\documentclass[12pt,english,a4paper]{book}
\usepackage[dvips]{graphicx}
\usepackage{fancyhdr}
\usepackage{latexsym}
\usepackage{amsfonts}
\usepackage{amssymb}
\usepackage{amsthm}
\usepackage{amsmath}
\usepackage{color}
\usepackage{colordvi}
\usepackage{array}
\usepackage{supertabular}
\usepackage{bm}
\usepackage{multirow}
\usepackage{axodraw}
\usepackage{lscape}
\usepackage[rightcaption]{sidecap}

\fancypagestyle{plain}{ \fancyhead{}   }

\fancypagestyle{normal}{
\fancyhead[LE,RO]{\bfseries\thepage}
\fancyhead[LO]{\bfseries\rightmark}
\fancyhead[RE]{\bfseries\leftmark}

\addtolength{\headheight}{0.5pt}}

\fancypagestyle{introduccio}{
\fancyhead[LE,RO]{\bfseries\thepage}
\fancyhead[LO]{\bfseries Introducci\'o}
\fancyhead[RE]{\bfseries Introducci\'o}

\addtolength{\headheight}{0.5pt}}

\fancypagestyle{index}{
\fancyhead[LE,RO]{\bfseries\thepage}
\fancyhead[LO]{\bfseries Contents}
\fancyhead[RE]{\bfseries Contents}

\addtolength{\headheight}{0.5pt}}

\fancypagestyle{OPERA}{
\fancyhead[LE,RO]{\bfseries\thepage}
\fancyhead[LO]{\bfseries \thesection ~Can OPERA help in constraining neutrino NSI?}
\fancyhead[RE]{\bfseries\leftmark}

\addtolength{\headheight}{0.5pt}}

\fancypagestyle{decoherence}{
\fancyhead[LE,RO]{\bfseries\thepage}
\fancyhead[LO]{\bfseries \thesection ~Decoherence in SN
  neutrino transformations$\dots$}
\fancyhead[RE]{\bfseries\leftmark}

\addtolength{\headheight}{0.5pt}}

\fancypagestyle{densematter}{
\fancyhead[LE,RO]{\bfseries\thepage}
\fancyhead[LO]{\bfseries \thesection ~Role of dense matter in
  collective SN neutrino$\dots$}
\fancyhead[RE]{\bfseries\leftmark}

\addtolength{\headheight}{0.5pt}}

\fancypagestyle{bibliografia}{
\fancyhead[LE,RO]{\bfseries\thepage}
\fancyhead[LO]{\bfseries Bibliography}
\fancyhead[RE]{\bfseries Bibliography}

\addtolength{\headheight}{0.5pt}}

\fancypagestyle{appendixa}{
\fancyhead[LE,RO]{\bfseries\thepage}
\fancyhead[LO]{\bfseries Appendix A: Equations of motion}
\fancyhead[RE]{\bfseries Appendix A: Equations of motion}

\addtolength{\headheight}{0.5pt}}

\fancypagestyle{conclusions}{
\fancyhead[LE,RO]{\bfseries\thepage}
\fancyhead[LO]{\bfseries Conclusions}
\fancyhead[RE]{\bfseries Conclusions}

\addtolength{\headheight}{0.5pt}}

\fancypagestyle{agraiments}{
\fancyhead[LE,RO]{\bfseries\thepage}
\fancyhead[LO]{\bfseries Agra\"{i}ments}
\fancyhead[RE]{\bfseries Agra\"{i}ments}

\addtolength{\headheight}{0.5pt}}

\newcommand{\Eps}{\varepsilon}

\newcommand{\I}{{\rm i}}

\newcommand{\D}{{\rm d}}

\newcommand{\be}{\begin{equation}}
\newcommand{\ee}{\end{equation}}
\newcommand{\bea}{\begin{eqnarray}}
\newcommand{\eea}{\end{eqnarray}}

\begin{document}

\pagestyle{empty}

\thispagestyle{empty}

\begin{center}

\includegraphics[scale=0.5]{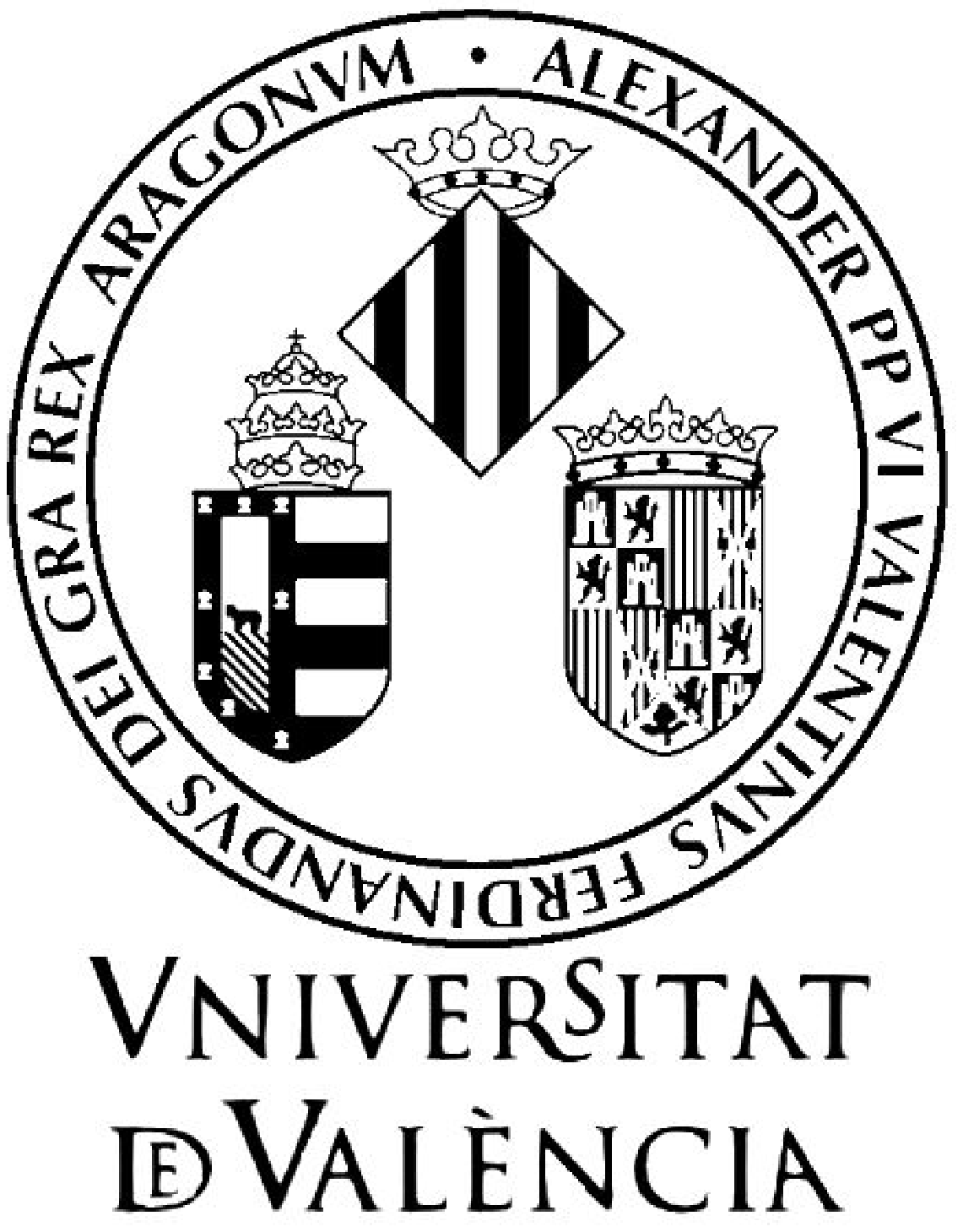}\\
\vspace{2cm}

\begin{huge}
{\bf
Supernoves com a laboratoris\\
de propietats de neutrins
}\\
\end{huge}

\vspace{0.5cm}

\begin{large}
{\bf
(Supernovae as laboratories for
neutrino properties)
}\\
\end{large}
\vspace{4cm}

\begin{large}
{\bf
Andreu Esteban Pretel 
}
\vspace{0.15cm}

Institut de F\'{i}sica Corpuscular (IFIC),\\
Departament de F\'{i}sica Te\`orica
\vspace{0.15cm}

Tesi Doctoral, Maig de 2009
\vspace{0.6 cm}

Directors de Tesi: 
\vspace{0.15cm}

Jose W. Furtado Valle i Ricard Tom\`as Bayo

\end{large}
\end{center}

\cleardoublepage

\cleardoublepage

\vspace*{2.cm}

Dr. Jose Wagner Furtado Valle, Professor d'Investigaci\'o del Consejo
Superior de Investigaciones Cient\'{\i}ficas, adscrit a l'Institut de
F\'{\i}sica Corpuscular - Departament de F\'{\i}sica Te\`orica,
Universitat de Val\`encia, i

\bigskip

Dr. Ricard Tom\`as Bayo, investigador postdoctoral al II. Institut
f\"ur Theoretische Physik, Universit\"at Hamburg,

\bigskip

\begin{quote}
CERTIFIQUEM:

Que la present mem\`oria ``{\it Supernoves com a laboratoris de
  propietats de neutrins}'' ha sigut realitzada sota la nostra
direcci\'o al Departament de F\'{i}sica Te\`orica de la Universitat de
Val\`encia, per N'Andreu Esteban Pretel i constitueix la seua Tesi per
tal d'optar al grau de Doctor en F\'{i}sica.
\end{quote}

\bigskip

I perqu\`e aix\'{i} conste, en compliment de la legislaci\'o vigent,
presentem davant del Departamet de F\'{i}sica Te\`orica de la
Universitat de Val\`encia la referida Tesi.

\vspace{1.5cm}

\begin{flushright}
A Val\`encia, a 25 de Maig de 2009.
\vspace{2cm}

Signat: Jose Wagner Furtado Valle \hspace{2cm} Signat: Ricard Tom\`as Bayo

\end{flushright}

\vfill

\cleardoublepage

\pagestyle{index}
\tableofcontents
\cleardoublepage

\pagestyle{normal}
\chapter*{Prefaci}
\addcontentsline{toc}{chapter}{Prefaci-Preface}
\markboth{\bf Prefaci}{}

La f\'{i}sica de neutrins ha experimentat un avan\c c espectacular en
els darrers deu anys. En juny de 1998, la col$\cdot$laboraci\'o
Super-Kamiokande~\cite{Fukuda:1998mi} don\`a el primer pas important
en aquest sentit en observar una forta evid\`encia de conversi\'o de
sabor en els neutrins atmosf\`erics, produ\"its en la
col$\cdot$lisi\'o de raigs c\`osmics amb l'atmosfera. No obstant,
aquest resultat no fou una sorpresa completament, ja que durant les
dues d\`ecades anteriors s'havia obtingut indicacions en favor
d'aquesta hip\`otesi. En experiments amb neutrins solars i dades
pr\`evies de neutrins atmosf\`erics s'observava respectivament un
d\`eficit de neutrins electr\`onics i neutrins mu\`onics en relaci\'o
als predits pels models te\`orics. Discrep\`ancies conegudes com el
problema dels neutrins solars i l'anomalia dels neutrins
atmosf\`erics. En 2002 es va confirmar l'oscil$\cdot$laci\'o de
neutrins, amb un esquema de massa i mescla, com el mecanisme correcte
per a explicar el problema del d\`eficit de neutrins solars. Per
demostrar-ho, van ser prou les primeres dades obtingudes per la
col$\cdot$laboraci\'o KamLand~\cite{Eguchi:2002dm}, experiment
terrestre amb neutrins generats en reactors nuclears. En eixe mateix
sentit, i tamb\'e en 2002, va ser resolta l'anomalia dels neutrins
atmosf\`erics fent \'us de les dades de neutrins d'accelerador
obtingudes en l'experiment K2K~\cite{Ahn:2002up}. M\'es tard
MINOS~\cite{Michael:2006rx} no nom\'es confirmaria aquest resultat,
sin\'o que augmentaria, i continua fent-ho, la precisi\'o en la
determinaci\'o dels corresponents par\`ametres d'oscil$\cdot$laci\'o.

La prova experimental de l'oscil$\cdot$laci\'o de neutrins demostrava
doncs que aquests tenen massa i, essent part\'{i}cules sense massa
dins del Model Est\`andard (SM) de les interaccions electrofebles,
aix\`o suposava alhora la primera evid\`encia robusta de f\'{i}sica
m\'es enll\`a del SM. Amb els experiments de neutrins a les portes de
l'era de la precisi\'o~\cite{McDonald:2004dd}, la determinaci\'o de
les propietats dels neutrins i el seu impacte te\`oric \'es un dels
principals objectius per als f\'{i}sics d'astropart\'{i}cules i
d'altes energies~\cite{Schwetz:2008er}. Aix\'{i} doncs els principals
esfor\c cos se centren actualment, per una banda en la identificaci\'o
de la seua naturalesa, Dirac o Majorana, i per una altra banda en la
determinaci\'o precisa dels par\`ametres d'oscil$\cdot$laci\'o i, de
forma complement\`aria, en la verificaci\'o de possibles efectes
subdominants no oscil$\cdot$latoris, tals com la conversi\'o
d'esp\'{i}n i sabor~\cite{Schechter:1981hw,Akhmedov:1988uk} o
possibles interaccions no est\`andard (NSI d'ac\'{i} endavant) dels
neutrins~\cite{Wolfenstein:1977ue}. La determinaci\'o d'aquests
obriria una finestra \'unica per a explorar f\'{i}sica m\'es enll\`a
del SM.

La tesi que ac\'{i} es presenta pret\'en ser una an\`alisi de diversos
aspectes de fenomenologia de neutrins en dos escenaris diferents. D'un
costat, es tracta l'estudi de les NSI en experiments terrestres
d'accelerador i reactor. D'altre costat es discuteix la propagaci\'o
de neutrins de supernova (SN), tenint en compte els nous descobriments
que mostren la import\`ancia que el propi fons de neutrins t\'e en
l'evoluci\'o d'aquests. Aquest efecte, menyspreat durant molt de
temps, pot ser de vital import\`ancia a l'hora d'entendre el senyal
que una possible explosi\'o de SN a la nostra gal\`axia donaria als
detectors de neutrins. L'an\`alisi dels neutrins de SN es presenta
tant en abs\`encia com en pres\`encia de NSI.

La pres\`encia de NSI pot afectar dr\`asticament a la propagaci\'o
dels neutrins en mat\`eria, aix\'{i} doncs \'es important discutir les
implicacions d'incloure aquestes interaccions en les an\`alisis dels
experiments terrestres previstos de neutrins d'accelerador i
reactor. D'aquesta manera s'ha considerat l'efecte que les NSI poden
tindre en els experiments MINOS, OPERA i Double
Chooz~\cite{EstebanPretel:2008qi}, i per tant els l\'{i}mits que
d'aquests es pot obtenir. La motivaci\'o del treball \'es doble, d'un
costat tots tres experiments estaran funcionant durant els propers
anys, per tant \'es una informaci\'o de la que disposarem en un futur
pr\`oxim. D'altre costat, OPERA mesurar\`a per primera vegada
l'oscil$\cdot$laci\'o $\nu_\mu \to \nu_\tau$ detectant directament els
$\nu_\tau$ i t\'e, a m\'es a m\'es, una relaci\'o distancia-energia
($L/E$) molt diferent a la de MINOS, dues caracter\'{i}stiques que
\'es sabut que ajuden en l'estudi de les NSI. Els l\'{i}mits que
d'aquest estudi i d'altres previs s'extrauen seran utilitzats en
l'an\`alisi de l'efecte de les NSI en la propagaci\'o de neutrins de
SN.

L'estudi de neutrins de SN ve motivat principalment per dues raons. En
primer lloc, si una explosi\'o de SN es don\'es a la nostra gal\`axia,
el nombre d'esdeveniments que s'observarien en els experiments
presents i futurs seria enorme, de l'ordre de $\mathcal{O}(10^4$--$10^5)$. En
segon lloc, les condicions extremes que els neutrins travessarien des
que s\'on creats al nucli de la SN fins ser detectats a la terra,
tindria un efecte dr\`astic en la seua propagaci\'o. En aquest estudi
farem especial \`emfasi en l'efecte que els propis neutrins tenen en
la seua evoluci\'o. Al dens flux de neutrins que emergeix del nucli
d'una SN la refracci\'o neutr\'{i}-neutr\'{i} causa un fenomen de
conversi\'o de sabor no lineal que \'es completament diferent a
qualsevol efecte indu\"{i}t per la mat\`eria
ordin\`aria~\cite{Pastor:2002we, Sawyer:2005jk, Fuller:2005ae,
  Hannestad:2006nj, Raffelt:2007cb, Fogli:2007bk,
  EstebanPretel:2007ec, EstebanPretel:2007yq, Dasgupta:2008cd,
  EstebanPretel:2008ni}. El principal efecte \'es un mode
col$\cdot$lectiu de transformaci\'o de parells de la forma $\nu_e
\bar\nu_e \to \nu_x \bar\nu_x$, on $\nu_x$ correspon a una
superposici\'o de $\nu_\mu$ i $\nu_\tau$. Per tindre aquest proc\'es
es necessari una gran densitat de neutrins i un exc\'es de parells
d'algun sabor. Ambdues condicions estan presents als models t\'{i}pics
de SN.

D'un altre costat, com ja s'ha comentat, es presentar\`a l'efecte que
pot tindre l'exist\`encia de NSI en l'evoluci\'o de neutrins de SN,
estudi que es presentar\`a tant en
abs\`encia~\cite{EstebanPretel:2007yu}, com en
pres\`encia~\cite{EstebanPretel:2009is} d'autointeracci\'o dels
neutrins. Una explosi\'o de SN \'es un bon escenari per a fer aquesta
an\`alisi, ja que els efectes de NSI xicotetes poden veure's
amplificats degut a les condicions extremes de densitat de mat\`eria
que troben els neutrins en la seua propagaci\'o.

Aix\'{i} doncs, la present tesi est\`a organitzada de la seg\"uent
forma. Al Cap\'{i}tol~\ref{chapter:supernova} resumim els coneixements
actuals sobre la f\'{i}sica de les SNe que esclaten per
col$\cdot$lapse gravitatori del nucli. Al
Cap\'{i}tol~\ref{chapter:oscillations} discutim el fenomen
d'oscil$\cdot$laci\'o de neutrins tant per a dos com per a tres
sabors, en buit i en mat\`eria, i sempre en abs\`encia d'un fons de
neutrins. Al final d'aquest cap\'{i}tol apliquem el formalisme descrit
anteriorment per a estudiar l'evoluci\'o de neutrins en l'embolcall
d'una SN. Al Cap\'{i}tol~\ref{chapter:NSI} analitzem com la inclusi\'o
de les NSI afecta a l'evoluci\'o dels neutrins i recordem els
l\'{i}mits actuals sobre els par\`ametres que les caracteritzen. A
m\'es a m\'es, estudiem el que es pot aprendre d'aquestes amb els
resultats dels experiments MINOS, OPERA i Double Chooz. Els
Cap\'{i}tols~\ref{chapter:coll2flavors} i \ref{chapter:coll3flavors}
els dediquem a l'estudi dels efectes de l'autointeracci\'o dels propis
neutrins en una SN. Al primer d'aquests cap\'{i}tols, despr\'es de
resumir el coneixement que del fenomen de transformaci\'o
col$\cdot$lectiva de neutrins es t\'e actualment, centrem la
discussi\'o en un escenari de dos sabors i tractem la q\"uesti\'o dels
efectes multiangulars en aquest fenomen. El
Cap\'{i}tol~\ref{chapter:coll3flavors} el dediquem a l'an\`alisi de
possibles efectes caracter\'{i}stics de tres neutrins. Al
Cap\'{i}tol~\ref{chapter:SN_NSI} estudiem les conseq\"u\`encies de les
NSI dels neutrins en la seua evoluci\'o en una SN. Per tal de
distingir els efectes derivats de les NSI d'aquells que v\'enen de
l'autointeracci\'o dels neutrins, comencem menyspreants aquests
\'ultims. A la segona part del cap\'{i}tol incloem ja tots els
ingredients i discutim el seu efecte conjunt. Per \'ultim, al
Cap\'{i}tol~\ref{chapter:conclusions} fem un breu resum de tot all\`o
analitzat en aquesta tesi doctoral remarcant el punts que considerem
m\'es importants del treball.

Tots els resultats originals discutits en aquesta tesi doctoral han
sigut publicats en les
Refs.~\cite{EstebanPretel:2008qi,EstebanPretel:2007ec,EstebanPretel:2007yq,EstebanPretel:2008ni,EstebanPretel:2007yu,EstebanPretel:2009is}:
\begin{itemize}
\item A.~Esteban-Pretel, R.~Tom\`as and J.~W.~F.~Valle, ``Probing
  non-standard neutrino interactions with supernova neutrinos,''
  Phys.\ Rev.\ D {\bf 76} (2007) 053001 [arXiv:0704.0032 [hep-ph]].
\item A.~Esteban-Pretel, S.~Pastor, R.~Tom\`as, G.~G.~Raffelt and
  G.~Sigl, ``Decoherence in supernova neutrino transformations
  suppressed by deleptonization,'' Phys.\ Rev.\ D {\bf 76} (2007)
  125018 [arXiv:0706.2498 [astro-ph]].
\item A.~Esteban-Pretel, S.~Pastor, R.~Tom\`as, G.~G.~Raffelt and
  G.~Sigl, ``Mu-tau neutrino refraction and collective three-flavor
  transformations in supernovae,'' Phys.\ Rev.\ D {\bf 77} (2008)
  065024 [arXiv:0712.1137 [astro-ph]].
\item A.~Esteban-Pretel, J.~W.~F.~Valle and P.~Huber, ``Can OPERA help
  in constraining neutrino non-standard interactions?,'' Phys.\ Lett.\
  B {\bf 668} (2008) 197 [arXiv:0803.1790 [hep-ph]].
\item A.~Esteban-Pretel, A.~Mirizzi, S.~Pastor, R.~Tom\`as,
  G.~G.~Raffelt, P.~D.~Serpico and G.~Sigl, ``Role of dense matter in
  collective supernova neutrino transformations,'' Phys.\ Rev.\ D {\bf
    78} (2008) 085012 [arXiv:0807.0659 [astro-ph]].
\item A.~Esteban-Pretel, R.~Tom\`as and J.~W.~F.~Valle, ``Interplay
  between collective effects and non-standard neutrino interactions of
  supernova neutrinos,'' arXiv:0909.2196 [hep-ph].
\end{itemize}

\cleardoublepage

\chapter*{Preface}
\markboth{\bf Preface}{}

Neutrino physics has experienced a spectacular breakthrough in the
last ten years. In June 1998, the Super-Kamiokande
Collaboration~\cite{Fukuda:1998mi} gave the first important step in
this direction when they observed a strong evidence of flavor
conversion for atmospheric neutrinos, those produced from the
collision of cosmic rays with the atmosphere. However, this result was
not a complete surprise. For two decades indications favoring this
hypothesis had been obtained. A deficit of electron and muon
neutrinos, compared to the prediction of theoretical models, was
observed in solar neutrino experiments and previous atmospheric
neutrino data, respectively. These discrepancies are known as the
solar neutrino problem and the atmospheric neutrino anomaly. In 2002,
flavor neutrino oscillation, within a scheme of mass and mixing, was
confirmed as the correct mechanism to explain the solar neutrino
deficit problem. The first data obtained by the
KamLAND~\cite{Eguchi:2002dm} collaboration, a terrestrial experiment
detecting reactor neutrinos, were enough to demonstrate this
oscillation scenario. In the same direction, and also in 2002, the
atmospheric neutrino anomaly was explained using the accelerator
neutrino data obtained in the K2K~\cite{Ahn:2002up} experiment. Later
on, MINOS~\cite{Michael:2006rx} would not only confirm this result but
would also increase, and continues to do so, the precision in the
determination of the corresponding oscillation parameters.

The experimental evidence of neutrino oscillations proved that they
have mass. Therefore, neutrinos being massless within the electro-weak
Standard Model (SM), it also represented the first robust evidence of
physics beyond the SM. With neutrino experiments at the threshold of
the precision era~\cite{McDonald:2004dd}, the determination of
neutrino properties and their theoretical impact is one of the main
goals for astroparticle and high energy
physicists~\cite{Schwetz:2008er}. Most of the effort is nowadays
focused on the precise determination of the oscillation parameters
and, in a complementary way, on the verification of possible
sub-leading non-oscillation effects, such as spin and flavor
conversions~\cite{Schechter:1981hw,Akhmedov:1988uk} or possible
non-standard neutrino interactions (NSI from now
on)~\cite{Wolfenstein:1977ue}. Their determination would open a unique
window to explore physics beyond the SM.

The present thesis aims to be an analysis of various aspects of
neutrino phenomenology in two different scenarios. On the one hand, we
address the study of neutrino NSI in accelerator and reactor
terrestrial experiments. On the other hand, we discuss the propagation
of supernova (SN) neutrinos, taking into account the recent
developments showing the importance that neutrino background may have
in their evolution. This effect, neglected for a long time, may be of
capital importance when trying to understand the neutrino signal from
a future galactic SN. Our SN neutrino analysis is presented both in
absence and presence of NSI.

The presence of NSI may drastically affect the propagation of
neutrinos through matter. Thus, it is important to discuss the
implications of including these new interactions in the future planned
neutrino terrestrial experiments. In this sense, we have considered
the effects that NSI may induce in MINOS, OPERA and Double Chooz
experiments~\cite{EstebanPretel:2008qi}, and the bounds that can be
obtained from them. The motivation of the work is twofold, on the one
hand all three experiments will be taking data during the next years,
providing valuable information in the near future. On the other hand,
OPERA will measure for the first time the oscillation channel $\nu_\mu
\to \nu_\tau$, detecting directly the $\nu_\tau$. Moreover, it has a
very different distance-energy ratio ($L/E$) than MINOS. Both factors
are expected to help in disentangling NSI from pure oscillations. The
bounds obtained in this study and the previous ones will be used in
the analysis of NSI effects in the propagation of SN neutrinos.

The study of SN neutrinos is motivated mainly by two reasons. First,
if a future SN explosion takes place in our galaxy, an enormous number
of neutrino events are expected in the present and future planned
detectors, $\mathcal{O}(10^4$--$10^5)$. Second, the extreme conditions
under which neutrinos travel since they are created in the SN core
until they are detected at the Earth, would have a dramatic effect in
their propagation. In our study we pay special attention to the effect
of the neutrino background itself. In the dense neutrino flux emerging
from a SN core, neutrino-neutrino refraction causes non-linear flavor
conversion phenomena that are unlike anything produced by ordinary
matter~\cite{Pastor:2002we, Sawyer:2005jk, Fuller:2005ae,
  Hannestad:2006nj, Raffelt:2007cb, Fogli:2007bk,
  EstebanPretel:2007ec, EstebanPretel:2007yq, Dasgupta:2008cd,
  EstebanPretel:2008ni}. The crucial phenomenon is a collective mode
of pair transformations of the form $\nu_e\bar\nu_e\to\nu_x\bar\nu_x$
where $x$ represents a suitable superposition of $\nu_\mu$ and
$\nu_\tau$. Collective pair transformations require a large neutrino
density and a pair excess of a given flavor. Both conditions are
present in typical SN models.

On the other hand, as it has already been commented, we will discuss
the effect that the existence of NSI would have in the evolution of SN
neutrinos. This study will be done both in
absence~\cite{EstebanPretel:2007yu}, and
presence~\cite{EstebanPretel:2009is} of neutrino-neutrino
interactions. A SN explosion is an attractive scenario to study NSI,
since the effect of small NSI can be amplified due to the extreme
conditions found in the SNe.

The present thesis is therefore organized as follows. In
Chapter~\ref{chapter:supernova} we review the current knowledge on
core collapse SN physics. In Chapter~\ref{chapter:oscillations} we
discuss the neutrino oscillation phenomenon for two and three flavors,
in vacuum and matter, but always in the absence of a neutrino
background. At the end of this chapter we apply the formalism
previously described to study the evolution of neutrinos through the
SN envelope. In Chapter~\ref{chapter:NSI} we analyze how the inclusion
of NSI affects the evolution of neutrinos and review the current
bounds on the parameters that characterize them. Furthermore, we study
what we can learn about NSI from the results of MINOS, OPERA and
Double Chooz experiments. Chapters~\ref{chapter:coll2flavors} and
\ref{chapter:coll3flavors} are devoted to the study of
neutrino-neutrino interactions in the SN context. In the first one,
after reviewing the current knowledge on the collective neutrino
transformation phenomenon, we center the discussion in the two flavor
scenario and address the question of multi-angle effects in this
phenomenon. In Chapter~\ref{chapter:coll3flavors} we analyze the
possible three flavor characteristic effects. In
Chapter~\ref{chapter:SN_NSI} we study the consequences that NSI would
have in neutrino evolution through the SN envelope. In order to
separate the effects of NSI from those of neutrino-neutrino
interaction, we start by neglecting the latter.  In the second part of
the chapter we include all the ingredients, and discuss the interplay
among them. Finally, in chapter~\ref{chapter:conclusions} we briefly
summarize the topics discussed in the present Ph.D. thesis emphasizing
the points that we consider of most interest.

All the original results presented in this Ph.D. thesis have been
published in
Refs.~\cite{EstebanPretel:2008qi,EstebanPretel:2007ec,EstebanPretel:2007yq,EstebanPretel:2008ni,EstebanPretel:2007yu,EstebanPretel:2009is}:
\begin{itemize}
\item A.~Esteban-Pretel, R.~Tom\`as and J.~W.~F.~Valle, ``Probing
  non-standard neutrino interactions with supernova neutrinos,''
  Phys.\ Rev.\ D {\bf 76} (2007) 053001 [arXiv:0704.0032 [hep-ph]].
\item A.~Esteban-Pretel, S.~Pastor, R.~Tom\`as, G.~G.~Raffelt and
  G.~Sigl, ``Decoherence in supernova neutrino transformations
  suppressed by deleptonization,'' Phys.\ Rev.\ D {\bf 76} (2007)
  125018 [arXiv:0706.2498 [astro-ph]].
\item A.~Esteban-Pretel, S.~Pastor, R.~Tom\`as, G.~G.~Raffelt and
  G.~Sigl, ``Mu-tau neutrino refraction and collective three-flavor
  transformations in supernovae,'' Phys.\ Rev.\ D {\bf 77} (2008)
  065024 [arXiv:0712.1137 [astro-ph]].
\item A.~Esteban-Pretel, J.~W.~F.~Valle and P.~Huber, ``Can OPERA help
  in constraining neutrino non-standard interactions?,'' Phys.\ Lett.\
  B {\bf 668} (2008) 197 [arXiv:0803.1790 [hep-ph]].
\item A.~Esteban-Pretel, A.~Mirizzi, S.~Pastor, R.~Tom\`as,
  G.~G.~Raffelt, P.~D.~Serpico and G.~Sigl, ``Role of dense matter in
  collective supernova neutrino transformations,'' Phys.\ Rev.\ D {\bf
    78} (2008) 085012 [arXiv:0807.0659 [astro-ph]].
\item A.~Esteban-Pretel, R.~Tom\`as and J.~W.~F.~Valle, ``Interplay
  between collective effects and non-standard neutrino interactions of
  supernova neutrinos,'' arXiv:0909.2196 [hep-ph].
\end{itemize}

\pagestyle{normal}
\chapter{Core Collapse Supernovae}\label{chapter:supernova}

One of the most spectacular cosmic events is a core collapse supernova
(SN) explosion. It means the death of a massive star and gives birth
to the most exotic states of matter known, neutron stars and black
holes. SN explosions determine also the evolution of galaxies, since
most of the heavy elements in nature, with mass number $A>70$, are
thought to be created through the s- and r- (slow and rapid)
processes. It has long been thought that the latter take place in this
kind of events. Elements that will afterwards serve as raw material in
the creation of new stars and planets.

Such an event involves as much instantaneous power as all the rest of
the luminous visible Universe combined, it releases about $10^{52}$
erg s$^{-1}$ ($\sim 10^{45}$ J s$^{-1}$) during some tens of
seconds. Around 99\% of this energy is emitted as neutrinos. They are,
therefore, expected to play a crucial role in the SN evolution.

There exist, though, different types of SNe, and not all of them are
the consequence of the collapse of a massive star core.

\section{Supernova types}

\begin{figure}
  \begin{center}
    \includegraphics[width=0.65\textwidth]{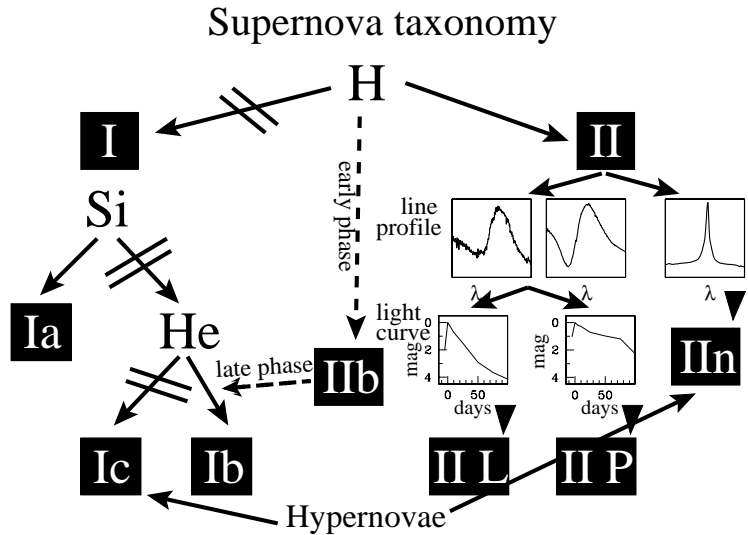}
  \end{center}
  \caption{\small SN classification according to observational
    criteria~\cite{Cappellaro:2000ez}.}
  \label{fig:taxonomy}
\end{figure}

The work of SN classification started with Minkowski~\cite{Minkowski}
(1941) who divided them in two types, whether they did (type II) or
did not (type I) show hydrogen lines in their spectra. Nevertheless, a
detailed classification of SNe according to observational criteria
needs, besides the identification of the characteristics in the
spectrum, an analysis of the line profile, luminosities and spectral
evolution. In Fig.~\ref{fig:taxonomy} we show this classification
schematically.

Another criterion used in the classification of SNe is according to
the explosion mechanism, under which we can distinguish two big
groups: thermonuclear and core collapse explosions.

\subsection{SNIa: Thermonuclear explosion}

SNIa are quite homogeneous events with similar luminosity and spectral
evolutions. Indeed until the 1990s it was commonly accepted that all
SNIa explosions were identical, and that the observed differences came
from observational errors. Thanks to this homogeneity this kind of SNe
are used as \emph{standard candles} in determining distances to far
galaxies. Type Ia SNe also provide a strong evidence for an
accelerating Universe, and are the best single tool for directly
measuring the density of dark energy.

From an observational point of view, their spectrum is characterized
by the absence of hydrogen lines and the presence of silicon lines,
see Fig.~\ref{fig:taxonomy}. The emission of elements like oxygen is
not very important, which means that the progenitor star was not very
massive. They have been observed in elliptical and spiral galaxies
formed by old stars.

According to these observational features, the standard scenario for
the SNIa explosion consists of a binary system where one of the stars
is a white dwarf which accretes matter from the secondary star. The
increment in mass leads to an increment in temperature for the central
region of the star, until the threshold temperature of the carbon
burning is reached. The high degeneracy of the stellar material turns
the combustion regime unstable and triggers the thermonuclear
explosion. After the explosion the progenitor star is completely
destroyed, resulting in an expanding nebula without a central compact
object.

The total energy released in type Ia SNe is approximately
$3\times10^{51}$ erg. Neutrinos carry just the 1\% of this energy, and
thus do not seem to play an important role in thermonuclear SNe. Since
neutrinos are the main subject of this thesis, we will not further
discuss this type of SN.

\subsection{SNIb, SNIc, SNII: Core collapse explosion}

The second group of SNe is much more heterogeneous. On the one hand,
SNIb and SNIc, just as SNIa, do not have hydrogen lines in their
spectra, but contrary to these, SNIb and SNIc also show an absence of
silicon lines. Finally, SNIb present helium lines, while SNIc do not.

On the other hand we have type II SNe which contain hydrogen in the
spectrum. These are also subclassified: SNII-P, in which after a
maximum the luminosity curve remains fairly constant for 2--3 months,
forming a \emph{plateau} (e.g.~SN 1987A), see Fig.~\ref{fig:taxonomy};
SNII-L, in which the luminosity falls rather linearly with
time. Nevertheless, there is no clear separation between these two SN
types and lots of intermediate cases exist. A third subtype of SNII is
known as IIn, where the ``n'' stands for narrow-line, since their
spectra show narrow components on top of the broader emissions. They
are very bright and show a slow evolution. Some SNe seem to change
along their evolution from a SNII, in the early phases, to a SNIb in
the nebular phase, and are therefore classified as SNIIb. Finally, one
very interesting development in the field of SNe has been the
discovery of a very energetic type of them in the late 1990s, known as
Hypernovae. The kinetic energy of such an event exceeds $10^{52}$ erg,
which is 10 times larger than that of a usual SN explosion. These high
energetic SNe have been observed as type Ic and IIn, and some hints
exist that they might be related to $\gamma$-ray
bursts~\cite{Nomoto:2002rc}.

This kind of explosions have been observed mainly in regions populated
by young stars and presenting high stellar formation activity, like
the arms of spiral galaxies. This is due to the fact that the
evolution of this type of SN is much faster than the SNIa type. They
are also less luminous than these and present a heterogeneous behavior
on every aspect. More specifically the luminosity curves are different
for each case, depending on the structure of the progenitor star. They
are therefore not useful as standard candles.

The differences in the SN spectra are due to the loss of different
envelope layers at some point of the evolution: SNIb ejects the
hydrogen layer while SNIc also loses the helium layer. In spite of
these spectral differences, SNIb, SNIc and SNII are all the result of
the same explosion mechanism, related to the death of massive stars
($M \gtrsim 8M_{\odot}$). At the end of their life, massive stars
accumulate iron in their center after several nuclear burning
stages. When the iron core reaches a certain mass it becomes unstable
and the collapse starts. According to the so called delayed explosion
mechanism, the collapse is inverted into an explosion and a shock wave
traverses the star, expelling the material found in its way. This
explosion is accompanied with the formation of a neutron star or a
black hole.

The total energy released in this kind of explosion is of the order of
$10^{53}$ erg, from which only the $1$\% is released as kinetic energy
of the expelled material and approximately $0.01$\% as light. The rest
of the energy is emitted in the form of neutrinos, which will,
therefore, be very important in the core collapse SN explosions, and
may play a determinant role in the effective realization of the
process.

The detection of neutrinos coming from such an event would be crucial
for different reasons:
\begin{itemize}
\item Neutrinos, contrary to photons, emerge from the deepest regions
  of the star, since they are much more weakly interacting particles
  than photons. Neutrino detection would therefore be the only way,
  apart from gravitational waves, to obtain information of the inner
  layers of the star, highly important for the understanding of the
  explosion mechanism.
\item While neutrinos escape the star seconds after the collapse,
  photons remain trapped and are only emitted from the envelope with a
  delay of several hours with respect to them. Therefore, neutrinos
  are the first signal expected from the explosion and could serve as
  an early warning system for galactic SNe.
\item It can occur that the SN is optically obscured, or that the
  stellar collapse does not produce an explosion and creates a black
  hole. In such cases, the detection of neutrinos would be very
  important if not the only observable effect of the SN.
\end{itemize}

\section{Core collapse supernova dynamics}

The main topic of the present thesis is SNe as neutrino sources, thus
we will study the type of SN where neutrinos play an important
role, namely core collapse SNe (SNIb, SNIc and SNII). In order to
understand the physical processes involved in the SN, it is useful to
distinguish four stages in the phenomenon:

\begin{itemize}

\item The life of the progenitor star
\item Stellar core collapse
\item Deleptonization and cooling
\item Supernova explosion

\end{itemize}

\subsection{The life of the progenitor star}

During its life a star must keep an equilibrium between two possible
fatal effects working in opposite directions. On the one hand the
gravitational force tends to collapse the star, on the other hand the
thermal pressure tends to expand it. In order to maintain this
equilibrium the star will undergo a series of nuclear burning stages.

Initially the hydrogen in the star will be transformed into helium
through nuclear fusion reactions. When there is no more hydrogen in
the center of the star this process can no longer compensate the
gravitational force. As the star contracts, the density and
temperature increase, and eventually the conditions for the helium
burning are reached, stabilizing the system once more. This situation
repeats itself, leading to carbon, neon, oxygen and silicon burning
stages, where each time the new fuel is the product of the previous
reactions. The energy released at each new step of the chain is
smaller every time, while the energy losses are larger, leading to
shorter and shorter periods of burning. In this way, while the
hydrogen combustion can last for hundreds of millions of years the
silicon sustains the star typically for no longer than days, see
Table~\ref{tab:evSN}.

\begin{table}[t]
  \caption{\small Evolution of a 15 $M_\odot$ star
    \cite{Woosley:2006ie}. Here ``s.u.''~stands for solar units, and
    the luminosity and neutrino losses are given normalized to the
    ones of the Sun: $L_\odot~=~
    3.839~\times~10^{33}$~erg/s and the Sun emits $2\times 10^{38}$
    neutrinos per second.}
\label{tab:evSN}
\begin{center}
{\footnotesize
\begin{tabular}{||c|c|c|c|c|c|c|c||}
\hline
\hline
Stage & Time & Fuel & Ash or &  Central $T$ &Central $\rho$ & $L$ (s.u.) & Neutrino \\
      & scale &     & Product & ($10^9$ K)& (g/cm$^3$)&     & losses (s.u.)\\
\hline
H & 11 My & H & He & 0.035 & 5.8 & $28000$ & 1800\\
\hline
He & 2 My & He & C,O & 0.18 & 1390 & $44000$ & 1900\\
\hline
C & 2000 y & Ne & Ne,Mg & 0.81 & $2.8 \times 10^7$ & $72000$ & $3.7 \times 10^5$ \\
\hline
Ne & 0.7 y & Ne & O,Mg & 1.6 & $1.2 \times 10^7$ & $75000$ & $1.4 \times 10^8$ \\
\hline
O & 2.6 y & O,Mg & Si,S, & 1.9 & $8.8 \times 10^6$ & $75000$ & $9.1 \times 10^8$ \\
  &       &      & Ar,Ca &     &                   &         &                   \\
\hline
Si & 18 d & Si,S, & Fe,Ni,& 3.3 & $4.8 \times 10^7$ & $75000$ & $1.3 \times 10^{11}$ \\
  &       & Ar,Ca &Cr,Ti,\dots&  &                 &         &                   \\
\hline
Fe & $\sim 1$ s & Fe,Ni,& Neutron  & $> 7.1$ & $> 7.3 \times 10^9$ & $75000$ & $> 3.6 \times 10^{15}$ \\
  &            & Cr,Ti,\dots & Star &    &               &         &                   \\
\hline
\hline
\hline
\end{tabular}
}
\end{center}
\end{table}

\renewcommand{\arraystretch}{1}

\begin{figure}[t]
  \begin{center}
    \includegraphics[width=0.6\textwidth]{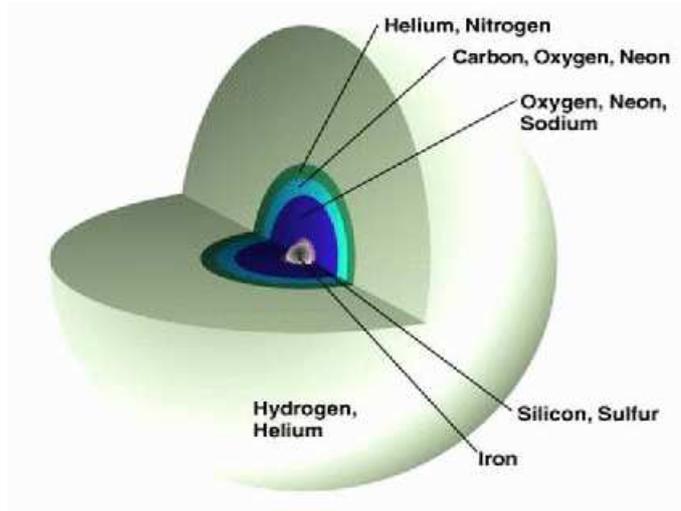}
  \end{center}
  \caption{\small Schematic picture of the onion structure developed
    in a typical progenitor.}
  \label{fig:ceba}
\end{figure}

After several burning stages the initial composition of hydrogen and
helium turns into the onion-shell structure, as schematically shown in
Fig.~\ref{fig:ceba}. Stars with masses above $11 M_{\odot}$ will have
a core mainly composed of Fe and Ni, while stars under that limit will
have an O-Ne-Mg core. The ignition of the iron core will never occur,
since it is the nucleus with the largest binding energy. At this point
the fate of the star will depend on whether the iron core reaches or
not the Chandrasekhar mass ($M_{\rm Ch} = 5.8 Y^2_L M_{\odot} \simeq
1.2$--$1.5M_{\odot}$, where $Y_L$ is the lepton fraction, i.e.~number
of leptons per baryon).

The Chandrasekhar limit is a stability criterion for compact objects
like white dwarfs or the iron core of a massive star. Such objects are
stabilized by electron degeneracy pressure. Depending on its mass,
whether being above or below the Chandrasekhar limit, the star will
follow different paths: $M < M_{\rm Ch}$, electrons become
non-relativistic, stabilizing the star again; $M > M_{\rm Ch}$ (SN
case), the degeneration of electrons cannot compensate the
gravitational pressure and the star collapses.

\subsection{Stellar core collapse}

As soon as the last stage of nuclear burning begins at the center of
the star, it starts developing a degenerate core formed by iron group
elements, covered by a silicon crust. The iron core will continuously
grow as the nuclear reactions at the border with the silicon layer add
new material to it. Since the iron ignition will never occur this
situation will not be stable for a long time. We have an inert sphere
under a huge pressure, which is a configuration similar to that of a
white dwarf. The stationary equilibrium is obtained thanks to the
electron degeneration pressure, which is subjected to the
Chandrasekhar limit, and is in this case around $1.2$--$1.5M_{\odot}$.
\vspace{0.5cm}\\
\textbf{A) Start of the collapse}\vspace{0.2cm}\\
We end up then with an iron white dwarf of around $1.5M_{\odot}$, a
central density of about $3.7 \times 10^9$ g cm$^{-3}$, a central
temperature of around $0.69$ MeV and an electron fraction (number of
electrons per baryon) of $Y_e \simeq 0.42$. This last stage developes
very fast and the iron core reaches the Chandrasekhar limit in
days. At this point the electron degeneration pressure can no longer
sustain the star and the collapse starts, lasting less than a second.

The rise in density and temperature originated by the collapse leads
to new processes that accelerate the infall. There will be different
processes involved, depending on the progenitor mass. The main ones
are:

\begin{itemize}

\item \textbf{Electron capture} ($15$--$20 M_{\odot}$). Due to the high
  densities attained during the collapse we will obtain electron
  capture by heavy elements, \be e^- (Z,A) \to \nu_e (Z-1,A)~, \ee
  such as,
  \bea
  ^{24}{\rm Mg} + e^- \to ~^{24}{\rm Na} + \nu_e~, \\
  ^{56}{\rm Fe} + e^- \to ~^{56}{\rm Mn} + \nu_e~. 
  \eea
  These processes have not only the effect of reducing the electron
  degeneration pressure by removing free electrons from the medium,
  but are also responsible for a huge energy loss in form of
  neutrinos. This is the first of the four neutrino emission phases.

\item \textbf{Nuclei photodissociation} ($\gtrsim 20 M_{\odot}$). In such
  massive stars the high temperatures reached make the iron nuclei
  photodissociation to be an important source of energy loss. The iron
  atoms are disintegrated in $\alpha$ particles due to the absorption
  of high energy photons,
  \be
  ^{56}{\rm Fe} + \gamma \to 13~^4{\rm He} + 4n - 124.4~{\rm MeV}~.
  \ee
  The fast contraction of the star releases a big amount of
  gravitational energy, most of which is absorbed in the
  photodissociation of the iron atoms. Part of the energy required in
  these processes is obtained from the electrons, leading to a
  reduction of their pressure.

\end{itemize}

The net result in both cases is the acceleration of the collapse.\vspace{0.5cm}\\
\textbf{B) Neutrino trapping}\vspace{0.2cm}\\
We could say that the first stage of the collapse comes to an end when
the density of the stellar core reaches a value of about $10^{12}$ g
cm$^{-3}$. This is by no means the maximum density the core will
register, since it continues to contract. Nevertheless, it marks an
important point in the SN evolution: at this density matter becomes
opaque to neutrinos, contrary to the initial moments of the collapse,
where neutrinos can freely escape the star. The dominant opacity
source for neutrinos in the collapse is neutral current interactions
with heavy elements. The coherent scattering cross section for these
processes is proportional to $A^2$.

The confinement is not permanent and after several interactions the
neutrino would eventually escape the core. The diffusion time, though,
is longer than the dynamic time of the collapse, leading to an
effective confinement. One can define the \emph{neutrino sphere}
($R_{\nu}$) as the surface where the optical depth of neutrinos,
$\int_{r}^{\infty}\lambda_{\nu}^{-1}{\rm d}r'$, becomes unity. This radius is
shown in Fig.~\ref{fig:janka} as a function of time with a dotted
line. One can approximately consider the region inside the neutrino
sphere opaque to neutrinos and the exterior transparent.

The main consequences of the neutrino trapping are:

\begin{itemize}

\item Lepton fraction conservation, $Y_L = Y_e + Y_{\nu} \simeq 0.35$,
  during the collapse. Once neutrinos are trapped, they become
  degenerate, as the electrons, and reach $\beta$ equilibrium,
  \be
  e^- + p \longleftrightarrow n + \nu_e~.
  \ee

\item Change in the nuclear state. The degeneration of neutrinos leads
  to a suppression of the neutronization process (protons convert to
  neutrons through electron capture), since the neutrino emission
  derived from it is forbidden by the Pauli exclusion
  principle. Therefore heavy nuclei do not melt into free nucleons
  until the density approaches the nuclear density.

\end{itemize}
\textbf{C) Core bounce and shock wave formation}\vspace{0.2cm}\\
The electron captures produced at the first stages of the collapse
will not only reduce the electron degeneration pressure but also the
electron fraction, and therefore $M_{\rm Ch}$. On the other hand,
there is a change in the role played by this parameter in the SN
dynamics. While initially it represented the largest amount of mass
that could be supported by the electron pressure, it now becomes the
largest amount of mass that can collapse homologously. As a
consequence the collapsing core is considered to be composed of two
parts (stage 1 of Fig.~\ref{fig:etapes}):

\begin{figure}
  \begin{center}
    \includegraphics[width=0.9\textwidth]{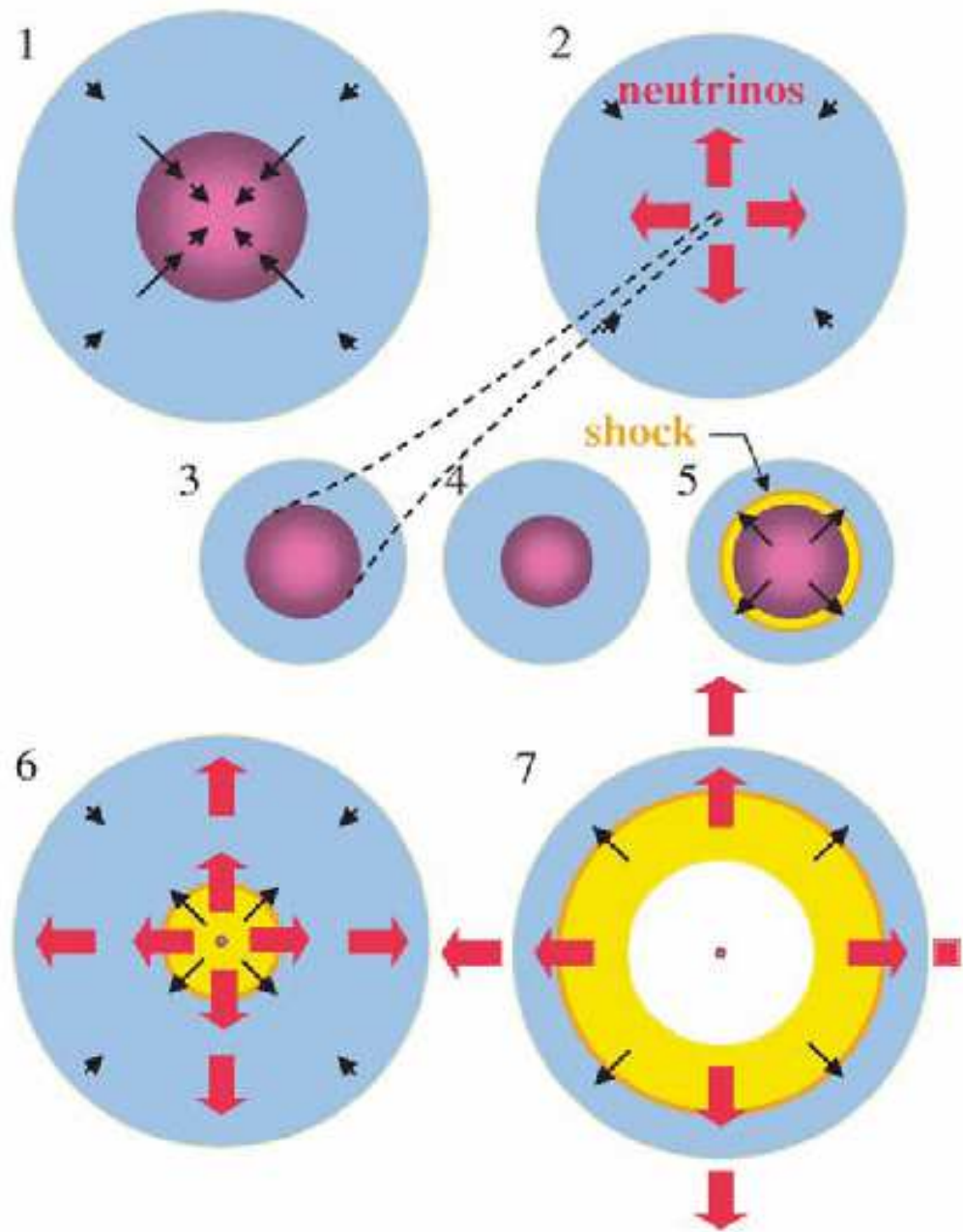}
  \end{center}
  \caption{\small Different stages of the SN evolution. The core
    separates into an inner, subsonically collapsing core and an
    outer, supersonically collapsing core. When the core is no longer
    compressible it bounces, generating a shock wave which propagates
    outwards and ultimately will produce the SN
    explosion~\cite{Mezzacappa:2005ju}.}
  \label{fig:etapes}
\end{figure}

\begin{itemize}

\item inner core ($R_{\rm ic}$ in Fig. \ref{fig:janka}), collapsing
  homologously and subsonically.

\item outer core ($R_{\rm Fe}$ in Fig. \ref{fig:janka}), collapsing
  supersonically in a free fall.

\end{itemize}

\begin{figure}[t]
  \begin{center}
    \includegraphics[width=0.7\textwidth]{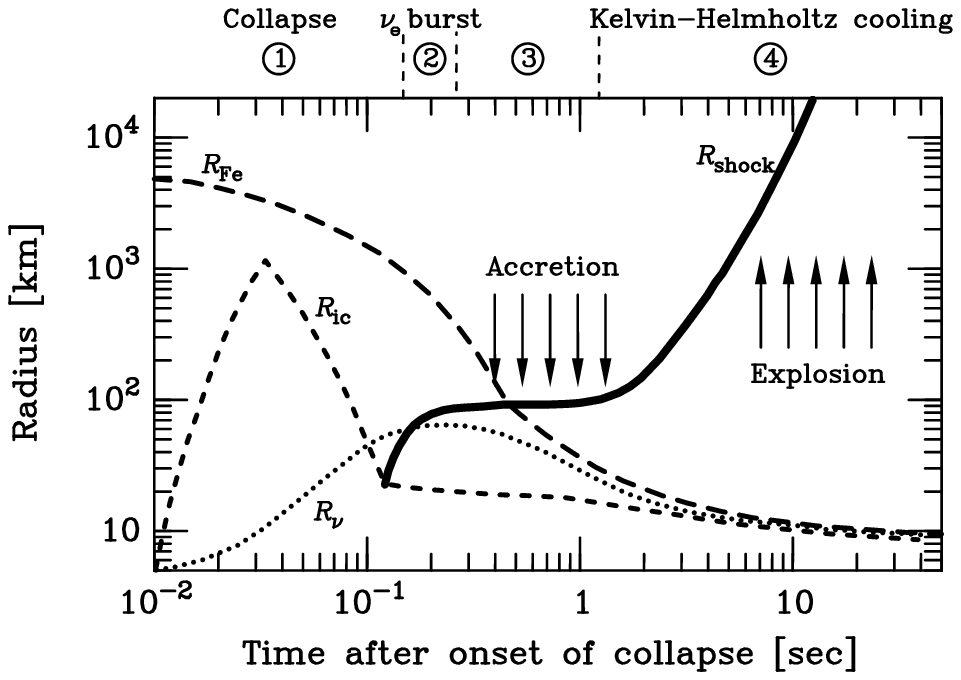}
  \end{center}
  \caption{\small Schematic picture of stellar core collapse,
    formation of the neutron star remnant and the start of the SN
    explosion. The figure shows particular radial positions in the
    star's central region as they evolve in time. The evolution can be
    divided in four stages: 1.~collapse phase, 2.~neutronization
    burst, 3.~matter accretion phase, 4.~protoneutron star cooling
    phase. $R_{\rm Fe}$ is the radius of the stellar iron core,
    $R_{\rm ic}$ is the inner core, $R_{\nu}$ is the neutrino sphere
    and $R_{\rm shock}$ indicates the shock wave
    position~\cite{jankavulcano}.}
  \label{fig:janka}
\end{figure}

The collapse continues and the central density keeps growing until
nuclear matter densities are reached ($\sim 1$--$3 \times 10^{14}$ g
cm$^{-3}$). At this point the nucleons and nuclei in the inner core
merge to form a macroscopic state of nuclear matter. Due to Fermi
effects and the repulsive nature of the nucleon-nucleon interactions
at short distances, there is a dramatic rise in pressure. Consequently
the inner core becomes incompressible and rebounds (stages 3--5 of
Fig.~\ref{fig:etapes}). The core bounce generates sound waves that
start propagating radially out of the inner core. They will not get
very far though, since the material in the outer core is falling
supersonically, forcing them to accumulate at the sonic point (border
between the subsonically infalling inner core and the supersonically
infalling outer core). The net effect is the formation of a density,
pressure, and velocity discontinuity in the flow, i.e., a shock wave,
which acquires more and more energy and almost immediately propagates
to the outer part of the iron core.

\subsection{Deleptonization and cooling}
\textbf{A) Neutronization burst}\vspace{0.2cm}\\
Once the core bounces and the shock wave is created, it starts
propagating outwards, dissociating nuclei into free nucleons (stages
5--7 of Fig.~\ref{fig:etapes}). Since the electron capture cross
section on free protons is larger than the one on nuclei, an enormous
amount of electron neutrinos are created through the neutronization
reaction $e^- + p \to n + \nu_e$, in those regions affected by the
shock wave. As already explained, neutrinos are initially trapped due
to the large density of the medium. The situation changes when the
shock wave gets to the neutrino sphere dissociating the iron nuclei,
some of the pressure is relieved and neutrinos can freely escape. This
sudden neutrino emission leads to a momentary rise in the luminosity
up to $10^{54}$ erg s$^{-1}$, known as \emph{neutronization burst} or
\emph{prompt neutrino burst}, and constitutes the second stage of the
neutrino emission. The duration of this peak is $\lesssim 10$ ms, and
is shown in Fig.~\ref{fig:etapemisionu}.

The two processes here described, namely nuclei dissociation and
neutrino emission, are responsible for an energy loss in the shock
wave which gets stalled in the iron core at a few hundred km. Its
revival is one of the most important issues currently discussed in the
theory of
gravitational core collapse SNe.\vspace{1.5cm}\\
\textbf{B) Matter accretion and mantle cooling}\vspace{0.2cm}\\
\begin{figure}[t]
  \begin{center}
    \includegraphics[width=0.6\textwidth]{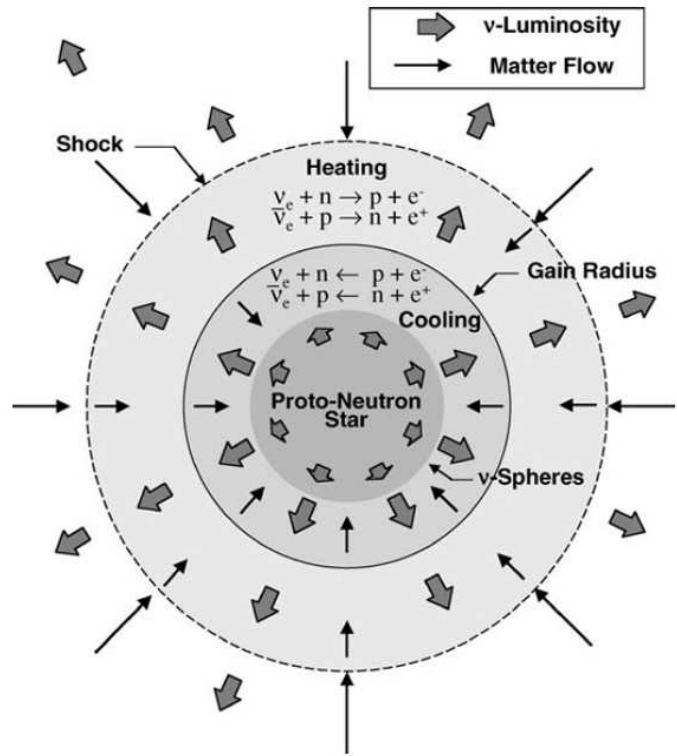}
  \end{center}
  \caption{\small Schematic representation of the exploding star after
    the shock wave passage in the reheating phase. Below the neutrino
    sphere we have the central radiating proto-neutron star while
    above it, but below the stalled shock wave, there is a net cooling
    region and a net heating region, mediated by electron neutrino and
    antineutrino absorption and emission~\cite{Mezzacappa:2005ju}.}
  \label{fig:protoestel}
\end{figure}
At this point the situation is as represented in
Fig.~\ref{fig:protoestel}. Under the shock wave remains a central
radiating object, the proto-neutron star (PNS), which will go on to
form a neutron star or a black hole. The PNS has a relatively cold
inner part, below the point where the shock wave was formed, composed
of neutrons, protons, electrons and neutrinos ($Y_L \simeq
0.35$). Surrounding this region there is a hot mantle formed by
``shocked'' nuclear material with low lepton number.

Since this mantle is hot ($T \gtrsim \mathcal{O}(10)$ MeV) and has a relatively
low density, the electrons are not quite degenerate and relativistic
thermal positrons can be created. Their presence will give rise to the
appearance of neutrinos through $e^+ + n \to \bar{\nu}_e + p$ and $e^+
+ e^- \to \nu + \bar{\nu}$ reactions. This point marks the third phase
in the neutrino emission. In contrast to the neutronization burst,
where only electron neutrinos are emitted, here all three flavors of
neutrinos and antineutrinos are created and emitted as the mantle
cools and contracts in the Kelvin-Helmholtz cooling phase.

On top of that, the external core accretes material over the PNS. The
gravitational energy released in this process is transformed into
thermal energy and emitted as thermal neutrinos. This stage lasts
between 10 ms and 1 s, and the neutrino luminosities remain in an
average value of $\sim 10^{52}$ erg s$^{-1}$ thanks to the accreted
material. We therefore obtain that the cooling and the
neutronization/deleptonization take place for the shocked
outer regions earlier than for the inner regions.\vspace{1.5cm}\\
\textbf{C) Proto-neutron star}\vspace{0.2cm}\\
In this stage (also known as Kelvin-Helmholtz cooling phase), the part
of the star that has not been ejected during the explosion evolves
from a hot and lepton rich configuration (PNS) to a cold and
deleptonized neutron star.

This stage represents the fourth and last of the neutrino emission
phases. Once the explosion starts, after the accretion phase, there is
a dramatic decrease in luminosity. As shown in
Fig.~\ref{fig:etapemisionu}, we observe an exponential fall in the
neutrino luminosity characteristic of the neutron star formation and
its cooling.

\begin{figure}[t]
  \begin{center}
    \includegraphics[width=0.82\textwidth]{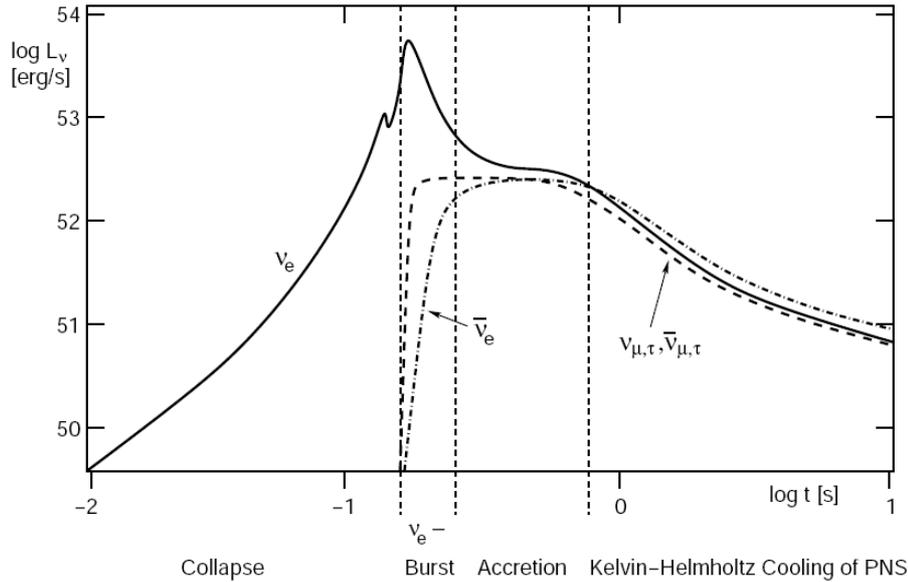}
  \end{center}
  \caption{\small Schematic neutrino luminosity curves for the
    different emission phases, corresponding to $\nu_e$ (solid line),
    $\bar\nu_e$ (dot-dash line) and each of the non-electron neutrinos
    (dashed line)~\cite{jankaKarlsruhe}.}
  \label{fig:etapemisionu}
\end{figure}

\subsection{Supernova Explosion}

The simplest scenario for a SN to take place would be that where the
shock wave has enough energy to go beyond all the infalling material
and blow up the star. In less than a second it would leave the iron
core and a moment later would eject the remaining outer layers,
producing a purely hydrodynamical explosion in a time scale of about
10 ms. This is the so called prompt explosion
scenario~\cite{Colgate:1960zz} and in order to be successful one needs
a sufficiently small and cold core and a soft equation of
state. Nevertheless, in general, numerical simulations do not seem to
confirm such a simple scenario as the one chosen by nature to carry
out the final SN explosion.

The shock wave undergoes a series of processes, resulting in energy
losses which progressively weakens it and ultimately stops its
progression. On the other hand SNe are not a theoretical hypothesis,
but take place in the Universe.
This is why a lot of effort has been
focused in determining the way the shock wave is revived and the
explosion is obtained in the delayed mechanism.

Different ingredients are being considered as possible contributions
to the phenomenon, and a combination of them may actually be involved
in the SN explosion mechanism: heating of the postshock region by
neutrinos, multi-dimensional hydrodynamic instabilities of the
accretion shock, in the postshock region, and in the PNS, rotation,
PNS pulsations, magnetic fields, and nuclear burning.

Three SN explosion mechanisms are centering nowadays the
discussion~\cite{Janka:2006fh}:
\begin{itemize}
\item The neutrino mechanism, where the shock wave is reactivated by
  the electron neutrinos and antineutrinos coming out the PNS. Part of
  these are absorbed by protons and neutrons behind the shock wave,
  providing the energy required.
%
%
  This mechanism was proposed by Wilson and Bethe~\cite{Bethe:1984ux},
  and although the energy released in form of neutrinos is by far
  larger than the energy needed to drive the explosion, it is very
  difficult to clarify the role played by the neutrino heating in the
  SN explosion mechanism.

\item The magneto-rotational or magneto-hydrodynamic
  (MHD)~\cite{Takiwaki:2004kf,Kotake:2004jt} mechanism, where the
  needed energy would be obtained from the rapid rotation of the
  collapsing stellar core and the amplification of magnetic fields
  through compression and wrapping. Such scenarios seem plausible in
  the cases of hypernovae, leading to jet-like explosions, but are
  disfavored for ordinary SNe because of the slow rotation of their
  progenitors predicted by stellar evolution calculations.

\item The acoustic mechanism, recently proposed by Burrows et
  al.~\cite{Burrows:2005dv,Burrows:2006uh}, relies on the acoustic
  power generated in the core of the PNS. According to this mechanism,
  the energy produced in the large-amplitude core motions would be
  transported via strong sound waves to the postshock region and
  deposited there, eventually triggering the late explosions at
  $\gtrsim$ 1 s after bounce. This mechanism appears to be
  sufficiently robust to blow up even the most massive and extended
  progenitors, but has so far not been confirmed by other
  groups. Although the acoustic modes are also found in other
  numerical simulations, like the ones performed by the Garching
  group~\cite{Janka:2006fh}, their amplitude seem to be much smaller,
  leading to no practical effects.
\end{itemize}

\section{Expected neutrino signal}

Independently of the concrete SN explosion mechanism, presumably there
are several characteristics regarding neutrinos that must result from
such an event. Let us here review the most important ones.

\subsection{General properties}

The energy released in such an event comes from the gravitational
binding energy of the compact star born after the collapse
\footnote{All the estimates given in this section are calculated using
  Newtonian physics.}
\bea
E_{\rm b} = \Delta E_{\rm G} & = & -\frac{3}{5}\left(\frac{G_{\rm N}
    M_{\rm core}^2}{R_{\rm Fe~core}}-\frac{G_{\rm N} M_{\rm
      core}^2}{R_{\rm Neutron~star}} \right)\nonumber  \\ 
& \approx & 1.60 \times 10^{53}{\rm erg}\left(\frac{M_{\rm core}}{M_{\odot}} \right)^2 \left(\frac{10~{\rm km}}{R_{\rm NS}}\right)~.
\eea
It seems reasonable to assume approximate equipartition of this energy
among the different neutrino flavors, receiving $E_{\rm b}/6$ each
of the 6 degrees of freedom that conform the standard (anti)neutrinos.

These neutrinos are trapped inside the PNS due to its huge density,
being released only after several collisions from a surface of $r
\approx 10$--$20$ km. The gravitational pressure of this compact
object is sustained by the thermal pressure, as long as matter near
its surface is not degenerate. We can make use of the virial theorem
in order to obtain the mean kinetic energy of a typical nucleon near
the PNS surface,
\be
\langle E_{\rm kin} \rangle = - \frac{1}{2} \langle E_{\rm G}\rangle \approx \frac{1}{2} G_{\rm N} \frac{M m_{\rm N}}{R}~,
\ee
leading to a typical value for the temperature of order 10 MeV, which
will therefore characterize the thermal neutrinos released.

As for the duration of the emission, it should be a multiple of the
typical diffusion time, 
\be
t_{\rm diff} \approx R_{\rm NS}^2/\lambda~,
\ee
where $\lambda$ stands for the mean free path. In order to give an
estimate of $t_{\rm diff}$ we use the scattering cross section off
non-relativistic nucleons, $\sigma \approx G_F^2 E_{\nu}^2/\pi =1.7
\times 10^{-42}$ cm$^2(E_{\nu}/10~{\rm MeV})^2$ and an approximate
density of $\rho_0 \approx 3 \times 10^{14}$ g cm$^{-3}$. Using
characteristic values of the involved quantities we obtain a $\lambda
\approx 300$ cm for neutrinos of 30 MeV, which leads to a diffusion
time of order $t_{\rm diff} = \mathcal{O}(1~{\rm s})$.


\subsection{Energies and spectra}

We have then some generic features of neutrinos coming out of the SN
core, confirmed by numerical simulations of neutrino
transport. Furthermore it is obvious that the nature of the scenario
we are dealing with will give rise to substantial differences among
neutrino flavors. Let us try to analyze some of them.

On top of the average neutrino energy of 10 MeV previously motivated,
all numerical simulations seem to obtain the same hierarchy for the
specific flavor average energies. According to the simulations,
electron neutrinos would start their free streaming above the neutrino
sphere with a lower average energy than electron antineutrinos, which
in turn have a lower average energy than muon and tau (anti)neutrinos,
$\langle E_{\nu_e}\rangle \lesssim \langle E_{\bar\nu_e}\rangle
\lesssim \langle E_{\nu_x}\rangle$ ($\nu_x \equiv
\nu_\mu,\nu_\tau,\bar\nu_\mu,\bar\nu_\tau$). This can be understood by
using some simple arguments.

\begin{figure}[t]
  \begin{center}
    \includegraphics[width=0.84\textwidth]{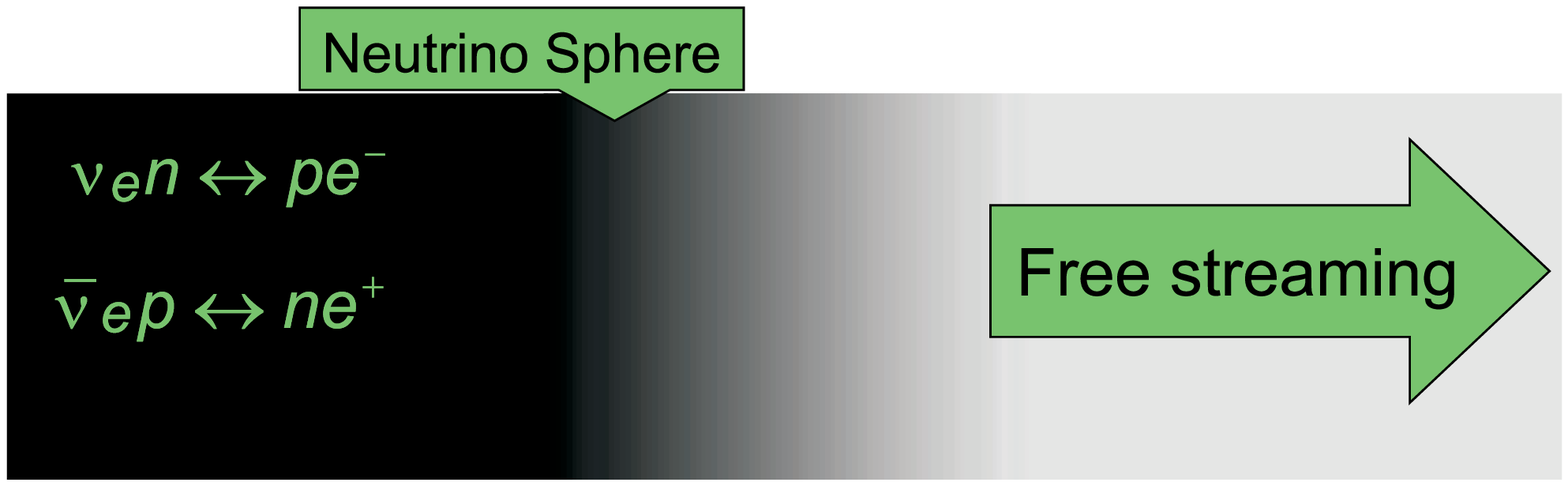}
    \includegraphics[width=0.84\textwidth]{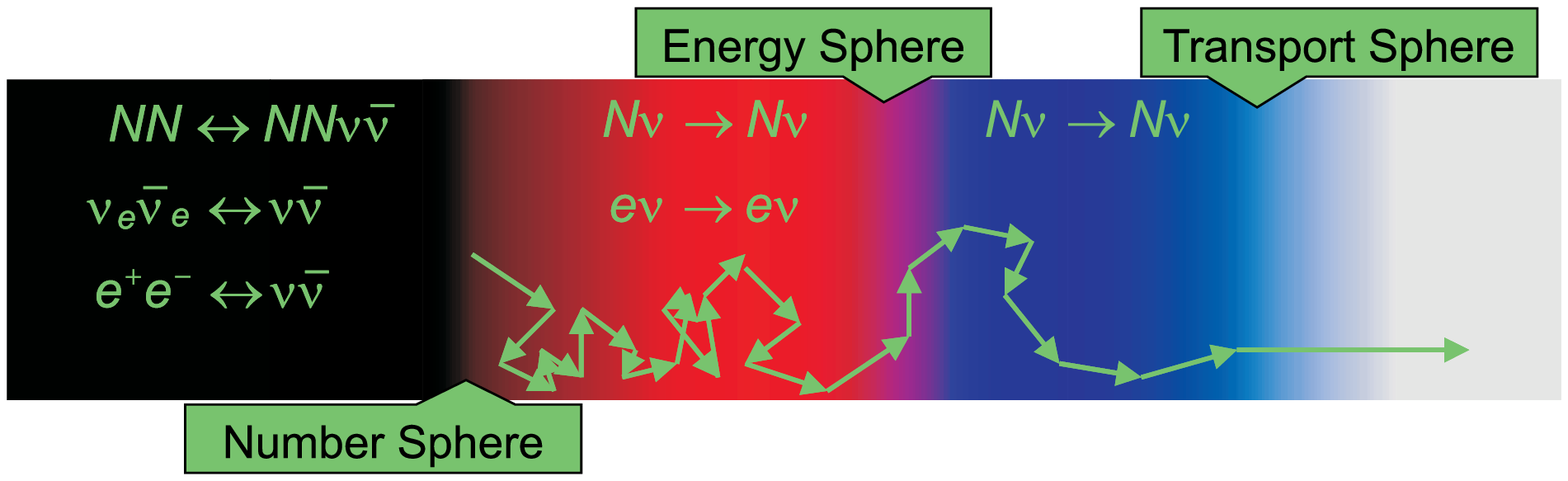}
  \end{center}
  \caption{\small Schematic representation of the neutrino evolution
    through the star, showing the different neutrino spheres. The top
    panel correspond to $\nu_e$ and $\bar\nu_e$ and the bottom one to
    non-electron neutrinos \cite{Keil:2003sw}.}
  \label{fig:nuesferes}
\end{figure}

The top panel of Fig.~\ref{fig:nuesferes} shows schematically the
propagation of $\nu_e$ and $\bar\nu_e$ throughout the SN. The main
reactions responsible for keeping them trapped inside the core in
thermal equilibrium are $\beta$ processes: neutron capture and proton
capture, respectively. The energy dependence of these reactions is
exactly the same, leading in principle to equal energy spectra for
both types of neutrinos. Nevertheless, this is not the whole story,
since the core of the star, in its way of becoming a neutron star,
contains more neutrons than protons, and this difference will only
grow with time. As a consequence $\nu_e$'s have a higher absorption
rate than $\bar\nu_e$'s, which is translated into a deeper $\nu_e$
neutrino sphere. The radius where neutrinos decouple from the PNS will
determine their energy, higher radius means lower densities and
temperatures. Since this argument applies for all neutrino energies,
the mean energy of the emitted $\bar\nu_e$ will always be larger than
the mean energy of the $\nu_e$.

Concerning the remaining part of the energy hierarchy relation, the
bottom panel of Fig.~\ref{fig:nuesferes} shows schematically the
non-electron neutrino transport. Since there are no $\mu$ nor $\tau$
leptons in the medium and therefore $\nu_x$ neutrinos do not
experience charged currents, we would naively expect them to decouple
deeper inside the PNS, compared to electron flavor neutrinos. The
situation is a bit more complicated though. It is true that there are
no charged current interactions for $\nu_x$, but they suffer from a
variety of neutral currents that must be taken into account. According
to the dominant interactions taking place we can distinguish four
regimes of evolution separated by different spheres: number sphere,
energy sphere and transport sphere.

In the innermost region the non-electron neutrinos are kept in thermal
equilibrium by energy exchanging scattering processes and the
following pair processes: Bremsstrahlung $NN \leftrightarrow NN \nu
\bar\nu$, neutrino-pair annihilation $\nu_e \bar\nu_e \leftrightarrow
\nu \bar\nu$ and electron-positron-pair annihilation $e^+ e^-
\leftrightarrow \nu \bar\nu$. The radius where these interactions
become inefficient defines the number sphere.

Beyond this point $\nu_x$'s are no longer in thermal equilibrium,
although they still exchange energy with the medium via scattering
reactions: $N\nu \to N\nu$ and $e^{\pm}\nu \to e^{\pm}\nu$. However,
the two processes are qualitatively very different. Scattering on
e$^{\pm}$ is less frequent since the interaction cross section is
smaller and there are fewer e$^{\pm}$ than nucleons. On the other hand
the amount of energy exchanged in each interaction with e$^{\pm}$ is
very large compared to the small recoil of nucleons.  At the radius,
where scattering on e$^{\pm}$ freezes out lies the energy sphere. A
diffusive regime starts, where neutrinos only scatter on nucleons and
therefore exchange little energy in each reaction. This regime is
terminated by the transport sphere, defined by the radius at which
also scattering on nucleons becomes ineffective and the $\nu_x$ start
streaming freely.

Due to its dependence on the square of neutrino energy the nucleon
scattering cross section has a filter effect, because it tends to
scatter high energy neutrinos more
frequently~\cite{Raffelt:2001kv}. The position of the number sphere
determines the flux, because neutrino creation is not effective beyond
that radius. The $\nu_x$ flux that passes the number sphere is
conserved. On the other hand, the mean energy of $\nu_x$ in this area
is still significantly lowered due to scattering processes before the
$\nu_x$ leave the star. The mean energy of $\nu_x$ emerging the star
is usually found to be larger than that of~$\bar\nu_e$.

Typical values for the mean energies obtained in numerical simulations
are:
\be
\langle E_{\nu} \rangle = \left\{\begin{array}{cl} 10\mbox{--}12~{\rm MeV} & \nu_e \\ 14\mbox{--}18~{\rm MeV} & \bar\nu_e \\ 18\mbox{--}24~{\rm MeV} & \nu_{\mu,\tau},\bar\nu_{\mu,\tau}\\ \end{array} \right.
\ee

As for the number fluxes, there is again a hierarchy relation among
them. The non-electron flavor ones are smaller than those of
$\bar\nu_e$ because the energy is found to be approximately
equipartitioned between the flavors. Similarly, since the lepton
number is carried away in $\nu_e$'s, their number flux is larger than
that of $\bar\nu_e$, so that again the energy is approximately
equipartitioned between $\nu_e$ and $\bar\nu_e$. In summary, we obtain
the hierarchy in the number fluxes: $F_{\nu_e} > F_{\bar\nu_e} >
F_{\nu_x} = F_{\bar\nu_x}$.

Throughout the literature one can find different forms of
parameterizing the non-thermal spectra of the neutrino fluxes. Two of
them are the most used. The first one is the quasi Fermi-Dirac
distribution:
\begin{equation}
  F^0_{\nu}(E) = 
  \frac{\Phi_{\nu} 
  }{T_{\nu}^3f_2(\eta_{\nu})} 
  \frac{E^2}{e^{E/T_{\nu}-\eta_{\nu}}+1}, 
\label{eq:flux-FD}
\end{equation}
where $E$ is the neutrino energy, and $T_{\nu}$ and $\eta_{\nu}$
denote an effective temperature and degeneracy parameter (chemical
potential), respectively.  The distribution is normalized so that
$\Phi_{\nu}$ stands for the total number of $\nu$ emitted.
The function $f_n(\eta_{\nu})$ is defined as 
\begin{equation}
f_n(\eta_{\nu}) \equiv \int_0^\infty 
\frac{x^n}{{\rm e}^{x-\eta_{\nu}}+1}{\rm d}x \,.
\end{equation}
The mean energy is consequently $\langle E_{\nu}\rangle =\left[
  f_3(\eta_{\nu}) / f_2(\eta_{\nu}) \right] T_{\nu}$, and the total
energy released is $E_{\nu}^{\rm tot} = \Phi_{\nu} \langle
E_{\nu}\rangle$.

\begin{figure}[t]
  \begin{center}
    \includegraphics[width=0.55\textwidth]{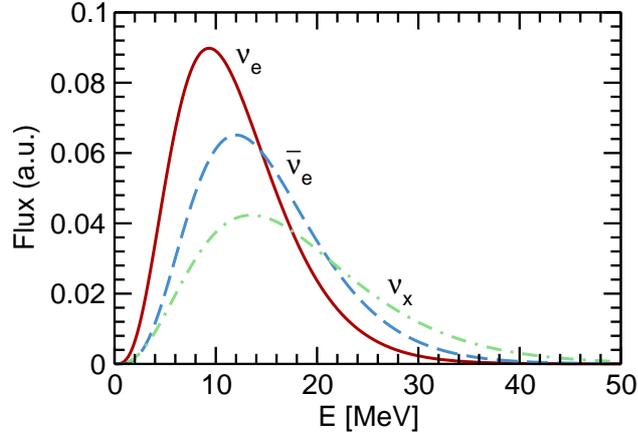}
  \end{center}
  \caption{\small Neutrino fluxes given by
    Eq.~(\ref{eq:spectralform}), with $\langle E_{\nu_e}\rangle = 12$,
    $\langle E_{\bar\nu_e}\rangle = 15$, $\langle E_{\nu_x}\rangle =
    18$, $\alpha_{\nu_e} = 3.5$, $\alpha_{\bar\nu_e} = 4$ and
    $\alpha_{\nu_e} = 3$. The flux is given in arbitrary units (a.u.)
    so that $\Phi_{\nu_e} = 1.15$, $\Phi_{\bar\nu_e} = 1$ and
    $\Phi_{\nu_x} = 0.85$.}
  \label{fig:fluxKeil}
\end{figure}

The second parametrization, shown in Fig.~\ref{fig:fluxKeil}, has been
recently introduced by the Garching group and fits better their
numerical results~\cite{Keil:2003sw}:
\begin{equation}\label{eq:spectralform}
  F^0_{\nu}(E)=
  \frac{\Phi_{\nu}}{\langle E_{\nu} \rangle}\,\frac{(1+\alpha)^{1+\alpha}}{\Gamma(1+\alpha)}  
  \left(\frac{E}{\langle E_{\nu} \rangle}\right)^\alpha 
  \exp\left[-(\alpha+1)\frac{E}{\langle E_{\nu} \rangle}\right] \,,
\end{equation}
where $\alpha$ describes a possible deformation with respect to a
Fermi-Dirac distribution.

\begin{figure}[t]
  \begin{center}
    \includegraphics[width=0.7\textwidth]{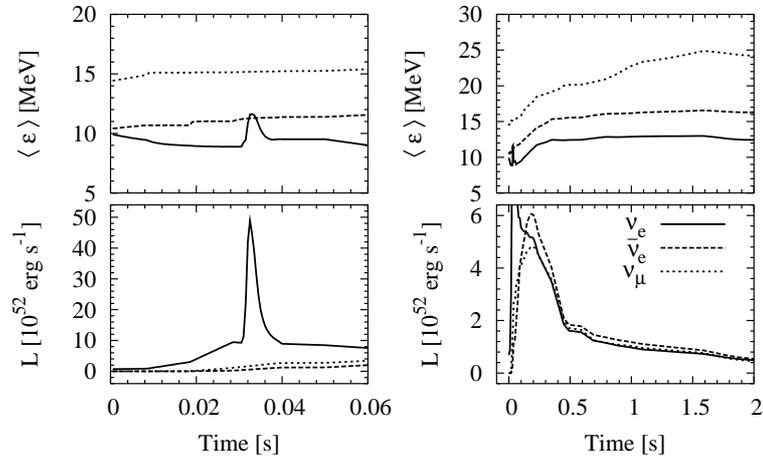}
  \end{center}
  \caption{\small Numerical result of the first two seconds after the
    core rebound. \emph{Top left:} Only the mean energy of the
    $\nu_e$'s is affected by the neutronization burst. \emph{Top
      right:} The evolution of the mean energies shows the hierarchy
    discussed in the text. \emph{Bottom left:} The neutronization
    explosion lasts for a few ms. \emph{Bottom right:} Temporal
    evolution of luminosities~\cite{Totani:1997vj}.}
  \label{fig:emisioliver}
\end{figure}

Fig.~\ref{fig:emisioliver} shows a numerical simulation where we can
see the mean energies (denoted by $\epsilon$) and the luminosities. In
this simulation we can identify all the features we have been
discussing in this section. 

\cleardoublepage

\pagestyle{normal}
\chapter{Standard Neutrino Oscillations}\label{chapter:oscillations}

In the last years, it has been widely demonstrated that neutrinos
oscillate from one flavor to another. This phenomenon occurs when the
interaction and mass bases do not coincide, meaning that the particles
which propagate are not the same as the ones that are created or
detected. This has been proven to be the case for neutrinos, created
via charged current weak interactions as one of the weak eigenstates
$\nu_e$, $\nu_\mu$ and $\nu_\tau$, different in general from the
propagating mass eigenstates $\nu_1$, $\nu_2$ and $\nu_3$, since the
mass matrix in the flavor basis is not diagonal. This, as we will
later show, leads inevitably to neutrino oscillations.

The first one to address the question of oscillation in the neutrino
system was Pontecorvo~\cite{Pontecorvo:1957cp} in 1957. Inspired in
the well known $K^0 \leftrightarrow \bar K^0$ oscillations, Pontecorvo
initially proposed neutrino-antineutrino oscillations. The precise
realization of the idea in terms of mass and mixing was introduced by
Maki, Nakagawa and Sakata~\cite{Maki:1962mu} in 1962 and later
developed by Pontecorvo~\cite{Pontecorvo:1967fh} in 1967.

This phenomenon will affect neutrino propagation through the SN
envelope and therefore has to be taken into account when studying SN
neutrinos. This problem can be attacked in two different ways. Either
we assume we have under control the part involving neutrino properties
(masses, mixing and CP-violating phases) and try to learn about SN
physics (explosion dynamics, SN neutrino fluxes and spectra, etc),
or we assume we have a good enough understanding of the SN physics and
try to improve our knowledge on the neutrino parameters. Throughout
this thesis we will mainly follow this second approach, trying to gain
some insight in the neutrino properties, by making some assumptions in
the SN models.

In this chapter we will introduce the basics of neutrino oscillations
in different steps. We will first treat vacuum oscillations in two and
three neutrino scenarios. After that we will discuss the effect of
neutrino interactions with matter, starting with a constant density
medium and later consider the varying density case. Finally we will
apply the formalism discussed in these sections to analyze the
evolution of neutrinos through SN envelopes.

\section{Vacuum oscillations}

The problem we want to discuss is the probability of observing a
neutrino flavor eigenstate different than the one created after a
certain time. While the natural basis to study neutrino interactions
is the flavor one, their evolution in vacuum is much simpler in the
mass eigenstate basis. We can always define a unitary transformation,
linking both neutrino basis and express the flavor eigenstates
$\nu_{\alpha}$ ($\alpha = e,\mu,\tau$) as a linear combination of the
mass eigenstates $\nu_i$ ($i = 1,2,3$),
\be
|\nu_{\alpha}\rangle=U^*_{\alpha i}\,|\nu_i\rangle\,,
\label{eq:alfaUi}
\ee
where we are summing over repeated indexes up to the number of
neutrino species. The relation for antineutrinos is exactly the same
but complex conjugating the elements in the mixing matrix $U$,
i.e.~$|\bar{\nu}_{\alpha}\rangle=U_{\alpha
  i}\,|\bar{\nu}_i\rangle\,$. Therefore, only for a complex $U$
(including a CP violating phase) we would observe a difference in the
evolution of neutrinos and antineutrinos in vacuum.

Using Eq.~(\ref{eq:alfaUi}), the initial neutrino state at $t=0$ can
be written as $|\nu(0)\rangle=|\nu_{\alpha}\rangle =U^*_{\alpha
  i}|\nu_i\rangle$. After a time $t$ the mass eigenstates just acquire
a phase, leading to
\be
|\nu(t)\rangle=U^*_{\alpha i}\,e^{-iE_i t}|\nu_i\rangle \,.
\label{eq:evol1}
\ee
If we now project this state onto a flavor eigenstate we find the
probability amplitude of finding the initial neutrino in that
particular state,
\be 
A(\nu_\alpha \to \nu_\beta)= \langle \nu_\beta|\nu(t)\rangle =
U^*_{\alpha i}\,e^{-iE_i t}\,\langle \nu_{\beta}|\nu_i \rangle =
U_{\beta j} U^*_{\alpha i}\,e^{-iE_i t}\,\langle \nu_j|\nu_i\rangle =
U_{\beta i}\,e^{-iE_i t}\, U^*_{\alpha i} \,.
\label{eq:evol2}
\ee
And finally, squaring this amplitude we obtain what we were looking
for, the probability of finding $|\nu_{\beta}\rangle$ at $t$ when
creating $|\nu_{\alpha}\rangle$ at $t=0$,
\be
P(\nu_{\alpha}\to \nu_{\beta})=
|A(\nu_{\alpha} \to \nu_{\beta})|^2=|U_{\beta i}\,e^{-iE_i t}\, U^*_{\alpha i}|^2 \,.
\label{eq:evol2.5}
\ee
Expanding this last expression we obtain
\be
P(\nu_{\alpha}\to \nu_{\beta})=\sum_i|U_{\beta i}|^2|U^*_{\alpha
  i}|^2+2{\rm Re}\left[\sum_{i\neq j} U_{\beta i}U^*_{\alpha i}U^*_{\beta j}U_{\alpha
    j}\,e^{-i(E_i-E_j)t}\right]\,,
\label{eq:evol2.6}
\ee
where we have explicitly written the summation. In all cases of
interest to us, the neutrinos are relativistic, so that we can
approximate,
\be
E_i=\sqrt{p^2+m_i^2}\simeq p+\frac{m_i^2}{2p}\simeq p+\frac{m_i^2}{2E}\,, 
\ee
and rewrite Eq.~(\ref{eq:evol2.6}) as
\be
P(\nu_{\alpha}\to \nu_{\beta})=\sum_i|U_{\beta i}|^2|U^*_{\alpha
  i}|^2+2{\rm Re}\left[\sum_{i\neq j} U_{\beta i}U^*_{\alpha i}U^*_{\beta j}U_{\alpha
    j}\,e^{-i\frac{\Delta m^2_{ij}}{2E}t}\right]\,,
\label{eq:evol3}
\ee
in terms of the neutrino squared mass differences, $\Delta m^2_{ij} =
m^2_i-m^2_j$.

\subsection{Two flavor case}

Let us take a closer look at Eq.~(\ref{eq:evol3}) in the two flavor
scenario, i.e.~we will only consider for the moment $\nu_e$ and
$\nu_{\mu}$. The mixing matrix connecting the flavor and interaction
basis takes the simple form
\be
U=\left(\begin{array}{cc}
\cos\theta_0 & \sin\theta_0 \\
-\sin\theta_0 & \cos\theta_0 \end{array} \right)\,,
\label{eq:U2}
\ee
where $\theta_0$ is the mixing angle. Making use of
Eqs.~(\ref{eq:evol3}) and~(\ref{eq:U2}) we obtain the oscillation
probabilities in two flavors
\be
P(\nu_e\to\nu_\mu)=P(\nu_\mu\to\nu_e)=
\sin^2 2\theta_0 \,\sin^2 \left (\frac{\Delta m^2}{4E}L\right)\,, 
\label{eq:prob1}
\ee
where $\Delta m^2=m_2^2-m_1^2$ and $L\simeq t$ (for relativistic
neutrinos) is the distance between the source and the
detector. Unitarity assures that the survival probabilities are
$P(\nu_e\to\nu_e) = P(\nu_\mu\to\nu_\mu) = 1 -
P(\nu_e\to\nu_\mu)$. Since $U$ is real in the two flavor scenario, the
same expressions are obtained for antineutrino survival and
oscillation probabilities. Another convenient way of expressing the
transition probability is given by
\be
P(\nu_e\to\nu_\mu)=
\sin^2 2\theta_0\,\sin^2 \left(1.27 \Delta m^2\frac{L}{E}\right)\,,
\label{eq:prob3}
\ee
where $L$ is in m and $E$ in MeV or $L$ is in km and $E$ in GeV.

There are several remarkable features in these expressions. The first
one is the oscillatory behavior in $L/E$, explaining why we call them
neutrino oscillations. In Eq.~(\ref{eq:prob1}) we can distinguish two
factors: a constant amplitude, $\sin^2 2\theta_0$, and an oscillatory
term, $\sin^2 (\frac{\Delta m^2}{4E} L)$. If we first focus in the
amplitude we note that a non-zero mixing angle is required to obtain
oscillations. On the other extreme, the maximum in the amplitude
corresponds to a mixing angle of $\theta_0=45^\circ$, maximal
mixing. Paying now attention to the oscillatory term we observe that
no flavor transitions would occur for massless neutrinos. Summarizing,
neutrino oscillations require both mass and mixing to take place.

Furthermore, $(\Delta m^2/4E)L$ must be of order unity if we want to
observe the oscillatory pattern. We can explicitly define the
oscillation length,
\be
L_{osc}=\frac{4\pi E}{\Delta m^2}\,\simeq \,2.48\;{\rm m}\,\frac{E\,\mbox{(MeV)}}
{\Delta m^2\, (\mbox{eV}^2)}=\,2.48\;{\rm km}\,\frac{E\,\mbox{(GeV)}}
{\Delta m^2\, (\mbox{eV}^2)}\,,
\ee
which will help us in this argument. For $L \ll L_{osc}$ no
oscillations have developed yet, the phase in Eq.~(\ref{eq:prob1}) is
very small, leading to no visible effect. On the other hand, for a very
large phase, $L \gg L_{osc}$, the transition probability experiences
very fast oscillations, translated at the detector in the averaged
probability over distance
\be
\overline{P(\nu_e\to \nu_\mu)}=\overline{P(\nu_\mu\to \nu_e)}
=\frac{1}{2}\, \sin^2 2\theta_0\,.
\label{aver}
\ee

It is also remarkable how the oscillation probability depends on the
neutrino mass, through the squared mass splittings $\Delta m^2$. The
unfortunate consequence is that it will not be possible to access the
information about the absolute individual neutrino masses through
oscillation experiments, but only the squared mass differences.

\subsection{Three flavor case}

The number of active neutrino species can be indirectly determined
through the invisible width of the $Z^0$
decay~\cite{Amsler:2008zzb}. LEP measured experimentally this quantity,
obtaining $N_\nu = 2.984\pm 0.008$, and proving the existence of only
three active neutrino flavors, $\nu_e$, $\nu_{\mu}$ and
$\nu_{\tau}$. In the three-neutrino scenario the flavor eigenstates
are related to the mass eigenstates through
\be
\left(\begin{array}{c}
\nu_{e} \\
\nu_{\mu}\\   
\nu_{\tau}
\end{array} \right)=
\left(\begin{array}{ccc}
U_{e1} & U_{e2} & U_{e3}  \\
U_{\mu 1} & U_{\mu 2} & U_{\mu 3}  \\
U_{\tau 1} & U_{\tau 2} & U_{\tau 3}  
\end{array} \right)
\left(\begin{array}{c}
\nu_{1} \\
\nu_{2}\\   
\nu_{3}
\end{array} \right)\,. 
\label{eq:mix2}
\ee
The simplest unitary form of the lepton mixing matrix, for the case of
Dirac neutrinos, is given in terms of three mixing angles
$\theta_{12}$, $\theta_{13}$ and $\theta_{23}$ and one CP-violating
phase, $\delta$. The case of Majorana neutrinos is slightly more
complicated, adding two more phases, $\varphi_1$ and $\varphi_2$,
although they will not affect neutrino oscillations. The resulting
leptonic mixing matrix $U$, also known as PMNS matrix, can be
factorized following e.g.~the Particle Data
Group~\cite{Amsler:2008zzb}, into four different matrices,
\be 
{\setlength\arraycolsep{3pt}
U=V_{23} W_{13} V_{12} D\equiv V D\,,
\label{Unew}
}
\ee
with 
\be
{\setlength\arraycolsep{3pt}
V_{12}=\left(\begin{array}{ccc}
c_{12}    &  s_{12}  & 0     \\
-s_{12}   &  c_{12}    & 0     \\
0         &    0       & 1
\end{array} \right),~
W_{13}=\left(\begin{array}{ccc}
c_{13}                  &    0       & s_{13}\,e^{-i\delta} \\
0                       &    1       & 0       \\
-s_{13}\,e^{i\delta}    &    0       & c_{13}
\end{array} \right),~
V_{23}=\left(\begin{array}{ccc}
1         &    0       & 0        \\
0         &  c_{23}    & s_{23}     \\
0         & -s_{23}    & c_{23}
\end{array} \right),
}
\ee
where $c_{ij}=\cos \theta_{ij}$, $s_{ij}=\sin\theta_{ij}$, and $D={\rm
  diag}(e^{-i\varphi_1},\, 1,\,e^{-i\varphi_2})$. Since the Majorana
phases do not have any effect on neutrino oscillations, we can omit
the $D$ factor, resulting in the following expression for
Eq.~(\ref{Unew})
\be
U =
\left(\begin{array}{ccc}
c_{12}\,c_{13}   & s_{12}\, c_{13}    & s_{13}\, e^{-i\delta} \\
-s_{12}\,c_{23}-c_{12}\,s_{23}\,s_{13}\,e^{i\delta} &
c_{12}\,c_{23}-s_{12}\,s_{23}\,s_{13}\,e^{i\delta} &
s_{23}\,c_{13}   \\
s_{12}\,s_{23}-c_{12}\,c_{23}\,s_{13}\,e^{i\delta} &
-c_{12}\,s_{23}-s_{12}\,c_{23}\,s_{13}\,e^{i\delta} &
c_{23}\,c_{13}
\end{array}
\right)\,.
\label{eq:U3}
\ee
Replacing this matrix into Eq.~(\ref{eq:evol3}), we obtain the
corresponding neutrino oscillation formulas in three flavors. Contrary
to the two-flavor case, the neutrino and antineutrino formulas do not
coincide, unless $\delta = 0$ or $\pi$. Even though there are no
simple expressions in this case, there are several approximations in
terms of the two flavor ones, that apply for practical purposes, see
for instance~\cite{Akhmedov:1999uz}.

In the three-neutrino scenario there exist two independent squared mass
differences, $\Delta m^2_{21}$ and $\Delta m^2_{31}$, that will
determine the evolution of neutrinos, while $\Delta m^2_{32}$ can be
easily reexpressed in terms of the other two.

\begin{figure}
  \begin{center}
    \includegraphics[width=0.6\textwidth]{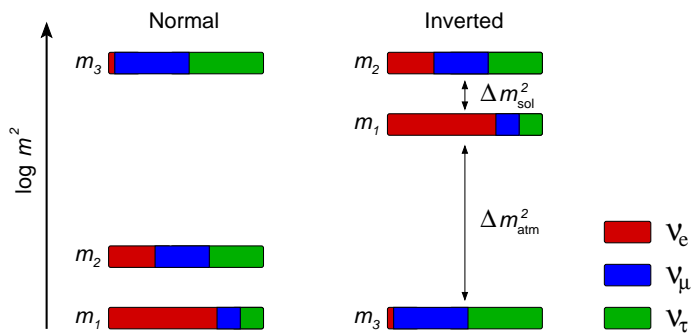}
  \end{center}
  \caption{\small Possible neutrino mass hierarchy patterns. The left
    one shows the normal case, while the right one corresponds to
    inverted hierarchy. The colors show the amount of a given flavor
    eigenstate in each mass eigenstate, according to current best fits
    for the mixing angles.}
  \label{fig:hierarchy}
\end{figure}

\subsection{Present status of three-flavor neutrino oscillations}

Four of the six neutrino oscillation parameters are rather well
determined by the oscillation data, the so called atmospheric
($|\Delta m^2_{31}|$, $\theta_{23}$) and solar ($\Delta m^2_{21}$,
$\theta_{12}$) neutrino parameters, while $\theta_{13}$, $\delta$ and
the sign of $\Delta m^2_{31}$ remain unknown~\cite{Maltoni:2004ei,
  Schwetz:2008er, Fogli:2008ig, Bandyopadhyay:2008va,
  GonzalezGarcia:2007ib}.

The status of the atmospheric neutrino parameters is determined by the
combination of different analyses. On the one hand, of course, we have
the atmospheric neutrino measurements from
Super-Kamiokande~\cite{Ashie:2005ik}, which give the most stringent
bound on the 23-mixing angle. On the other hand, the determination of
$|\Delta m^2_{31}|$ is dominated by accelerator experiments, mainly
MINOS data~\cite{Adamson:2008zt}, while K2K~\cite{Aliu:2004sq}
basically has no impact any more. The complementarity of these
experiments leads to the following best fit point and $1\sigma$
errors~\cite{Schwetz:2008er}:
\begin{equation}
\sin^2\theta_{23} = 0.50^{+0.07}_{-0.06} \,,\qquad
|\Delta m^2_{31}| = 2.40^{+0.12}_{-0.11} \times 10^{-3}\,{\rm eV^2} \,.
\end{equation}
Although we have quite a good measurement of $|\Delta m^2_{31}|$, it
is not possible to determine the hierarchy of neutrino masses,
i.e.~the sign of $\Delta m^2_{31}$, with the current data.

The determination of the solar neutrino parameters comes from the
combination of KamLAND reactor experiment~\cite{:2008ee} and
SNO~\cite{Aharmim:2008kc}, Super-Kamiokande~\cite{Fukuda:2002pe},
Borexino~\cite{Galbiati:2008zz} and Gallex/GNO~\cite{gallex-nu08:07}
solar neutrino experiments. Just as before, the determination of each
parameter is dominated by one type of experiment. Thus, $\theta_{12}$
is mostly constrained by solar experiments (mainly SNO), while $\Delta
m^2_{21}$ is basically determined by KamLAND. Nevertheless, KamLAND is
also starting to help on the lower limit of $\theta_{12}$. The
resulting parameters from this analysis are (at
$1\sigma$)~\cite{Schwetz:2008er}:
\begin{equation}\label{eq:solar}
\sin^2\theta_{12} = 0.304^{+0.022}_{-0.016} \,,\qquad
\Delta m^2_{21} = 7.65^{+0.23}_{-0.20} \times 10^{-5}\,{\rm eV^2} \,.
\end{equation}

Concerning the 13-mixing angle, at this moment we only have upper
bounds coming from null results of the short-baseline CHOOZ reactor
experiment~\cite{Apollonio:2002gd} with some effect also from solar
and KamLAND data, especially at low $\Delta m^2_{31}$ values. At
90\%~CL ($3\sigma$) the following limits are
obtained~\cite{Schwetz:2008er}:
\begin{equation}\label{eq:th13}
  \sin^2\theta_{13} \le \left\lbrace \begin{array}{l@{\qquad}l}
      0.060~(0.089) & \text{(solar+KamLAND)} \\
      0.027~(0.058) & \text{(CHOOZ+atm+K2K+MINOS)} \\
      0.035~(0.056) & \text{(global data)}
    \end{array} \right.
\end{equation}

Finally, no limit at all has been yet obtained for the CP violating
phase in neutrino oscillation experiments.

From solar experiments we know that $\Delta m^2_{21}$, also known as
solar squared mass difference ($\Delta m^2_{\rm sol}$), is positive,
while the sign of $\Delta m^2_{31}$ or $\Delta m^2_{\rm atm}$, is yet
unknown. This sign determines what is called the neutrino mass
hierarchy, $\Delta m^2_{31} = |\Delta m^2_{31}|$ corresponds to normal
hierarchy, and $\Delta m^2_{31} = -|\Delta m^2_{31}|$ to inverted
hierarchy. As we will later see it will be crucial in the evolution of
neutrinos inside the SN. Fig.~\ref{fig:hierarchy} shows the two
possible configurations for the hierarchy of neutrino masses.

A summary of the current knowledge of the neutrino parameters is given
in Table~\ref{tab:summary}. For a more detailed description of these
fits see Refs.~\cite{Schwetz:2008er,Maltoni:2004ei}.

\begin{table}[t]\centering
    \catcode`?=\active \def?{\hphantom{0}}
    
\begin{tabular}{|@{\quad}>{\rule[-2mm]{0pt}{6mm}}l@{\quad}|@{\quad}c@{\quad}|@{\quad}c@{\quad}|@{\quad}c@{\quad}|}
        \hline
        \hline
        parameter & best fit $\pm 1\sigma$ & 2$\sigma$ & 3$\sigma$ 
        \\
        \hline
        $\Delta m^2_{21}\: [10^{-5}{\rm eV^2}]$
        & $7.65^{+0.23}_{-0.20}$  & 7.25--8.11 & 7.05--8.34 \\[2mm]
        $|\Delta m^2_{31}|\: [10^{-3}{\rm eV^2}]$
        & $2.40^{+0.12}_{-0.11}$  & 2.18--2.64 & 2.07--2.75 \\[2mm]
        $\sin^2\theta_{12}$
        & $0.304^{+0.022}_{-0.016}$ & 0.27--0.35 & 0.25--0.37\\[2mm]
        $\sin^2\theta_{23}$
        & $0.50^{+0.07}_{-0.06}$ & 0.39--0.63 & 0.36--0.67\\[2mm]
        $\sin^2\theta_{13}$
        & $0.01^{+0.016}_{-0.011}$  & $\leq$ 0.040 & $\leq$ 0.056 \\
        \hline
        \hline
        \hline
\end{tabular}
\caption{\small Best-fit values with 1$\sigma$ errors,
  and 2$\sigma$ and 3$\sigma$ intervals (1 d.o.f.) for the
  three-flavor neutrino oscillation parameters from global data,
  including solar, atmospheric, reactor (KamLAND and CHOOZ) and
  accelerator (K2K and MINOS) experiments~\cite{Schwetz:2008er}.}
\label{tab:summary} 
\end{table}

\section{Neutrino oscillations in matter}

In most experiments, neutrinos travel through matter before being
detected. Solar neutrinos are emitted from deep inside the Sun and
have to travel not only through it before hitting the detector but
also through part of the Earth at night. Geometry tells us that
accelerator neutrinos must cross the Earth before being detected, as
well as most of the atmospheric neutrinos, depending on the original
interaction point. And of course the ones of most interest to us, SN
neutrinos, which traverse the whole envelope of the star before
finding vacuum. It is therefore of the utmost importance to determine
the effect of matter in neutrino oscillations.

\subsection{Neutrino interactions with matter}

The Standard Model is built over the gauge group $SU(3)_C \otimes
SU(2)_L \otimes U(1)_Y$, and describes the strong, weak and
electromagnetic interactions of matter. According to this model,
neutrinos are $SU(3)_C \otimes U(1)_{em}$ singlets, and therefore
interact only via charged (CC) and neutral (NC) weak currents,
described by the following Lagrangians:
\bea
\label{eq:L_CC} {\mathcal L}^{\rm CC} & = & \frac{g}{\sqrt{2}}\sum_{\alpha}\bar e_{\alpha L} \gamma^{\mu} \nu_{\alpha L} W_{\mu}^- + {\rm h.c.}~,\\
\label{eq:L_NC} {\mathcal L}^{\rm NC} & = & \frac{g}{2 \cos\theta_W} \sum_{\alpha}\bar\nu_{\alpha L} \gamma^{\mu} \nu_{\alpha L} Z_{\mu}^0~,
\eea
where $g = e/\sin\theta_W$, with $e$ the electron charge and
$\theta_W$ the weak angle, $f_{\alpha L} =
\frac{1}{2}(1-\gamma_5)f_\alpha$ correspond to the left-handed fermion
fields, with $\alpha = e, \mu, \tau$, $\gamma^\mu$ are the Dirac
matrices, and $W_{\mu}^-$ and $Z_{\mu}^0$ are the gauge boson fields.

\begin{figure}
\begin{center}
\begin{picture}(359,80)(-5,-40)
\ArrowLine(40,50)(70,25)
\ArrowLine(40,-50)(70,-25)
\Photon(70,25)(70, -25)3 4
\Text(95,0)[r]{\small$W^\pm$}
\Text(95,35)[r]{\small$e$}
\Text(100,-35)[r]{\small$\nu_e$}
\Text(40,-35)[l]{\small$e$}
\Text(40,35)[l]{\small$\nu_e$}
\ArrowLine(70,25)(100,50)
\ArrowLine(70,-25)(100,-50)
\ArrowLine(240,50)(270,25)
\ArrowLine(240,-50)(270,-25)
\Photon(270,25)(270, -25)3 4
\Text(290,0)[r]{\small$Z^0$}
\Text(325,35)[r]{\small$\nu_{e, \mu,\tau}$}
\Text(325,-35)[r]{\small$p, n, e$}
\Text(220,-35)[l]{\small$p,n,e$}
\Text(220,35)[l]{\small$\nu_{e,\mu,\tau}$}
\ArrowLine(270,25)(300,50)
\ArrowLine(270,-25)(300,-50)
\end{picture}
\end{center}
\caption{\small Neutrino scattering diagrams. \label{fig:WZ}}
\end{figure}
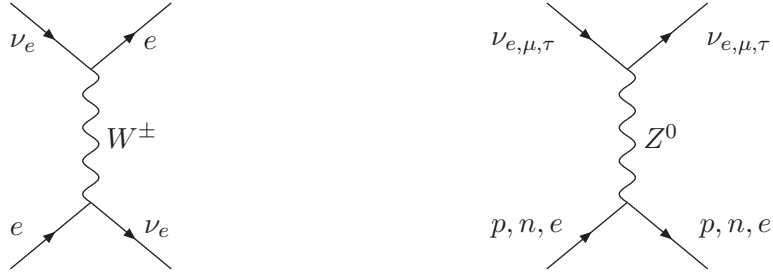

The dominant source of neutrino interactions in a medium is coherent
forward elastic scattering, under which the medium remains
unchanged. 
As first discussed by Wolfenstein~\cite{Wolfenstein:1977ue} in 1978,
the effect of this coherent process on neutrinos can be parameterized
as an effective potential affecting their evolution. Ordinary matter
is composed of electrons, protons and neutrons, but not $\mu$ or
$\tau$ leptons (in Chapter~\ref{chapter:coll2flavors} we will study
the effect of considering neutrinos as a background medium).  As a
consequence only $\nu_e$'s participate in CC mediated by the $W^{\pm}$
exchange, while all neutrino species have equal NC interactions on
$n$, $p$ and $e^-$ mediated by $Z^0$ bosons, see
Fig.~\ref{fig:WZ}.

Let us start by considering the effective potential induced by CC
interactions. Its only contribution comes from elastic scattering of
$\nu_e$'s on electrons, which in the effective low energy limit gives
the following term to the interaction Hamiltonian,
\bea
{\mathcal H}^{\rm CC}_{\rm eff} & = & \frac{G_{\rm F}}{\sqrt{2}}\{\bar e \gamma^{\mu}(1-\gamma_5)\nu_e\} \{\bar\nu_e \gamma_{\mu} (1-\gamma_5)e\}\\
& = & \frac{G_{\rm F}}{\sqrt{2}}\{\bar\nu_e  \gamma^{\mu}(1-\gamma_5)\nu_e\} \{\bar e \gamma_{\mu} (1-\gamma_5)e\}~,
\eea
where $G_{\rm F}/\sqrt{2} = g^2/8M_W^2$, and we have applied the Fierz
rearrangement in the second line. The effective potential $V_{\rm CC}$ can
be calculated as the matrix element of this interaction Hamiltonian:
\be
V_{\rm CC} = \langle \Psi|
{\mathcal H}^{\rm CC}_{\rm eff}| \Psi\rangle ~,
\label{matel}
\ee
where $\Psi$ is the wave function of the system of neutrino and
medium. We define the vector of polarization of electrons as 
\be
\vec{\lambda}_e \equiv \omega_e^{\dagger}\vec{\sigma}\omega_e~,  
\ee 
where $\omega_e$ is the two component spinor.  Suppose electrons have
some density distribution over the momentum, $\vec{p}_e$ and
polarization $\vec{\lambda}_e$:
\be
\frac{f(\vec{\lambda}_e,\vec{p}_e)}{(2\pi)^3}~.
\ee
Then the  total number density  of electrons, $N_e$, equals
\be
\label{ne}
N_e = \sum_{\vec{\lambda}}
\int  \frac{d^3 p_e}{(2\pi)^3} ~f(\vec{\lambda_e},\vec{p}_e)~.
\ee
The average polarization of electrons is defined as
\be
\langle \vec{\lambda}_{e}\rangle =
\frac{1}{N_e}\
\sum_{\vec{\lambda}}
\int  \frac{d^3 p_e}
{(2\pi)^3} \vec{\lambda} ~
f(\vec{\lambda}_e,\vec{p}_e)~. 
\label{avpol}
\ee
The matrix element of Eq.~(\ref{matel}) can be calculated as
\be 
\sum_{\vec{\lambda}} \int  \frac{d^3 p_e}
{(2\pi)^3}  ~
f(\vec{\lambda}_e,\vec{p}_e) 
\langle e_{p, \lambda}|\bar{e} \gamma_{\mu}(1 - \gamma_5) e  
     | e_{p, \lambda}\rangle~.
\label{polave} 
\ee
In the case of an unpolarized medium, $\vec{\lambda}_e = 0$, only the
vector current contributes to the potential:
\be
V_{\rm CC} = V^V_{\rm CC}(\vec{p}_e)  = \sqrt{2} G_{\rm F}\, \frac{f_e (\vec{p}_e)}{(2\pi)^3}
\left(1  - \frac{\vec{p}_e \cdot \widehat{k}_{\nu}}{E_e} \right) ,
\label{unpol}
\ee
where $\widehat{k}_{\nu} \equiv \vec{p}_{\nu}/|\vec{p}_{\nu}|$ with
$\vec{p}_{\nu}$ being the neutrino momentum, $E_e$ is the energy of
electrons. In the case of a moving medium both $\gamma_0$ and the space
components of the vector current, $\vec{\gamma}$, give non-zero
contribution. The former gives the electron density, $N_e$, while the
latter is also proportional to the velocity of electrons in the
medium: $ \langle \psi_e | \vec{\gamma}|\psi_e\rangle \propto \vec{v}
$ and~\cite{Nunokawa:1997dp,Langacker:1982ih}
\be
V_{\rm CC}(v_e) = \sqrt{2} G_{\rm F} N_e ( 1 - v\cdot \cos \beta)~,
\label{vpot}
\ee
where $\beta$ is the angle between the momenta of the electrons and
neutrinos.

If such unpolarized medium is composed of non-relativistic electrons
or ultra-relativistic electrons from an isotropic distribution, the
only non-zero contribution to the potential comes from the $\gamma_0$
term. Therefore the second term in Eq.~(\ref{vpot}) disappears and we
obtain \cite{Wolfenstein:1977ue}\footnote{For a detailed calculation
  of the effective potential in media with different properties
  see~\cite{Smirnov:1998cr}.}
%
%
\be
V_{\rm CC}  = \sqrt{2} G_{\rm F} N_e~.
\label{eq:VCC}
\ee


We can determine in an analogous way the effective potential due to
NC interactions that affect all neutrino flavors. The result is the
following:
\be
V_{\rm NC} = -\sqrt{2} G_{\rm F} N_n/2\,.
\ee

It is important to notice here that, besides these tree level
contributions to the matter potential, another one arises from
radiative corrections to neutral-current of $\nu_\mu$ and $\nu_\tau$
scattering. Although there are no $\mu$ nor $\tau$ leptons in normal
matter, they can appear as virtual states, causing a shift between
$\nu_\mu$ and $\nu_\tau$ due to the difference in the $\mu$ and $\tau$
lepton masses. As a consequence, the following effective potential
must be added to $\nu_\tau$~\cite{Botella:1986wy}:
\be\label{eq:Vmt}
V_{\mu \tau} \approx \frac{3\sqrt{2}G_{\rm F}
  m_{\tau}^2}{(2\pi)^2 Y_e} \left[\ln
  \left(\frac{m_W^2}{m_{\tau}^2}\right) - 1 +
  \frac{Y_n}{3}\right]V_{\rm CC} \equiv \frac{Y_\tau^{\rm eff}}{Y_e}~V_{\rm CC}~,
\ee
where $m_{\tau}$ is the $\tau$ mass and $m_W$ is the $W$-boson mass,
and $Y_e$ and $Y_n$ are the electron and neutron fraction numbers
respectively, i.e.~$Y_e = \frac{N_e}{N_p+N_n}$ and $Y_n =
\frac{N_n}{N_p+N_n}$. We have defined an effective $\tau$ fraction of
the medium, $Y_\tau^{\rm eff}$. If we assume $Y_e = Y_n = 0.5$
(typical in SN), we obtain $Y_\tau^{\rm eff} \approx 2.7 \times
10^{-5}$ and $V_{\mu \tau}\approx 5.4\times 10^{-5} ~V_{\rm CC}$.

Summarizing, the effects of the SM neutrino interactions in matter can
be parameterized for a neutrino of flavor $\alpha$, $\nu_\alpha$,
using the effective potentials $V_\alpha$:
\bea
\label{eq:Ve} &&V_e = V_{\rm CC} + V_{\rm NC} = \sqrt{2} G_{\rm F} \left(N_e - \frac{N_n}{2} \right)~,\\
\label{eq:Vmu} &&V_{\mu} = V_{\rm NC} = \sqrt{2} G_{\rm F} \left(-\frac{N_n}{2}
\right)~,\\
\label{eq:Vmutau} &&V_{\tau} = V_{\rm NC} + V_{\mu \tau} = \sqrt{2}
G_{\rm F} \left(- \frac{N_n}{2}+ 5\times10^{-5}N_e \right)~.
\eea
In the case of antineutrinos propagating in matter, the effective
potentials are identical but with opposite sign. These potentials can
be reexpressed in terms of the medium density $\rho$:
\bea
\label{eq:Vcc_Ye}V_{\rm CC} &=& \sqrt{2} G_{\rm F} N_e \simeq 7.6 \times 10^{-14} Y_e \rho({\rm g/cm^3})~{\rm eV} = V_0 Y_e \rho({\rm g/cm^3}) ~,\\
\label{eq:Vnc_Ye}V_{\rm NC} &=& \frac{\sqrt{2}}{2} G_{\rm F} N_n \simeq -3.8 \times 10^{-14} Y_n \rho({\rm g/cm^3})~{\rm eV}\nonumber \\
&=& -\frac{V_0}{2} (1-Y_e) \rho({\rm g/cm^3})~,
\eea
where we have defined $V_0 \equiv 7.6 \times 10^{-14}$ eV. Three
important examples are:
\begin{itemize}
\item At the Earth core: $\rho \sim 10$ g/cm$^3$ and $V_{\rm CC} \sim 10^{-13}$ eV.
\item At the Sun core $\rho \sim 100$ g/cm$^3$ and $V_{\rm CC} \sim 10^{-12}$ eV.
\item At a SN core $\rho \sim 10^{14}$ g/cm$^3$ and $V_{\rm CC}
  \sim$ eV.
\end{itemize}

\subsection{Evolution equation}\label{sec:evoleq}

It is convenient to derive the evolution equations in matter using the
weak eigenstate basis, since these are the neutrinos that interact and
feel the potential, and thus enter diagonally in the Hamiltonian.

Let us start again by considering the two flavor case. The evolution
equation in vacuum in the mass eigenstate basis is given by the
equation of Schr\"odinger:
\be
i\frac{d}{dt}|\nu_i\rangle=H_m|\nu_i\rangle~,
\ee
where $H_m={\rm diag}(E_1,\,E_2)$. In the flavor basis the resulting
equation of Schr\"odinger is:
\be
\label{eq:Schr_fl}i\frac{d}{dt}|\nu_{\alpha}\rangle=H_W|\nu_{\alpha}\rangle = U H_m U^\dagger |\nu_{\alpha}\rangle~.
\ee
For relativistic neutrinos we can use $E_i\simeq p+m_i^2/2E$, and
therefore obtain,
\be
{\setlength\arraycolsep{2pt}
i\frac{d}{dt}\left(\begin{array}{c}
\nu_e \\
\nu_\mu   
\end{array} \right)=
\left(\begin{array}{cc}
\left(p+\frac{m_1^2+m_2^2}{4E}\right)-\frac{\Delta m^2}{4E}\cos 2\theta_0 & 
\frac{\Delta m^2}{4E}\sin 2\theta_0 \\
\frac{\Delta m^2}{4E}\sin 2\theta_0 &
\left(p+\frac{m_1^2+m_2^2}{4E}\right)+\frac{\Delta m^2}{4E}\cos 2\theta_0
\end{array} \right)
\left(\begin{array}{c}
\nu_{e} \\
\nu_{\mu}   
\end{array} \right)\,. 
\label{eq:ev1}
}
\ee
Common terms in the diagonal elements of the effective Hamiltonian can
only add a common phase to all neutrino states, and therefore do not
have any effect in neutrino oscillations, where only relative phases
matter. We can then subtract any multiple of the identity matrix
without affecting neutrino oscillations. If we remove the terms in
brackets in the Hamiltonian we end up with the following evolution
equation in vacuum: \renewcommand{\arraystretch}{1.3} \be
i\frac{d}{dt}\left(\begin{array}{c}
    \nu_e \\
    \nu_\mu
\end{array} \right)=
\left(\begin{array}{cc}
-\frac{\Delta m^2}{4E}\cos 2\theta_0 & 
\frac{\Delta m^2}{4E}\sin 2\theta_0 \\
\frac{\Delta m^2}{4E}\sin 2\theta_0 &
\frac{\Delta m^2}{4E}\cos 2\theta_0
\end{array} \right)
\left(\begin{array}{c}
\nu_{e} \\
\nu_{\mu}   
\end{array} \right)
\equiv H_W^{\rm kin}\left(\begin{array}{c}
\nu_{e} \\
\nu_{\mu}   
\end{array} \right)\,. 
\label{eq:ev2}
\ee
\renewcommand{\arraystretch}{1}

If we want to study the effect of matter in the propagation of
neutrinos, we have to add a new term, $H^{\rm int}_W$, to the
Hamiltonian,
\be
\label{eq:hamiltonia}H_W=H_W^{\rm kin}+H_W^{\rm int}~,
\ee
with the corresponding effective potentials, given in Eqs.~(\ref{eq:Ve})
and~(\ref{eq:Vmu}), in the diagonal elements. Omitting again the
common terms due to NC interactions, we find the evolution equation in
matter,
\be
\renewcommand{\arraystretch}{1.3}
i\frac{d}{dt}\left(\begin{array}{c}
\nu_e \\
\nu_\mu   
\end{array} \right)=
\left(\begin{array}{lc}
-\frac{\Delta m^2}{4E}\cos 2\theta_0 + V_{\rm CC} ~~ & 
\frac{\Delta m^2}{4E}\sin 2\theta_0 \\
~~~\frac{\Delta m^2}{4E}\sin 2\theta_0 &
\frac{\Delta m^2}{4E}\cos 2\theta_0
\end{array} \right)
\left(\begin{array}{c}
\nu_{e} \\
\nu_{\mu}   
\end{array} \right)\,. 
\label{eq:ev3}
\renewcommand{\arraystretch}{1}
\ee
This general expression applies both for constant and varying
density. The physics, though, will be different and we will treat
these cases separately.

\subsection{Constant density case}

Let us start assuming a scenario with constant density. This case,
although unrealistic, is of particular interest, since it is a very
good approximation used for accelerator neutrinos. In this kind of
experiments neutrinos travel through part of the Earth, but rarely
leave the mantle, which at first order of approximation can be
considered to have constant density.

We define the matter eigenstates, $\nu^m_i$, as the eigenstates of the
effective Hamiltonian given in Eq.~(\ref{eq:ev3}), which for a varying
density medium depend on time/position, but are constant for the case
now under study. The relation with the interaction basis is given by
the unitary transformation:
\be
\left(\begin{array}{c} \nu_e \\ \nu_\mu \end{array} \right) = U(\theta_m) \left(\begin{array}{c} \nu^m_1 \\ \nu^m_2 \end{array}\right) =\left(\begin{array}{cc} \cos \theta_m & \sin \theta_m \\ -\sin \theta_m & \cos \theta_m \end{array} \right) \left(\begin{array}{c} \nu^m_1 \\ \nu^m_2 \end{array}\right)\,,
\label{eq:mateig}
\ee 
where the effective mixing angle, $\theta_m$, is of course different
from the vacuum mixing angle, $\theta_0$. It is obtained from the
diagonalization of the Hamiltonian in Eq.~(\ref{eq:ev3}), and is given
by
\be
\tan 2\theta_m=\frac{\frac{\Delta m^2}{2E}\sin 2\theta_0}{\frac{\Delta
m^2}{2E}\cos 2\theta_0-V_{\rm CC}}\,.
\label{eq:mix3}
\ee
The difference of neutrino eigenenergies in matter is
\be
E^m_1-E^m_2=\sqrt{\left(\frac{\Delta m^2}{2E}\cos 2\theta_0-V_{\rm CC}
\right)^2+\left(\frac{\Delta m^2}{2E}\right)^2 \sin^2 2\theta_0}\;. 
\label{eq:E1E2}
\ee

With these redefined ingredients it is easy to understand that the
evolution of neutrinos in a medium of constant density is just as in
vacuum with some effective mixing angle and masses. The oscillation
probability of $\nu_e \leftrightarrow \nu_\mu$ is therefore given,
analogously to Eq.~(\ref{eq:prob3}), by:
\be
P(\nu_e\to\nu_\mu)=
\sin^2 2\theta_m\,\sin^2 \left(\pi \frac{L}{L_{m}}\right)\,,
\label{eq:pr1}
\ee
where the oscillation length in matter is defined as,
\be
L_m=\frac{2\pi}{E^m_1-E^m_2}=\frac{2\pi}{\sqrt{\left(\frac{\Delta m^2}{2E}
\cos 2\theta_0-V_{\rm CC} \right)^2+\left(\frac{\Delta m^2}{2E}
\right)^2 \sin^2 2\theta_0}}\,,
\label{eq:lm}
\ee
and the oscillation amplitude is,
\be
\sin^2 2\theta_m =\frac{\left(\frac{\Delta m^2}{2E}\right)^2\sin^2 2\theta_0}
{\left(\frac{\Delta m^2}{2E}\cos 2\theta_0-V_{\rm CC}\right)^2
+\left(\frac{\Delta m^2}{2E}\right)^2\sin^2 2\theta_0}\,. 
\label{eq:amp}
\ee
The most important point here is that this oscillation amplitude is no
longer limited by the vacuum mixing angle, and even for a very small
$\theta_0$ we can obtain substantial oscillatory
transitions. Furthermore, it presents a typical resonant behavior,
acquiring its maximal value, $\sin^2 2\theta_m=1$, when the so called
MSW (Mikheyev-Smirnov-Wolfenstein)~\cite{Mikheev:1986gs} resonance
condition is satisfied:
\be
V_{\rm CC} =\frac{\Delta m^2}{2E}\cos 2\theta_0\,.
\label{eq:res1}
\ee
This condition, which leads to maximal mixing in matter ($\theta_m =
\pi/4$), will be fulfilled by either neutrinos or antineutrinos,
depending on the sign of $\Delta m^2$, i.e.~the true mass hierarchy
scheme, but never by both of them simultaneously. Assuming the usual
convention for the mixing angle, where $\cos 2\theta_0 > 0$, a
positive $\Delta m^2$ would take the resonance to the neutrino channel
(positive $V_{\rm CC}$), while a negative $\Delta m^2$ would take it
to the antineutrino channel (negative $V_{\rm CC}$). Therefore, the
neutrino-antineutrino symmetry present in vacuum when $\delta = 0$ or
$\pi$, is broken by the matter potentials.

\subsection{Varying density case}

The situation is more complicated when neutrinos propagate through a
non-constant density medium. The matter basis is no longer constant,
and $\nu^m_i$ and therefore the unitary transformation $U(\theta_m)$
depend on time/position. If we derive Eq.~(\ref{eq:mateig}) with
respect to time we find
\begin{equation}
\frac{\partial}{\partial t}\left(\begin{array}{c}{ \nu_e}\\{ \nu_\mu} \end{array}
\right)={\dot{U}(\theta_m)}\left(\begin{array}{c}{\nu^m_1}\\ {\nu_2^m}
\end{array}\right)+{{U}(\theta_m)}\left(\begin{array}{c}
{ \dot{\nu}^m_1}\\ { \dot{\nu}_2^m} \end{array}\right)\;,
\end{equation}
where the dot stands for time derivative. Taking this expression to
the evolution equation in the flavor basis Eq.~(\ref{eq:Schr_fl}), we
obtain
\begin{equation}
i\left(\begin{array}{c}{\dot{\nu}^m_1}\\ { \dot{\nu}_2^m} \end{array}\right)=
{ {U^\dagger(\theta_m)}} { H_W} { {U}(\theta_m)}
\left(\begin{array}{c}{ \nu^m_1}\\ { \nu_2^m} \end{array}\right)
-i\;{ {U^\dagger}} { \dot{U}(\theta_m)}\left(\begin{array}{c}
{ \nu^m_1}\\ { \nu_2^m} \end{array}\right)\; .
\end{equation}
For a constant density medium the effective mixing angle $\theta_m$ is
constant and the second term on the right-hand side of this expression
vanishes. Therefore, we obtain a diagonal relation, where the
evolution of $\nu_1^m$ is only determined by $\nu_1^m$ and the same for
$\nu_2^m$, with no interference between them. But that is not the case
if the density is not constant, where we have
\begin{equation}
i\left(\begin{array}{c}{\dot{\nu}^m_1}\\ { \dot{\nu}_2^m}\end{array}\right)=
\left(\begin{array}{cc} E^m_1(t) & -i{\dot{\theta}_m(t)}
\\ i{\dot{\theta}_m(t)}  & E^m_2(t)\end{array}\right)
\left(\begin{array}{c}{{\nu}^m_1}\\ { {\nu}_2^m} \end{array}\right)~,
\label{eq:evoleq2}
\end{equation}
with $\dot{\theta}_m \equiv {\rm d}\theta_m/{\rm dt}$. The effective
Hamiltonian in the matter eigenstate basis is not diagonal in general,
meaning that $\nu^m_i$ mix in the evolution and are not energy
eigenstates. The importance of this effect will depend on the size of
the off-diagonal terms with respect to the diagonal ones, determining
two types of evolution. \vspace{0.5cm}\\
\textbf{A) Adiabatic case}\vspace{0.2cm}\\
The adiabatic approximation correspond to the case where the
off-diagonal terms are small, in the sense explained before,
i.e.~$|\dot{\theta}_m| \ll |E^m_1-E^m_2|$. In this case the
transitions between matter eigenstates are suppressed. This
suppression can be quantified with the adiabaticity condition,
\be
\gamma^{-1}(r) \ll 1\,, 
\label{eq:adiab_cond}
\ee
being $\gamma$ the adiabaticity parameter, defined as the relation
between the off-diagonal terms in Eq.~(\ref{eq:evoleq2}) and the
diagonal ones,
\be
\gamma^{-1}(r)\equiv \frac{2|\dot{\theta}_m|}{|E^m_1-E^m_2|}=\frac{\sin 2\theta_0
\frac{\Delta m^2}{2E}}
{|E^m_1-E^m_2|^{3}}\,
|\dot{V}_{\rm CC}|\,. 
\label{eq:adiab_param}
\ee
In this expression $E^m_1-E^m_2$ and $V_{\rm CC}$ are given by the
Eqs.~(\ref{eq:E1E2}) and (\ref{eq:VCC}) respectively. The parameter
$\gamma$ can also be expressed in terms of the elements of the
Hamiltonian matrix as
\begin{equation}
  \label{eq:gamma2_text}\gamma^{-1}(r) =2
  \frac{\dot{H}_{12}(H_{22}-H_{11}) - (\dot{H}_{22} -
  \dot{H}_{11})H_{12}}{ \left[(H_{22}-H_{11})^2+4
  H_{12}^2\right]^{3/2}}\,.
\end{equation}
When the condition in Eq.~(\ref{eq:adiab_cond}) applies, the
Hamiltonian in Eq.~(\ref{eq:evoleq2}) is basically diagonal, leading
to a very simple time evolution of the matter eigenstates, they just
acquire phase factors.

The adiabaticity condition has a simple physical meaning.  Let us
define the resonance width at half height $\delta r$ as the spatial
width of the region where the amplitude of neutrino oscillations in
matter is $\sin^2 2\theta_m\ge 1/2$. According to Eq.~(\ref{eq:amp}),
the limiting condition $\sin^2 2\theta_m(r+\delta r/2) = 1/2$ will be
satisfied at the point where $(\Delta m^2/2E~\cos 2\theta_0-V_{\rm
  CC}-\delta V_{\rm CC}/2)^2 =(\Delta m^2/2E~\sin^2
2\theta_0)^2$. Making use of Eq.~(\ref{eq:res1}), we find:
\be
\delta V_{\rm CC} = 2 \frac{\Delta m^2}{2E}\sin^2 2\theta_0\,,
\ee
which can be converted into a distance using
\be
\delta V_{\rm CC} = \left|\frac{{\rm d}V_{\rm CC}}{{\rm d}r}\right|_{\rm res}{\rm d}r~,
\ee
or,
\be
\delta r = \frac{\delta V_{\rm CC}}{\left|\frac{{\rm d}V_{\rm CC}}{{\rm
        d}r}\right|_{\rm res}}~.
\ee
Here, ``res'' denotes here the point where the resonance condition is
satisfied. We can define the density scale height at the resonance as
\be
\label{eq:denshs}L_{\rho} \equiv \left|\frac{1}{V_{\rm CC}} \frac{{\rm
      d}V_{\rm CC}}{{\rm d}r}\right|_{\rm res} = \left|\frac{1}{N_e}
  \frac{{\rm d}N_e} {{\rm d}r}\right|_{\rm res}\,,
\ee
and reexpress the resonance width as
\be
\delta r = \frac{2\tan 2\theta_0}{L_{\rho}}\,.
\ee

Using Eq.~(\ref{eq:lm}) we find that the oscillation length at the
resonance is given by $(L_m)_{\rm res}=2\pi/|E^m_1-E^m_2|_{\rm
  res}=(4\pi E/\Delta m^2)/\sin 2\theta_0$. Therefore the adiabaticity
parameter at the resonance can be rewritten as
\be
\gamma_{\rm res} =\pi\frac{\delta r}{(L_m)_{\rm res}}\,,
\label{eq:adiab3}
\ee
i.e.~the adiabaticity condition $\gamma_{\rm res} > \pi$ is just the
condition that at least one oscillation length fits into the resonance
region.

Let us explicitly discuss under these adiabatic circumstances the
evolution of an electron neutrino created at time $t = t_i$, as a
superposition of matter eigenstates:
\be
\nu(t_i)=\nu_e=\cos\theta^i_m \,\nu^m_1+\sin\theta^i_m\, \nu^m_2\,.
\label{eq:init}
\ee
At a later time $t_f$, the evolution in the adiabatic approximation,
where no transition $\nu_1^m \to \nu_2^m$ can occur, takes the
neutrino state to
\be
\nu(t_f)=
\cos\theta^i_m\, e^{-i\int_{t_i}^{t_f} E^m_1(t')dt'}\,\nu^m_1 +\sin\theta^i_m\,
e^{-i\int_{t_i}^{t_f} E^m_2(t')dt'}\,\nu^m_2\,.
\label{eq:fin}
\ee
Taking into account that the mixing angle $\theta(t_f) \equiv
\theta^f_m$ changes with time and therefore is different from
$\theta^i_m$, we find the transition probability to be
\be
P(\nu_e\to\nu_\mu)=\frac{1}{2}-\frac{1}{2}\cos 2\theta^i_m\cos 2\theta^f_m-
\frac{1}{2}\sin 2\theta^i_m\sin 2\theta^f_m \cos \Phi\,,
\label{eq:pr3}
\ee
where
\be
\Phi=\int_{t_i}^{t_f}(E^m_1-E^m_2) dt'\,.
\label{eq:phi}
\ee
The second term in Eq.~(\ref{eq:pr3}) is a smooth function of $t_f$,
while the third term oscillates with time. If the matter density at
the neutrino production point is far above the MSW resonance one,
$N^i_e \gg N^{res}_e$, the initial mixing angle is $\theta^i_m \approx
\pi/2$ and the third term is strongly suppressed due to the $\sin 2
\theta^i_m$ factor. As the neutrinos travel toward lower density
regions, the effective mixing angle decreases as well down to
$\theta^f_m=\theta_0$ for vanishing matter. On the way it passes
through maximal mixing, $\theta_m = \pi/4$, at the resonance. In this
case, the neutrino probability is
$P(\nu_e\to\nu_\mu)=\cos^2\theta^f_m$, which for a low final density
becomes $P(\nu_e\to\nu_\mu)=\cos^2\theta_0$. In particular, if
$\theta_0$ is small the conversion between $\nu_e$ and $\nu_\mu$ is
almost complete, contrary to the vacuum case. This amplification of
the conversion probability in matter is known as the MSW
effect~\cite{Wolfenstein:1977ue,Mikheev:1986gs}.

\begin{figure}
  \begin{center}
    \includegraphics[width=0.6\textwidth]{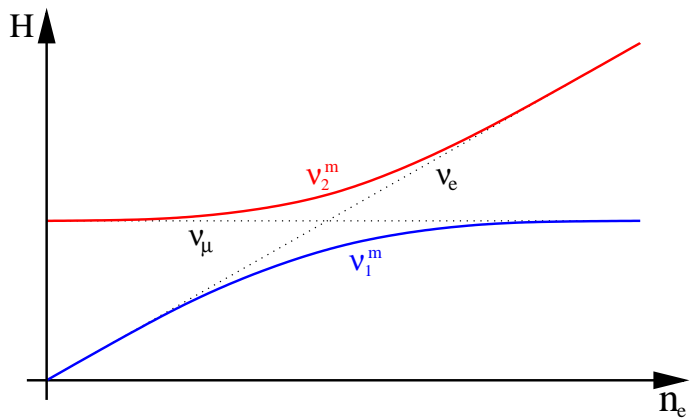}
  \end{center}
  \caption{\small Level crossing scheme in two flavors. Shown are the
    neutrino energy levels in matter as a function of the electron
    number density $N_e$. Dashed line in absence of mixing, solid line
    with mixing.}
  \label{fig:levcross}
\end{figure}

In Fig.~\ref{fig:levcross} we represent the diagram known as level
crossing scheme. It shows the energy levels of $\nu^m_1$ and $\nu^m_2$
along with those in absence of mixing (i.e.~$\nu_e$ and $\nu_\mu$) as
the function of the electron number density, $N_e$. In absence of
mixing the energy levels cross at the resonance point, but with
non-vanishing mixing the levels repel each other and the avoided
level crossing results. \vspace{0.5cm}\\
\textbf{B) Non-adiabatic case}\vspace{0.2cm}\\
The situation is different when the off-diagonal terms of
Eq.~(\ref{eq:evoleq2}) are comparable or larger than the diagonal
ones, $|\dot{\theta}_m| \gtrsim |E^m_1-E^m_2|$. The adiabatic
approximation does not apply anymore, and one has to take into account
possible transition between matter eigenstates driven by the violation
of the adiabaticity. We can generalize Eq.~(\ref{eq:pr3}) to include
this effect in the following way (omitting the oscillatory terms which
average to zero),
\be
\overline{P(\nu_e\to\nu_\mu)}\simeq \frac{1}{2}-\frac{1}{2}\cos
2\theta^i_m\cos 2\theta^f_m
(1-2P')\,,
\label{eq:pr4}
\ee
where $P'$ stands for the hopping probability and for small values of
the mixing angle is given by the Landau-Zener formula
\be
P'\simeq e^{-\frac{\pi}{2}\gamma_{\rm res}}\,,
\label{eq:hop}
\ee
where $\gamma_{\rm res}$ is the adiabaticity parameter computed at the
resonance point. As discussed in Ref.~\cite{Kuo:1988pn}, this
expression is valid as long as the density profile can be approximated
as linear around the resonance point and the mixing angle is
small. For an arbitrary density distribution and mixing angle the
general expression is
\be 
P_f = \frac{ {\rm exp}(-\frac{\pi}{2} \gamma F) - {\rm
    exp}(-\frac{\pi}{2} \gamma F/\sin^2 \theta)} {1 - {\rm
    exp}(-\frac{\pi}{2} \gamma F/\sin^2 \theta)}~~,
\label{exact-pf}
\ee
where $F$ depends on the density profile and the mixing angle. This
expression has to be computed at the resonance point. There are other
more general formalisms where the analysis is independent of the
resonance and the adiabaticity parameter can be calculated at an
arbitrary point~\cite{Kachelriess:2001bs}. Another useful expression
for the jumping probability, which applies for an arbitrary mixing
angle, is
\be
P'\simeq \frac{{\rm exp}(-2\pi L^{-1}_\rho \frac{\Delta m^2}{2E}\cos^2\theta)-1}{{\rm exp}(-2\pi L^{-1}_\rho \frac{\Delta m^2}{2E})-1}\,,
\label{eq:hop2}
\ee
where $L_\rho$ is defined in Eq.~(\ref{eq:denshs}). The whole
expression can be easily reinterpreted in terms of $\gamma_{\rm res}$.

In the adiabatic limit $\gamma_{\rm res} \gg 1$ and $P' \simeq 0$,
reproducing Eq.~(\ref{eq:pr3}). In the non-adiabatic limit
$\gamma_{\rm res} \ll 1$, and as discussed in
Ref.~\cite{Friedland:2000rn} the jumping probability is given by $P'
\simeq \sin^2(\theta_m - \theta_m^i)$, where $\theta_m^i$ is the
effective mixing angle at the neutrino production point. If we assume
the matter density at that point to be far above the resonance
density, we obtain in the limit of $\theta_m \to \theta_0$ a crossing
probability for neutrinos of $P' \simeq \cos^2\theta_0$, which in the
case of a small mixing angle reduces to $P' \simeq 1$. This situation
leads to an interchange between the survival and transition
probabilities. If we consider then the case of a very small vacuum
mixing angle and $N_e^f\ll N_e^{\rm res} \ll N_e^i$ (or $N_e^i\ll
N_e^{\rm res}\ll N_e^f$), Eq.~(\ref{eq:pr4}) becomes the simple
expression
\be
\overline{P(\nu_e\to\nu_\mu)}\simeq 1-P'\,,
\label{pr5}
\ee
with $P'$ given by Eq.~(\ref{eq:hop2}).

\section{Neutrino oscillations in the SN envelope}

After having discussed the generalities of neutrino oscillation
physics, we are going to study the actual three flavor situation in
the particular case of SN neutrinos.  The evolution of neutrinos
inside the SN can be divided in two zones: inside the core, dominated
by neutrino collisions with matter, and through the envelope, where
neutrinos participate only in elastic forward scattering, the border
being defined as the neutrino sphere. In this thesis we do not
consider what happens inside the core, and parameterize its effect as
distribution functions for the neutrinos. We take this distribution
functions as input and study their evolution outside the neutrino
sphere, as discussed in Chapter~\ref{chapter:supernova}.

\subsection{Supernova matter profiles}\label{sec:SN_profiles}

\begin{figure}
  \begin{center}
    \includegraphics[width=0.5\textwidth]{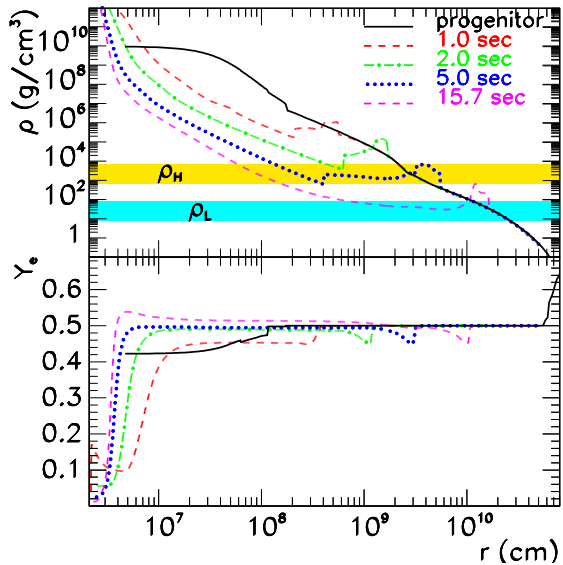}
  \end{center}
  \caption{\small Density (upper panel) and electron fraction (bottom panel)
    profiles for the SN progenitor and at different instants after the
    core bounce, from Ref.~\cite{Tomas:2004gr}. The regions where the
    $H$- (yellow) and the $L$- (cyan) resonance take place are also
    indicated.}
  \label{fig:snprofiles}
\end{figure}

The main conclusion of the previous sections is that in order to
determine the evolution of neutrinos in a medium with varying density
we need to know both the neutrino parameters and the properties of the
medium, especially the matter and chemical profiles. In a SN scenario,
these exhibit an important time dependence during the explosion.
Figure~\ref{fig:snprofiles} shows the density $\rho(t,r)$ and the
electron fraction $Y_e(t,r)$ profiles for a typical SN progenitor as
well as at different times post-bounce.

\begin{figure}[t]
  \begin{center}
    \includegraphics[width=0.62\textwidth]{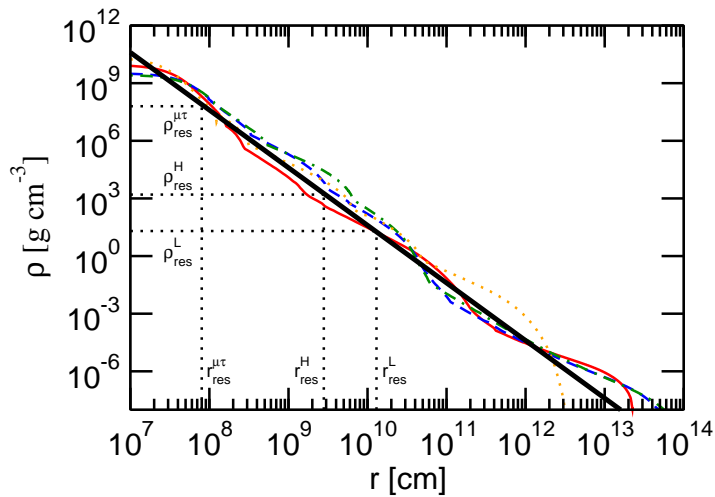}
  \end{center}
  \caption{\small Matter density profiles for three different SN
    progenitor masses: 11 (red solid), 20 (blue dashed) and 30 (green
    dot-dashed) $M_{\odot}$ from Woosley~\cite{Wo}, one from
    Nomoto~\cite{No} for the progenitor of SN1987A (orange dotted) and
    the expression $\rho = 4\times10^4{\rm gcm^{-3}}(10^9{\rm
      cm}/r)^3$ (black thick solid line). The regions around the
    points where the $\mu\tau$, $H$ and $L$-resonances take place are
    also shown with dotted lines, $r_{\rm res}^{\mu\tau}$, $r_{\rm
      res}^H$ and $r_{\rm res}^L$.\label{fig:SN_profiles}}
\end{figure}

The density in the envelope of a SN depends on the characteristics of
the progenitor star (mass, metallicity, $\dots$). However, it can be
reasonably well approximated by a power law of the type:
\begin{equation}\label{eq:rho_prof}
\rho(r) = \rho_0 \left(\frac{r_0}{r}\right)^n~,
\end{equation}
where $\rho_0 \sim 10^4$~g/cm$^3$, $r_0\sim 10^9$~cm, and $n\sim
3$. In Fig.~\ref{fig:SN_profiles} we can see how well several profiles
adjust to this expression of the density. The electron fraction
profile varies depending on the matter composition of the different
layers.  For instance, typical values of $Y_e$ between 0.42 and 0.45
in the inner regions are found in stellar evolution
simulations~\cite{Woosley:2002zz}. In the intermediate regions,
where the MSW $H$- and $L$-resonances take place $Y_e\approx 0.5$.
This value can further increase in the most outer layers of the SN
envelope due to the presence of hydrogen.

After the SN core bounce the matter profile is affected in several ways.
First note that a front shock wave starts to propagate outwards and
eventually ejects the SN envelope. The evolution of the shock wave
will strongly modify the density profile and therefore the neutrino
propagation~\cite{Schirato:2002tg,Fogli:2003dw}. Following
Ref.~\cite{Tomas:2004gr} we have represented in
Fig.~\ref{fig:snprofiles} a more complicated structure of the shock
wave, where an additional ``reverse wave'' appears due to the
collision of the neutrino-driven wind and the slowly moving material
behind the forward shock.

On the other hand, the electron fraction is also affected by the time
evolution as the SN explosion proceeds. Let us discuss how it changes
during the different stages of the explosion.
Once the collapse starts the core density grows so that the neutrinos
become eventually effectively trapped within the neutrino sphere. At
this point the trapped electron fraction has decreased until values of
the order of 0.33~\cite{Cardall:2007dy}. When the inner core reaches
nuclear density it can not contract any further and bounces. As a
consequence a shock wave forms in the inner core and starts
propagating outwards.  When the newly formed SN shock reaches
densities low enough for the initially trapped neutrinos to begin
streaming faster than the shock propagates~\cite{Bethe:1980gq}, a
breakout pulse of $\nu_e$ is launched.
In the shock-heated matter, which is still rich of electrons and
completely disintegrated into free neutrons and protons, a large
number of $\nu_e$ are rapidly produced by electron captures on
protons. They follow the shock on its way out until they are released
in a very luminous flash, the breakout burst, at about the moment when
the shock penetrates the neutrino sphere and the neutrinos can escape
essentially unhindered.  As a consequence, the lepton number in the
layer around the neutrino sphere decreases strongly and the matter
neutronizes. The value of $Y_e$ steadily decreases in these layers
until values of the order of $\mathcal{O}(10^{-2})$.  Outside the
neutrino sphere there is a steep rise until $Y_e\approx 0.5$. This is
a robust feature of the neutrino-driven baryonic wind. Neutrino
heating drives the wind mass loss and causes $Y_e$ to rise within a
few $10$~km from low to high values, between 0.45 and
0.55~\cite{private}, see bottom panel of Fig.~\ref{fig:snprofiles}.
Inspired in the numerical results of Ref.~\cite{Tomas:2004gr} we can
parameterized the behavior of the electron fraction near the neutrino
sphere phenomenologically as,
\begin{equation}
Y_e = a + b\arctan[(r-r_0)/r_s]~,
\label{eq:Ye}
\end{equation}
where $a\approx 0.23$--$0.26$ and $b\approx 0.16$--$0.20$. The
parameters $r_0$ and $r_s$ describe where the rise takes place and how
steep it is, respectively. As can be seen in Fig.~\ref{fig:snprofiles}
both decrease with time.

\subsection{Factorization of the evolution}

In the SN there exists such extreme conditions that neutrinos travel
from regions with densities of the order $10^{12}$ g/cm$^3$, right
behind the neutrino sphere, to basically zero density at the
atmosphere of the star. As a consequence, contrary to the case of the
Sun, neutrinos will undergo three kinematic resonant conversions. Two
of them come from $V_{\rm CC}$, the $H$-resonance taking place at high
densities and defined by $V_{\rm CC}(r^H_{\rm res}) = \Delta m^2_{\rm
  31}/(2E)\cos 2 \theta_{13}$, and the $L$-resonance occurring at low
densities where $V_{\rm CC}(r^L_{\rm res}) = \Delta m^2_{21}/(2E)\cos
2 \theta_{12}$. The third one is due to $V_{\mu\tau}$, and is
determined by $V_{\mu\tau}(r^{\mu\tau}_{\rm res}) = \Delta
m^2_{31}/(2E)\cos 2 \theta_{23}$. These three conditions are
represented in the density profiles of Fig.~\ref{fig:SN_profiles} and
the level crossing diagrams for the three flavor scenario shown in
Fig.~\ref{fig:generic}. Since the sign of $\Delta m^2_{31}$ is
undetermined both hierarchies are still possible. The left panel of
this figure corresponds to normal hierarchy ($\Delta m^2_{31} > 0$),
which after Eq.~(\ref{eq:res1}) translates in the $H$-resonance
occurring for neutrinos, while the right panel is done for the
inverted hierarchy case ($\Delta m^2_{31} < 0$) and the resonance
takes place in the antineutrino channel. The $\mu\tau$-resonance
depends not only on the hierarchy, but also on the octant of
$\theta_{23}$ as shown in Fig.~\ref{fig:generic}.

The interesting point here is that the large difference between the
two mass splittings ($\Delta m_{21}^2/\Delta m_{31}^2\sim 10^{-2}$),
and therefore the densities at the resonance layers, allows us to
factorize the evolution of neutrinos. We can then treat the three
possible transitions as an approximate two neutrino problem. In the
$H$-resonance region the mixing $U^m_{e2}$ associated to $\Delta
m^2_{21}$ is suppressed by matter by more than two orders of magnitude,
\be
\frac{U_{e2}^m}{U_{e2}} \sim \frac{\rho_L}{\rho_H} 
\lesssim 10^{-2}~.
\ee

The situation is similar at the $L$-resonance region, where the
effects of $\Delta m^2_{31}$ basically go as the mixing at vacuum,
\be
U_{e3}^m = U_{e3} [1 + {\cal O}(\xi)]~,~~
~~~~~ \xi \approx \frac{\rho_L}{\rho_H} 
\lesssim 10^{-2}~.
\ee

Finally, the same argument stands for the $\mu\tau$-resonance, where
the mixing between $\nu_\mu$ or $\nu_\tau$ with the third state is
suppressed at the resonance by the huge hierarchy between
$V_{\mu\tau}$ and $V_{\rm CC}$, ${\cal O}(10^{-4})$.

This means that by an appropriate redefinition of the states we can
always disentangle one of the neutrinos, reducing the three neutrino
problem to an effective two neutrino one.

\subsection{Level crossing schemes}

\begin{figure}[t]
  \begin{center}
    \includegraphics[width=0.45\textwidth]{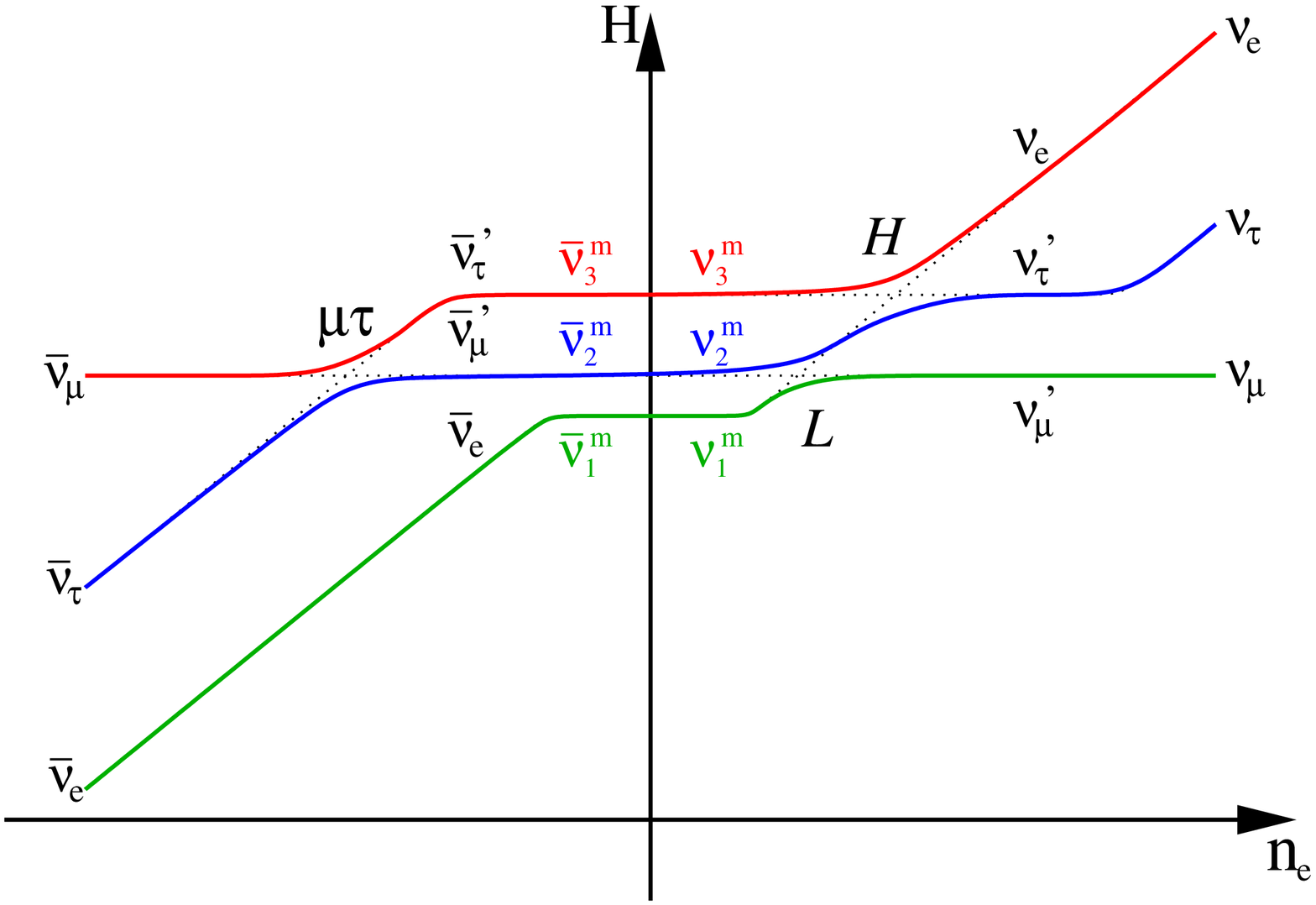}
    \includegraphics[width=0.45\textwidth]{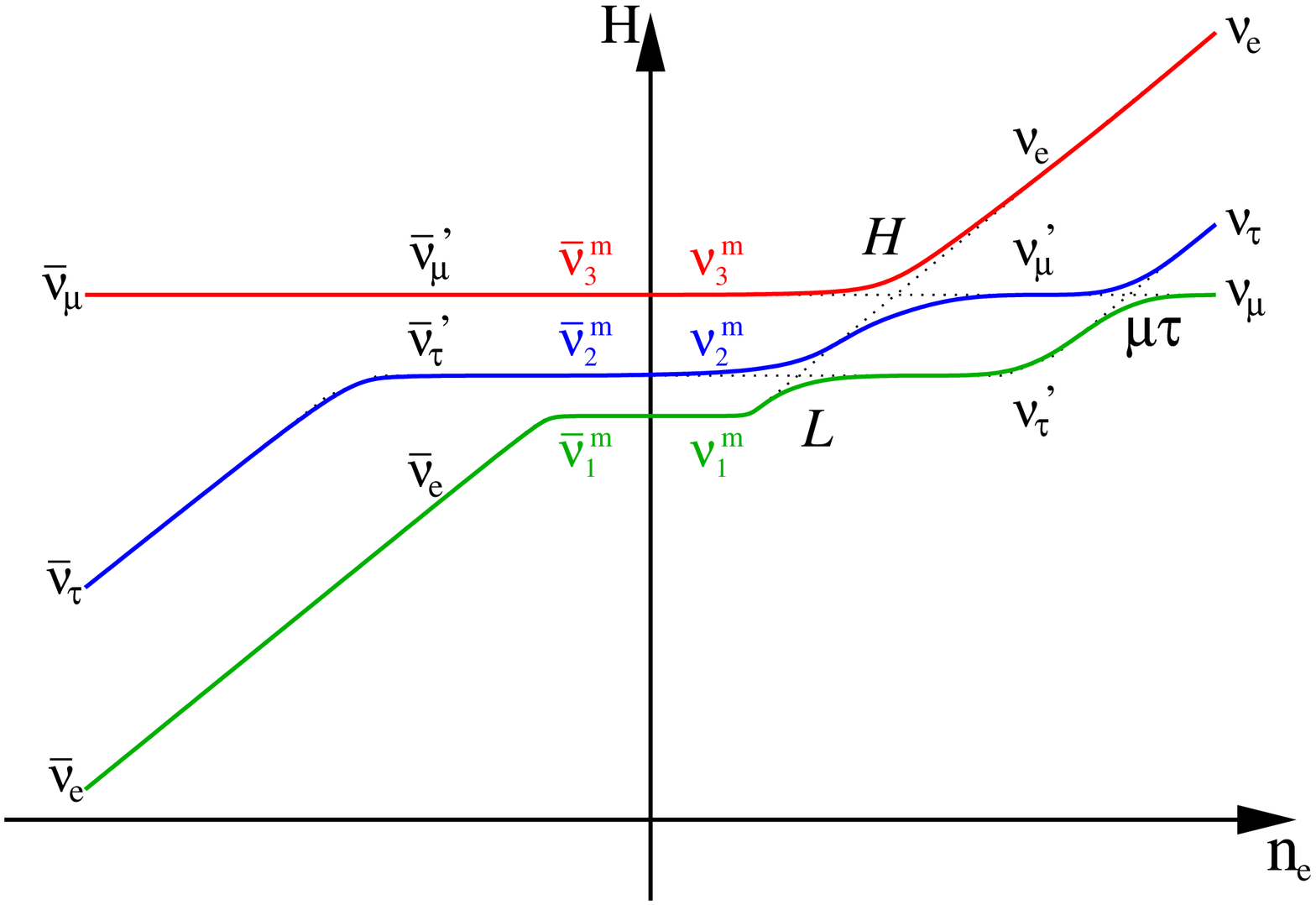}
    \includegraphics[width=0.45\textwidth]{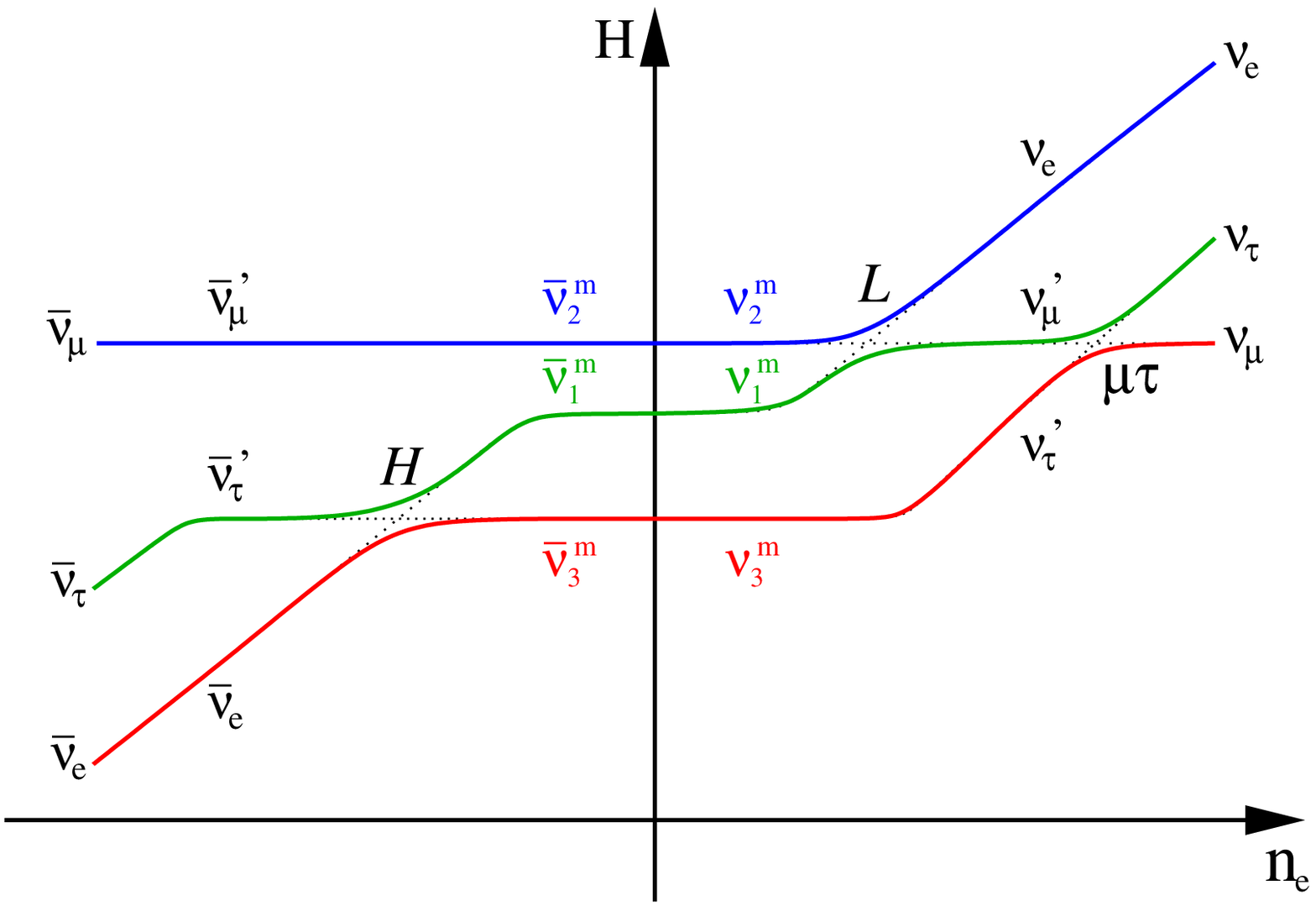}
    \includegraphics[width=0.45\textwidth]{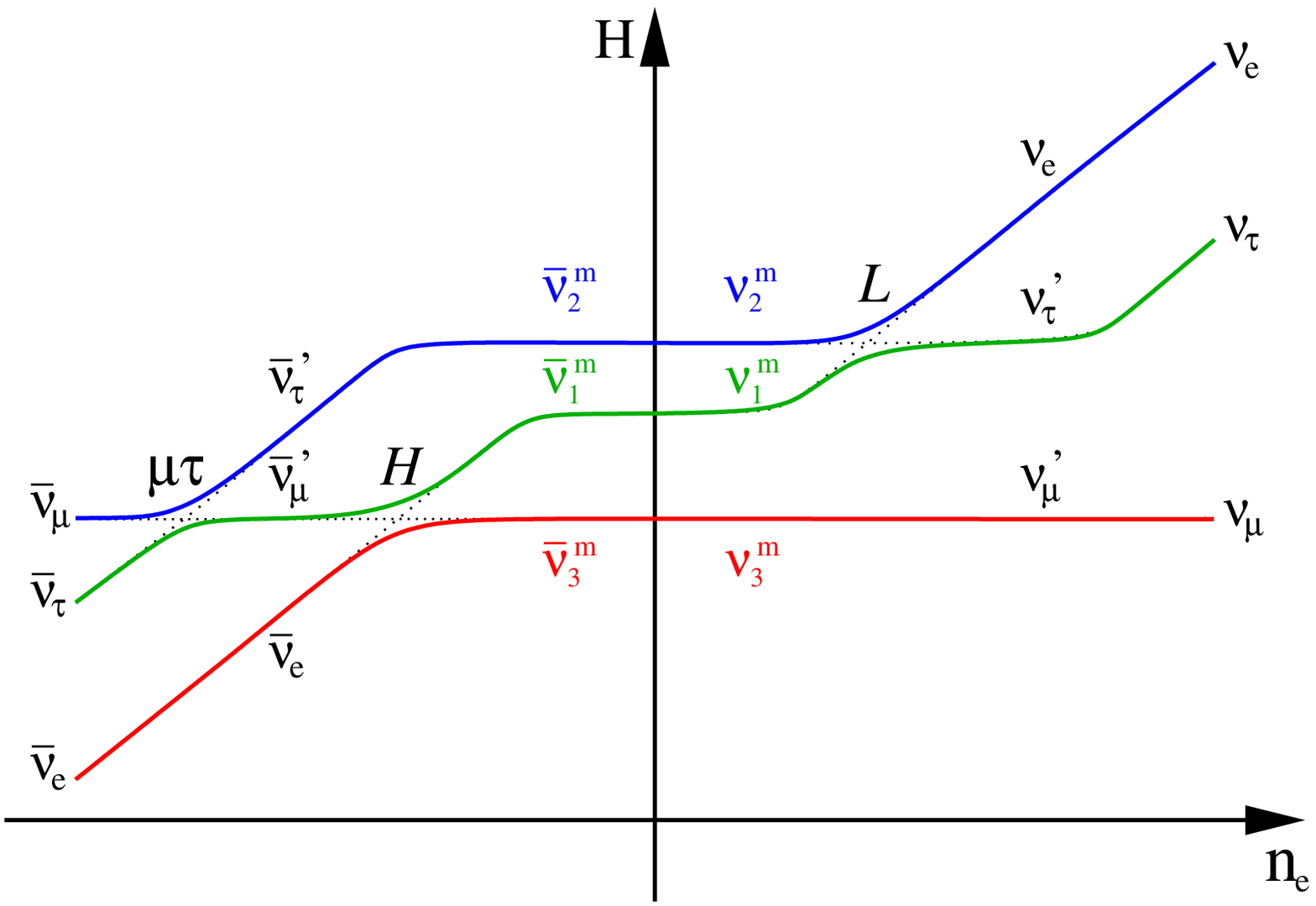}
  \end{center}
  \caption{\small Level crossing diagrams for normal (top panels) and
    inverted mass hierarchy (bottom panels). The left panels
    correspond to $\theta_{23}<\pi/4$ (first octant) and the right
    ones to $\theta_{23}>\pi/4$ (second octant). Solid lines show the
    eigenvalues of the effective Hamiltonian as a function of the
    electron number density. Dashed lines correspond to energies of
    the flavor levels $\nu_e$, $\nu'_{\mu}$ and $\nu'_{\tau}$. The
    part of the plot with $n_e < 0$ corresponds to the antineutrino
    channel.
  \label{fig:generic}}
\end{figure}

As discussed in Sec.~\ref{sec:evoleq}, the evolution of neutrinos can
be described using the equation of Schr\"odinger,
Eq.~(\ref{eq:Schr_fl}). Far enough from the neutrino sphere, so that
no effect is to be expected from neutrino-neutrino interactions, the
Hamiltonian involved in the flavor basis is given by
\be
H =  H_{\rm kin} + H_{\rm int} = \frac{1}{2 E} \left( 
\begin{array}{lcc} 
m_{ee}^2 + 2 E V_{\rm CC} & m_{e\mu}^2 & m_{e\tau}^2  \\
m_{e\mu}^2 & m_{\mu\mu}^2 & m_{\mu\tau}^2  \\
m_{e\tau}^2 & m_{\mu\tau}^2 & m_{\tau\tau}^2  + 2 E V_{\mu\tau}
\end{array}
\right)~,
\label{eq:hamilt}
\ee
where the $m^2_{\alpha\beta}$ come from the kinetic term,
\begin{equation}
\label{eq:Hkin}H_{\rm kin} = 
U\frac{1}{2E} \left( \begin{array}{ccc} m_1^2 & 0 & 0 \\ 0
  & m_2^2 & 0 \\ 0 & 0 & m_3^2 \end{array} \right)
U^\dagger~,
\end{equation}
with $U$ given in Eq.~(\ref{eq:U3}), and $H_{\rm int} = {\rm
  diag}(V_{\rm CC},0,V_{\mu\tau})$ is the resulting interaction
Hamiltonian after removing the NC common terms. By using
Eq.~(\ref{eq:hamilt}) it is easy to reproduce the level crossing
schemes of Fig.~\ref{fig:generic}. Its diagonal terms basically
determine the energies of the flavor states, represented by dotted
lines, while its eigenvalues, represented by the colored solid lines,
give the actual neutrino path.

For very high densities, where $V_{\mu\tau}$ and obviously $V_{\rm CC}$
are larger than the kinetic terms, the whole Hamiltonian leading the
evolution of neutrinos can be effectively reduced to the interaction
one,
\be
H \approx \left( 
\begin{array}{lcc} 
V_{\rm CC} & 0 & 0  \\
0 & m_{\mu\mu}^2  & 0  \\
0 & 0 & V_{\mu\tau}
\end{array}
\right)~.
\label{eq:hamilt-int}
\ee
As a consequence, the matter and flavor eigenstate coincide, and the
eigenvalues will go as shown in the high density region of
Fig.~\ref{fig:generic}: $\nu_e \sim V_{\rm CC}$, $\nu_\tau \sim
V_{\mu\tau}$ and $\nu_\mu \sim m_{\mu\mu}^2$, and the same with
opposite sign in the potentials for antineutrinos. The $V_{\mu\tau}$
potential will be important if
\be
V_{\mu \tau} \gtrsim \Delta m^2_{31}/2E \approx  2 m_{\mu \tau}^2/2E \rightarrow \rho_{\mu \tau} \gtrsim 10^{7} - 10^{8}~ \mbox{g/cm}^3.
\ee

In the interval of densities where $V_{\mu \tau} \ll \Delta
m^2_{31}/2E \ll V_{\rm CC}$, the $V_{\mu \tau}$ potential can be
neglected so that we obtain the Hamiltonian
\be
H =  \frac{1}{2 E} \left( 
\begin{array}{lcc} 
m_{ee}^2 + 2 E V_{\rm CC} & m_{e\mu}^2 & m_{e\tau}^2  \\
m_{e\mu}^2 & m_{\mu\mu}^2 & m_{\mu\tau}^2  \\
m_{e\tau}^2 & m_{\mu\tau}^2 & m_{\tau\tau}^2
\end{array}
\right)~.
\label{eq:hamilt-noVmt}
\ee
Now, by diagonalizing the 23-sector,
\be
H = \frac{1}{2E}\left( 
\begin{array}{lcc} 
m_{ee}^2 + 2 EV_{\rm CC} & m_{e \mu'}^2 & m_{e\tau'}^2  \\
m_{e\mu'}^2 & m_{\mu' \mu'}^2 & 0  \\
m_{e\tau'}^2 & 0 & m_{\tau' \tau'}^2 
\end{array}
\right)~,
\label{eq:hamilt-rot}
\ee
we will be basically diagonalizing the whole Hamiltonian, since
$V_{\rm CC} \gg \Delta m^2_{31}/2E$. For this density range $H \approx
{\rm diag}(V_{\rm CC},m_{\mu' \mu'}^2,m_{\tau' \tau'}^2)$, and the
resulting matter basis is ($\nu_e,\nu_{\mu'},\nu_{\tau'}$).
Therefore, in the region $V_{\mu\tau} \sim \Delta m^2_{31}/2E$ the
level crossing occurs, taking place for neutrinos or antineutrinos
depending on the hierarchy and the octant of $\theta_{23}$, as shown
in Fig.~\ref{fig:generic}. For instance, for normal hierarchy and
first octant (top left panel of Fig.~\ref{fig:generic}) neutrinos have
undergone the transitions $\bar\nu_{\mu} \rightarrow \bar\nu_{\tau}'$,
$\bar\nu_{\tau} \rightarrow \bar\nu_{\mu}'$, $\nu_{\mu} \rightarrow
\nu_{\mu}'$ and $\nu_{\tau} \rightarrow \nu_{\tau}'$.

As the density decreases, the potential reaches the different mass
scales, producing the level crossings at the resonances. The sign of
$\Delta m^2_{21}$ is known to be positive, and therefore the
$L$-resonance will always take place for neutrinos, but depending on
the hierarchy in the neutrino mass scheme, the $H$-resonance will
occur for neutrinos (normal hierarchy, $\Delta m^2_{31} > 0$) or
antineutrinos (inverted hierarchy, $\Delta m^2_{31} < 0$), as shown in
Fig.~\ref {fig:generic}. The positive values of the density (x-axis)
in the level crossings represent the evolution of neutrinos, while
that of antineutrinos is shown in the negative $N_e$ values. This is a
simple trick, that allows us to represent the level crossing schemes
in a more compact form, and is motivated by the fact that the
effective potential for antineutrinos has the opposite sign than the
one for neutrinos. Thus, antineutrinos can be visualized as neutrinos
travelling through matter with effectively negative $N_e$. Neutrinos
will start their evolution from the far right side of the figure,
while antineutrinos will be born at the far left side. Both will
thereafter evolve towards vacuum, represented as $N_e = 0$ at the
center.

The evolution of neutrinos in the SN envelope will be determined at
the resonances, the points where the flavor transitions can occur. Let
us then analyze what happens at the different resonances. 

\subsection{Mu-tau-resonance}

The crossing of $\nu_\mu$ and $\nu_\tau$ levels, the
$\mu\tau$-resonance, is a consequence of the potential that arises at
one loop level due to the difference in the masses of the $\mu$ and
$\tau$ leptons~\cite{Botella:1986wy}. Although $V_{\mu\tau}$ is almost
five orders of magnitude smaller than $V_{\rm CC}$, the huge densities in
the inner layers of the SN may make it important.

In principle, one could think that this possible crossing does not
really matter, since the initial $\nu_\mu$ and $\nu_\tau$ fluxes are
expected to be identical, and therefore no potential effect between
them would lead to any observable features. However, the situation
changes when we include neutrino-neutrino interactions into the
picture. As we will discuss in Chapter~\ref{chapter:coll3flavors},
these could lead to a transition right before the $\mu\tau$-resonance
moving the flux difference to the $\nu_\mu$ and $\nu_\tau$
branches. It will be then important to understand what happens at the
resonance.

The conversion due to the $V_{\mu\tau}$ potential will take place in
the dense inner regions, where $\nu_e$ is completely decoupled, since
$V_{\rm CC}$ is much larger than the elements $H_{1i}$ (i=2,3) in the
Hamiltonian of Eq.~(\ref{eq:hamilt}). Consequently, we obtain an
effective two neutrino problem. The $\mu\tau$-resonance is governed by
the atmospheric mixing angle and mass splitting, and the resonance
condition is given by
\be
V_{\mu\tau} \simeq \frac{\Delta m^2_{31}}{2E} \cos 2\theta_{23}\,.
\label{eq:mutau_res}
\ee
Thus, it will take place in the neutrino or antineutrino channel
depending on both the hierarchy scheme and the octant of
$\theta_{23}$, due to the combination of signs of $\Delta m^2_{31}$,
$\cos 2\theta_{23}$ and $V_{\mu\tau}$.
%
%
Numerically, one gets $\rho_{\mu \tau} \gtrsim 10^{7} - 10^{8}
~\mbox{g/cm}^3$. Applying Eq.~(\ref{eq:hop2}) at the resonance point,
it can be obtained~\cite{Akhmedov:2002zj}
\bea
&&P'_{\mu\tau} = \frac{e^{\chi\cos^2\theta_{23}}-1}{e^{\chi}-1}\,,\\
&&\chi \equiv -2\pi \frac{\Delta m^2_{31}}{2E} \left|\frac{1}{V_{\mu\tau}} \frac{{\rm d}V_{\mu\tau}}{{\rm d}r}\right|^{-1}_{res}\,.
\eea
Substituting into this expression the potential profile and
parameters, we find $\chi \approx 500$--$900$, which corresponds to a
very good adiabaticity: $P'_{\mu\tau} = 0$.

\subsection{{\em H}-resonance}

Let us move now to the resonances due to the $V_{\rm CC}$, and let us
start with the high density one, $H$-resonance. In order to reduce the
three-neutrino problem to a two-neutrino one we should diagonalize the
submatrix ($\nu_\mu,\nu_\tau$). However, this transformation leads to
very complicated expressions for the Hamiltonian and no easy
analytical formulae can be derived. As an approach to the problem,
though, we can undo the rotation $V_{23}$ of Eq.~(\ref{eq:U3}) which
to a good approximation diagonalizes the desired
submatrix~\cite{Kuo:1986sk}. After performing such rotation the
Hamiltonian takes the following form:
\bea
H^H & = & V^{\dagger}_{23} (U H_{\rm kin} U^{\dagger} + H_{\rm int}) V_{23} = \nonumber \\
& = & \left(\setlength\arraycolsep{2pt}
\begin{array}{ccc}
  \Delta_{\rm atm}s^2_{13} - \Delta_{\odot}c^2_{13}c^2_{12} + V_{\rm CC} & \frac{1}{2}\Delta_{\odot}c_{13}s2_{12} & \frac{1}{2} \left(\Delta_{\rm atm}s2_{13} + \Delta_{\odot}c^2_{12}s2_{13} \right) \\
  \frac{1}{2}\Delta_{\odot}c_{13}s2_{12} & - \Delta_{\odot}s^2_{12} & - \frac{1}{2}\Delta_{\odot}s2_{12}s_{13} \\
  \frac{1}{2} \left( \Delta_{\rm atm}s2_{13} + \Delta_{\odot}c^2_{12}s2_{13} \right) & - \frac{1}{2}\Delta_{\odot}s2_{12}s_{13} & \Delta_{\rm atm}c^2_{13} - \Delta_{\odot}c^2_{12}s^2_{13} \end{array} \right)~, \nonumber \\
\eea
where $\Delta_{\rm atm} \equiv (m^2_3-m^2_1)/2E$, $\Delta_{\odot}
\equiv (m^2_2-m^2_1)/2E$, $s2_{ij} \equiv \sin{2\theta_{ij}}$,
$c2_{ij} \equiv \cos{2\theta_{ij}}$, $s^2_{ij} \equiv
\sin^2{\theta_{ij}}$ and $c^2_{ij} \equiv \cos^2{\theta_{ij}}$. In the
approximation where $\Delta_{\odot} \ll \Delta_{\rm atm}$ this
Hamiltonian reduces to the one given in Eq.~(\ref{eq:hamilt-rot}) and
the evolution of neutrinos breaks to a decoupled one ($\nu'_\mu$) and
a two-neutrino mass matrix of the type of Eq.~(\ref{eq:ev3}).

The resonance point will be now obtained using the condition
$H^H_{33}-H^H_{11} = 0$, which gives $V_{\rm CC} = (\Delta_{\rm
  atm}+\Delta_{\odot}c^2_{12})c2_{13}$. Applying Eq.~(\ref{eq:Vcc_Ye}) we
find that the density at the resonance point is given by
\be\label{eq:rhoH_res}
\rho_H = \frac{(\Delta_{\rm atm}+\Delta_{\odot}c^2_{12})c2_{13}}{V_0 Y_e}~.
\ee
Assuming that $Y_e \simeq 0.5$ around these layers of the SN, we find
the density range for the resonance region $5 \times
10^4$ g/cm$^3 \gtrsim \rho_H \gtrsim 10^3$ g/cm$^3$ for the typical
energy range of the SN neutrinos, 1 MeV $\lesssim E \lesssim$ 50 MeV.

We can now calculate the adiabaticity parameter at the resonance using
Eq.~(\ref{eq:gamma2_text}). Taking into account that
$H^H_{33}-H^H_{11} = 0$, we obtain:
\be\label{eq:gamma_H}
\gamma^{-1}_H = \left|\frac{\dot{V}_{\rm CC}}{(\Delta_{\rm atm}s2_{13} + \Delta_{\odot}c^2_{12}s2_{13})^2}\right|_{\rm res}~.
\ee
Making use of Eqs.~(\ref{eq:Vcc_Ye}),~(\ref{eq:rho_prof})
and~(\ref{eq:rhoH_res}) we can reexpress the adiabaticity parameter at
the $H$-resonance as
\be\label{eq:gamma_H_2}
\gamma^{-1}_H = \left|\frac{-5.9\times 10^{-5}
    (c2_{13})^{4/3}}{\left((\Delta_{\rm atm} + \Delta_{\odot}c^2_{12})^2 V_0
    Y_e \rho_0\right)^{1/3}(s2_{13})^2 r_0}\right|~,
\ee
where $\Delta_{\rm atm}$ and $\Delta_{\odot}$ are given in eV,
$\rho_0$ in g/cm$^{3}$ and $r_0$ in cm.

\begin{figure}[t]
  \begin{center}
    \includegraphics[angle=0,width=0.6\textwidth]{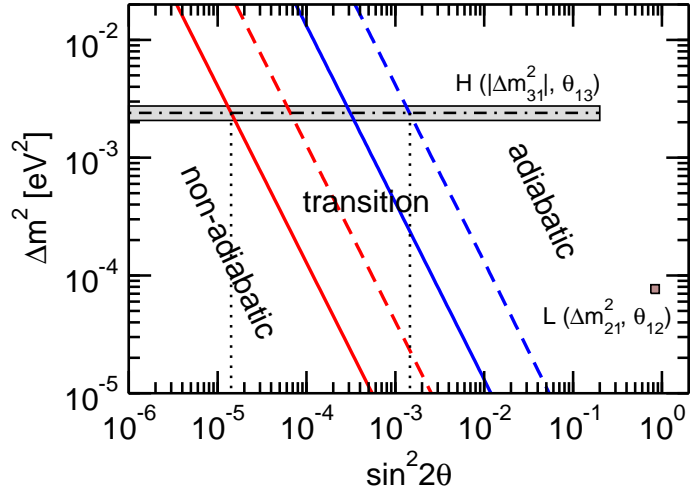}
  \end{center}
  \caption{\small Contours of constant hopping probability $P'$. The
    solid lines correspond to a neutrino energy of $E = 5$ MeV, the
    left one (red) stands for $P' = 0.9$ (highly non-adiabatic
    transition) and the right one (blue) to $P' = 0.1$ (adiabatic
    transition). The dashed lines are done for neutrinos with $E = 50$
    MeV. The gray band shows the current allowed region at 3$\sigma$
    for the parameters involved in the $H$-resonance, for which the
    vertical lines mark the borders for adiabatic and non-adiabatic
    conversions. The small brown band shows the current allowed region
    for the $L$-parameters, lying in the adiabatic region.}
  \label{fig:adresHL}
\end{figure}

In Fig.~\ref{fig:adresHL} we show the corresponding jumping
probability contours for the whole $\Delta m^2-\sin^22\theta$ space of
parameters. To the right of the dashed blue line we obtain an
adiabatic evolution, while to the left of the solid red line the
evolution is highly non-adiabatic. The gray band shows the allowed
region for the parameters involved in the $H$-resonance at 3$\sigma$,
as given in Table~\ref{tab:summary}. The dotted vertical lines in the
figure mark the adiabaticity borders for this case. In the figure we
see, and can be calculated from Eq.~(\ref{eq:gamma_H_2}), that for
$\sin^2\theta_{13} \gtrsim 10^{-3}$ the evolution is adiabatic, while
$\sin^2\theta_{13} \lesssim 10^{-5}$ leads to a non-adiabatic
evolution.

\subsection{{\em L}-resonance}

In order to properly study the adiabaticity at the $L$-resonance, we
undo the rotations $V_{23}$ and $W_{13}$ in the flavor basis around
the resonance point~\cite{Kuo:1986sk}. The corresponding Hamiltonian
is given by
\bea
H^L & = & W^{\dagger}_{13} V^{\dagger}_{23} (U H_{\rm kin} U^{\dagger} + H_{\rm std}) V_{23} W_{13} = \nonumber \\
& = & \left(\begin{array}{ccc}
  - \Delta_{\odot}c^2_{12} + V_{\rm CC}c^2_{13} & \frac{1}{2}\Delta_{\odot}s2_{12} & V_{\rm CC}s2_{13} \\
  \frac{1}{2}\Delta_{\odot}s2_{12} & - \Delta_{\odot}s^2_{12} & 0 \\
  V_{\rm CC}s2_{13} & 0 & \Delta_{\rm atm} + V_{\rm CC}s^2_{13} \end{array} \right)~.
\eea
From this equation it is easy to see that the third neutrino decouples
($\Delta_{\rm atm} \gg V_{\rm CC}s2_{13}$, for the lower resonance), leaving
us with the effective two-neutrino evolution governed by the 1-2
submatrix in $H^L$.

The resonance point will be given by $H^L_{22}-H^L_{11} = 0$, thus
$V_{\rm CC}c^2_{13} = \Delta_{\odot}c2_{12}$. Just as before, we can
obtain the density at the resonance point
\be\label{eq:rhoL_res}
\rho_L = \frac{\Delta_{\odot}c2_{12}}{V_0 Y_e c^2_{13}}~,
\ee
and what we were looking for, the adiabaticity parameter,
\be\label{eq:gamma_L}
\gamma^{-1}_L =
\left|\frac{\dot{V}_{\rm CC}c^2_{13}}{(\Delta_{\odot}s2_{12})^2}\right|_{\rm
  res}~.
\ee
Again, making use of Eqs.~(\ref{eq:Vcc_Ye}),~(\ref{eq:rho_prof})
and~(\ref{eq:rhoL_res}) we can reexpress the adiabaticity parameter at
the $L$-resonance as
\be\label{eq:gamma_L_2}
\gamma^{-1}_L = \left|\frac{-5.9\times 10^{-5}
    (c2_{12})^{4/3}}{\left((\Delta_{\odot}^2 V_0
    Y_e \rho_0\right)^{1/3}(s2_{12})^2 r_0}\right|~.
\ee
We can see how the adiabaticity parameter at the $L$-resonance has
essentially the same expression as the one at the $H$-resonance,
Eq.~(\ref{eq:gamma_H_2}), but changing $\Delta_{\rm atm} \to
\Delta_{\odot}$ and $\theta_{13} \to \theta_{12}$, provided that
$\Delta_{\odot}/\Delta_{\rm atm} \ll 1$ and $\sin^2\theta_{13} \ll
1$. This allows us to use the same contours of constant jumping
probability of Fig.~\ref{fig:adresHL}, if we reinterpret the
oscillation parameters as $\Delta m^2_{21}$ and
$\sin^22\theta_{12}$. The small brown band in the figure shows the
current allowed region at 3$\sigma$ for these parameters given in
Table~\ref{tab:summary}. Since both neutrino properties are rather
well constrained, we can conclude that the $L$-resonance is always
adiabatic, and takes place in the neutrino channel.

\cleardoublepage

\pagestyle{normal}
\chapter{Non-Standard Neutrino Interactions}\label{chapter:NSI}

\section{Introduction}

The confirmation of the neutrino oscillation interpretation of solar
and atmospheric neutrino data by reactor~\cite{Araki:2004mb} and
accelerator~\cite{Ahn:2006zza,Michael:2006rx} neutrino experiments
provides a unique picture of neutrino physics in terms of
three-neutrino oscillations~\cite{Schwetz:2008er}, leaving little room
for other non-standard neutrino
properties~\cite{Pakvasa:2003zv}. Nevertheless, it has long been
recognized that any gauge theory of neutrino mass generation
inevitably introduces dimension-6 non-standard neutrino interaction
(NSI) terms.  Such sub-weak strength operators arise in the broad
class of seesaw-type models, due to the non-trivial structure of
charged and neutral current weak interactions~\cite{Schechter:1980gr},
and may be
sizeable~\cite{Mohapatra:1986bd,Bernabeu:1987gr,Branco:1989bn,Rius:1989gk,Deppisch:2004fa}.
Alternatively, neutrino NSI may arise in models where neutrino masses
are radiatively ``calculable''~\cite{Zee:1980ai,Babu:1988ki}. Finally,
in some supersymmetric unified models, the strength of neutrino NSI
may arise from renormalization and/or threshold
effects~\cite{Hall:1985dx}.

We stress that NSI strengths are highly model dependent.  In some
models NSI strengths are too small to be relevant for neutrino
propagation, because they are either suppressed by some large mass
scale or restricted by limits on neutrino masses, or both.  However,
this need not be the case, and there are many theoretically attractive
scenarios where moderately large NSI strengths are possible and
consistent with the smallness of neutrino masses. In fact one can show
that NSI may exist even in the limit of massless
neutrinos~\cite{Mohapatra:1986bd,Bernabeu:1987gr,Branco:1989bn,Rius:1989gk,Deppisch:2004fa}.
This may also occur in the context of fully unified models like
$SO(10)$~\cite{Malinsky:2005bi}.

We argue that, in addition to the precision determination of the
oscillation parameters, it is necessary to test for sub-leading
non-oscillation effects that could arise from neutrino NSI. These are
natural outcome of many neutrino mass models and can be of two types:
flavor-changing (FC) and non-universal (NU).
These are constrained by existing experiments (see
Sec.~\ref{sec:limits}) and, with neutrino experiments now entering a
precision phase~\cite{McDonald:2004dd}, an improved determination of
neutrino parameters and their theoretical impact constitute an
important goal in astroparticle and high energy
physics~\cite{Schwetz:2008er}.

One of the objectives of this thesis is the study of neutrino NSI,
since they would open a new window to physics beyond the Standard
Model. We have performed several analysis in the subject using
neutrinos from different scenarios. In this chapter, after introducing
the NSI parameterization and summarizing the current limits on the
parameters, we will discuss what we can learn about the possible NSI
using the combination of MINOS and OPERA, two experiments currently
taking data. In Chapter~\ref{chapter:SN_NSI} we will present the
possibility of probing these NSI using SN neutrinos.

\section{Parameterization}

\begin{figure}
\begin{center}
\includegraphics[width=0.5\textwidth]{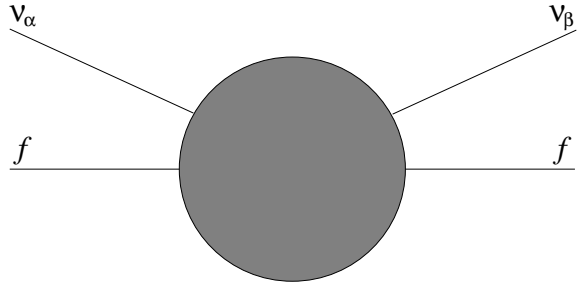}
\caption{\small Schematic diagram of the neutrino NSI described by the
  effective low-energy four-fermion operator given in
  Eq.~(\ref{eq:Lnsi}).}
\label{fig:NSI_diagram} 
\end{center}
\end{figure}

A large class of NSI may be parameterized with the effective
low-energy four-fermion operator:
\begin{equation}
\label{eq:Lnsi}
\mathcal{L}_{\rm nsi} = -\varepsilon^{fP}_{\alpha\beta}
2\sqrt{2}G_F(\bar\nu_\alpha\gamma_\mu L \nu_\beta) (\bar f\gamma^\mu P f)~.
\end{equation}
where $P=L$ or $R$ and $f$ is a first generation fermion:
$e,~u,~d$. The coefficients $\varepsilon^{fP}_{\alpha\beta}$ denote
the strength of the NSI between the neutrinos of flavors $\alpha$ and
$\beta$ and the $P$-handed component of the fermion $f$. This
Lagrangian describes the elastic scattering processes
\be
\nu_{\alpha} ~f \to \nu_{\beta} ~f ~~~~\alpha,\beta=e,\mu,\tau~,
\ee
schematically represented in Fig.~\ref{fig:NSI_diagram}.

Just as in the SM case, we can calculate the potential induced in the
evolution of neutrinos due to this new interactions. We can rewrite
Eq.~(\ref{eq:Lnsi}) as
\bea
\mathcal{L}_{\rm nsi} & = & \frac{1}{\sqrt{2}}G_F\{\bar\nu_{\alpha}  \gamma^{\mu}(1-\gamma_5)\nu_{\beta}\} \left( \varepsilon^{fL}_{\alpha \beta} \{\bar f \gamma_{\mu} (1-\gamma_5) f\} + \varepsilon^{fR}_{\alpha \beta} \{\bar f \gamma_{\mu} (1+\gamma_5) f\}\right) = \nonumber \\
& = &\frac{1}{\sqrt{2}}G_F\{\bar\nu_{\alpha}  \gamma^{\mu}(1-\gamma_5)\nu_{\beta}\} \left\{\bar f \gamma_{\mu} \left[(\varepsilon^{fL}_{\alpha \beta}+\varepsilon^{fR}_{\alpha \beta})-(\varepsilon^{fL}_{\alpha \beta}-\varepsilon^{fR}_{\alpha \beta})\gamma_5\right] f\right\}~.
\eea
The effective axial vector coupling, $(\varepsilon^{fL}_{\alpha
  \beta}-\varepsilon^{fR}_{\alpha \beta})$, does not contribute to the
coherent elastic scattering of neutrinos with the particles in an
unpolarized medium, as we discussed in the calculation of $V_{\rm CC}$
and $V_{\rm NC}$. The vector coupling, $\varepsilon^{fV}_{\alpha
  \beta}\equiv(\varepsilon^{fL}_{\alpha
  \beta}+\varepsilon^{fR}_{\alpha \beta})$, on the other hand does
affect neutrino propagation in matter, and although no positive signs
of them have yet been obtained there are limits on their magnitude as
we will later discuss.

Finally, the presence of NSI like the ones described in the Lagrangian
of Eq.~(\ref{eq:Lnsi}) induces the following potentials on neutrinos:
\bea
V^{\rm FC}_{\alpha \beta} = \sum_{f} \sqrt{2} G_F N_f \varepsilon^{fV}_{\alpha \beta} ~, & & \alpha \ne \beta ~,\\
V^{\rm NU}_{\alpha \alpha} = \sum_{f} \sqrt{2} G_F N_f \varepsilon^{fV}_{\alpha \alpha} ~,
\eea
where $V^{\rm FC}_{\alpha \beta}$ correspond to the scattering
amplitudes in the flavor changing processes $\nu_{\alpha} ~f \to
\nu_{\beta} ~f$, while $V^{\rm NU}_{\alpha \alpha}$ represents the
difference between the non-standard and the standard components in the
elastic scattering amplitudes of $\nu_{\alpha} ~f \to \nu_{\alpha}
~f$. In general, we can write:
\be
\label{eq:V_alfabet}V^{\rm nsi}_{\alpha \beta} = \sum_{f} \sqrt{2} G_F N_f \varepsilon^{fV}_{\alpha \beta} ~,
\ee
where in principle we sum over all three fundamental fermions found in
ordinary matter. However, due to the complexity of a global analysis,
usually the studies are done for each one of them separately. The
number density for the different fermions will be given by:
\be
N_f = \left\{
\begin{array}{ll} 
N_e, & f = e\\
N_p + 2 N_n, & f = d\\
2 N_p + N_n, & f = u
\end{array}
\right.
\ee

If we add this new part to the Hamiltonian discussed in
Chapter~\ref{chapter:oscillations}, with its kinetic and standard
matter interaction terms, we obtain,
\begin{equation}
{\setlength\arraycolsep{3pt}
\label{eq:hamiltonian}
H=\frac{1}{2E}U\left(\begin{array}{ccc}
0&0&0\\
0&\Delta m^2_{21}&0\\
0&0&\Delta m^2_{31}
\end{array}\right)U^\dagger +
\left(\begin{array}{ccc}
V_{\rm CC}&0&0\\
0&0&0\\
0&0&0
\end{array}\right)+\left(\begin{array}{ccc}
V^{\rm nsi}_{ee}&V^{\rm nsi}_{e\mu}&V^{\rm nsi}_{e\tau}\\
V^{\rm nsi*}_{e\mu}&V^{\rm nsi}_{\mu\mu}&V^{\rm nsi}_{\mu\tau}\\
V^{\rm nsi*}_{e\tau}&V^{\rm nsi*}_{\mu\tau}&V^{\rm nsi}_{\tau\tau}
\end{array}\right) \,,
}
\end{equation}
where $V^{\rm nsi}_{\alpha\beta}$ are given in Eq.~(\ref{eq:V_alfabet}),
and $V^{\rm nsi*}_{\alpha\beta} = V^{\rm nsi*}_{\beta\alpha}$ for Hermiticity
of the Hamiltonian.

\section{Current limits}\label{sec:limits}

Current bounds on the NSI parameters can be obtained from different
sources: laboratory experiments, solar, reactor, atmospheric and
accelerator neutrino experiments, cosmology and SN neutrinos. The
latter case will be discussed in detail in
Chapter~\ref{chapter:SN_NSI}, while we will here review the present
limits coming from the rest of the experiments. Note that these limits
have been obtained by varying each $\varepsilon_{\alpha\beta}$ one at
a time. In general, when correlations among different
$\varepsilon_{\alpha\beta}$ are included the bounds become weaker.

\subsection{Laboratory}
\label{sec:laboratory}

The measurement of $\nu e$ and $\nu q$ elastic cross sections in
neutrino scattering
experiments~\cite{Auerbach:2001wg,Daraktchieva:2003dr,Dorenbosch:1986tb,Vilain:1994qy,Zeller:2001hh}
provide the following bounds on the NSI parameters,
$|\varepsilon^{fP}_{\mu\mu}| \lesssim 10^{-3}$--$10^{-2},~
|\varepsilon^{fP}_{ee}|\lesssim
10^{-1}$--$1,~|\varepsilon^{fP}_{\mu\tau}|\lesssim 0.1,~
|\varepsilon^{fP}_{e\tau}|\lesssim 0.5$ at 90 \%
C.L~\cite{Barger:1991ae,Davidson:2003ha,Barranco:2005ps}.
On the other hand the analysis of the $e^+e^-\to \nu\bar\nu\gamma$
cross section measured at LEP II  leads to a
bound on $|\varepsilon^{eP}_{\tau\tau}|\lesssim
0.5$~\cite{Berezhiani:2001rs}.
Future prospects to improve the current limits imply the measurement
of $\sin^2\theta_W$ leptonically in the scattering off electrons in
the target, as well as in neutrino deep inelastic scattering in a
future neutrino factory. The main improvement would be in the case of
$|\varepsilon^{fP}_{ee}|$ and $|\varepsilon^{fP}_{e\tau}|$, where
values as small as $10^{-3}$ and $0.02$, respectively, could be
reached~\cite{Davidson:2003ha}.

The search for flavor violating processes involving charged leptons is
expected to restrict the corresponding neutrino interactions, to the
extent that the $SU(2)$ gauge symmetry is assumed. However, this can
at most give indicative order-of-magnitude restrictions, since we know
$SU(2)$ is not a good symmetry of nature.
Using radiative corrections it has been argued that, for example,
$\mu$-$e$ conversion on nuclei like in the case of Ti$\mu$ also
constrains $|\varepsilon^{qP}_{e\mu}|\lesssim 7.7\times
10^{-4}$~\cite{Davidson:2003ha}. The rather good bound
$|\varepsilon^{eP}_{e\mu}|\lesssim 5\times 10^{-4}$ is also obtained
because of the strong experimental limit on $\mu \to 3e$
transitions. Finally, only an $\mathcal{O}(1-10)$ limit can be imposed
on $|\varepsilon^{eP}_{\tau\tau}|$ from $\tau$ decays.

A detailed relation of the commented limits is given in
Table~\ref{tab:nsi}~\cite{Maltoni:2008mu}.

\renewcommand{\arraystretch}{1.2}
\begin{table}[t]
  \caption{\small Present bounds at 90\% CL on the NC-like NSI couplings
    $\Eps_{\alpha\beta}^{fP}$ from non-oscillation experiments.
    Limits have been obtained by varying each $\Eps_{\alpha\beta}$
    one at a time, with all the others set to zero. Table taken
    from~\cite{Maltoni:2008mu}.}
  \begin{center}
    {\footnotesize
      \begin{tabular}{ccccc}
	\hline
	\hline
	Left-handed & Right-handed & Process & Experiment & Reference \\
	\hline
	\hline
	$-0.03 < \Eps_{ee}^{eL} < 0.08$ & $0.004 < \Eps_{ee}^{eR} < 0.15$
	& $\begin{aligned} \nu_e e &\to \nu e \\[-2mm]
	\bar{\nu}_e e &\to \bar{\nu} e \end{aligned}$
	& $\begin{aligned} &\text{LSND} \\[-2mm] &\text{Reactors} \end{aligned}$
	& \cite{Barranco:2005ps, Barranco:2007ej}
	\\[1mm]
	$-1 < \Eps_{ee}^{uL} < 0.3$ & $-0.4 < \Eps_{ee}^{uR} < 0.7$
	& $\nu_e q \to \nu q$ & CHARM
	& \cite{Davidson:2003ha}
	\\[1mm]
	$-0.3 < \Eps_{ee}^{dL} < 0.3$ & $-0.6 < \Eps_{ee}^{dR} < 0.5$
	& $\nu_e q \to \nu q$ & CHARM & \cite{Davidson:2003ha}
	\\
	\hline
	$|\Eps_{\mu\mu}^{eL}| < 0.03$ & $|\Eps_{\mu\mu}^{eR}| < 0.03$
	& $\nu_\mu e \to \nu e$ & CHARM II
	& \cite{Davidson:2003ha, Barranco:2007ej}
	\\[1mm]
	$|\Eps_{\mu\mu}^{uL}| < 0.003$ & $-0.008 < \Eps_{\mu\mu}^{uR} < 0.003$
	& $\nu_\mu q \to \nu q$ & NuTeV
	& \cite{Davidson:2003ha}
	\\[1mm]
	$|\Eps_{\mu\mu}^{dL}| < 0.003$ & $-0.008 < \Eps_{\mu\mu}^{dR} < 0.015$
	& $\nu_\mu q \to \nu q$ & NuTeV
	& \cite{Davidson:2003ha}
	\\
	\hline
	$-0.5 < \Eps_{\tau\tau}^{eL} < 0.2$ & $ -0.3 < \Eps_{\tau\tau}^{eR} < 0.4$
	& $e^+ e^- \to \nu \bar{\nu} \gamma$ & LEP
	& \cite{Berezhiani:2001rs, Barranco:2007ej}
	\\[1mm]
	$|\Eps_{\tau\tau}^{uL}| < 1.4$ & $|\Eps_{\tau\tau}^{uR}| < 3$
	& rad.~corrections & $\tau$ decay
	& \cite{Davidson:2003ha}
	\\[1mm]
	$|\Eps_{\tau\tau}^{dL}| < 1.1$ & $|\Eps_{\tau\tau}^{dR}| < 6$
	& rad.~corrections & $\tau$ decay
	& \cite{Davidson:2003ha}
	\\
	\hline
	\hline
	$|\Eps_{e\mu}^{eL}| < 0.0005$ & $|\Eps_{e\mu}^{eR}| < 0.0005$
	& rad.~corrections & $\mu\to 3e$
	& \cite{Davidson:2003ha}
	\\[1mm]
	$|\Eps_{e\mu}^{uL}| < 0.0008$ & $|\Eps_{e\mu}^{uR}| < 0.0008$
	& rad.~corrections & $\text{Ti}\, \mu \to \text{Ti}\, e$
	& \cite{Davidson:2003ha}
	\\[1mm]
	$|\Eps_{e\mu}^{dL}| < 0.0008$ & $|\Eps_{e\mu}^{dR}| < 0.0008$
	& rad.~corrections & $\text{Ti}\, \mu \to \text{Ti}\, e$
	& \cite{Davidson:2003ha}
	\\
	\hline
	$|\Eps_{e\tau}^{eL}| < 0.33$ & $|\Eps_{e\tau}^{eR}| < 0.28$
	& $\nu_e e \to \nu e$ & LEP+LSND+Rea
	& \cite{Berezhiani:2001rs, Barranco:2007ej}
	\\[1mm]
	$|\Eps_{e\tau}^{uL}| < 0.5$ & $|\Eps_{e\tau}^{uR}| < 0.5$
	& $\nu_e q \to \nu q$ & CHARM
	& \cite{Davidson:2003ha}
	\\[1mm]
	$|\Eps_{e\tau}^{dL}| < 0.5$ & $|\Eps_{e\tau}^{dR}| < 0.5$
	& $\nu_e q \to \nu q$ & CHARM
	& \cite{Davidson:2003ha}
	\\
	\hline
	$|\Eps_{\mu\tau}^{eL}| < 0.1$ & $|\Eps_{\mu\tau}^{eR}| < 0.1$
	& $\nu_\mu e \to \nu e$ & CHARM II
	& \cite{Davidson:2003ha, Barranco:2007ej}
	\\[1mm]
	$|\Eps_{\mu\tau}^{uL}| < 0.05$ & $|\Eps_{\mu\tau}^{uR}| < 0.05$
	& $\nu_\mu q \to \nu q$ & NuTeV
	& \cite{Davidson:2003ha}
	\\[1mm]
	$|\Eps_{\mu\tau}^{dL}| < 0.05$ & $|\Eps_{\mu\tau}^{dR}| < 0.05$
	& $\nu_\mu q \to \nu q$ & NuTeV
	& \cite{Davidson:2003ha}
	\\
	\hline
	\hline
      \end{tabular}
    }
  \end{center}
\label{tab:nsi}
\end{table}    

\renewcommand{\arraystretch}{1}

\subsection{Solar and reactor}
\label{sec:solar-reactor}

NSI can also affect neutrino propagation through matter, probed in
current neutrino oscillation experiments.  The bounds so obtained
apply to the vector coupling constant of the NSI,
$\varepsilon^{fV}_{\alpha\beta} = \varepsilon^{fL}_{\alpha\beta} +
\varepsilon^{fR}_{\alpha\beta}$, since only this combination appears
in neutrino propagation in matter, as previously discussed.

The role of neutrino NSIs as subleading effects on the solar neutrino
oscillations and KamLAND has been considered in
Refs.~\cite{Friedland:2004pp,Guzzo:2004ue,Miranda:2004nb} with the
following bounds at 90 \% CL for $\varepsilon \equiv -\sin \theta_{23}
\varepsilon^{dV}_{e\tau}$ with the allowed range $-0.93\lesssim
\varepsilon \lesssim 0.30$, while for the diagonal term
$\varepsilon'\equiv
\sin^2\theta_{23}\varepsilon^{dV}_{\tau\tau}-\varepsilon^{dV}_{ee}$,
the only forbidden region is $[0.20,0.78]$~\cite{Miranda:2004nb}.
Only in the ideal case of infinitely precise determination of the
solar neutrino oscillation parameters, the allowed range would ``close
from the left'' for negative NSI parameter values, at $-0.6$ for
$\varepsilon$ and $-0.7$ for $\varepsilon'$.

\subsection{Atmospheric and accelerator neutrinos}
\label{sec:atmosph-accel-neutr}

NSI involving muon neutrinos can be constrained by atmospheric
neutrino experiments as well as accelerator neutrino oscillation
searches at K2K and MINOS.  In Ref.~\cite{Fornengo:2001pm}
Super-Kamiokande and MACRO observations of atmospheric neutrinos were
considered in the framework of two neutrinos.  The limits obtained
were $-0.05\lesssim \varepsilon^{dV}_{\mu\tau}<0.04$ and
$|\varepsilon^{dV}_{\tau\tau}-\varepsilon^{dV}_{\mu\mu}|\lesssim 0.17$
at 99 \% CL. The same data set together with K2K were recently
considered in Refs.~\cite{Friedland:2004ah,Friedland:2005vy} to study
the nonstandard neutrino interactions in a three generation scheme
under the assumption
$\varepsilon_{e\mu}=\varepsilon_{\mu\mu}=\varepsilon_{\mu\tau}=0$. The
most recent analysis including also accelerator experiments
\cite{GonzalezGarcia:2007ib} gives at 3$\sigma$,
$|\varepsilon_{\mu\tau}^{eV}| \leq 0.058 \,,
|\varepsilon_{\tau\tau}^{eV} - \varepsilon_{\mu\mu}^{eV}| \leq 0.19
\,, |\varepsilon_{\mu\tau}^{uV}| \leq 0.019 \,,
|\varepsilon_{\tau\tau}^{uV} - \varepsilon_{\mu\mu}^{uV}| \leq 0.061
\,, |\varepsilon_{\mu\tau}^{dV}| \leq 0.019 \,,
|\varepsilon_{\tau\tau}^{dV} - \varepsilon_{\mu\mu}^{dV}| \leq 0.060$.
%
%


\subsection{Cosmology}
\label{sec:cosmology}

If NSI with electrons were large they might also lead to important
cosmological and astrophysical implications. For instance, relic
neutrinos could have been kept in thermal contact with electrons and
positrons for a longer time than in the standard case, hence they
would share a larger fraction of the entropy release from $e^\pm$
annihilations.  This would affect the predicted features of the cosmic
background of neutrinos, in particular, it could enhance the radiation
content of the Universe.  As pointed out in
Ref.~\cite{Mangano:2006ar}, the required NSI couplings to observe this
kind of effects with cosmological mesurements are larger than the
current laboratory bounds.

\pagestyle{OPERA}

\section{Can OPERA help in constraining neutrino non-standard interactions?}\label{sec:OPERA}

The interplay of oscillation and neutrino NSI was studied
in~\cite{Grossman:1995wx} and subsequently it was
shown~\cite{Huber:2001de,Huber:2002bi} that in the presence of NSI it
is very difficult to disentangle genuine oscillation effects from
those coming from NSI. The latter may affect production, propagation
and detection of neutrinos and in general these three effects need not
be correlated.  It has been shown that in this case cancellations can
occur which make it impossible to separate oscillation from NSI
effects.  Subsequently it was discovered that the ability to detect
$\nu_\tau$ may be crucial in order to overcome that
problem~\cite{Campanelli:2002cc}, though this method requires
sufficiently large beam energies to be applicable.  Barring the
occurrence of fine-tuned cancellations, NSI and oscillations have very
different $L/E$ dependence. Therefore, combining different $L/E$ can
be very effective in probing the presence of NSI. The issue of NSI and
oscillation in neutrino experiments with terrestrial sources has been
studied in a large number of publications~\cite{Bergmann:1998ft,
  Ota:2001pw, Ota:2002na, Honda:2006gv, Kitazawa:2006iq,
  Friedland:2006pi, Blennow:2007pu, GonzalezGarcia:2001mp,
  Huber:2001zw, Gago:2001xg, Huber:2002bi, Campanelli:2002cc,
  Bueno:2000jy, Kopp:2007mi, Adhikari:2006uj, Ribeiro:2007ud,
  Kopp:2007ne}.  In Ref.~\cite{Blennow:2007pu} it was shown that
MINOS~\cite{Michael:2006rx} on its own is not able to put new
constraints on NSI parameters. On the other hand, in
Ref.~\cite{Friedland:2006pi} the combination of atmospheric data with
MINOS was proven to be effective in probing at least some of the NSI
parameters. Since matter effects are relatively small in MINOS, its
main role in that combination is to constrain the vacuum mixing
parameters.

The question we would like to address here is whether the combination
of MINOS and OPERA~\cite{Guler:2000bd} can provide useful information
on NSI. OPERA has recently seen the first events in the emulsion cloud
chamber~\cite{OPERAfirst} and hence it appears timely to ask this
question. The idea is that OPERA will be able to detect $\nu_\tau$ and
has a very different $L/E$ than MINOS.  Both factors are known to help
distinguish NSI from oscillation effects.  Clearly, much larger
improvements on existing sensitivities are expected from superbeam
experiments like T2K~\cite{Itow:2001ee} and
NO$\nu$A~\cite{Ayres:2004js} especially in combination with reactor
neutrino experiments like Double
Chooz~\cite{Ardellier:2004ui,Ardellier:2006mn} or Daya
Bay~\cite{Guo:2007ug}, see Ref.~\cite{Kopp:2007ne}.  We will here
focus on the simple case where NSI only affects neutrino propagation.

\subsection{Basic setup}

\textbf{A) Evolution Hamiltonian}\vspace{0.2cm}\\
The evolution of neutrinos will be governed by the Hamiltonian given
in Eq.~(\ref{eq:hamiltonian})
\begin{equation}
{\setlength\arraycolsep{3pt}
\label{eq:OP_ham}
H=\frac{1}{2E}U\left(\begin{array}{ccc}
0&0&0\\
0&\Delta m^2_{21}&0\\
0&0&\Delta m^2_{31}
\end{array}\right)U^\dagger +
V_{\rm CC}\left(\begin{array}{ccc}
1&0&0\\
0&0&0\\
0&0&0
\end{array}\right)+ V_{\rm CC}\left(\begin{array}{ccc}
0&0&\varepsilon_{e\tau}\\
0&0&0\\
\varepsilon_{e\tau}&0&\varepsilon_{\tau\tau}
\end{array}\right) \,,
}
\end{equation}
where we have assumed the $\varepsilon$'s to be real for
simplicity\footnote{The inclusion of phases has been considered in the
  literature, see, e.g., Refs.~\cite{Kopp:2007mi,Blennow:2008ym}}. We
have also made use of the fact that all $\varepsilon_{x\mu}$ are
fairly well constrained and hence are expected not to play a
significant role at leading order. The effect of $\varepsilon_{ee}$ is
a re-scaling of the matter density and all experiments considered here
are not expected to be sensitive to standard matter effects. Hence we
will set $\varepsilon_{ee}= 0$. Note, that the $\varepsilon$ as
defined here, are effective parameters. At the level of the underlying
Lagrangian describing the NSI, the NSI coupling of the neutrino can be
either to electrons, up or down quarks. From a phenomenological point
of view, however, only the (incoherent) sum of all these contributions
is relevant. For simplicity, we chose to normalize our NSI to the
electron abundance.  This introduces a relative factor of 3 compared
to the case where one normalizes either to the up or down quark
abundance (assuming an isoscalar composition of the Earth), i.e.~the
NSI coupling to only up or down quark would need to be 3 times as
strong to produce the same effect in oscillations. Since both
conventions can be found in the literature, care is required in making
quantitative comparisons.

There are two potential benefits beyond adding statistics from
combining the data from MINOS and OPERA: First, OPERA can detect
$\nu_\tau$ which, in principle, allows to directly access any effect
from $\varepsilon_{x\tau}$. Moreover, although the baseline is the same,
the beam energies are very different $\langle E \rangle \simeq 3
\,\mathrm{GeV}$ for MINOS, whereas $\langle E \rangle \simeq 17
\,\mathrm{GeV}$ for OPERA.\vspace{0.5cm}\\
\textbf{B) Experiments}\vspace{0.2cm}\\
All numerical simulations have been done using the GLoBES (General
Long Baseline Experiment Simulator)
software~\cite{Huber:2004ka,Huber:2007ji}, a package especially
designed for the simulation of long baseline neutrino oscillation
experiments. In order to include the effects of the NSI we have
customized the package by adding a new piece to the Hamiltonian as
shown in Eq.~(\ref{eq:OP_ham}). We have considered three different
experiments: MINOS, OPERA and Double Chooz, the main characteristics
of which are summarized in table~\ref{tab:exp}.

\renewcommand{\arraystretch}{1.2}
\begin{table}
\begin{center}
{\footnotesize
  \begin{tabular}{l@{\hspace{0.5cm}}l@{\hspace{0.5cm}}r@{\hspace{0.5cm}}c@{\hspace{0.5cm}}c@{\hspace{0.5cm}}l}
    \hline
    \hline
    Label & $L$ & $\langle E_\nu \rangle$ & Power & $t_{\rm run}$ & Channel \\
    \hline
    \hline
    MINOS$_2$ (M2) & 735 km & 3 GeV & $5\times 10^{20}$ pot/yr & 5 yr
    & $\nu_\mu \to \nu_{e,\mu}$ \\
    \hline
    OPERA (O) & 732 km & 17 GeV & $4.5\times 10^{19}$ pot/yr & 5 yr &
    $\nu_\mu \to \nu_{e,\mu,\tau}$ \\
    \hline
    \multirow{2}{*}{Double Chooz (DC)} & 0.2 km (near) &
    \multirow{2}{*}{4 MeV} & \multirow{2}{*}{$8.4$ GW} &
    \multirow{2}{*}{5 yr} & \multirow{2}{*}{$\bar\nu_e \to \bar\nu_e$} \\
    & 1.05 km (far) & & & \\
    \hline
    \hline
  \end{tabular}
}
\end{center}
\caption{\small Main parameters of the experiments under study.}
\label{tab:exp}
\end{table}

\renewcommand{\arraystretch}{1}

MINOS is a long baseline neutrino oscillation experiment using the
NuMI neutrino beam, at FNAL. It uses two magnetized iron calorimeters.
One serves as near detector and is located at about $1\,\mathrm{km}$
from the target, whereas the second, larger one is located at the
Soudan Underground Laboratory at a distance of $735\,\mathrm{km}$ from
the source. The near detector is used to measure the neutrino beam
spectrum and composition. The near/far comparison also mitigates the
effect of cross section uncertainties and various systematical errors.
In our simulations, based
on~\cite{Huber:2004ug,Ables:1995wq,NUMIL714}, we have used a running
time of 5 years with a statistics corresponding to a primary proton
beam of $5 \times 10^{20}$ per year, giving a total of $2.5 \times
10^{21}$, the maximum reachable value reported by the MINOS
collaboration. The mean energy of the neutrino beam is $\langle E
\rangle \simeq 3 \,\mathrm{GeV}$.

The OPERA detector is located at Gran Sasso and gets its beam from
CERN (CNGS). OPERA consists of two parts: a muon tracker and an
emulsion cloud chamber. The latter one is the part which is able to
discern a $\nu_\tau$ charged current interaction by identifying the
subsequent $\tau$-decay. The baseline is $732\,\mathrm{km}$.
Following~\cite{Huber:2004ug,Guler:2000bd,Komatsu:2002sz} we assume a
5 year run with a nominal beam intensity of $4.5 \times
10^{19}\,\mathrm{pot}$ per year. The CNGS neutrino beam has an average
energy of $\langle E \rangle \simeq 17 \,\mathrm{GeV}$. Since both
MINOS and OPERA have the same baseline we use the same matter density
which we take constant and equal to its value at the Earth's crust,
that is $\rho = 2.7$ g/cm$^3$.

Finally, Double Chooz is a reactor experiment, to be located in the old
site of CHOOZ, in France. The experiment consists of a pair of nearly
identical near and far detectors, each with a fiducial mass of
$10.16\,\mathrm{t}$ of liquid scintillator. The detectors are located
at a distance of $0.2\,\mathrm{km}$ and $1.05\,\mathrm{km}$
respectively. As considered in~\cite{Huber:2006vr} we assume the
thermal power of both reactor cores to be 4.2 GW and a running time of
5 years. The neutrinos mean energy is $\langle E \rangle \simeq 4
\,\mathrm{MeV}$.

Concerning the neutrino oscillation parameters used to calculate the
simulated event rates, we have taken the best fit values as given in
Ref.~\cite{Maltoni:2004ei}, unless stated otherwise:
\begin{equation}
  \begin{array}{rclrcl}
    \sin^2\theta^{\rm true}_{12} & = & 0.32, & (\Delta m^2_{21})^{\rm true} & = & +7.6 \times 10^{-5} ~\mathrm{eV^2},\\
    \sin^2\theta^{\rm true}_{23} & = & 0.5,  & (\Delta m^2_{31})^{\rm
      true} & = & +2.4 \times 10^{-3} ~\mathrm{eV^2},\\
    \sin^2\theta^{\rm true}_{13} & = & 0, &\delta^{\rm
      true}_{CP} & = & 0.
  \end{array}
\end{equation}
Note the positive sign assumed for $(\Delta m^2_{31})^{\rm true}$
which corresponds to the case of normal hierarchy. Since, none of the
experiments considered here is very sensitive to ordinary matter
effects, our results would be very similar when choosing as true
hierarchy, the inverted one.

\subsection{Results} 

\begin{figure}
\begin{center}
\includegraphics[width=0.55\textwidth]{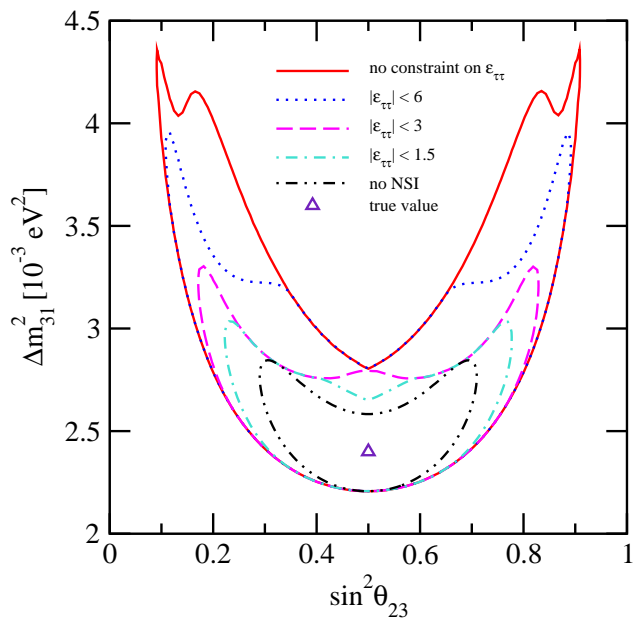}
\caption{\small Allowed region in the $\sin^2\theta_{23}$-$\Delta
  m^2_{31}$-plane at 95\% CL (2 d.o.f.). In this fit $\theta_{13}$,
  $\varepsilon_{e\tau}$ and $\varepsilon_{\tau\tau}$ are left
  free. The different lines correspond to different priors for the
  $\varepsilon_{\tau\tau}$ values as explained in the
  legend~\cite{EstebanPretel:2008qi}.}
\label{fig:atm} 
\end{center}
\end{figure}

\textbf{A) Disappearance: probing NU NSI ($\bm{\varepsilon_{\tau\tau}}$)}\vspace{0.2cm}\\
As it has been previously shown in~\cite{Friedland:2006pi,
  Blennow:2007pu} the presence of NSI, notably ${\sf 
  \varepsilon_{\tau\tau}}$,
substantially degrades the goodness of the determination of the
``atmospheric'' neutrino oscillation parameters from
experiment. Indeed as shown in Fig.~\ref{fig:atm} our calculation
confirms the same effect, showing how the allowed region in the
$\sin^2\theta_{23}$-$\Delta m^2_{31}$-plane increases in the presence
of NSI.

This figure is the result of a combined fit to simulated OPERA and
MINOS data in terms of the ``atmospheric'' neutrino oscillation
parameters, leaving the mixing angle $\theta_{13}$ to vary freely. The
inner black dot-dashed curve corresponds to the result obtained in the
pure oscillation case (no NSI).  As displayed in the figure, allowing
for a free nonzero strength for NSI parameters $\varepsilon_{\tau\tau}$
and $\varepsilon_{e\tau}$ the allowed region grows substantially, as seen
in the solid, red curve.
Intermediate results assuming different upper bounds on
$|\varepsilon_{\tau\tau}|$ strengths are also indicated in the figure,
and given in the legend.
One sees that the NSI effect is dramatic for large NSI magnitudes.  A
similar result has been obtained in Fig.~2 of
Ref.~\cite{Blennow:2007pu}.  However, such large values are in conflict
with atmospheric neutrino
data~\cite{Fornengo:2001pm,Friedland:2006pi}.
In contrast, for lower NSI strengths allowed by the atmospheric +
MINOS data combination~\cite{Friedland:2006pi}, say
$|\varepsilon_{\tau\tau}|=1.5$, the NSI effect becomes much smaller.
Clearly beam experiments currently can not compete with atmospheric
neutrino data in constraining $\varepsilon_{\tau\tau}$. The reason for
the good sensitivity of atmospheric data to the presence of NSI is the
very large range in $L/E$, especially the very high energy events are
crucial in constraining NSI~\cite{Fornengo:2001pm}.

In summary, the inclusion of OPERA data helps only for very large
values of $\varepsilon_{\tau\tau}$ as can be seen also from the first
line of table~\ref{tab:00}. These large values, however are already
excluded by the combination of MINOS and atmospheric
results~\cite{Friedland:2006pi}.  We checked that doubling the OPERA
exposure does not change this conclusion. The slight improvement by
OPERA is exclusively due to the $\nu_\mu$ sample in the muon tracker
and the results do not change if we exclude the $\nu_\tau$ sample from
the analysis. The usefulness of the $\nu_\mu$ sample stems from the
very different value of $L/E$ compared to MINOS. These results are not
too surprising, since even a very high energy neutrino factory will
not be able to improve the bound on $\varepsilon_{\tau\tau}$ in
comparison to
atmospheric neutrino data~\cite{Huber:2001zw}.\vspace{0.3cm}\\
\textbf{B) Appearance: probing FC NSI ($\bm{\varepsilon_{e\tau}}$)}\vspace{0.2cm}\\
It is well known that, in the presence of NSI, the determination of
$\theta_{13}$ exhibits a continuous degeneracy~\cite{Huber:2001de}
between $\theta_{13}$ and $\varepsilon_{e\tau}$, which leads to a
drastic loss in sensitivity in $\theta_{13}$.  A measurement of only
$P_{e\mu}$ and $P_{\bar\mu \bar e}$ at one $L/E$ cannot disentangle
the two and will only yield a constraint on a combination of
$\theta_{13}$ and $\varepsilon_{e\tau}$.  In this context, it was
shown in Ref.~\cite{Campanelli:2002cc}, that even a very rudimentary
ability to measure $P_{\mu\tau}$ may be sufficient to break this
degeneracy. Therefore, it seems natural to ask whether OPERA can
improve upon the sensitivity for $\varepsilon_{e\tau}$ that can be
reached only with MINOS. The latter has been studied in
Ref.~\cite{Friedland:2006pi} in combination with atmospheric neutrinos
and on its own in Ref.~\cite{Blennow:2007pu}. The result, basically,
was that MINOS will not be able to break this degeneracy, and hence a
possible $\theta_{13}$ bound from MINOS will, in reality, be a bound
on a combination of $\varepsilon_{e\tau}$ and $\theta_{13}$.

In Tables~\ref{tab:00} and \ref{tab:01} we display our results for a
true value of $\theta_{13}=0$ and no NSI\footnote{Note that the values
  given in our tables are obtained from the projected $\chi^2$ and for
  1 degree of freedom only. Moreover, the resulting projected $\chi^2$
  is strongly non-Gaussian.}. The allowed range for
$\varepsilon_{e\tau}$ shrinks only very little by the inclusion of
OPERA data. As in the case of $\varepsilon_{\tau\tau}$ we checked that
this result is not due to the $\nu_\tau$ sample in OPERA but is
entirely due to the different $L/E$ compared to MINOS. Also a two-fold
increase of the OPERA exposure does not substantially alter the
result.

\begin{figure}[t]
\begin{center}
\includegraphics[width=0.55\textwidth]{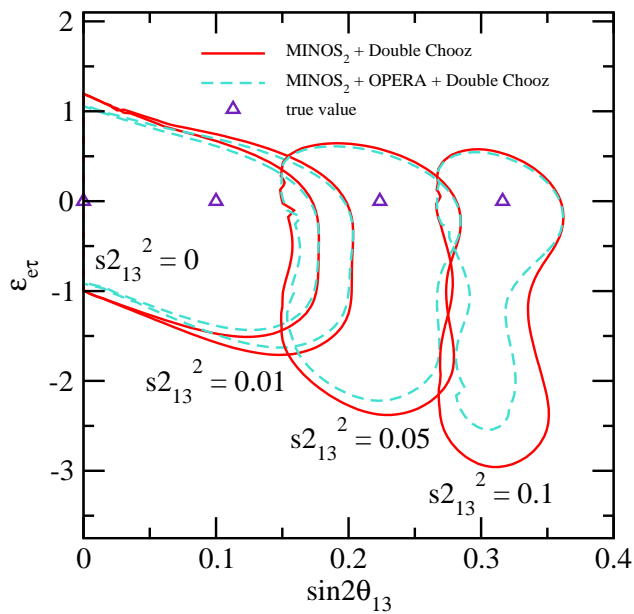}
\caption{\small Allowed regions in the
  $\sin2\theta_{13}$-$\varepsilon_{e\tau}$-plane at 95\% CL (2
  d.o.f.).  $\Delta m^2_{31}$, $\theta_{23}$ and
  $\varepsilon_{\tau\tau}$ are left free in this fit. The solid lines
  correspond to the combination of MINOS$_2$ and Double Chooz while
  the dashed lines also include OPERA in the analysis. Each set of
  lines correspond to different true values for $\sin^22\theta_{13}$,
  from left to right: 0, 0.01, 0.05 and
  0.1~\cite{EstebanPretel:2008qi}.}
\label{fig:th13} 
\end{center}
\end{figure}

In order to improve the sensitivity to NSI and to break the degeneracy
between $\theta_{13}$ and $\varepsilon_{e\tau}$ it will be necessary
to get independent information on either $\varepsilon_{e\tau}$ or
$\theta_{13}$. An improvement of direct bounds on
$\varepsilon_{e\tau}$ is in principle possible by using a very high
energy $\nu_e$ beam and a close detector, but this would require
either a neutrino factory or a high $\gamma$ beta beam. Both these
possibilities are far in the future and will therefore not be here
considered any further. Thus, we focus on independent information on
$\theta_{13}$. Reactor experiments are very sensitive to $\theta_{13}$
but do not feel any influence from $\varepsilon_{e\tau}$ since the
baseline is very short and the energy very low which leads to
negligible matter effects. This is true for standard MSW-like matter
effects as well as non-standard matter effects due to
NSI~\cite{Valle:1987gv}. We consider here as new reactor experiment
Double Chooz~\cite{Ardellier:2006mn}, but for our discussion Daya
Bay~\cite{Guo:2007ug} or RENO~\cite{Joo:2007zzb} would work equally
well. In Fig.~\ref{fig:th13} we show the allowed regions in the
$\sin2\theta_{13}$-$\varepsilon_{e\tau}$ plane for the combinations of
MINOS and Double Chooz (red solid curves) and of MINOS, Double Chooz
and OPERA (blue dashed curves) for four different input values of
$\sin^22\theta_{13}$ indicated in the plot. As expected, the effect of
Double Chooz in all four cases is to constrain the allowed
$\sin2\theta_{13}$ range. The impact of OPERA, given by the difference
between the solid and dashed lines, is absent for very small true
values of $\sin2\theta_{13}$ and increases with increasing true
values. For the largest currently permissible values of
$\theta_{13}\simeq0.16$, OPERA can considerably reduce the size of the
allowed region and help to resolve the degeneracy. In that parameter
region a moderate increase in the OPERA exposure would make it
possible to constrain large negative values of
$\varepsilon_{e\tau}$. Again, this effect has nothing to do with
$\nu_\tau$ detection and, in this case, is based on the different
$L/E$ in $\nu_e$-appearance channel.

\subsection{Conclusion}

We have here studied how OPERA can help in improving the sensitivities
on neutrino non-standard contact interactions of the third family of
neutrinos.  In our analysis we considered a combined OPERA fit
together with high statistics MINOS data, in order to obtain
restrictions on neutrino oscillation parameters in the presence of
NSI.  Due to its unique ability of detecting $\nu_\tau$ one would
expect that the inclusion of OPERA data would provide new improved
limits on the universality violating NSI parameter
$\varepsilon_{\tau\tau}$. We found, however, that the $\nu_\tau$ data
sample is too small to be of statistical significance. This holds even
if we double the nominal exposure of OPERA to
$9\times10^{19}\,\mathrm{pot}$.  OPERA also has a $\nu_\mu$ sample,
which can help constraining NSI.  Here the effect is due to the very
different $L/E$ of OPERA compared to MINOS. This makes the OPERA
$\nu_\mu$ sample more sensitive to NSI. However, the improvement is
small and happens in a part of the NSI parameter space which is
essentially excluded by atmospheric neutrino data.

We have also studied the possibility of constraining the FC NSI
parameter $\varepsilon_{e\tau}$. For this purpose it is crucial to have a
good knowledge of $\theta_{13}$.  Therefore, we included future
Double Chooz data, since reactor neutrino experiments are insensitive
to the presence of NSI of the type considered here. Therefore, reactor
experiments can provide a clean measurement of $\theta_{13}$, which in
turn can be used in the analysis of long baseline data to probe the
NSI. Double Chooz is only the first new reactor experiment and more
precise ones like Daya Bay or Reno will follow. Our result would be
qualitatively the same if we would have considered those, more
precise, experiments, but clearly the numerical values of the obtained
bounds would improve. The conclusion for $\varepsilon_{e\tau}$ with
respect to the $\nu_\tau$ sample is the same as before: the sample is
very much too small to be of any statistical significance. OPERA's
different $L/E$ again proves to be its most important feature and
allows to shrink the allowed region on the
$\sin^2\theta_{13}$-$\varepsilon_{e\tau}$ plane for large $\theta_{13}$
values. Here a modest increase in OPERA exposure would allow to
completely lift the $\theta_{13}$-$\varepsilon_{e\tau}$ degeneracy and
thus to obtain a unique solution.

\renewcommand{\arraystretch}{1.2}
\begin{landscape}
\begin{table}[b]
\begin{center}
{\footnotesize
  \begin{tabular}{|c|c|c|c|c|c|c|}
   
    \hline
    \hline
    & \multicolumn{2}{c|}{M2} & \multicolumn{2}{c|}{O} & \multicolumn{2}{c|}{M2+O} \\
    \hline
    & 90\% C.L. & 95\% C.L. & 90\% C.L. & 95\% C.L. & 90\% C.L. & 95\% C.L. \\
    \hline
    \hline
    $\varepsilon_{\tau\tau}$ & [-10.8,10.8] & [-11.8,11.8] & [-10.4,10.4] & [-11.0,11.0] & [-8.5,8.5] & [-9.2,9.2] \\
    \hline
    $\varepsilon_{e\tau}$ & [-1.9,0.9] & [-2.3,1.0] & [-2.1,1.4] & [-2.5,1.6] & [-1.6,0.9] & [-2.0,1.0] \\
    \hline
    $\Delta m^2_{31}$ [10$^{-3}$ eV$^2$] & [2.3,4.5] & [2.2,4.9] & [2.0,5.0] & [2.0,5.3] & [2.3,3.8] & [2.2,4.0] \\
    \hline
    $\sin^2\theta_{23}$ & [0.08,0.92] & [0.07,0.93] & [0.08,0.92] & [0.07,0.93] & [0.12,0.88] & [0.11,0.89] \\
    \hline
    \hline
  \end{tabular}
}
\end{center}
\caption{\small 90\% and 95\% C.L. allowed regions for
  $\varepsilon_{\tau\tau}$, $\varepsilon_{e\tau}$, $\Delta m^2_{31}$
  and $\sin^2\theta_{23}$ for different sets of experiments. Each row
  is obtained marginalizing over the remaining parameters in the table, 
  plus $\theta_{13}$. The true value for $\sin^22\theta_{13}$ is  $0$~\cite{EstebanPretel:2008qi}. }
\label{tab:00}
\end{table}

\begin{table}
\begin{center}
{\footnotesize
  \begin{tabular}{|c|c|c|c|c|c|c|c|c|}
    \hline
    \hline
    & \multicolumn{2}{c|}{M2} & \multicolumn{2}{c|}{O} & \multicolumn{2}{c|}{M2+O} & \multicolumn{2}{c|}{M2+O+DC} \\
    \hline
    & 90\% C.L. & 95\% C.L. & 90\% C.L. & 95\% C.L. & 90\% C.L. & 95\% C.L. & 90\% C.L. & 95\% C.L. \\
    \hline
    \hline
    $\varepsilon_{\tau\tau}$ & [-10.1,11.0] & [-11.2,12.0] & [-10.1,10.3] & [-10.8,11.0] & [-7.9,9.0] & [-8.7,9.6] & [-5.1,5.3] & [-5.6,5.8] \\
    \hline
    $\varepsilon_{e\tau}$ & [-4.2,1.3] & [-4.5,1.5] & [-4.3,1.5] & [-5.0,1.8] & [-3.7,1.2] & [-4.1,1.4] & [-0.5,0.4] & [-0.7,0.5] \\
    \hline
    $\Delta m^2_{31}$ [10$^{-3}$ eV$^2$] & [2.3,4.6] & [2.2,5.0] & [2.0,4.8] & [2.0,5.2] & [2.3,4.0] & [2.2,4.2] & [2.3,2.8] & [2.3,2.9] \\
    \hline
    $\sin^2\theta_{23}$ & [0.09,0.92] & [0.08,0.93] & [0.09,0.93] & [0.08,0.94] & [0.13,0.90] & [0.12,0.91] & [0.24,0.78] & [0.22,0.80] \\
    \hline
    \hline
  \end{tabular}
}
\end{center}
\caption{\small Same as table~\ref{tab:00} with true value $\sin^22\theta_{13}$ of $0.1$~\cite{EstebanPretel:2008qi}. }
\label{tab:01}
\end{table}
\end{landscape}

\renewcommand{\arraystretch}{1}

\cleardoublepage

\pagestyle{normal}
\chapter{Collective Supernova Neutrino Transformations in Two Flavors}\label{chapter:coll2flavors}

It is well known that neutrino interactions with the medium have a
strong impact in their evolution, through the induced effective matter
potentials $V_{\rm CC},~V_{\rm NC}$ and $V_{\mu\tau}$, as discussed in
Chapter~\ref{chapter:oscillations}. What is not really well understood
is the effect of neutrinos themselves, as a background medium. It was
first pointed by Pantaleone~\cite{Pantaleone:1992eq} that the
inclusion of neutrino-neutrino interaction into the problem would
introduce an off-diagonal refractive index. The oscillation equations
become then non-linear, leading sometimes to surprising collective
phenomena in very dense environments such as the early Universe or
core-collapse SNe.

Initially most of the attention was centered on the context of the
early Universe, first in a series of papers by Samuel, Kosteleck\'y
and Pantaleone~\cite{Samuel:1993uw, Kostelecky:1993yt,
  Kostelecky:1993dm, Kostelecky:1993ys, Kostelecky:1994dt,
  Kostelecky:1995xc, Samuel:1995ri,Kostelecky:1996bs,
  Pantaleone:1998xi}, and later by other
authors~\cite{Pastor:2001iu,Lunardini:2000fy, Dolgov:2002ab,
  Wong:2002fa, Abazajian:2002qx}. But it was not until very recently
that the community has realized the importance this ingredient might
have in the frame of SN neutrinos~\cite{Pastor:2002we, Sawyer:2005jk,
  Fuller:2005ae, Duan:2005cp, Duan:2006an, Duan:2006jv,
  Hannestad:2006nj, Balantekin:2006tg, Duan:2007mv, Raffelt:2007yz,
  EstebanPretel:2007ec, Raffelt:2007cb, Raffelt:2007xt, Duan:2007fw,
  Fogli:2007bk, Duan:2007bt, Duan:2007sh, Dasgupta:2008cd,
  EstebanPretel:2007yq, Dasgupta:2007ws, Duan:2008za, Dasgupta:2008my,
  Sawyer:2008zs, Duan:2008eb, Chakraborty:2008zp, Dasgupta:2008cu}.

In the dense neutrino flux emerging from a SN core, neutrino-neutrino
refraction causes non-linear flavor oscillation phenomena that are
unlike anything produced by ordinary matter. There are several aspects
that have to be taken into account when studying neutrino-neutrino
interactions in the context of SN. The main two are multi-angle and
multi-energy effects. On the one hand, neutrinos are not emitted from
a single point but a spherical surface, the neutrino sphere. As a
consequence, and due to the current-current nature of the weak
interaction, neutrinos moving in different directions will experience
a different refractive effect caused by the other neutrinos. On the
other hand, the effect is energy dependent, and since SN neutrinos are
emitted with a broad spectrum, this dependency could be an issue.

In this chapter we will address the question of multi-angle effects in
SN neutrino transformations in a two flavor scenario, for both
single and multi-energy configurations. But before we will review the
main characteristics of the collective phenomena caused by the dense
neutrino background.

\section{Introduction}\label{sec:5.intro}

In order to study this complicated non-linear problem it is convenient
to use the density matrix formalism. Mixed neutrinos are described by
matrices of density $\varrho_{\bf p}$ and $\bar\varrho_{\bf p}$ for
each (anti)neutrino mode, where overbarred quantities refer to
antineutrinos here and now on. The diagonal entries are the usual
occupation numbers whereas the off-diagonal terms encode phase
information. The equations of motion (EOMs) are
\begin{equation}
\I\partial_t\varrho_{\bf p}=[{\sf H}_{\bf p},\varrho_{\bf p}]\,,
\label{eq:gen_eoms}
\end{equation}
where the Hamiltonian is~\cite{Sigl:1992fn}
\begin{equation}
 {\sf H}_{\bf p}=\Omega_{\bf p}
 +{\sf V}+\sqrt{2}\,G_{\rm F}\!
 \int\!\frac{\D^3{\bf q}}{(2\pi)^3}
 \left(\varrho_{\bf q}-\bar\varrho_{\bf q}\right)
 (1-{\bf v}_{\bf q}\cdot{\bf v}_{\bf p})\,,
\label{eq:hamiltonianrho}
\end{equation}
${\bf v}_{\bf p}$ being the velocity of a neutrino mode with momentum
${\bf p}$. The matrix of vacuum oscillation frequencies is $\Omega_{\bf
  p}={\rm diag}(m_1^2,m_2^2,m_3^2)/2|{\bf p}|$ in the mass basis. The
matter effect is represented, in the weak interaction basis, by ${\sf
  V}=\sqrt{2}\,G_{\rm F}n_B\,{\rm diag}(Y_e,0,Y_\tau^{\rm eff})$, as
defined in Chapter~\ref{chapter:oscillations}. The factor $(1-{\bf
  v}_{\bf q}\cdot{\bf v}_{\bf p})=(1-\cos\theta_{\bf pq})$ represents
the current-current nature of the weak interaction. For antineutrinos
the only difference is $\Omega_{\bf p}\to-\Omega_{\bf
  p}$\footnote{This convention is equivalent to the one used in the
  previous chapters, where antineutrinos had opposite sign for the
  matter potential with respect to neutrinos, i.e.~$V \to -V$.}, i.e.,
in vacuum antineutrinos oscillate ``the other way around''.

In the $\nu_e$-$\nu_x$ two-flavor system, the density matrices can be
reduced to polarization vectors by using the Pauli matrices and the
unit matrix. Therefore the EOMs can be reexpressed in terms of the
polarization vectors using:
\begin{equation}
\varrho = \frac{1}{2}\left[P_0+{\bf P}\cdot{\boldsymbol \sigma}\right]~,
\end{equation}
where ${\boldsymbol \sigma}$ is the vector of Pauli matrices. In this
formalism the survival probability of $\nu_e$ is given at time $t$ by
$\frac{1}{2}[1+P_z(t)]$, and Eqs.~(\ref{eq:gen_eoms}) and
(\ref{eq:hamiltonianrho}) become
\begin{equation}
\dot{\bf P}_{\bf p}={\bf H}_{\bf p}\times{\bf P}_{\bf p}\,,
\label{eq:eomsP}
\end{equation}
\begin{equation}
 {\bf H}_{\bf p}=\omega_{\bf p}{\bf B}+\lambda{\bf L}+
 \mu_0\int\frac{\D^3{\bf q}}{(2\pi)^3}\,
 \left({\bf P}_{\bf q}-\bar{\bf P}_{\bf q}\right)
 \left(1-{\bf v}_{\bf q}\cdot{\bf v}_{\bf p}\right)\,,
\label{eq:hamiltonianP}
\end{equation}
where $\omega_{\bf p}=|\Delta m^2/2E|$ is the vacuum oscillation
frequency with $E=|{\bf p}|$ for relativistic neutrinos, ${\bf
  B}=(\sin2\theta,0,\pm\cos2\theta)$ is a unit vector in the mass
direction in flavor space where the ``$-$'' sign corresponds to normal
hierarchy and the ``$+$'' sign to inverted hierarchy, ${\bf L}$ is a
unit vector in the weak-interaction direction with ${\bf B}\cdot{\bf
  L}=\cos2\theta$ and $\theta$ the vacuum mixing angle. The effect of
a homogeneous and isotropic medium is parameterized by
\begin{equation}
\lambda=\sqrt2\,G_{\rm
  F}(n_{e^-}-n_{e^+})\,,
\label{eq:lambda}
\end{equation}
where $n_{e^{\pm}}$ is the electron/positron number density. Finally
the neutrino-neutrino term is given by
\begin{equation}\label{eq:mudefine}
  \mu_0=\sqrt2 G_{\rm F}\left(F_{\bar\nu_e}^{R_\nu}-F_{\bar\nu_x}^{R_\nu}\right)\,,
\end{equation}
where $F_{\bar\nu_e}^{R_\nu}$ and $F_{\bar\nu_x}^{R_\nu}$ are the
$\bar\nu_e$ and $\bar\nu_x$ fluxes, taken at the neutrino sphere with
radius~$R_\nu$.

\subsection{Single-angle and single-energy}\label{sec:single-single}

Let us begin with the simplest case, single-angle and
single-energy. We consider a two flavor system initially composed of
equal densities of pure $\nu_e$ and $\bar\nu_e$, all of them emitted
in the same direction with equal energies. The flavor content of these
ensembles will be given by the polarization vectors in flavor space
${\bf P}$ and $\bar{\bf P}$. By definition, we take these vectors to be
unitary.

In vacuum the general EOMs given by Eq.~(\ref{eq:eomsP}) are reduced
for this system to
\begin{eqnarray}\label{eq:eomvac}
 \partial_t{\bf P}&=&\left[+\omega{\bf B}
 +\mu\left({\bf P}-\bar{\bf P}\right)\right]\times{\bf P}\,,
 \nonumber\\
 \partial_t\bar{\bf P}&=&\left[-\omega{\bf B}
 +\mu\left({\bf P}-\bar{\bf P}\right)\right]\times\bar{\bf P}\,.
\end{eqnarray}

\begin{figure}
\begin{center}
\includegraphics[angle=0,width=0.55\textwidth]{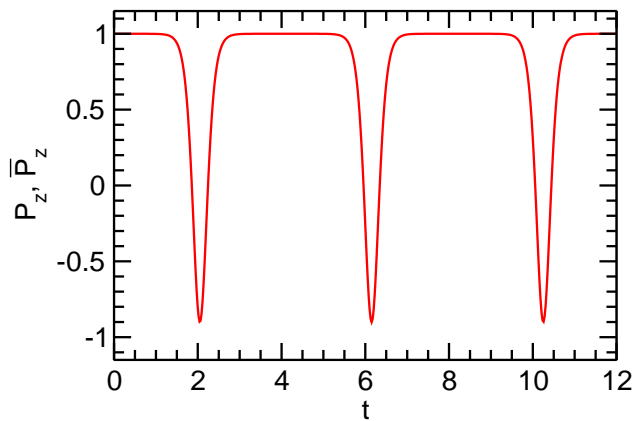}
\end{center}
\caption{\small Evolution of $P_z$ and $\bar{P}_z$ for the system of
  equations Eq.~(\ref{eq:eomvac}) for inverted hierarchy, with
  $\theta=0.01$, $\omega=1$ km$^{-1}$, and a constant
  neutrino-neutrino interaction $\mu=10$
  km$^{-1}$.}\label{fig:fexample}
\end{figure}

In Fig.~\ref{fig:fexample} we show the evolution of $P_z$ and
$\bar{P}_z$, given by Eq.~(\ref{eq:eomvac}), for a small vacuum mixing
angle, inverted mass hierarchy and a constant $\mu$. Initially, both
${\bf P}$ and ${\bf \bar{P}}$ stay put. After some time, though, they
flip completely, but immediately return to their original state,
leading to periodic motion. This behavior, translated to flavor
language, means that we obtain complete and simultaneous conversion of
$\nu_e$ and $\bar\nu_e$ to $\nu_\mu$ and $\bar\nu_\mu$ and back. For
the normal hierarchy on the other hand, nothing visible happens.

This evolution was proved in Ref.~\cite{Hannestad:2006nj} to be
equivalent to a pendulum in flavor space, by reducing the EOMs to
those of a spherical pendulum. Making use of this analogy it is
possible to explain why we only obtain bipolar conversion in the
inverted hierarchy case. For normal hierarchy the system initially
sets near the minimum of the pendulum potential and is therefore
trapped leading to no visible effects. In the inverted hierarchy,
conversely, the evolution starts with the system close to the maximum
of the potential in an unstable equilibrium. Therefore it will always
move to the minimum, leading to this periodic bipolar conversion. The
bipolar period, i.e.~the separation between dips, is obtained to scale
logarithmically with the small vacuum mixing angle.

\begin{figure}
\begin{center}
\includegraphics[angle=0,width=0.48\textwidth]{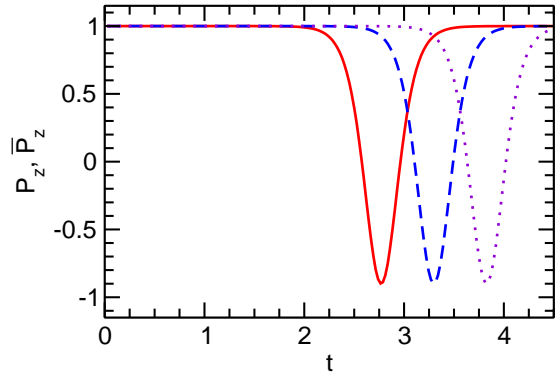}
\end{center}
\caption{\small Evolution of $P_z$ and $\bar{P}_z$ in several systems
  with background matter described by Eq.~(\ref{eq:eommatt}). We are
  using inverted hierarchy, $\sin2\theta=0.001$, $\omega=1$ km$^{-1}$,
  and $\mu=10$ km$^{-1}$ for all three systems.  The red/solid line
  has $\lambda=10^2$, the blue/dashed line $\lambda=10^3$, and the
  violet/dotted line $\lambda=10^4$.}\label{fig:pmatter}
\end{figure}

In the presence of a background medium of charged leptons the
EOMs, Eq.~(\ref{eq:eomsP}), become:
\begin{eqnarray}\label{eq:eommatt}
 \partial_t{\bf P}&=&\left[+\omega{\bf B}+\lambda{\bf L}
 +\mu\left({\bf P}-\bar{\bf P}\right)\right]\times{\bf P}\,,
 \nonumber\\
 \partial_t\bar{\bf P}&=&\left[-\omega{\bf B}+\lambda{\bf L}
 +\mu\left({\bf P}-\bar{\bf P}\right)\right]\times\bar{\bf P}\,.
\end{eqnarray}
As it was first pointed out by Duan et al.~\cite{Duan:2005cp}, by
going to the frame co-rotating around the ${\bf L}$-direction we can
reduce this equations to the ones in vacuum, Eq.~(\ref{eq:eomvac}),
with a time dependent ${\bf B}$,
\begin{equation}\label{eq:bt}
  {\bf B}=\left(
    \begin{array}{c}
      \sin(2\theta_0)\cos(-\lambda t)\\
      \sin(2\theta_0)\sin(-\lambda t)\\
      -\cos(2\theta_0)
    \end{array}\right).
\end{equation}
In this picture ${\bf B}$ rotates around the $z$-direction with
frequency $-\lambda$. If this rotation is faster than all other
frequencies one would na\"{\i}vely expect that the transverse
components of ${\bf B}$ average to zero, leaving us with $\langle {\bf
  B}\rangle$ along the $z$-axis, i.e., an effectively vanishing mixing
angle and no flavor conversion. However, this fast rotating transverse
components are still enough to trigger the conversion effect, with a
matter suppressed effective mixing angle, though. Thus, the presence
of matter in this simple picture is just to extend logarithmically the
bipolar period. This is illustrated in Fig.~\ref{fig:pmatter}. A more
detailed description of the consequences of adding an ordinary matter
background will be given in Sec.~\ref{sec:densematter} for the
multi-angle case.

\begin{figure}
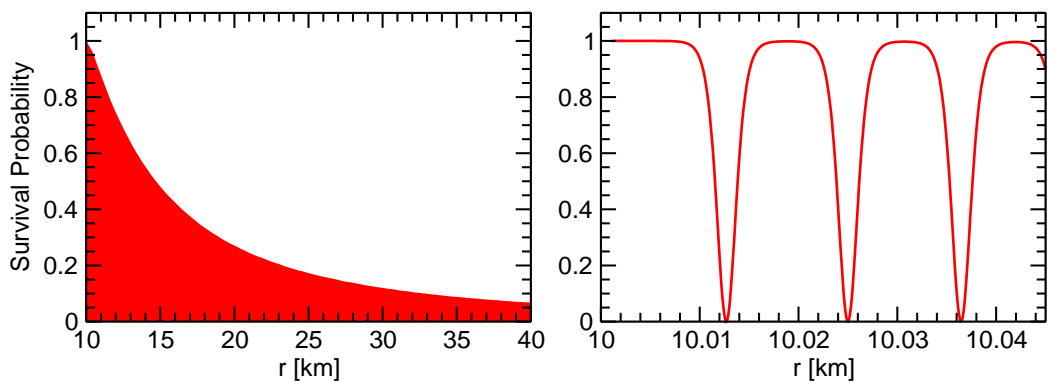

\begin{center}
\includegraphics[angle=0,height=0.33\textwidth]{./cap4/figures/noves/s-s_l0_0_s2t.001_w.3_m0_7e5.eps}
\includegraphics[angle=0,height=0.33\textwidth]{./cap4/figures/noves/s-s_l0_0_s2t.001_w.3_m0_7e5_r10-10.5.eps}
\end{center}
\caption{\small Survival probabilities for $\nu_e$ or $\bar\nu_e$ for
  a system with $\mu_0= 7\times 10^{5}$ km$^{-1}$, $\sin2\theta =
  0.001$ and symmetric initial conditions.}\label{fig:SNbimodal}
\end{figure}

In a realistic SN scenario the neutrino self-interaction strength is
not constant. There is a radial dependence for the $\mu$
parameter. From the neutrino flux dilution we obtain a $r^{-2}$
scaling, but there will also be another $r^{-2}$ factor due to the
collinearity suppression, coming from the $(1-{\bf v}_{\bf q}\cdot{\bf
  v}_{\bf p})$ factor in Eq.~(\ref{eq:hamiltonianrho}). These two
effects leave us essentially with an $r^{-4}$ decline for the
self-interaction term, see Appendix~A for a more detailed
description. We add now this neutrino density dependence,
normalized\footnote{This number, which we take here by definition,
  will be motivated in Sec.~\ref{sec:setup}} at the neutrino sphere to
$\mu_0\equiv\mu(R_\nu) = 7\times 10^{5}$ km$^{-1}$, to the previous
picture: initial equal numbers of $\nu_e$ and $\bar\nu_e$ alone,
inverted mass hierarchy and small mixing angle $\sin2\theta =
0.001$. We then find numerically the survival probability shown in
Fig.~\ref{fig:SNbimodal}. We obtain a decline of the oscillation
amplitude as a function of radius leading to a complete flavor
conversion caused by the bipolar effect. This result is to be compared
with the periodic bipolar oscillations obtained in
Fig.~\ref{fig:fexample}, with a constant $\mu$. This behavior can be
understood using the pendulum analogy,
Ref.~\cite{Hannestad:2006nj}. The basic idea is that by reducing $\mu$
the energy of the system is also being reduced, so that, after the
oscillation, the pendulum will not come back to its initial position
(the maximum of the potential) but to a lower one, which will be
closer and closer to the rest point as $\mu$ decreases.

Another important feature to be taken into account in our way to
simulate a realistic SN scenario, is the initial flux asymmetry
obtained in typical SN numerical simulations. As described in
Chapter~\ref{chapter:supernova}, one expects the hierarchy of neutrino
number fluxes $F^{R_\nu}_{\nu_e} > F^{R_\nu}_{\bar\nu_e} > F^{R_\nu}_{\nu_x} =
F^{R_\nu}_{\bar\nu_x}$. We take this into account by imposing the initial
conditions $\bar P_z = 1$ and $P_z = 1 + \epsilon$, where the
asymmetry parameter, $\epsilon$, represents the neutrino excess coming
from the deleptonization of the collapsed core, and is defined as
\begin{equation}\label{eq:epsdefine}
 \epsilon\equiv
 \frac{F^{R_\nu}_{\nu_e}-F^{R_\nu}_{\nu_x}}{F^{R_\nu}_{\bar\nu_e}-F^{R_\nu}_{\bar\nu_x}}-1
 =\frac{F^{R_\nu}_{\nu_e}-F^{R_\nu}_{\bar\nu_e}}{F^{R_\nu}_{\bar\nu_e}-F^{R_\nu}_{\bar\nu_x}}\,,
\end{equation}
where we have used $F^{R_\nu}_{\nu_x}=F^{R_\nu}_{\bar\nu_x}$. We can
assume $F^{R_\nu}_{\nu_x}=F^{R_\nu}_{\bar\nu_x}=0$ at the neutrino
sphere without loss of generality, since flavor oscillations do not
change those parts of the flavor fluxes that are already equal. Only
the transformation of the excess $\bar\nu_e$ flux over the $\bar\nu_x$
flux is observable, and likewise for neutrinos.

\begin{figure}
\begin{center}
\includegraphics[angle=0,width=0.52\textwidth]{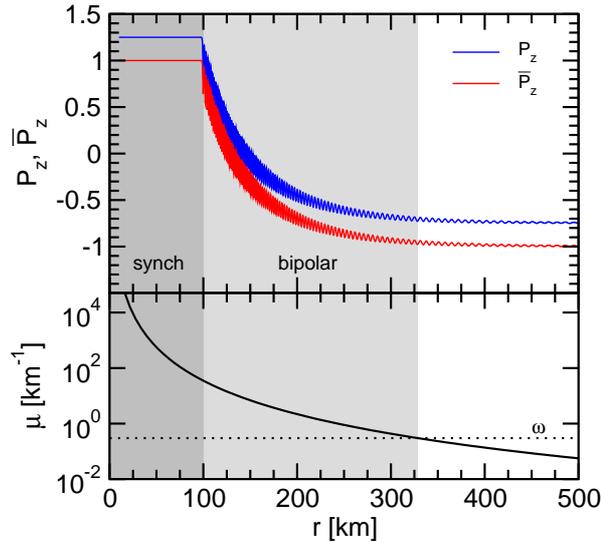}
\end{center}
\caption{\small $P_z$ (blue) and $\bar P_z$ (red) evolution in our toy
  SN model with 25\% more neutrinos than antineutrinos and
  $\sin2\theta=0.001$. The grey shaded bands show the synchronization
  (dark) and bipolar regions (light).}\label{fig:SN2}
\end{figure}

Choosing $\epsilon = 0.25$ and solving numerically the EOMs for this
system, given in Eq.~(\ref{eq:eomvac}), we obtain the results shown in
Fig.~\ref{fig:SN2}. We can distinguish different regimes of evolution,
corresponding to different relative strengths of $\mu$ with respect to
$\omega$. For large values of $\mu/\omega$, right behind the neutrino
sphere and up to about 100 km, we observe a first stage of
synchronized evolution where both neutrinos and antineutrinos stay in
their original flavor. For intermediate values of $\mu/\omega$, after
what we call $r_{\rm syn}$, we get a second stage of bipolar
conversion, where complete pair transformations $\nu_e\bar\nu_e \to
\nu_x\bar\nu_x$ are developed, keeping always the initial flavor
lepton asymmetry conserved. The nutations we observe during this
bipolar phase depend on the mixing angle, the smaller $\theta$ is, the
larger the depth of the nutations. Beyond this point, small
$\mu/\omega$, the neutrino-neutrino interactions die out and we obtain
ordinary oscillations, where of course, normal matter will play an
important role.

The limiting condition between synchronized oscillations and bipolar
conversions is found to be in Ref.~\cite{Hannestad:2006nj}\footnote{In
  this paper $\mu$ was normalized to the density of neutrinos and the
  results were expressed in terms of $\alpha=1/(1+\epsilon)$. Here we
  have normalized $\mu$ to the density of antineutrinos and use the
  picture of an excess of neutrinos, expressed by $\epsilon$. For the
  same physical system our $\mu$ is the one of
  Ref.~\cite{Hannestad:2006nj}, divided by $1+\epsilon$.}
\begin{equation}
\mu(r_{\rm syn}) \simeq \frac{2\omega}{(1-\sqrt{1+\epsilon})^2}\,.
\label{eq:synchcond}
\end{equation}
This means that for values of $\mu$ larger than this quantity
synchronized evolution is obtained, while for smaller values is the
turn of bipolar transformations. The end of this bipolar regime is
determined by the condition,
\begin{equation}
  \mu(r_{\rm bip}) \simeq \omega,
\end{equation}
thus for smaller values of $\mu$ no bipolar conversions are to be
expected. In the single-angle case we find from
Eq.~(\ref{eq:muvariation}) that the effective neutrino-neutrino
interaction strength varies at large distances as
\begin{equation}
\mu_{\rm eff}(r)=\mu_0\,\frac{R_\nu^4}{2r^4}\,.
\end{equation}
Therefore, the synchronization radius is
\begin{equation}\label{eq:r_synch}
 \frac{r_{\rm syn}}{R_\nu} = 
 \left(\frac{\sqrt{1+\epsilon}-1}{2}\right)^{1/2}
 \left(\frac{\mu_0}{\omega}\right)^{1/4} \approx \frac{\sqrt\epsilon}{2}\,
 \left(\frac{\mu_0}{\omega}\right)^{1/4}\,.
\end{equation}
The second line assumes $\epsilon\ll 1$. The bipolar radius on the
other hand is given by
\begin{equation}\label{eq:r_bip}
 \frac{r_{\rm bip}}{R_\nu} = \left(\frac{\mu_0}{2\omega}\right)^{1/4}\,.
\end{equation}
If we assume the values of the parameters used in Fig.~\ref{fig:SN2}:
$\omega=0.3~{\rm km}^{-1}$, $\mu_0=7\times10^5~{\rm km}^{-1}$,
$R_\nu=10$~km and $\epsilon=0.25$, we find $r_{\rm syn}=95$~km and
$r_{\rm bip}=330$~km corresponding well to the figure. For a typical
Fe core SN the $H$- and $L$-resonances take place far outside this
region, and therefore both effects are decoupled. Lower mass
progenitors, though, may collapse with a O-Ne-Mg core. In this kind of
SN the density profile may be such that both effects occur in the same
region~\cite{Duan:2007sh}. Throughout this thesis we will concentrate
on the former case.

\begin{figure}[t]
\begin{minipage}[b]{0.48\linewidth}
\centering
\includegraphics[angle=0,height=0.72\textwidth]{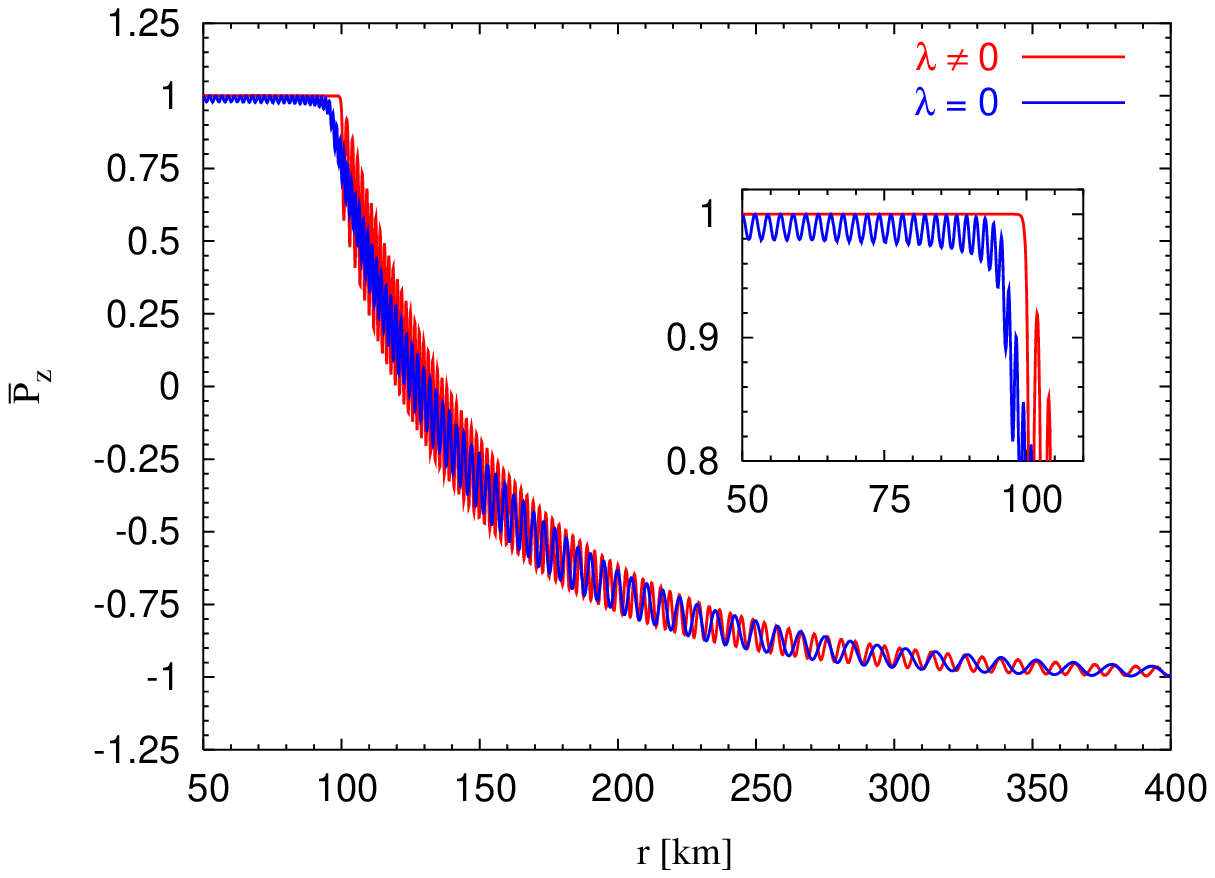}
\vskip12pt
\includegraphics[angle=0,height=0.7\textwidth]{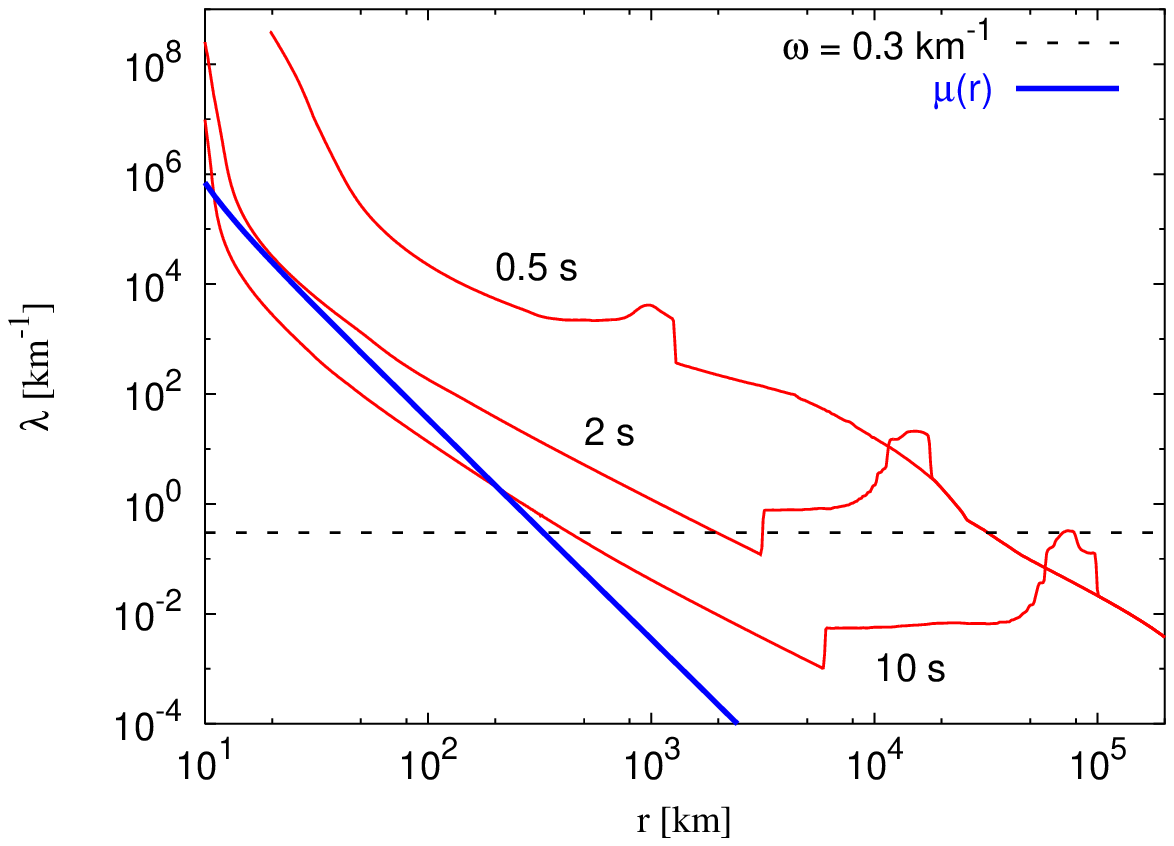}
\end{minipage}
\hspace{0.2cm}
\begin{minipage}[b]{0.48\linewidth}
\centering
\includegraphics[angle=0,height=0.716\textwidth]{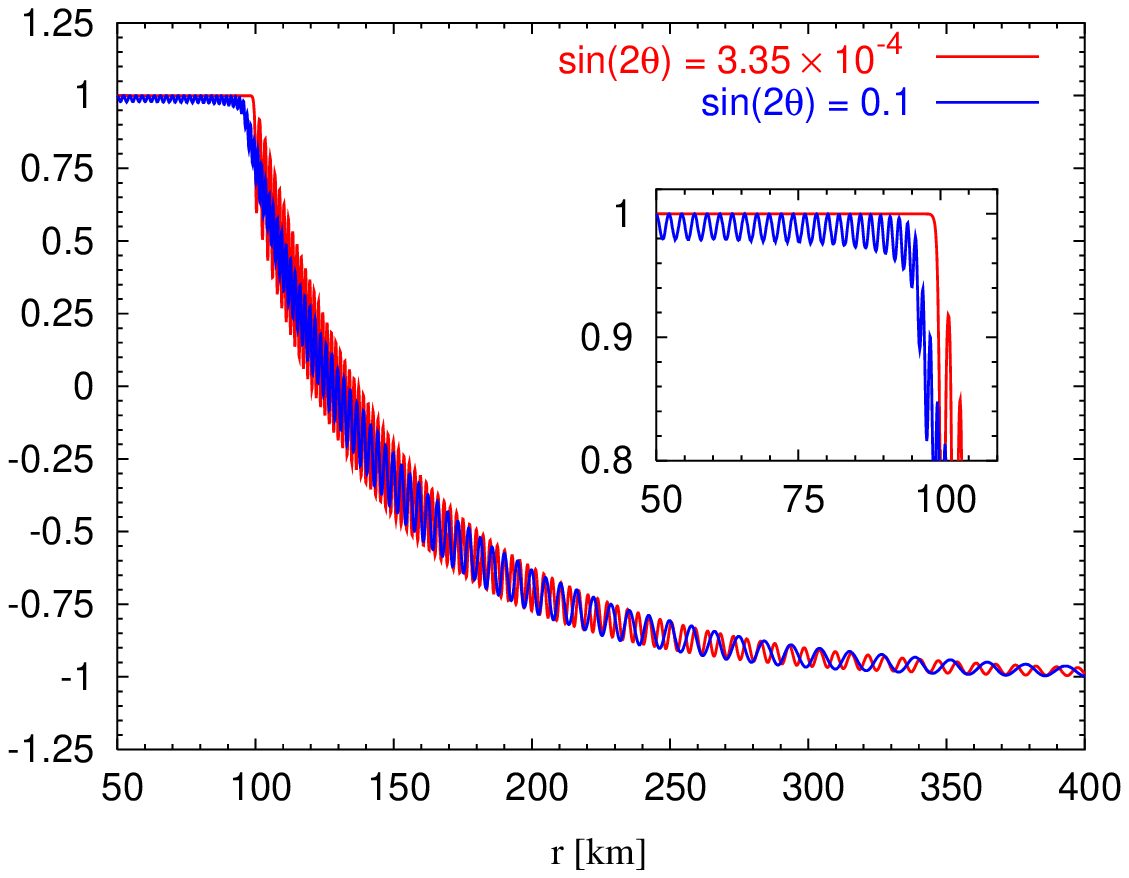}
\vskip10pt
\caption{\small Top: Same case as in Fig.~\ref{fig:SN2} but for large
  vacuum mixing angle of $\sin2\theta=0.1$ (blue in both panels),
  compared with different ways of suppressing the mixing angle
  (red). Left: ordinary matter according to the profile at $t=2\,$s in
  bottom panel, where typical profiles of Garching
  group~\cite{Arcones:2006uq} plus $\mu(r)$ are shown. Right: small
  vacuum mixing angle of
  $\sin2\theta=3.35\times10^{-4}$~\cite{EstebanPretel:2007ec}.\label{fig:matterangle}}
\end{minipage}
\end{figure}

Changing the vacuum mixing angle and adding normal matter causes only
the minor logarithmic changes discussed earlier.  We illustrate this
point in Fig.~\ref{fig:matterangle} with the evolution of $\bar P_z$
for the same case as in Fig.~\ref{fig:SN2}, but assuming now a large
vacuum mixing angle $\sin 2\theta=0.1$. In the synchronization region
one can now see oscillations. On the left panel of the top row, we
overlay this curve with $\bar P_z$, using a typical matter profile,
while on the right panel we superimpose the vacuum case with a reduced
mixing angle. This reduced mixing angle is chosen to be the effective
one at $r_{\rm syn}$ due to the matter profile used for the left
panel. As expected, matter has the effect of slightly delaying the
onset of pair transformations.

\subsection{Single-angle and multi-energy}\label{sec:single-multi}

In addition to the radial dependence of $\mu$ and the initial
neutrino-antineutrino asymmetry, another feature of SN neutrinos that
plays an important role in this analysis is their energy spectra. As
described in Chapter~\ref{chapter:supernova}, far from being
monochromatic, neutrinos are emitted from the star in a rather wide
energy range, typically from 1 to 50 MeV. We need then to study the
consequences of a multi-energy analysis.

\begin{figure}
\begin{center}
\includegraphics[angle=0,width=0.5\textwidth]{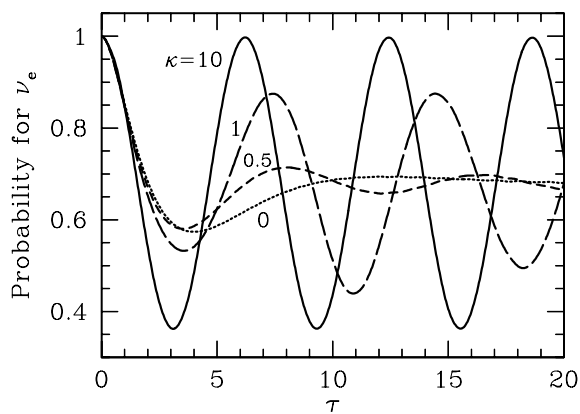}
\caption{\small Total $\nu_e$ survival probability as a function of
  time, where $\tau \equiv (\Delta m^2/2p_0)t$ and $p_0 = \langle
  p^{-1} \rangle^{-1} \simeq 2.2\,T$. The curves are for different
  values $\kappa\equiv\mu/(\Delta m^2/2p_0)$ of the neutrino
  self-coupling as indicated where $\kappa=0$ corresponds to vacuum
  oscillations~\cite{Pastor:2001iu}.}
\label{totalP}
\end{center}
\end{figure}

Let us start considering the evolution of an ensemble consisting of
only neutrinos with many momenta, $p_j$. In this case the general
EOMs, Eq.~(\ref{eq:eomsP}), can be written for each
mode as
\begin{equation}
  \partial_t{\bf P}_j = \left(\omega_j{\bf B} + \mu {\bf J}\right)\times{\bf P}_j\,,\label{eq:eommultiennu}
\end{equation}
analogously to Eq.~(\ref{eq:eomvac}), where $\omega_j \equiv \Delta
m^2/2p_j$. We have defined the polarization vector for the entire
ensemble as
\begin{equation}
{\bf J}\equiv\sum_{j=1}^{N_\nu} {\bf P}_j\,,
\end{equation}
$N_{\nu}$ being the number of neutrino modes. In the well known case
where we do not consider neutrino-neutrino interactions, or these are
too weak, each energy mode oscillates around ${\bf B}$ with its
particular frequency $\omega_j$. This behavior leads inevitably to a
quick decoherence. The effect is illustrated in Fig.~\ref{totalP} for
an ensemble of neutrinos with a thermal momentum distribution at
temperature $T$. The vacuum mixing angle was taken to be $\sin2 \theta
= 0.8$ and all neutrinos were originally in a pure $\nu_e$ state.

The situation changes, though, if we consider a sufficiently dense
neutrino ensemble~\cite{Pastor:2001iu}. In such a case all neutrino
modes oscillate together with a common $synchronized$ frequency. This
leads to a net oscillation effect for the whole neutrino ensemble, as
shown in Fig.~\ref{totalP}. In this case, we can no longer neglect the
second term in Eq.~(\ref{eq:eommultiennu}).

This effect can be intuitively understood by looking at the EOMs,
Eq.~(\ref{eq:eommultiennu}). We see two different oscillatory
components. On the one hand each individual mode precesses around
${\bf B}$ with a particular $\omega_j$, on the other hand all of them
precess around the global polarization vector ${\bf J}$ with a common
frequency. Of course, ${\bf J}$ is not a constant vector, but depends
on the behavior of ${\bf P}_j$. If we sum Eq.~(\ref{eq:eommultiennu})
over all modes we obtain
\begin{equation}
\dot{\bf J}=
{\bf B}\times\sum_{j=1}^{N_\nu}\omega_j{\bf P}_j.
\label{eq:motion1a}
\end{equation}
Therefore, if the precession frequency around ${\bf J}$ is much larger
than the one around ${\bf B}$, i.e.~if $\mu$ is sufficiently large and
changes sufficiently slowly (adiabatic limit~\cite{Raffelt:2007xt}),
the evolution of a given mode remains dominated by ${\bf J}$. The fast
precession around ${\bf J}$ implies that the projection of ${\bf P}_j$
on ${\bf J}$ is conserved while the transverse component averages to
zero on a fast time scale relative to the slow precession around ${\bf
  B}$. If ${\bf J}$ moves slowly, then the individual modes will
follow ${\bf J}$. Of course, if the external field is much larger than
the internal ones (dilute neutrino gas), then the modes will decouple
and precess individually around the external field with their separate
vacuum oscillation frequencies.

We can now derive the synchronization frequency that leads the
evolution of the compound system,
\begin{equation}
\dot{\bf J}=\omega_{\rm syn}\,{\bf B}\times{\bf J}.
\label{Jdot}
\end{equation}
Of the individual modes, the external field ``sees'' only the
projection along ${\bf J}$ because the transverse components average
to zero. One can define here an effective magnetic moment, ${\bf M}$,
of the system, obtaining that the contribution of mode $j$ to this
total magnetic moment is $\omega_j\,{\bf\hat J}\cdot{\bf P}_j$ so that
\begin{equation}
{\bf M}={\bf\hat J}\,\sum_{j=1}^{N_\nu}
\omega_j\,{\bf\hat J}\cdot{\bf P}_j,
\end{equation}
where ${\bf\hat J}$ is a unit vector in the direction of ${\bf J}$.
And we therefore obtain
\begin{equation}
\omega_{\rm syn}=\frac{|{\bf M}|}{|{\bf J}|}=
\frac{1}{|{\bf J}|}\sum_{j=1}^{N_\nu}
\omega_j\,{\bf\hat J}\cdot{\bf P}_j.
\label{eq:wsynchnu}
\end{equation}
In particular, if all modes started aligned (coherent flavor state)
then $|{\bf J}|=N_\nu$ and ${\bf\hat J}\cdot{\bf P}_j=1$ so that
\begin{equation}
  \omega_{\rm syn}=\left\langle\omega\right\rangle
  =\frac{1}{N_\nu}\sum_{j=1}^{N_\nu}\frac{\Delta m^2}{2p_j}=\frac{\Delta m^2}{2}\langle\frac{1}{p}\rangle.
\end{equation}

Let us study now the case where neutrinos and antineutrinos are
simultaneously present, with a continuum of energy modes. Therefore,
all the summations in the quantities defined previously will become
integrals. We can then define ${\bf J}=\int_0^\infty d\omega\,{\bf
  P}_\omega$ and $\bar{\bf J}=\int_0^\infty d\omega\,\bar{\bf
  P}_\omega$ and introduce ${\bf D} \equiv {\bf J}-\bar{\bf J}$,
representing the net lepton number. The equivalent EOMs to those of
Eq.~(\ref{eq:eomvac}) are now
\begin{eqnarray}\label{eq:eom1}
 \partial_t{\bf P}_\omega&=&\left(+\omega{\bf B} +\mu {\bf D}\right)\times{\bf P}_\omega\,,
 \nonumber\\*
 \partial_t\bar{\bf P}_\omega&=&\left(-\omega{\bf B} +\mu {\bf D} \right)
 \times\bar{\bf P}_\omega\,.
\end{eqnarray}
As discussed earlier, the only difference between the equations for
neutrinos and antineutrinos is the sign in the vacuum term. This
suggests the use of a more compact notation, where we use only ${\bf
  P}_\omega$ by extending it to negative frequencies such that
$\bar{\bf P}_{\omega} = {\bf P}_{-\omega}$ ($\omega > 0$). The EOMs
take then the very same form as Eq.~(\ref{eq:eommultiennu})
\begin{equation}\label{eq:eom1a}
 \dot{\bf P}_\omega=\left(\omega{\bf B}
 +\mu {\bf D}\right)\times{\bf P}_\omega\,.
\end{equation}
The difference vector is then redefined as
\begin{equation}
\label{def-d}
{\bf D}= \int_{-\infty}^{+\infty}d\omega\,s_\omega\,{\bf P}_\omega,
\end{equation}
where $s_\omega \equiv {\rm
  sign}(\omega)=\omega/|\omega|$. Integrating both sides of
Eq.~(\ref{eq:eom1a}) over $s_\omega d\omega$ provides
\begin{equation}\label{eq:eomD}
 \dot{\bf D} = {\bf B} \times{\bf M}\,,
\end{equation}
where
\begin{equation}
{\bf M} \equiv
 \int_{-\infty}^{+\infty} d\omega\, s_\omega \omega\,
 {\bf P}_\omega
\end{equation}
is again the effective magnetic moment of the system.

We have obtained a completely analogous situation to the neutrino-only
case. Here both neutrinos and antineutrinos behave in the same way if
$\mu$ is sufficiently large, and therefore ${\bf M} \propto {\bf
  D}$. As a consequence, and making use of Eq.~(\ref{eq:eomD}), we see
how ${\bf D}$ precesses around ${\bf B}$ with the synchronization
frequency
\begin{equation}\label{eq:wsynch}
 \omega_{\rm syn}=\frac{|{\bf M}|}{|{\bf J}|}=
 \frac{\int_{-\infty}^{+\infty} d\omega\,s_\omega\omega\,P_\omega}
 {\int_{-\infty}^{+\infty} d\omega\,P_\omega}\,,
\end{equation}
which is just the generalization of Eq.~(\ref{eq:wsynchnu}).

So, just as in the neutrino-only case, we obtain that a multi-energy
ensemble behaves as if all neutrinos had a common vacuum oscillation
frequency given by Eq.~(\ref{eq:wsynch}). Therefore the evolution of a
neutrino and antineutrino ensemble with an energy spectrum will be
basically the same discussed in Sec.~\ref{sec:single-single}, with a
vacuum oscillation frequency $\omega_{\rm syn}$.

\begin{figure}
\begin{center}
\includegraphics[angle=0,width=0.5\textwidth]{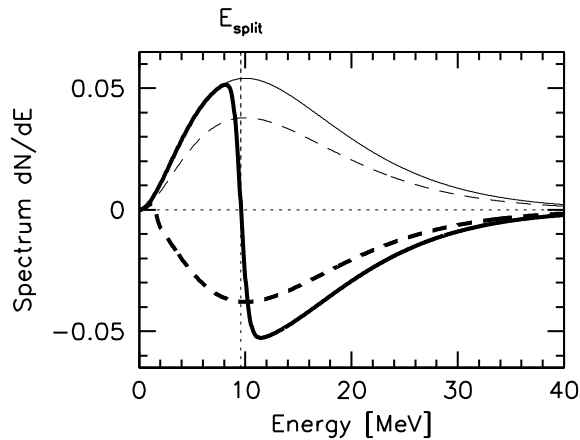}
\caption{\small Neutrino spectra at the neutrino sphere (thin lines)
  and beyond the dense-neutrino region (thick lines) for the schematic
  SN model described in the text. Solid: neutrinos. Dashed:
  antineutrinos. Positive spectrum: electron (anti)neutrinos.
  Negative spectrum: $x$ (anti)neutrinos.\label{fig:snspectrum}}
\end{center}
\end{figure}

After this generic discussion on the multi-energy case, let us
concentrate again on the specific scenario of interest, a SN with
varying $\mu$ and flavor asymmetry $\epsilon$. Although the evolution
of the system is basically the same discussed for the single-energy
situation, as we have just deduced, there is one special feature
characteristic of such a system, namely the spectral split. This
effect was first numerically observed by Duan et
al.~\cite{Duan:2006an,Duan:2006jv} and later on discussed in the
context of two flavors by Raffelt and
Smirnov~\cite{Raffelt:2007cb,Raffelt:2007xt}, Fogli et
al.~\cite{Fogli:2007bk}, and in three flavors by Duan et
al.~\cite{Duan:2008za} and Dasgupta et al.~\cite{Dasgupta:2008cd}. The
main result is shown in Fig.~\ref{fig:snspectrum}, taken from
Ref.~\cite{Raffelt:2007xt}. Represented with thin lines are the
thermal $\nu_e$ and $\bar\nu_e$ flux spectra with an average energy of
15~MeV produced in the neutrino sphere of a SN core. The mean energies
are chosen to be equal while the $\nu_e$ flux is 25$\%$ larger than
the $\bar\nu_e$ flux and the other fluxes are completely ignored. With
thick lines are shown the emerging spectra, once neutrino-neutrino
interactions can be neglected. On the one hand we observe how the
antineutrino spectra (dashed line) has completely flipped, converting
to the x-flavor. On the other hand a twofold behavior is observed for
neutrinos (solid), all of them above a characteristic energy ($E_{\rm
  split}$) convert also to the x-flavor, while the ones below $E_{\rm
  split}$ stay in their original state.

To understand this result we will follow the description given in
Ref.~\cite{Raffelt:2007xt}. We rewrite the EOMs for this system in
terms of an effective Hamiltonian
\begin{equation}
\label{eq:eom2}
 \partial_t{\bf P}_\omega={\bf H}_\omega\times{\bf P}_\omega\,,
\end{equation}
where
\begin{equation}
\label{eq:eom2_H}
 {\bf H}_\omega = \omega{\bf B} +\mu {\bf D}.
\end{equation}
As described previously, in the adiabatic limit ${\bf P}_\omega$ will
precess around ${\bf H}_\omega$ following at the same time its
movement. We assume the usual situation in which all neutrinos are
initially in the same state, all of ${\bf P}_\omega$ pointing in the
direction of ${\bf H}_\omega$ due to the large initial
$\mu$. Therefore, in the adiabatic limit they stay aligned with ${\bf
  H}_\omega$ for the entire evolution:
\begin{equation}\label{eq:hialigned}
 {\bf P}_\omega (\mu) = \hat{\bf H}_\omega(\mu)\,P_\omega\,,
\end{equation}
which solves the EOMs. Here $P_\omega \equiv |{\bf P}_\omega|$ and
$\hat{\bf H}_\omega \equiv {\bf H}_\omega/|{\bf H}_\omega|$ is a unit
vector. As usual we assume an excess flux of neutrinos over
antineutrinos, implying that initially ${\bf P}_\omega$ and ${\bf D}$
are collinear and $D_z>0$.

In the adiabatic limit then, all ${\bf P}_\omega$ will have the same
direction as ${\bf H}_\omega$, which in turn will be confined to the
${\bf B}-{\bf D}$ plane, as given in Eq.~(\ref{eq:eom2_H}). As a
consequence ${\bf M}$ will also stay in that plane, allowing us to
redefine it as
\begin{equation}\label{decompose}
 {\bf M} = b\, {\bf B} + \omega_{\rm c} {\bf D}
\end{equation}
and rewrite the EOM of Eq.~(\ref{eq:eomD}) as
\begin{equation}
\label{eq:eomD2}
\partial_t{\bf D} = \omega_{\rm c}\,{\bf B} \times{\bf D}.
\end{equation}

The system is therefore evolving simultaneously in two ways: a fast
precession around ${\bf B}$ determined by $\omega_{\rm c} =
\omega_{\rm c}(\mu)$ and a drift in the co-rotating plane caused by
the explicit $\mu(t)$ variation. If we go to the co-rotating frame,
the individual Hamiltonians become
\begin{equation}\label{eq:Hi}
 {\bf H}_\omega=(\omega-\omega_{\rm c})\,{\bf B}
 +\mu {\bf D}\,.
\end{equation}

We can now see how the system evolves with the decrease of
$\mu$. Initially ($\mu\to\infty$) the oscillations are synchronized,
$\omega_{\rm c}^{\infty}=\omega_{\rm syn}$, and all ${\bf P}_\omega$
form a collective~${\bf P}$. As $\mu$ decreases, the ${\bf P}_\omega$
zenith angles spread out while remaining in a single co-rotating
plane. In the end ($\mu\to 0$) the co-rotation frequency is
$\omega_{\rm c}^0$ and Eqs.~(\ref{eq:hialigned}) and~(\ref{eq:Hi})
imply that all final ${\bf H}_\omega$ and therefore all ${\bf
  P}_\omega$ with $\omega > \omega_{\rm c}^0$ are aligned with ${\bf
  B}$, the others anti-aligned: a spectral split is inevitable with
$\omega_{\rm split}\equiv\omega_{\rm c}^0$ being the split
frequency. The lengths $P_\omega=|{\bf P}_\omega|$ are conserved and
eventually all ${\bf P}_\omega$ point in the $\pm{\bf B}$
directions. Therefore the conservation of flavor-lepton number gives
us $\omega_{\rm split}$, for $D_z>0$, by virtue of
\begin{equation}\label{spl}
 D_z=\int_{-\infty}^0 P_\omega\,d\omega
 -\int_{0}^{\omega_{\rm split}}P_\omega\,d\omega
 +\int_{\omega_{\rm split}}^{+\infty} P_\omega\,d\omega\,.
\end{equation}
In general, $\omega_{\rm split}=\omega_{\rm c}^0 \neq \omega_{\rm
c}^\infty=\omega_{\rm syn}$.

It appears naturally from this explanation that the spectral split
will take place only for neutrinos or antineutrinos, depending on
which of them has the largest flux. As a particular case, there would
be no spectral split for equal neutrino and antineutrino fluxes.

\section{Multi-angle analysis. Equations of motion}

As described in the introduction, the question we want to address in
this chapter is the consequences of the multi-angle analysis in the
evolution of SN neutrinos, for both single and multi-energy
cases. This kind of neutrinos are not emitted from a single point, but
from the neutrino sphere, which can be approximated as a spherical
surface. The EOMs will therefore be affected by the geometry of the
problem, including an angular dependence.

Our fundamental quantities are the flux matrices in flavor space ${\sf
  J}_r$ that depend on the radial coordinate $r$
(Appendix~A). The diagonal entries represent the total
neutrino number fluxes through a sphere of radius~$r$. In the absence
of oscillations, ${\sf J}_r$ would not depend on the radius at
all. The flux matrices are represented by polarization vectors ${\bf
  P}_r$ in the usual way,
\begin{eqnarray}\label{eq:pol}
 {\sf J}_r&=&
 \frac{F^{R_\nu}_{\nu_e}+F^{R_\nu}_{\nu_x}}{2}
 +\frac{F^{R_\nu}_{\bar\nu_e}-F^{R_\nu}_{\bar\nu_x}}{2}\,
 {\bf P}_r\cdot\bm{\sigma}\,,
 \nonumber\\
 \bar{\sf J}_r&=&
 \frac{F^{R_\nu}_{\bar\nu_e}+F^{R_\nu}_{\bar\nu_x}}{2}
 +\frac{F^{R_\nu}_{\bar\nu_e}-F^{R_\nu}_{\bar\nu_x}}{2}\,
 \bar{\bf P}_r\cdot\bm{\sigma}\,,
\end{eqnarray}
where $\bm\sigma$ is the vector of Pauli matrices. The number fluxes
$F^{R_\nu}_{\nu}$ are understood at the neutrino sphere, as always. In
both equations the term proportional to the polarization vector is
normalized to the antineutrino flux. As a consequence, at the neutrino
sphere we have the normalization
\begin{equation}
 P=|{\bf P}|=1+\epsilon
 \hbox{\quad and\quad}
 \bar P=|\bar{\bf P}|=1\,.
\end{equation}
where $\epsilon$ is the asymmetry parameter defined in
Eq.~(\ref{eq:epsdefine}). In this way, we treat the excess flux from
deleptonization as an adjustable parameter without affecting the
baseline flux of antineutrinos,.

The diagonal part of the flux matrices is conserved and irrelevant for
flavor oscillations. The polarization vector ${\bf P}_r$ only captures
the {\it difference\/} between the flavor fluxes. For this reason we
have defined the asymmetry $\epsilon$ in terms of the flux
differences.

\begin{figure}
\begin{center}
\includegraphics[angle=0,width=0.8\textwidth]{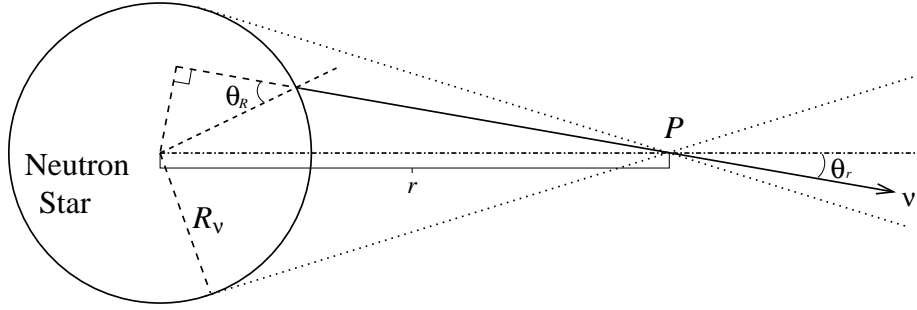}
\caption{\small Schematic neutrino emission in a spherical symmetric
  system. The solid line represents a neutrino emitted from the
  neutrino sphere ($R_\nu$) at an angle $\theta_R$ relative to the
  normal direction. This neutrino intersects the radial axis at point
  $P$ at distance $r$ from the center of the star, forming an angle
  $\theta_r$ with the axis. The system therefore satisfies
  $R_\nu\sin\theta_R = r\sin\theta_r$. Point $P$ sees only neutrinos
  traveling within the cone delimited by the dotted lines. Figure
  adapted from~\cite{Duan:2006an}. \label{fig:sphegeom}}
\end{center}
\end{figure}

We want to study here multi-angle effects. For the geometry we are
considering, it is convenient to label the different angular modes
with
\begin{equation}
u=\sin^2\vartheta_R\,,
\end{equation}
where $\vartheta_R$ is the zenith angle at the neutrino sphere $r=R_\nu$
of a given mode relative to the radial direction, see
Fig.~\ref{fig:sphegeom}. The parameter $u$ is fixed for every
trajectory whereas the physical zenith angle $\vartheta_r$ at distance
$r$ varies. Therefore, using the local zenith angle to label the modes
would complicate the equations.

We will consider two generic angular distributions for the modes. In
the multi-angle case we assume that the neutrino radiation field is
``half isotropic'' directly above the neutrino sphere, i.e., all
outward moving modes are equally occupied as expected for blackbody
emission. This implies (Appendix~A)
\begin{equation}
{\bf P}_{u,r}=\D {\bf P}_{r}/\D u=  {\rm const.}
\end{equation}
at $r=R_\nu$ for $0\leq u\leq 1$. Note that $u=0$ represents radial
modes, $u=1$ tangential ones. The other generic distribution is the
single-angle case where all neutrinos are taken to be launched at
$45^\circ$ at the neutrino sphere so that $u=1/2$ for all
neutrinos. This is a very interesting case, because as we will later
see it seems to catch the whole physics of the problem. Moreover, from
a more practical point of view, this approximation saves a lot of
numerical effort.

For a monochromatic energy distribution, the EOMs in spherical
symmetry are (Appendix~A)
\begin{eqnarray}\label{eq:eom5}
 \partial_r{\bf P}_{u,r}&=&
 +\frac{\omega {\bf B}\times{\bf P}_{u,r}}{v_{u,r}}
 +\frac{\lambda_r{\bf L}\times{\bf P}_{u,r}}{v_{u,r}}
 +\nonumber\\*
 &&\mu_0\,\frac{R_\nu^2}{r^2}
 \left[\left(\int_0^1\D u'\,
 \frac{{\bf P}_{u',r}-\bar{\bf P}_{u',r}}{v_{u',r}}\right)
 \times\left(\frac{{\bf P}_{u,r}}{v_{u,r}}\right)
 -({\bf P}_r-\bar{\bf P}_r)\times{\bf P}_{u,r}\right]\,,
 \nonumber\\*
 \partial_r\bar{\bf P}_{u,r}&=&
 -\frac{\omega {\bf B}\times\bar{\bf P}_{u,r}}{v_{u,r}}
 +\frac{\lambda_r{\bf L}\times\bar{\bf P}_{u,r}}{v_{u,r}}
 +\\*
 &&\mu_0\,\frac{R_\nu^2}{r^2}
 \left[\left(\int_0^1\D u'\,
 \frac{{\bf P}_{u',r}-\bar{\bf P}_{u',r}}{v_{u',r}}\right)
 \times\left(\frac{\bar{\bf P}_{u,r}}{v_{u,r}}\right)
 -({\bf P}_r-\bar{\bf P}_r)\times\bar{\bf P}_{u,r}\right]\,,\nonumber
\end{eqnarray}
where the radial velocity of mode $u$ at radius $r$ is
\begin{equation}
v_{u,r}=\sqrt{1-u\,R_\nu^2/r^2}\,.
\end{equation}



The radial velocity in Eq.~(\ref{eq:eom5}) introduces a relative
strength factor for the different angular modes. As a consequence, one
would expect two sources of kinematical decoherence, the
self-interaction term and the matter term. We do not consider the
vacuum term as a possible source for decoherence because by the time
this term is dominant neutrinos are basically collinear. It seems
important then to study the self-induced neutrino decoherence and the
role of dense matter in collective SN neutrino transformations.

\pagestyle{decoherence}
\section{Decoherence in supernova neutrino transformations suppressed
  by deleptonization}\label{sec:decoherence}

The current-current nature of the weak interaction causes the
interaction energy to depend on $(1-\cos\theta)$ for two trajectories
with relative angle $\theta$. Therefore, neutrinos emitted in
different directions from a SN core experience different refractive
effects~\cite{Duan:2006an, Duan:2006jv}. As a result, one would expect
that their flavor content evolves differently, leading to kinematical
decoherence between different angular modes~\cite{Sawyer:2005jk}. In
the SN context, however, it has been numerically observed that the
evolution is similar to the single-angle (or the isotropic)
case~\cite{Duan:2006an, Duan:2006jv}. We analyze here up to which
extent this approximation is valid and study the role of the different
model parameters on the decoherence coming from multi-angle effects.

In Ref.~\cite{Raffelt:2007yz} it was shown that this multi-angle
decoherence is indeed unavoidable in a ``symmetric gas'' of equal
densities of neutrinos and antineutrinos. Moreover, this effect is
self-accelerating in that an infinitesimal anisotropy is enough to
trigger an exponential run-away towards flavor equipartition, both for
the normal and inverted hierarchies. Therefore, the observed
suppression of multi-angle decoherence must be related to the
$\nu_e\bar\nu_e$ asymmetry that is generated by SN core
deleptonization.

\begin{figure}
\begin{center}
\includegraphics[width=0.95\textwidth]{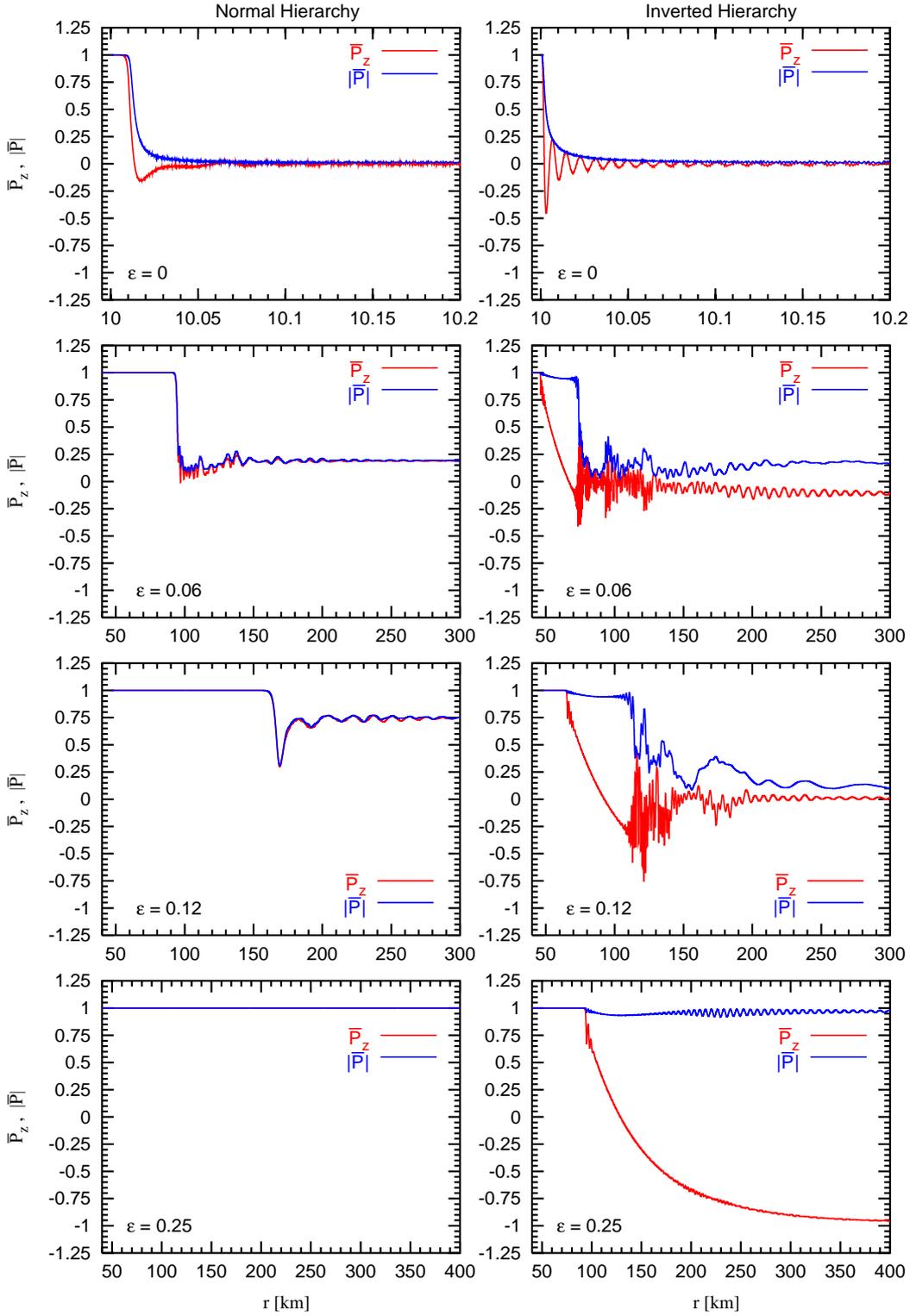}
\caption{\small Radial evolution of $\bar P_z$ in a schematic SN model
  as in Fig.~\ref{fig:SN2}, but now for multi-angle neutrino emission
  at the neutrino sphere ($R_\nu=10$~km). In addition we show the
  length $\bar P=|\bar {\bf P}|$ as a measure of kinematical
  coherence. Left: normal hierarchy. Right: inverted hierarchy.  From
  top to bottom: $\epsilon=0$, 0.06, 0.12 and~0.25, where $\epsilon$
  is defined in
  Eq.~(\ref{eq:epsdefine})~\cite{EstebanPretel:2007ec}.\label{fig:example2}}
\end{center}
\end{figure}

To illustrate this point we show in Fig.~\ref{fig:example2} a few
examples along the lines of Fig.~\ref{fig:SN2}, but now for
multi-angle emission from the neutrino sphere that is again taken at
10~km. We consider different values of the asymmetry parameter. The
left panels are for the normal hierarchy, the right panels for the
inverted hierarchy. Since $P_z(r)-\bar P_z(r)=\epsilon$ is constant,
it is sufficient to show $\bar P_z(r)$ alone. However, the length
$\bar P=|\bar {\bf P}|$ is no longer preserved: Complete kinematical
decoherence among the angular modes would cause $\bar P=0$. On the
other hand, if $\bar P=1$ remains fixed, this signifies that all modes
evolve coherently with each other. We use $\bar{\bf P}$ rather than
${\bf P}$ because the former measures what happens to the
$\nu_e\bar\nu_e$ pairs, whereas the latter also includes the conserved
$\nu_e$~excess.

In the top row we use $\epsilon=0$ (symmetric case). The flavor
content decoheres quickly in agreement with
Ref.~\cite{Raffelt:2007yz}. Both the length and the $z$-components of
${\bf P}$ and $\bar{\bf P}$ shrink to zero within about 20~meters of
the nominal neutrino sphere.

On the other extreme, we show in the bottom row the same for
$\epsilon=0.25$. In the normal hierarchy, nothing visible happens,
in analogy to the single-angle case.  In the inverted hierarchy, the
transformation is similar, but not identical, to the single-angle
case. The nutations wash out quickly. Shortly after exiting from the
synchronization phase, the length $\bar P$ shrinks a bit, but stays
almost constant thereafter. Clearly, some sort of multi-angle effect
has happened as we will discuss further in
Sec.~\ref{sec:coh-decoh}, but multi-angle decoherence has
certainly not occurred.

In the two middle rows we show intermediate cases with
$\epsilon=0.06$ and 0.12, respectively. For the inverted hierarchy,
these examples are qualitatively equivalent. The evolution is at first
similar to the single-angle case and analogous to $\epsilon=0.25$.
The nutations are washed out and the length $\bar P$ shrinks a
little bit after the synchronization radius. At some larger radius,
however, something new happens in that $\bar P$ suddenly shrinks
significantly, although not to zero, and there is a distinct feature
in the evolution of the $z$-component. Now we obtain partial
decoherence. The final flavor content is very different from the
single-angle case.

In the normal hierarchy, and for $\epsilon=0.06$, we obtain large
decoherence that begins abruptly at some radius far beyond $r_{\rm
  syn}$. For the larger asymmetry $\epsilon=0.12$, the length $\bar P$
also shrinks, but closely tracks $\bar P_z$. As we will see, this case
is somewhat like the second stage of the inverted-hierarchy case,
i.e., a certain amount of shrinking of the length of $\bar P$ and thus
a clear multi-angle effect, but no real decoherence.

Depending on the deleptonization flux, here represented by the
asymmetry parameter $\epsilon$, the system behaves very differently.
In particular, for the inverted hierarchy it is striking that there
are either two or three distinct phases. We always have the initial
synchronized phase at large neutrino densities. Next, there is
always the quasi single-angle pair-transformation phase at distances
larger than $r_{\rm syn}$. Just beyond this radius, the global
polarization vectors quickly shrink by a small amount, but then
stabilize immediately. Finally, if $\epsilon$ is below some critical
value, there is a sharp transition to a third phase where the
different angular modes decohere significantly, but not completely.
The practical outcome for the flavor fluxes emerging from the
dense-neutrino region is very different depending on $\epsilon$. The
transition between these regimes is abrupt, a small change of
$\epsilon$ is enough to cause one or the other form of behavior.

\subsection{Setup of the problem}                       \label{sec:setup}

In the previous section we have presented the different regimes of
evolution due to multi-angle effects. Let us now try to understand
this behavior.\vspace{0.5cm}\\
\textbf{A) Schematic supernova model}\vspace{0.2cm}\\
We will here consider a two-flavor oscillation scenario driven by the
atmospheric $\Delta m^2=1.9$--$3.0\times10^{-3}~{\rm eV}^2$.  Assuming
$\langle E_\nu\rangle=15$~MeV, the oscillation frequency is
$\omega=\hbox{0.3--0.5}~{\rm km}^{-1}$. To be specific, we use
\begin{equation}\label{eq:omegadefine}
\omega=\left\langle\frac{\Delta m^2}{2E}\right\rangle=
0.3~{\rm km}^{-1}
\end{equation}
as a benchmark value in the monochromatic model.

The total energy output of a SN is around $3\times10^{53}~{\rm
erg}$, corresponding to $0.5\times10^{53}~{\rm erg}$ in each of the
six neutrino species if we assume approximate equipartition of the
emitted energy. If this energy is emitted over 10~s, the average
luminosity per flavor would be $0.5\times10^{52}~{\rm erg/s}$.
However, at early times during the accretion phase, the luminosity
in the $\bar\nu_e$ flavor can exceed $3\times10^{53}~{\rm
erg/s}$~\cite{Raffelt:2003en}. As our baseline estimate we use
\begin{equation}\label{eq:muestimate}
  \mu_0=7\times10^5~{\rm km}^{-1} \times
  \left(\frac{L_{\bar\nu_e}}{\langle E_{\bar\nu_e}\rangle}
    -\frac{L_{\bar\nu_x}}{\langle E_{\bar\nu_x}\rangle}\right)
  \frac{15~{\rm MeV}}{10^{52}~{\rm erg/s}}\,
  \left(\frac{10~{\rm km}}{R_\nu}\right)^2\,.
\end{equation}
Unless otherwise stated, throughout this section we always use the
benchmark values for the different parameters summarized in
Table~\ref{tab:benchmark}.
\begin{table}
\caption{\small Default values for our model parameters.}
\label{tab:benchmark}
\begin{center}
\begin{tabular}{lll}
\hline
\hline
Parameter&Standard value&Definition\\
\hline
$\epsilon$&0.25&Eq.~(\ref{eq:epsdefine})\\
$\mu_0$&$7\times10^5~{\rm km}^{-1}$&Eq.~(\ref{eq:mudefine})\\
$\omega$&$0.3~{\rm km}^{-1}$&Eq.~(\ref{eq:omegadefine})\\
$\sin2\theta$&$10^{-3}$&---\\
\hline
\hline
\end{tabular}
\end{center}
\end{table}
In our calculations we always take the neutrino sphere at the radius
$R_\nu=10$~km. Of course, the physical neutrino sphere is not a
well-defined concept. Therefore, the radius $R_\nu$ simply represents
the location where we fix the inner boundary condition. However,
essentially nothing happens until the synchronization radius $r_{\rm
syn}\gg R_\nu$ because the in-medium mixing angle is extremely small
and both neutrinos and antineutrinos simply precess around~${\bf
B}$. Therefore, as far as the vacuum and matter oscillation terms
are concerned, it is almost irrelevant where we fix the inner
boundary condition.

Not so for the neutrino-neutrino term because we also fix the
angular distribution at $r=R_\nu$. While the $r^{-2}$ scaling
from flux dilution is unaffected by the radius for the inner
boundary condition, the ``collinearity suppression'' also scales as
$(R_\nu/r)^{2}$ for $r\gg R_\nu$. If we fix a half-isotropic distribution or
a single angle of $45^\circ$ at a larger radius $R_\nu'$, the new inner
boundary condition essentially amounts to
$\mu_0\to\mu'_0=\mu_0\,(R_\nu'/R_\nu)^2$. In the early phase after
bounce $R_\nu'=30$~km could be more realistic, leading to a $\mu_0$ value
almost an order of magnitude larger. Evidently, $\mu_0$ is a rather
uncertain model parameter that can differ by orders of magnitude
from our benchmark value.

However, collective pair conversions only begin at $r_{\rm syn}$
where $\mu$ is so small that synchronization ends. Therefore, the main
impact of a modified $\mu_0$ is to change $r_{\rm syn}$ and thus to
push the collective pair conversions to larger radii. In any event,
according to Eq.~(\ref{eq:r_synch}) if $\mu_0$ is taken to be uncertain
by two orders of magnitude, $r_{\rm syn}$ only changes by a factor
of 3.

The total electron lepton number emitted from a collapsed SN core is
about $3\times10^{56}$. On the other hand, assuming that each neutrino
species carries away $0.5\times10^{53}~{\rm erg}$ with an average
energy of 15~MeV, the SN core emits about $2\times10^{57}$ neutrinos
in each of the six species. In this simplified picture, the SN emits
on average about 15\% more $\nu_e$ than $\bar\nu_e$.  However, in the
oscillation context we need the excess of
$F^{R_\nu}_{\nu_e}-F^{R_\nu}_{\nu_x}$ relative to the same quantity
for antineutrinos as defined in Eq.~(\ref{eq:epsdefine}). The true
value of $\epsilon$ thus depends sensitively on the detailed fluxes
and spectra of the emitted neutrinos. The asymmetry parameter is large
when the first hierarchy in $F^{R_\nu}_{\nu_e} > F^{R_\nu}_{\bar\nu_e}
> F^{R_\nu}_{\bar\nu_x}=F^{R_\nu}_{\nu_x}$ is large and/or the second
hierarchy is small. Even if $F^{R_\nu}_{\bar\nu_x}$ is as small as
half of $F^{R_\nu}_{\bar\nu_e}$, the asymmetry $\epsilon$ would be as
large as 30\%, even when $F^{R_\nu}_{\nu_e}$ exceeds
$F^{R_\nu}_{\bar\nu_e}$ by only
15\%.\vspace{0.5cm}\\
\textbf{B) Numerical multi-angle decoherence and the inner boundary
condition}\vspace{0.2cm}\\
One important and somewhat confusing complication of numerically
solving the EOMs is the phenomenon of numerical multi-angle
decoherence. In order to integrate Eq.~(\ref{eq:eom5}) one needs to
work with a finite number of angular modes, equivalent to
coarse-graining the phase space of the system. If the number of
angular bins is chosen smaller than some critical number $N_{\rm
  min}$, multi-angle decoherence occurs for $r<r_{\rm syn}$, where
physically it is not possible and does not occur for a fine-grained
calculation. This phenomenon is shown, for example, in Fig.~3 of
Ref.~\cite{Duan:2006an}. It is not caused by a lack of numerical
precision, but a result of the coarse-graining of phase space. A
related phenomenon is recurrence as discussed in the context of
multi-angle decoherence in Ref.~\cite{Raffelt:2007yz}.

In other words, a coarsely grained multi-angle system behaves
differently than a finely grained one. A smaller mixing angle
reduces $N_{\rm min}$, a larger neutrino-neutrino interaction
strength increases it.

Starting the integration at $r=R_\nu$ is doubly punishing because the
fast oscillations of individual modes caused by a large $\mu$
requires many radial steps for the numerical integration and
avoiding numerical decoherence requires a large number of angular
modes. On the other hand, in this region nothing but fast
synchronized oscillations take place that have no physical effect if
the mixing angle is small. Using a larger radius as a starting point
for the integration avoids both problems and does not modify the
overall flavor evolution at larger distances.

From the physical perspective, the ``neutrino sphere'' is not a
well-defined concept because different energy modes and different
species decouple at different radii, and in any case, each
individual neutrino scatters last at a different radius. If the
exact inner boundary condition would matter, we would need to solve
the full kinetic equations, including neutral-current and
charged-current collisions. It is the beauty of the
neutrino-neutrino flavor transformation problem that the real action
begins at $r_{\rm syn}$, significantly outside the neutrino
sphere. Our approach of reducing the equations of motion to the
refractive terms is only self consistent because the exact location
of the inner boundary condition is irrelevant.

In summary, the nominal neutrino sphere at $R_\nu=10$~km is nothing but
a point of reference where we normalize the fluxes and fix the
angular distribution. As a starting point for integration we
typically use $r_0=0.75\,r_{\rm syn}$. A few hundred angular modes
are then usually enough to avoid numerical decoherence.

We note, however, that the normal-hierarchy cases are more sensitive
to both the number of angular bins and the starting radius for the
integration. It can happen that a case that looks like the
$\epsilon=0.12$ example in Fig.~\ref{fig:example2}, which shows a
mild shrinking of the polarization vector, can become ``more
coherent'' by choosing a smaller starting radius which then may also
require a larger number of modes. For the normal hierarchy, the
different multi-angle cases are less cleanly separated from each other
than in the inverted hierarchy in that the transition is less abrupt
as a function of $\epsilon$.

When physical multi-angle decoherence occurs (e.g.~the middle rows
of Fig.~\ref{fig:example2}), a much larger number of modes is needed
to provide reproducible results. However, we are here not interested
in the exact final outcome, we are mostly interested in the range of
parameters that lead to decoherence. Therefore, massive computation
power is not needed for our study.

For those cases where we include a non-trivial spectrum of energies we
also need energy bins. A distribution of energies does not lead to
kinematical decoherence in the context of collective neutrino
oscillations~\cite{Hannestad:2006nj} so that the number of energy bins
is not a crucial parameter. Of course, to resolve the energy-dependent
behavior and especially the spectral splits, a sufficiently
fine-grained binning is required. It provides better resolution, but
not a qualitatively different form of behavior.

\subsection{Coherent evolution vs.\ decoherence}
\label{sec:coh-decoh}

\begin{figure}
\begin{center}
\includegraphics[width=\textwidth]{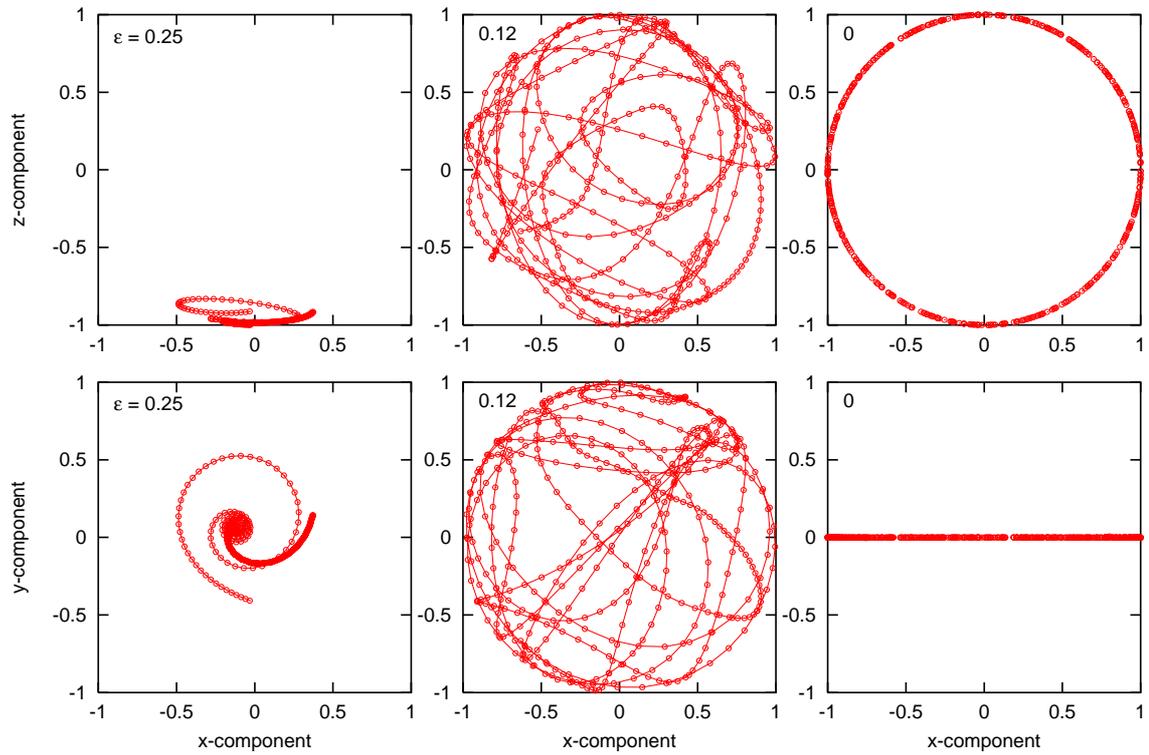}
\caption{\small Final location on the unit sphere of 500 antineutrino
  polarization vectors for our standard parameters and the inverted
  hierarchy. The top row is the ``side view'' ($x$-$z$-components),
  the bottom row the ``top view'' ($x$-$y$-components). Left:~quasi
  single-angle case ($\epsilon=0.25$). Middle: decoherent case
  ($\epsilon=0.12$). Right: symmetric system
  ($\epsilon=0$)~\cite{EstebanPretel:2007ec}.\label{fig:footprints1}}
\end{center}
\end{figure}

\textbf{A) Different forms of evolution}\vspace{0.2cm}\\
Before investigating the conditions for decoherence among angular
neutrino modes we first take a closer look at what happens in
the different cases shown in Fig.~\ref{fig:example2}. Considering
first the quasi single-angle case with the asymmetry
$\epsilon=0.25$, some insight is gained by looking at the final
state of the evolution at some large radius where the
neutrino-neutrino effects have completely died out and all modes
simply perform vacuum oscillations. In the left-hand panels of
Fig.~\ref{fig:footprints1} we show the end state of 500~polarization
vectors, representing modes uniformly spaced in the
angular coordinate $u$. In the upper
panel we show the final state in the $x$-$z$-plane (``side view''),
in the lower panel in the $x$-$y$-plane (``top view'').

Initially, all polarization vectors are aligned in the flavor
direction. At the beginning of the pair transformation phase at
$r_{\rm syn}$, some are peeled off, forming a spiral structure
that is easily gleaned from the left panels of
Fig.~\ref{fig:footprints1}. This structure continues to evolve
almost as in the single-angle case, i.e., once established it moves
almost like a rigid body and eventually orients itself in the
negative ${\bf B}$-direction. Of course, it continues to rotate
around the ${\bf B}$ direction even at large radii because of vacuum
oscillations.

\begin{figure}
\begin{center}
\includegraphics[width=\textwidth]{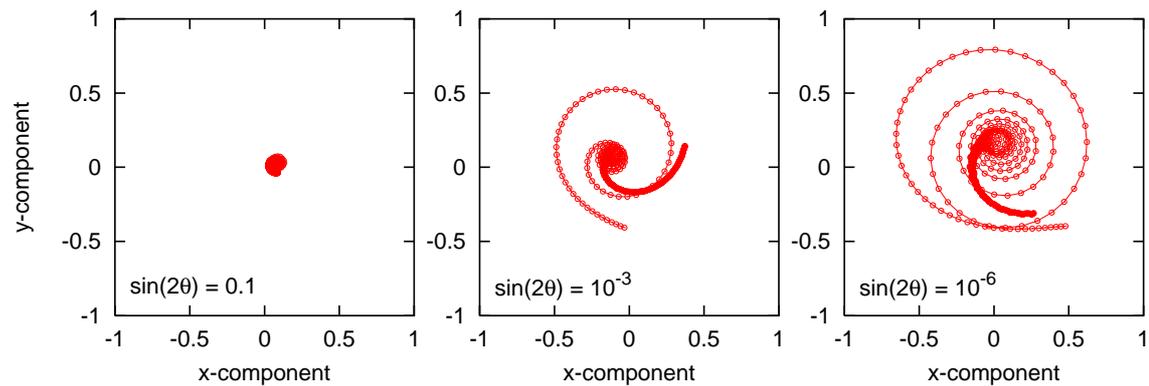}
\caption{\small Same as Fig.~\ref{fig:footprints1}, now only top views for
  quasi single-angle cases with the mixing angles $\sin2\theta=0.1$,
  $10^{-3}$ and $10^{-6}$ from left to right. The middle panel is
  identical with the bottom left panel of
  Fig.~\ref{fig:footprints1}~\cite{EstebanPretel:2007ec}.\label{fig:footprints4}}
\end{center}
\end{figure}

The spiral structure is different depending on the mixing angle. We
illustrate this in Fig.~\ref{fig:footprints4} where we show the top
view in analogy to the lower-left panel of Fig.~\ref{fig:footprints1}
for different choices of mixing angle. For a large $\sin2\theta$, the
polarization vectors stay close to each other. For a smaller
$\sin2\theta$, the spiral spreads over a larger solid angle and has
more windings. We recall that a smaller $\sin2\theta$ also has the
effect of causing a larger nutation depth of the flavor pendulum.

\begin{figure}[t]
\begin{center}
\includegraphics[width=0.35\textwidth]{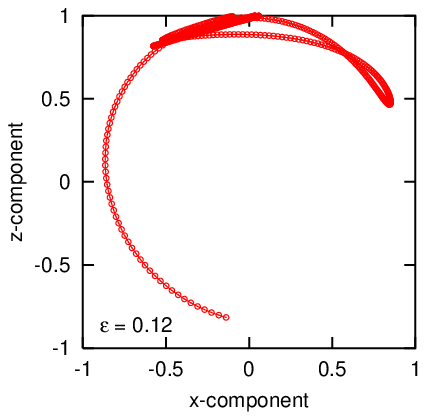}
\includegraphics[width=0.35\textwidth]{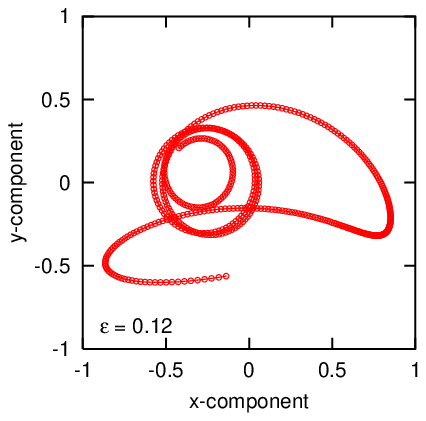}
\caption{\small Same as Fig.~\ref{fig:footprints1}, here for the normal
  hierarchy and
  $\epsilon=0.12$~\cite{EstebanPretel:2007ec}.\label{fig:footprints2}}
\end{center}
\end{figure}

Now turn to the quasi decoherent case with $\epsilon=0.12$.
Initially the same happens, but at the ``decoherence radius'' the
spiral structure dissolves almost instantaneously. The polarization
vectors enter a complicated structure as illustrated by the end
state (central panels of Fig.~\ref{fig:footprints1}). Moreover, they
are spread out all over the unit sphere, having both positive and
negative $z$-components. This structure looks different for
different choices of $\sin2\theta$ and $\epsilon$. However, once a
sufficient number of polarization vectors is used, it is
reproducible. For $\epsilon=0.06$ the picture would be qualitatively
similar.

Finally we show the fully symmetric case ($\epsilon=0$) in the
right-hand panels. Here decoherence is fast and complete. For a
small mixing angle, all polarization vectors are confined to the
$x$-$z$-plane. They distribute themselves on a circle in that plane.

\begin{figure}[t]
\begin{center}
\includegraphics[width=0.55\textwidth]{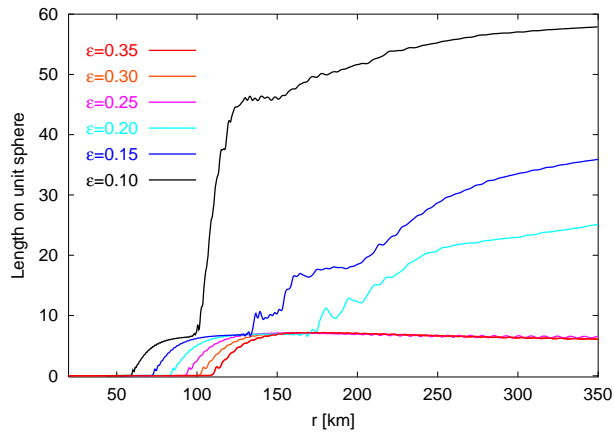}
\caption{\small Evolution of the length of the one-dimensional subspace
occupied by the polarization vectors for our standard inverted
hierarchy case, taking a series of different asymmetries
$\epsilon$. The length grows to larger values for smaller asymmetries~\cite{EstebanPretel:2007ec}.
\label{fig:phasespace}}
\end{center}
\end{figure}

For the normal hierarchy, we show in Fig.~\ref{fig:footprints2} as
an explicit example the $\epsilon=0.12$ case of
Fig.~\ref{fig:example2} that showed a clear multi-angle effect
without strong decoherence. Once more we find a spiral structure.
Most polarization vectors remain oriented roughly in their original
direction, but in this case also with a tail of a few polarization
vectors reversed. The quasi decoherent case ($\epsilon=0.06$) and
the symmetric system produce similar final pictures as the
corresponding cases of the inverted hierarchy.\vspace{0.5cm}\\
\textbf{B) Measures of decoherence}\vspace{0.2cm}\\
Even in the quasi-decoherent cases the unit sphere is not uniformly
filled with polarization vectors. Rather, in the mono-energetic case
considered here, the occupied phase space is a one-dimensional
subspace of the unit sphere. It is parameterized by the angular
variable $u$ and shows a clear line-like structure. This picture
suggests to use the length of this line on the unit sphere as
another global measure besides the length $\bar P$ to discriminate
between different modes of evolution~\cite{Raffelt:2007yz}. In a
numerical run with discrete angular bins, this quantity is simply
the sum of the angles between neighboring polarization vectors. In
Fig.~\ref{fig:phasespace} we show this quantity for the indicated
values of $\epsilon$ as a function of radius for our
inverted-hierarchy examples.

At the radius $r_{\rm syn}$ where the spiral forms, the length on
the unit sphere quickly increases from~0 to a value that is almost
independent of $\epsilon$, but depends on the mixing angle. For
smaller $\sin2\theta$ it is larger, corresponding to the spiral
having more windings as indicated earlier. Later, this length stays
practically constant, reflecting that the spiral structure, once
established, does not change much except tilting toward the negative
${\bf B}$-direction and precessing around it.

When $\epsilon$ is smaller than a critical value, at the
``decoherence radius'' a sudden second growth shoots up from
the plateau of these curves. For smaller $\epsilon$, the final
length is longer, representing a more ``phase-space filling'' line
on the unit sphere.

Note, however, that for $\epsilon$ close to zero, the line does not
fill the unit sphere, but essentially stays in a narrow band. In the
perfectly symmetric case, the motion of all polarization vectors is
essentially confined to the $x$-$z$-plane, i.e., the polarization
vectors distribute themselves over a great circle on the sphere as
shown in the right panels of Fig.~\ref{fig:footprints1}.

\subsection{Role of model parameters} \label{sec:parameters}

\textbf{A) Mixing angle}\vspace{0.2cm}\\
In the single-angle case, we have discussed that the mixing angle
affects only to the onset of the bipolar conversions. This discussion
suggests that, at least for the inverted hierarchy, the actual vacuum
mixing angle does not strongly influence the issue of multi-angle
decoherence because this effect happens when the global polarization
vector is tilted far away from the ${\bf B}$ direction. On the other
hand, we have already noted that the quasi-coherent spiral structure
that forms just beyond the synchronization radius has more windings
for a smaller mixing angle so that the system is not identical.

\begin{figure}
\begin{center}
\includegraphics[angle=0,width=0.55\textwidth]{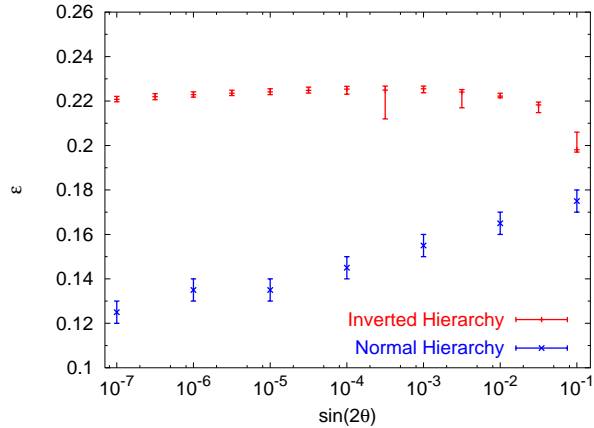}
\caption{\small Limiting $\epsilon$ for decoherence as a function of
  mixing angle for our standard example and both
  hierarchies~\cite{EstebanPretel:2007ec}.\label{fig:mixeps}}
\end{center}
\end{figure}

To clarify the role of the mixing angle we have used our standard
inverted-hierarchy case and have calculated the limiting asymmetry
$\epsilon$ for decoherence for a broad range of mixing angles. We show
the limiting contours in the plane of $\epsilon$ and $\sin 2\theta$ in
Fig.~\ref{fig:mixeps} for both hierarchies, above which
multi-angle decoherence does not appear.

We emphasize that the limiting $\epsilon$ shown in
Fig.~\ref{fig:mixeps} has a different meaning for the two hierarchies.
As discussed earlier, in the inverted hierarchy, $\bar P$ shortens
somewhat even in the quasi single-angle regime. Therefore, as a formal
criterion for distinguishing the regions of coherence and decoherence
we use that the final $\bar P$ has shortened to less than 0.85. The
exact choice is irrelevant because the transition between the
quasi-coherent and decoherent regimes is steep as a function
of~$\epsilon$.

Conversely, in the normal hierarchy, $\bar P$ need not visibly shorten
at all as illustrated by the example in the lower left panel of
Fig.~\ref{fig:example2}. Therefore, we here demand that $\bar P$ does
not visibly shorten in such a picture. We construct the demarcation
line by decreasing $\epsilon$ in steps of 0.01 until the polarization
vector for the first time shortens visibly. Finding this point
requires a significant amount of manual iterations with a modified
inner radius and number of angular bins to make sure the result does
not depend on these numerical parameters. The error bars represent our
confidence range for the true critical value.

We conclude that for the inverted hierarchy, multi-angle decoherence
is virtually independent from the value of $\sin2\theta$, except
that for very large $\theta$ a slightly smaller asymmetry is enough
to suppress decoherence. Assuming the presence of ordinary matter,
such large mixing angles seem irrelevant, except perhaps at late
times. Either way, it is conservative to assume a small mixing angle
and we will use $\sin2\theta=10^{-3}$ as a default value.

For the case of normal hierarchy we find a strong dependence of the
critical $\epsilon$ on $\log_{10}(\sin2\theta)$. For a smaller mixing
angle it is easier to suppress decoherence. The normal hierarchy is
very different from the inverted one in that for a small mixing angle,
all polarization vectors stay closely aligned with the $z$-direction
unless multi-angle decoherence takes place. Therefore, it is plausible
that for a smaller mixing angle, decoherence effects are
delayed.\vspace{0.5cm}\\
\textbf{B) Energy distribution}\vspace{0.2cm}\\
The neutrinos emitted from a SN core naturally have a broad energy
distribution. In Ref.~\cite{Raffelt:2007yz} it was noted that the
energy distribution of neutrinos and antineutrinos is largely
irrelevant for the question of decoherence as long as the
oscillations exhibit self-maintained
coherence~\cite{Kostelecky:1994dt}. The multi-angle transition to
decoherence typically occurs within the dense-neutrino region where
the synchronization of energy modes remains strong. Therefore, we
expect that multi-angle decoherence is not significantly affected by
the neutrino spectrum.

In order to compare a monochromatic system with one that has a broad
energy distribution, the crucial quantity to keep fixed is not the
average energy, but the average oscillation frequency
$\langle\omega\rangle=\langle\Delta m^2/2E\rangle$, as discussed in
Sec.~\ref{sec:single-multi}. If we assume that neutrinos and
antineutrinos have equal distributions, it is straightforward to
adjust, for example, the temperature of a thermal distribution such
that $\langle\omega\rangle$ is identical to our monochromatic standard
case $\omega_0=0.3~{\rm km}^{-1}$. If we assume different
distributions for neutrinos and antineutrinos, the equivalent
$\omega_0$ is somewhat more subtle, and we will have to use
Eq.~(\ref{eq:wsynch}).

We have studied several numerical examples of quasi single-angle
behavior and of multi-angle decoherence, taking different neutrino and
antineutrino energy spectra, such as flat or thermal and with equal or
different temperatures. We always found that the evolution of the
global polarization vectors is almost identical to the equivalent
monochromatic cases. We never observed that a broad energy spectrum
caused a significant deviation from the monochromatic behavior at
those radii that are relevant for decoherence.

\begin{figure}
\begin{center}
\includegraphics[angle=0,width=0.85\textwidth]{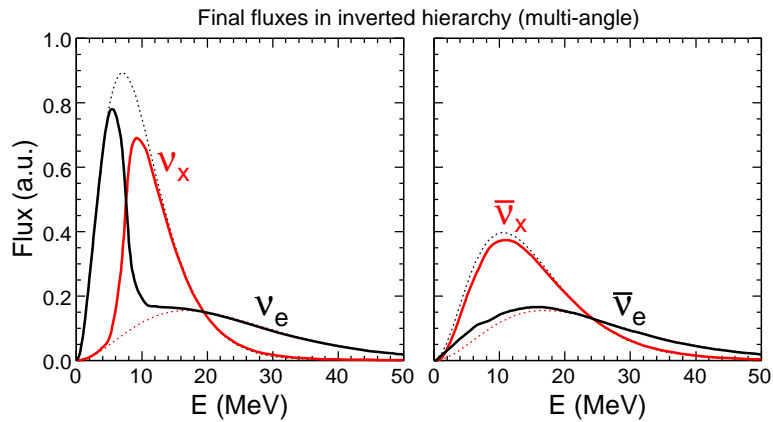}
\caption{\small Multi-angle simulation in inverted hierarchy: Final
  fluxes after the bipolar conversions for different neutrino species
  as a function of energy. Initial fluxes are shown as dotted lines to
  guide the eye~\cite{Fogli:2007bk}.\label{fig:multiansplit}}
\end{center}
\end{figure}

Of course, a multi-energy system is qualitatively different from a
monochromatic one in that the final energy distribution shows a
spectral split. Nevertheless, for sufficiently large asymmetries
$\epsilon$ where the multi-angle system evolves in the quasi
single-angle mode, there is no significant modification of the
spectral split. In Ref.~\cite{Fogli:2007bk} it is performed this
analysis for a fixed value of the $\epsilon$. They indeed obtain the
same kind of spectral split as in the single-angle case. Their result
is shown in Fig.~\ref{fig:multiansplit}. In the decoherent case, on
the other hand, the final spectra naturally are very different, but we
have not explored such cases systematically because multi-angle
decoherence does not seem to be generic for realistic SN scenarios.

\begin{figure}[t]
\begin{center}
\includegraphics[width=\textwidth]{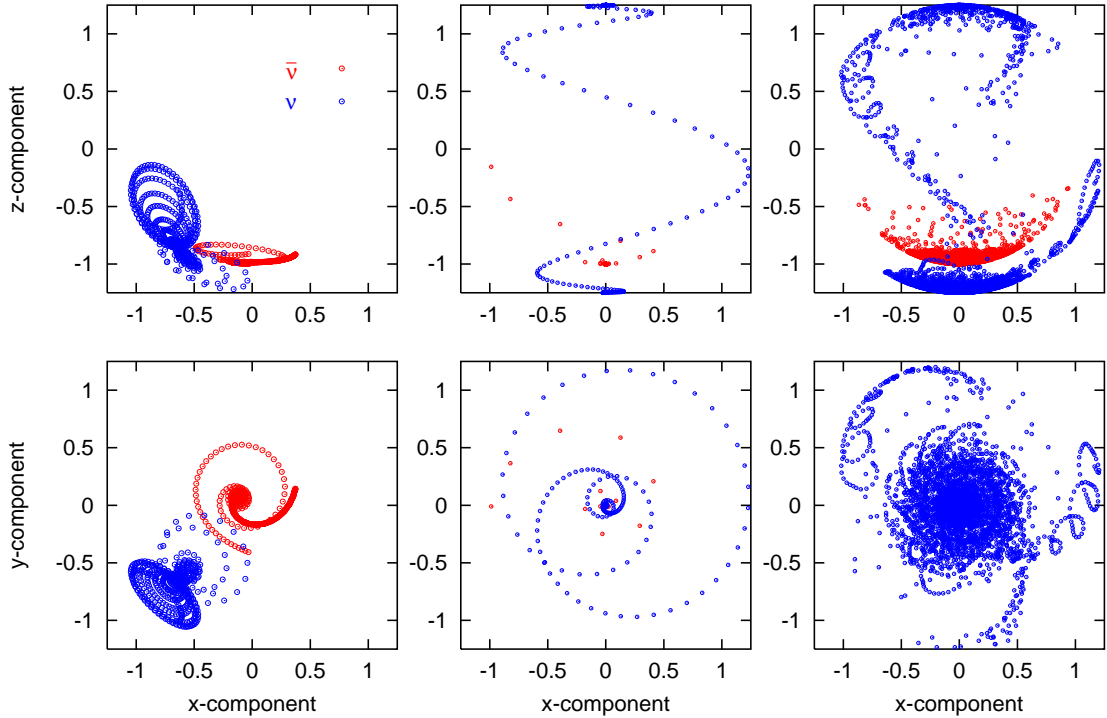}
\caption{\small Final state at a large radius of the polarization
  vectors for our standard parameters in analogy to
  Fig.~\ref{fig:footprints1}. The antineutrinos (red/light gray) are
  on the unit sphere, whereas the neutrinos (blue/dark gray) live on a
  sphere of radius $1+\epsilon=1.25$. Left: monochromatic multi-angle,
  the antineutrinos being identical with the left column of
  Fig.~\ref{fig:footprints1}. Middle: Box-like energy spectrum and
  single angle. Right: Box-like energy spectrum and multi angle. In
  the lower right panel we do not show the
  antineutrinos~\cite{EstebanPretel:2007ec}.\label{fig:footprints3}}
\end{center}
\end{figure}

To illustrate the modifications caused by an energy spectrum in a
different way from the previous literature, we show in
Fig.~\ref{fig:footprints3} the side and top views of the location of
neutrino and antineutrino polarization vectors on the unit sphere in
analogy to Fig.~\ref{fig:footprints1} for our standard parameter
values. In the left column we show the same monochromatic multi-angle
case that we already showed in the left column of
Fig.~\ref{fig:footprints1}, with 500 modes. In addition we include the
neutrinos (blue/dark gray) that here live on a sphere of radius
$1+\epsilon=1.25$. The neutrinos form a spiral structure similar to
the one of the antineutrinos, but in the final state this structure
cannot move to the negative ${\bf B}$ directions because of lepton
number conservation.

In the middle column we show a single-angle example with the same
parameters, now using a box-like spectrum of oscillation frequencies
where initially $\bar P_{\omega}^z=(2\omega_0)^{-1}$ and
$P_{\omega}^z=(1+\epsilon)(2\omega_0)^{-1}$ for
$0\leq\omega\leq2\omega_0$ so that $\langle\omega_\nu\rangle
=\langle\omega_{\bar\nu}\rangle=\omega_0$ and it is equivalent to
the original monochromatic case. We now see that most of the
antineutrinos have moved to the negative ${\bf B}$ direction as
before, whereas the neutrinos populate both the positive and
negative ${\bf B}$ direction, representing the spectral split. The
lack of full adiabaticity prevents the split from being complete,
leaving some polarization vectors not fully aligned or anti-aligned
with ${\bf B}$. At large radii when the neutrino-neutrino
interactions have died out, these modes precess with their different
vacuum oscillation frequencies so that they are found on a spiral
locus extending from the ``south pole'' to the ``north pole'' that
gets wound up further at larger radii. Note that here we have used
1000 energy modes in order to obtain a visible population occupying
these non-adiabatic final states. Still, only very few red dots
(antineutrinos) are visible, the vast majority being at the south
pole. Likewise for the neutrinos (blue dots), the spiral is
populated only by a small fraction of the 1000 modes. In other
words, the evolution is nearly adiabatic.

\begin{figure}[t]
\begin{center}
\includegraphics[width=0.55\textwidth]{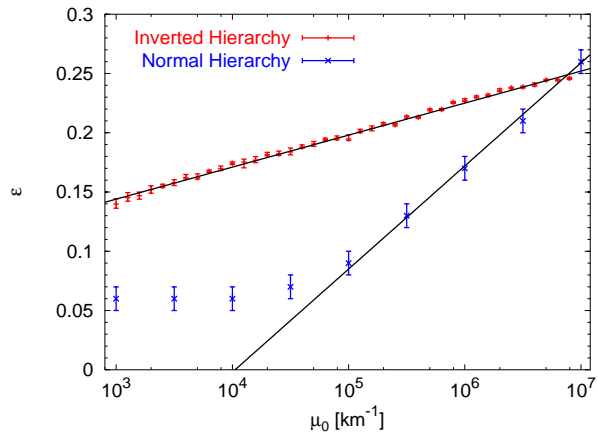}
\caption{\small Limiting $\epsilon$ for decoherence as a function of
  the effective neutrino-neutrino interaction strength $\mu_0$ for our
  standard parameters. The linear regressions are ``visual fits''
  represented by
  Eq.~(\ref{eq:epscontour})~\cite{EstebanPretel:2007ec}.\label{fig:regimes}}
\end{center}
\end{figure}

Finally we combine a box-like energy spectrum and a multi-angle
distribution (right panels). The antineutrinos all cluster around
the negative ${\bf B}$ direction and fill the ``southern polar cap''
more or less uniformly because at late times modes with different
energies precess with different frequencies. The neutrinos populate
both the northern and southern polar caps, representing the spectral
split. At intermediate latitudes we find coherent spiral structures.
They correspond to modes with different angles but equal $\omega$ so
that even at late times they do not dissolve by differential
precession.\vspace{0.5cm}\\
\textbf{C) Effective interaction strength}\vspace{0.2cm}\\
Besides the asymmetry $\epsilon$ itself, the most uncertain model
parameter is the effective neutrino-neutrino interaction strength
$\mu_0$ as defined in Eq.~(\ref{eq:mudefine}). In Fig.~\ref{fig:regimes}
we show the demarcation lines between coherence and decoherence for
both hierarchies in the $\mu_0$-$\epsilon$-plane, keeping all other
parameters at their standard values. The contours are constructed as
described in the discussion for the mixing angle at the beginning of
this section. The numerical contours are visually very well
approximated by linear regressions of the form
\begin{eqnarray}\label{eq:epscontour}
 \epsilon_{\rm IH}&\approx&0.225+0.027\,\log_{10}
 \left(\frac{\mu_0}{10^6~{\rm km}^{-1}}\right)\,,
 \nonumber\\*
 \epsilon_{\rm NH}&\approx&0.172+0.087\,\log_{10}
 \left(\frac{\mu_0}{10^6~{\rm km}^{-1}}\right)\,.
\end{eqnarray}
For the normal hierarchy, the linear regression would intersect
$\epsilon=0$ within the range of investigated $\mu_0$-values, but in
reality turns over and saturates around $\epsilon=0.06$.\vspace{0.5cm}\\
\textbf{D) Vacuum oscillation frequency}\vspace{0.2cm}\\
The average vacuum oscillation frequency $\omega$ depends on the
atmospheric $\Delta m^2$ that is quite well constrained, and a
certain average of the neutrino energies. Our standard value is
$\omega=0.3~{\rm km}^{-1}$. If we increase this to $1~{\rm
km}^{-1}$, the $\epsilon$-$\mu_0$-contour in Fig.~\ref{fig:regimes} is
essentially parallel-shifted to larger $\epsilon$ by about 0.035
(inverted hierarchy). This range of $\omega$ probably brackets the
plausible possibilities so that the uncertainty of $\omega$ does
not strongly influence the practical demarcation between the
regimes.

\begin{figure}[t]
\begin{center}
\includegraphics[width=0.55\textwidth]{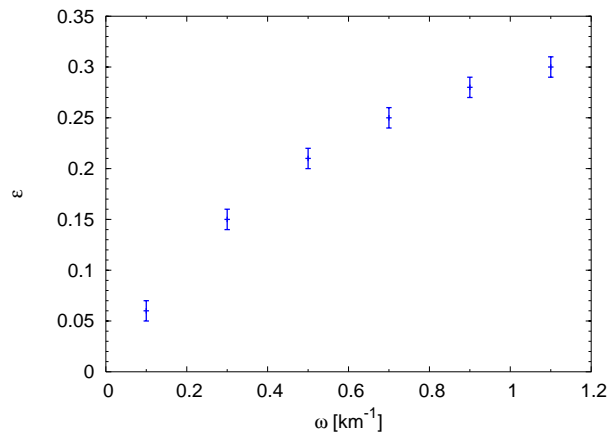}
\caption{\small Limiting $\epsilon$ for decoherence as a function of
  the vacuum oscillation frequency $\omega$ for our standard
  parameters and the normal
  hierarchy~\cite{EstebanPretel:2007ec}.\label{fig:nhomega}}
\end{center}
\end{figure}

The normal hierarchy is more sensitive to $\omega$.  In
Fig.~\ref{fig:nhomega} we show a contour for the coherence regime in
the $\epsilon$-$\omega$ plane, assuming otherwise our standard
parameter values. Changing $\omega$ from $0.3$ to $1~{\rm km}^{-1}$
increases the critical $\epsilon$ by almost~0.15.

\subsection{Concluding remarks}
\label{sec:conclusions}

We have here not attempted to develop further analytical insights, but
have taken a practical approach and explored numerically the range of
parameters where different forms of behavior dominate in a realistic
SN scenario.

To this end we have first clarified that ``multi-angle effects''
means one of two clearly separated forms of behavior. The flavor
content of the system can evolve in a quasi single-angle form. On
the level of the polarization vectors this means that they fill only
a restricted volume of the available phase space and maintain a
coherent structure. On the other hand, nearly complete flavor
equilibrium can arise where the available phase space is more or
less uniformly filled.

For realistic assumptions about SN and neutrino parameters, the
switch between these modes of evolution is set by the degree of
asymmetry between the neutrino and antineutrino fluxes. While this
asymmetry is caused by the deleptonization flux, the crucial parameter
$\epsilon$ is the asymmetry between
$F^{R_\nu}_{\nu_e}-F^{R_\nu}_{\nu_x}$ and the corresponding
antineutrino quantity as defined in Eq.~(\ref{eq:epsdefine}) because
for flavor oscillations the part of the density matrix that is
proportional to $F^{R_\nu}_{\nu_e}+F^{R_\nu}_{\nu_x}$ drops out. While
in a realistic SN on average $F^{R_\nu}_{\nu_e}$ is about 15\% larger
than $F^{R_\nu}_{\bar\nu_e}$, the asymmetry parameter as defined in
Eq.~(\ref{eq:epsdefine}) is typically much larger.

The critical value of $\epsilon$ that is enough to suppress
decoherence depends on the type of neutrino mass hierarchy, the
average energies, luminosities, and on the mixing angle. We have
found that for $\epsilon > 0.3$, decoherence is suppressed for the
entire range of plausible parameters, but a value smaller than 0.1
may be enough, depending on the combination of other parameters.

We conclude that the quasi single-angle behavior may well be typical
for realistic SN conditions, i.e., that the deleptonization flux is
enough to suppress multi-angle decoherence. To substantiate this
conclusion one should analyze the output of numerical simulations in
terms of our model parameters. Besides the flavor-dependent
luminosities and average energies, one needs the angular
distribution of the neutrino radiation field at some radius where
collisions are no longer important.

If our conclusion holds up in the light of realistic SN simulations,
a practical understanding of the effect of self-induced neutrino
flavor transformations quickly comes into reach. In the normal mass
hierarchy, nothing new would happen on a macroscopic scale. In the
inverted hierarchy, the final effect would be a conversion of
$\nu_e\bar\nu_e$ pairs and a split in the $\nu_e$ spectrum. These
phenomena are only mildly affected by multi-angle effects as long as
we are in the quasi single-angle regime.

If at late times the matter density profile contracts enough that an
MSW effect occurs in the dense-neutrino region, the situation becomes
more complicated as the neutrino-neutrino and ordinary matter effects
interfere and produce a richer structure of spectral
modifications~\cite{Duan:2006an, Duan:2006jv}.  Even then, numerical
simulations are much simpler if multi-angle decoherence is suppressed.

It is not obvious how $\epsilon$ evolves at late times. The
deleptonization of the core is probably faster than the cooling so
that one may think that $\epsilon$ becomes smaller. On the other hand,
the $\bar\nu_e$ can essentially only interact via neutral current
reactions and their flux and energy distribution should, therefore,
become very similar to the ones of $\nu_x$ and $\bar\nu_x$. Therefore,
it is not obvious if at late times the initial flux difference
$F^{R_\nu}_{\nu_e}-F^{R_\nu}_{\bar\nu_e}$ or
$F^{R_\nu}_{\bar\nu_e}-F^{R_\nu}_{\bar\nu_x}$ decreases more
quickly. We also note that there can be a cross-over in the sense that
at late times the flux hierarchy can become
$F^{R_\nu}_{\nu_x}=F^{R_\nu}_{\bar\nu_x} > F^{R_\nu}_{\nu_e} >
F^{R_\nu}_{\bar\nu_e}$ as in Ref.~\cite{Raffelt:2003en}, meaning that
we would have a pair excess flux of $\nu_x\bar\nu_x$ instead of a
$\nu_e\bar\nu_e$ excess.

Our results suggest that signatures of collective flavor
transformations are not erased by multi-angle decoherence and will
survive to the surface, modulated by the usual MSW flavor
conversions~\cite{Dighe:1999bi}. The survival of observable signatures
then also depends on the density fluctuations of the ordinary medium
that can be a source of kinematical flavor
decoherence~\cite{Fogli:2006xy,Friedland:2006ta}.

All authors in this field have relied on the simplifying assumption of
either homogeneity or exact spherical symmetry to make the equations
numerically tractable. The neutrino emission from a real SN is
influenced by density and temperature fluctuations of the medium in
the region where neutrinos decouple. Likewise, the neutrino fluxes
emitted from the accretion tori of coalescing neutron stars, the
likely engines of short gamma ray bursts, have fewer symmetries than
assumed here. It remains to be investigated if systems with more
general geometries behave qualitatively similar to the spherically
symmetric case or if deviations from spherical symmetry can provide a
new source of kinematical decoherence.

\pagestyle{densematter}
\section{Role of dense matter in collective supernova neutrino
  transformations}\label{sec:densematter}

Let us now analyze the other potential source of multi-angle
decoherence in the collective neutrino transformations, namely the
dense matter background. One of the many surprises of self-induced
flavor transformations has been that, in the single-angle
approximation, dense matter barely affects them. They are driven by an
instability in flavor space that is insensitive to matter because it
affects all neutrino and antineutrino modes in the same
way. Therefore, it can be transformed away by going to a rotating
frame in flavor space, see Eq.~(\ref{eq:bt}).

We here clarify, however, that the matter density can not be
arbitrarily large before it affects collective flavor conversions
after all. The matter term is ``achromatic'' only if we consider the
time--evolution of a homogeneous (but not necessarily isotropic)
neutrino ensemble on a homogeneous and isotropic matter background.
If the matter background is not isotropic, the current-current
nature of the neutrino-electron interaction already implies that
different neutrino modes experience a different matter effect.

It is more subtle that even without a current, matter still affects
different neutrino modes differently if we study neutrinos streaming
from a source. The relevant evolution is now the flavor variation of
a stationary neutrino flux as a function of distance. For a
spherically symmetric situation, ``distance from the source'' is
uniquely given by the radial coordinate $r$. Neutrinos reaching a
certain $r$ have travelled different distances on their trajectories
if they were emitted with different angles relative to the radial
direction. Therefore, at $r$ they have accrued different oscillation
phases even if they have the same vacuum oscillation frequency and
even if they have experienced the same matter background. In other
words, if we project the flavor evolution of different angular modes
on the radial direction, they have different effective vacuum
oscillation frequencies even if they have the same energy. The same
argument applies to matter that modifies the oscillation frequency
in the same way along each trajectory, but therefore acts
differently when expressed as an effective oscillation frequency
along the radial direction.

The neutrino-neutrino interaction, when it is sufficiently strong,
forces different modes to reach a certain $r$ with the same
oscillation phase. To achieve this ``self-main\-tained coherence''
the neutrino-neutrino term must overcome the phase dispersion that
would otherwise occur. Such a dispersion is caused not only by a
spectrum of energies, but also by a matter background.

\subsection{Homogeneous Ensemble}                 \label{sec:homogeneous}

\textbf{A) Isotropic background}\vspace{0.2cm}\\
Let us reanalyze the role of matter in collective neutrino
transformations. We begin with the EOMs in their simplest form,
relevant for a homogeneous (but not necessarily isotropic) gas of
neutrinos. We only consider two-flavor oscillations where the most
economical way to write the EOMs is in terms of the usual flavor
polarization vectors ${\bf P}_{\bf p}$ for each mode ${\bf p}$ and
analogous vectors $\bar{\bf P}_{\bf p}$ for the antineutrinos, as
shown in Eq.~(\ref{eq:hamiltonianP}). For simplicity we here assume
that initially only $\nu_e$ and $\bar\nu_e$ are present with an excess
neutrino density of $n_{\nu_e}=(1+\epsilon)\,n_{\bar\nu_e}$.  The
polarization vectors are initially normalized such that $|\int\D{\bf
  p}\, \bar{\bf P}_{\bf p}|=1$ and $|\int\D{\bf p}\, {\bf P}_{\bf
  p}|=1+\epsilon$.

The matter term is ``achromatic'' in that it affects all modes of
neutrinos and antineutrinos in the same way. As already discussed, one
may study the EOMs in a coordinate system that rotates around ${\bf
  L}$ with frequency $\lambda$ so that the matter term disappears. In
the new frame the vector ${\bf B}$ rotates fast around ${\bf L}$ so
that its transverse component averages to zero. Therefore, in the new
frame the rotation-averaged Hamiltonian is
\begin{equation}\label{eq:EOM2}
 \langle{\bf H}\rangle=\omega\cos(2\theta)\,{\bf L}+
 \mu({\bf P}-\bar{\bf P})\,,
\end{equation}
where for the moment we consider the even simpler case of an
isotropic and monochromatic neutrino ensemble where the entire system
is described by one polarization vector ${\bf P}$ for neutrinos and
one $\bar{\bf P}$ for antineutrinos.

A dense matter background effectively projects the EOMs on the
weak-interaction direction. In particular, the relevant vacuum
oscillation frequency is now $\omega\cos2\theta$. For a small mixing
angle, the case usually considered in this context, this projection
effect is not important. However, a large mixing angle would strongly
modify the projected $\omega$. Maximal mixing where $\cos2\theta=0$
would prevent any collective flavor transformations, an effect that is
easily verified in numerical examples\footnote{This effect and its
  consequences will be further discussed in the context of three
  flavors in Chapter~\ref{chapter:coll3flavors}.}.

One usually assumes that the (anti)neutrinos are prepared in
interaction eigenstates so that initially ${\bf P}$ and $\bar{\bf P}$
are oriented along ${\bf L}$. Therefore, the rotation-averaged EOMs
alone do not lead to an evolution. However, in the unstable case of
the inverted mass hierarchy, an infinitesimal disturbance is enough
to excite the transformation. The fast-rotating transverse ${\bf B}$
component that was left out from the EOM is enough to trigger the
evolution, but otherwise plays no crucial role~\cite{Duan:2005cp,
Hannestad:2006nj}.

If we consider a homogeneous system where $\mu$ is a slowly
decreasing function of time, one can find the adiabatic solution of
the EOM for the simple system consisting only of ${\bf P}$ and
$\bar{\bf P}$~\cite{Duan:2007mv, Raffelt:2007xt}. In vacuum, this is
a complicated function of $\epsilon$ and $\cos2\theta$. In dense
matter, however, we are effectively in the limit of a vanishing
mixing angle because the initial orientation of the polarization
vectors now coincides with the direction relevant for the
rotation-averaged evolution. The original vacuum mixing angle only
appears in the expression for the projected oscillation frequency
$\omega\cos2\theta$.

With $z=\bar P_z$ the adiabatic connection between $\bar P_z$ and
$\mu$ is now given by the inverse function of
\begin{equation}\label{eq:adiabatic}
 \frac{\omega\cos2\theta}{\mu}=\frac{\epsilon+2z}{2}
 -\frac{\epsilon+2z+(3\epsilon+2z)z}
 {2\sqrt{(1+z)(1+z+2\epsilon)}}\,,
\end{equation}
where $-1\leq z\leq+1$. The synchronization radius $r_{\rm syn}$
where the adiabatic curve begins its decrease is implied by $\bar
P_z=z=1$. One finds the familiar result of Eq.~(\ref{eq:synchcond})
\begin{equation}\label{eq:synch}
\frac{\omega\cos2\theta}{\mu}\Big|_{\rm
syn}=\frac{(\sqrt{1+\epsilon}-1)^2}{2}\,.
\end{equation}
For $\mu$ values larger than this limit, the polarization vectors are
stuck to the ${\bf L}$ direction.

Without matter one finds that $({\bf P}-\bar{\bf P})\cdot{\bf B}$ is
conserved. Here, the analogous conservation law applies to $({\bf
  P}-\bar{\bf P})\cdot{\bf L}$. Therefore, the adiabatic solution for
$P_z$ is such that $P_z-\bar P_z$ is conserved, i.e., $P_z=\bar
P_z+\epsilon$.

In summary, the presence of dense matter simplifies the EOMs and in
that the adiabatic solution is the one for a vanishing vacuum mixing
angle, provided one uses the projected vacuum oscillation frequency.\vspace{0.5cm}\\
\textbf{B) Background flux}\vspace{0.2cm}\\
As a next example we still consider a homogeneous system, but now
allow for a net flux of the background matter, assuming axial
symmetry around the direction defined by the flux. For simplicity we
consider a monochromatic ensemble with a single vacuum oscillation
frequency $\omega$. We characterize the angular neutrino modes by
their velocity component $v$ along the matter flux direction. The
EOMs are in this case
\begin{equation}
\dot{\bf P}_{v}={\bf H}_{v}\times{\bf P}_{v}\,.
\end{equation}
The Hamiltonian for the mode $v$ is
\begin{equation}
 {\bf H}_{v}=\omega{\bf B}+(\lambda-\lambda'v){\bf L}+
 \mu({\bf D}-v{\bf F})\,,
\end{equation}
where $\lambda'\equiv\lambda v_e$, $v_e$ being the net electron
velocity\footnote{This description is analogous to the one given in
  Chapter~\ref{chapter:oscillations}.}. Moreover,
\begin{eqnarray}
 {\bf D}&=&\int_{-1}^{+1}\D v\,
 \left({\bf P}_v-\bar{\bf P}_v\right)\,,
 \nonumber\\*
 {\bf F}&=&\int_{-1}^{+1}\D v\,v
 \left({\bf P}_v-\bar{\bf P}_v\right)
\end{eqnarray}
are the net neutrino density and flux polarization vectors. The
normalization is $|\int \D v\,\bar{\bf P}_v|=1$ and $|\int \D v\,{\bf
  P}_v|=1+\epsilon$.

Next, we transform the EOMs to a frame rotating with frequency
$\lambda$, allowing us to remove the matter term, but not the matter
flux,
\begin{equation}
 \langle{\bf H}_{v}\rangle=(\omega-\lambda'v){\bf L}+
 \mu({\bf D}-v{\bf F})\,.
\end{equation}
We assume $\theta$ to be small and thus use
$\omega\approx\omega\cos2\theta$. We now have a system where the
effective vacuum oscillation frequencies for neutrinos are uniformly
distributed between $\omega\pm\lambda'$ and for antineutrinos between
$-\omega\pm\lambda'$. Even after removing the average common
precession of all modes, their evolution is still dominated by the
matter-flux term if $\lambda'\gg\mu$. In other words, collective
behavior now requires $\mu > \lambda'$ and not only $\mu > \omega$.

The simplest example is the flavor pendulum where for $\epsilon=0$ and
an isotropic neutrino gas one obtains the well-known pendular motions
of the polarization vectors. Matter does not disturb this behavior,
except that it takes logarithmically longer for the motion to
start. However, a matter flux, if sufficiently strong, suppresses this
motion and the polarization vectors remain stuck to the ${\bf L}$
direction for both mass hierarchies. If the neutrino distribution is
not isotropic, the ensemble quickly decoheres
kinematically~\cite{Raffelt:2007yz}, an effect that is also suppressed
by a sufficiently strong matter flux.

We have verified these predictions in several numerical examples, but
have not explored systematically the transition between a ``weak''
and a ``strong'' matter flux because a homogeneous ensemble only
serves as a conceptual example where matter can have a strong
influence on self-induced transformations.

\subsection{Spherical Stream}                          \label{sec:stream}

The most general case of neutrino flavor evolution consists of an
ensemble evolving both in space and time. In practice, however, one
usually considers quasi-stationary situations where one asks for the
spatial flavor variation of a stationary neutrino flux streaming from
a source. The neutrino density decreases with distance so that one
can mimic this situation by a homogeneous system evolving in time
with a decreasing density, the expanding universe being a realistic
example. However, the analogy has important limitations because
collective oscillations introduce geometric complications into the
spatial-variation case.

The simplest non-trivial example is a perfectly spherical source (``SN
core'') that emits neutrinos and antineutrinos  like a
blackbody surface into space. The matter background is also taken to
be perfectly spherically symmetric, but of course varies with radius.
As a further simplification we consider monochromatic neutrinos and
antineutrinos that are all emitted with the same energy. In such a
system, we end up with the EOMs given in Eq.~(\ref{eq:eom5}), which we
can rewrite as
\begin{equation}
\partial_r{\bf P}_{u}={\bf H}_{u}\times{\bf P}_{u}\,,
\end{equation}
where the Hamiltonian is
\begin{equation}\label{eq:Ham1}
 {\bf H}_{u}=
 \frac{\omega {\bf B}+\lambda{\bf L}}{v_{u}}
 +\mu_r \left(\frac{{\bf D}}{v_{u}}-{\bf F}\right)\,.
\end{equation}
For antineutrinos we have, as always, $\omega\to-\omega$. Since the
polarization vectors describe the fluxes, the global density and flux
polarization vectors are
\begin{eqnarray}
 {\bf D}&=&\int_0^1\D u\,
 \frac{{\bf P}_{u}-\bar{\bf P}_{u}}{v_{u}}\,,
 \nonumber\\*
 {\bf F}&=&\int_0^1\D u\,
 \left({\bf P}_{u}-\bar{\bf P}_{u}\right)\,,
\end{eqnarray}
using the normalization $|\int_0^1 \D u\,\bar{\bf P}_{u}|=1$ and
$|\int_0^1 \D u\,{\bf P}_{u}|=1+\epsilon$. The matter coefficient
$\lambda=\sqrt2\,G_{\rm F}[n_{e^-}(r)-n_{e^+}(r)]$ encodes the
effective electron density at radius $r$ whereas
\begin{equation}
\mu_r=\mu_0\,\frac{R_\nu^2}{r^2}\,.
\end{equation}
Therefore, $\mu_r$ always varies as $r^{-2}$ due to the geometric flux
dilution, whereas $\lambda$ is given by the detailed matter profile
of a SN model~\cite{Schirato:2002tg,Fogli:2003dw,Tomas:2004gr}.



The variation of the polarization vectors with the common radial
coordinate $r$ now acquires dynamical significance in that the
polarization vectors evolve differently than they would in the
absence of neutrino-neutrino interactions. From Eq.~(\ref{eq:Ham1})
it is obvious that the matter term is no longer the same for all
modes and thus can not be transformed away by going to a rotating
frame. This behavior does not depend on the radial variation of
$\lambda$, even a homogeneous medium would show this multi-angle
matter effect.

For quasi single-angle behavior to occur, $\epsilon$ must not be too
small, a condition that is probably satisfied in a realistic SN.
Therefore, the synchronization radius implied by Eq.~(\ref{eq:synch})
is always much larger than the neutrino-sphere radius $R_\nu$, allowing
us to expand the EOMs in powers of $R_\nu/r\ll1$. Using
\begin{equation}
v_{u}^{-1}=1+\frac{u}{2}\,\frac{R_\nu^2}{r^2}
\end{equation}
we find
\begin{equation}
 {\bf H}_{u}=(\omega {\bf B}+\lambda{\bf L})
 \left(1+\frac{u}{2}\,\frac{R_\nu^2}{r^2}\right)
 +\mu_r\,\frac{R_\nu^2}{2r^2}
 \left({\bf Q}+u{\bf F}\right)\,,
\end{equation}
where
\begin{equation}
{\bf Q}=\int_0^1\D u\,u
\left({\bf P}_{u}-\bar{\bf P}_{u}\right)
\end{equation}
and ${\bf F}$ is the same as before. At large $r$ a small
correction to $\omega$ is not crucial and can be ignored. The radial
variation of $\lambda$ is slow compared to the precession, so we
can go to a frame that rotates with a different frequency $\lambda$
at each radius. Finally the rotation-averaged Hamiltonian is
\begin{equation}
 \langle{\bf H}_{u}\rangle=(\omega +u\,\lambda^*)\,{\bf L}
 +\mu^*\,\left({\bf Q}+u{\bf F}\right)\,,
\end{equation}
where we have assumed $\omega\cos2\theta\approx\omega$ and defined
\begin{eqnarray}\label{eq:mulambda}
 \lambda^*&=&\lambda\,\frac{R_\nu^2}{2r^2}\,,
 \nonumber\\*
 \mu^*&=&\mu_r\,\frac{R_\nu^2}{2r^2}=\mu_0\,\frac{R_\nu^4}{2r^4}\,.
\end{eqnarray}

The multi-angle matter effect can be neglected if in the collective
region beyond the synchronization radius we have
\begin{equation}
\lambda^*\ll \mu^*\label{eq:potcondition}
\end{equation}
equivalent to
\begin{equation}\label{eq:condition}
n_{e^-}-n_{e^+}\ll n_{\bar\nu_e}\,.
\end{equation}
In the opposite limit we expect that the large spread of effective
oscillation frequencies prevents collective oscillations. In this
case all polarization vectors remain pinned to the ${\bf L}$
direction and no flavor conversion occurs.

For intermediate values it is not obvious what will happen. One may
expect that the multi-angle matter effect triggers multi-angle
decoherence, destroying the quasi single-angle behavior. This indeed
occurs for the inverted hierarchy whereas in the normal hierarchy we
have not found any conditions where multi-angle decoherence was
triggered by the multi-angle matter effect. We recall that for a
sufficiently small $\epsilon$ multi-angle decoherence occurs even in
the normal hierarchy whereas no collective transformation arise for a
sufficiently large $\epsilon$~\cite{EstebanPretel:2007ec}.

\begin{figure}[!ht]
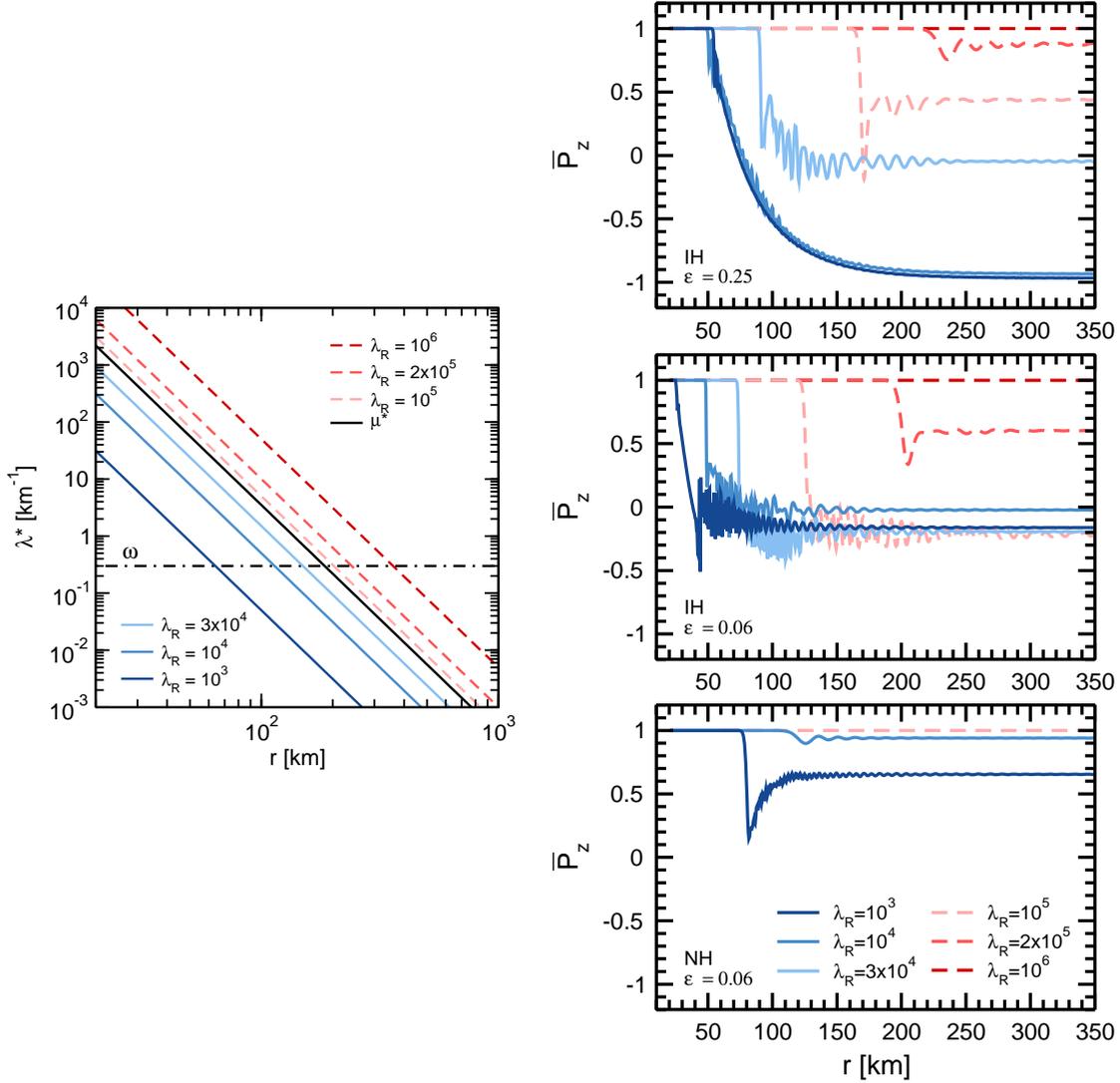

\begin{center}
\begin{tabular}{cc}
\vspace{3.4cm} & \multirow{3}{*}{\includegraphics[angle=0,width=0.5\textwidth]{./cap4/figures/0807.0659/fig2.eps}}\\
\includegraphics[angle=0,width=0.45\textwidth]{./cap4/figures/0807.0659/fig1.eps}
& \\
\vspace{3.5cm} & \\
\end{tabular}
\end{center}
\caption{\small On the left plot we show the radial variation of
  $\mu^*$ and $\lambda^*$ for our numerical examples with the
  indicated value of $\lambda_R\equiv\lambda_{R_\nu}$. On the right
  plots we have the radial variation of $\bar P_z$ for three different
  scenarios: IH and $\epsilon = 0.25$ (top panel), IH and $\epsilon =
  0.06$ (middle panel), and NH and $\epsilon = 0.06$ (bottom panel).
  In each panel different values of $\lambda_{R_\nu}$ have been
  assumed~\cite{EstebanPretel:2008ni}.} \label{fig:prof_trans}
\end{figure}



We illustrate these points with a numerical example where $R_\nu=10$~km,
$\omega=0.3~{\rm km}^{-1}$, $\theta=10^{-2}$, 
$\mu_0=7\times 10^4~{\rm km}^{-1}$ and
\begin{equation}
\lambda=\lambda_{R_\nu}\,\left(\frac{R_\nu}{r}\right)^n
\end{equation}
with $n=2$. This particular value of the power-law index leads to the
same radial dependence of $\mu^*$ and $\lambda^*$, see
Eq.~(\ref{eq:mulambda}).  In the left plot of Fig.~\ref{fig:prof_trans}
we show the radial variation $\mu^*$ and $\lambda^*$ for different
choices of $\lambda_{R_\nu}$, between $10^3$~km$^{-1}$ and
$10^6$~km$^{-1}$. Even for the smallest matter effect, the ordinary
MSW resonance, defined by the condition $\lambda=\omega$, stays
safely beyond the collective region.

In the right plots of Fig.~\ref{fig:prof_trans} we show the
corresponding variation of $\bar P_z$ for three different cases:
inverted mass hierarchy and $\epsilon = 0.25$ (top panel), inverted
mass hierarchy and $\epsilon = 0.06$ (middle panel), and normal mass
hierarchy and $\epsilon = 0.06$ (bottom panel).
In the top panel we observe the usual transformation for a small
matter effect, a complete suppression of transformations for a large
matter effect, and multi-angle decoherence for intermediate
cases. Repeating the same exercise for the normal mass hierarchy and
the same $\epsilon$ reveals no macroscopic influence of the matter
term.

For a sufficiently small $\epsilon$ one finds self-induced multi-angle
decoherence for both hierarchies. In the middle and bottom panels of
Fig.~\ref{fig:prof_trans} we show how with a sufficiently strong
matter effect the decoherence can be suppressed for both hierarchies.

\subsection{Discussion}

We have identified a new multi-angle effect in collective neutrino
transformations that is caused by a matter background. Previous
numerical studies of multi-angle effects had used a matter profile
that satisfies the condition Eq.~(\ref{eq:condition}) in the critical
region~\cite{Duan:2006an, Fogli:2007bk}. In other multi-angle studies
matter was entirely ignored~\cite{EstebanPretel:2007ec} and
otherwise, single-angle studies were performed. Therefore, the
multi-angle matter effect discussed here had escaped numerical
detection.

In many practical cases relevant for SN physics or in coalescing
neutron stars, the density of matter is probably small enough so that
this effect can be ignored. On the other hand, for iron-core SNe,
during the accretion phase the matter can be large enough to be
important.

If at early times the matter density profile is such that our
multi-angle effect is important, this will not be the case at later
times when the explosion has occurred and the matter profile
contracts toward the neutron star. In principle, therefore,
interesting time-dependent features in the oscillation probability
can occur.

A large matter effect can be ``rotated away'' from the EOMs when it is
identical for all modes. Here we have seen that even a perfectly
uniform medium provides a multi-angle variation of the matter effect.
We note that the matter fluxes would not be important, in contrast to
our first example of a homogeneous ensemble, because the relevant
quantity is the spread of the matter effect between different
modes. Therefore, whenever a flux term would be important, the matter
density term already provides a strong multi-angle effect.

In addition, the medium can have density variations caused by
convection and turbulence~\cite{Scheck:2007gw} that is known to affect
the MSW resonance under certain circumstances~\cite{Loreti:1995ae,
  Fogli:2006xy, Friedland:2006ta, Choubey:2007ga, Kneller:2007kg}.
Density variations in the transverse direction to the neutrino stream
lines may well cause important variations of the matter effect between
different modes. It remains to be investigated in which way collective
flavor transformations are affected.

\cleardoublepage

\pagestyle{normal}
\chapter{Collective Supernova Neutrino Transformations in Three
  Flavors}\label{chapter:coll3flavors}

We here extend our previous numerical analysis to the case of three
neutrino flavors. Our main results can be summarized as follows: (i)~A
two-flavor treatment indeed captures the full effect if one ignores
the mu-tau potential, $V_{\mu\tau}$, and if the ordinary MSW
resonances occur outside of the collective neutrino
region. (ii)~Including $V_{\mu\tau}$ can strongly modify the $\nu_e$
or $\bar\nu_e$ survival probabilities, influencing the neutrino signal
from the next galactic~SN. (iii)~The effect depends sensitively on a
possible deviation from maximal~$\theta_{23}$. (iv)~The inclusion of
$V_{\mu\tau}$ also affects the onset of the bipolar
conversions. (v)~Multi-angle matter decoherence possibly supresses
these effects.

\section{Introduction}                        \label{sec:introduction}

As it was discussed in Chapter~\ref{chapter:oscillations}, neutrinos
of different flavor suffer different refraction in matter. The energy
shift between $\nu_e$ and $\nu_\mu$ or $\nu_\tau$ is $
V_{\rm CC}=\sqrt{2}\,G_{\rm F}Y_e n_B$ with $G_{\rm F}$ the Fermi
constant, $n_B$ the baryon density, and $Y_e=n_e/n_B$ the electron
fraction. The potential $V_{\rm CC}$ is caused by the charged-current
$\nu_e$-electron interaction that is absent for $\nu_\mu$ and
$\nu_\tau$. For a matter density $\rho=1$ g cm$^{-3}$ we have
$\sqrt{2}\,G_{\rm F}n_B = 7.6\times10^{-14}$~eV, yet this small energy
shift is large enough to be of almost universal importance for
neutrino oscillation physics.

In normal matter, $\mu$ and $\tau$ leptons appear only as virtual
states in radiative corrections to neutral-current $\nu_\mu$ and
$\nu_\tau$ scattering, causing a shift $ V_{\mu\tau}= \sqrt{2}\,G_{\rm
  F}Y_\tau^{\rm eff}n_B$ between $\nu_\mu$ and~$\nu_\tau$. It has the
same effect on neutrino dispersion as real $\tau$ leptons with an
abundance~\cite{Botella:1986wy}
\begin{equation}
 Y_\tau^{\rm eff}=\frac{3\sqrt{2}\,G_{\rm F}m_\tau^2}{(2\pi)^2}
 \left[\ln\left(\frac{m_W^2}{m_\tau^2}\right)-1+\frac{Y_n}{3}\right]
 =2.7\times10^{-5}\,,
\label{eq:Ytau}
\end{equation}
as defined in Eq.~(\ref{eq:Vmt}), where $n_e=n_p$ was assumed. For the
neutron abundance we have used $Y_n=n_n/n_B=0.5$, but it provides only
a 2.5\% correction so that its exact value is irrelevant. As discussed
in Chapter~\ref{chapter:NSI}, a large non-standard contribution to
$Y_\tau^{\rm eff}$ can arise from radiative corrections in
supersymmetric models~\cite{Roulet:1995qb}, but we will focus in this
chapter\footnote{We will treat the non-standard case in
  Chapter~\ref{chapter:SN_NSI}.} on the standard-model effect alone.

This ``mu--tau matter effect'' modifies oscillations if $
V_{\mu\tau} > \Delta m^2/2E$. For propagation through the Earth and
for $\Delta m^2_{\rm atm}=2$--$3\times10^{-3}~{\rm eV}^2$, this occurs
for neutrino energies $E > 100$~TeV. The oscillation length then far
exceeds $r_{\rm Earth}$ so that $ V_{\mu\tau}$ is irrelevant for
the high-energy neutrinos that are searched for by neutrino
telescopes.

\begin{figure}
\begin{center}
\includegraphics[angle=0,width=0.55\textwidth]{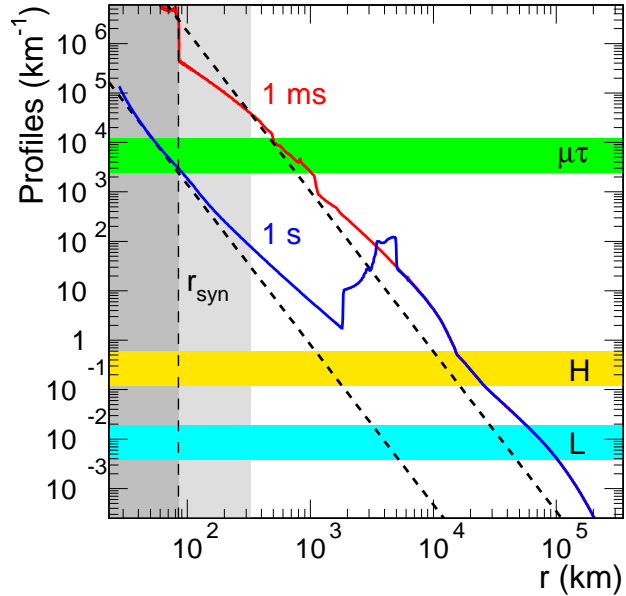}
\caption{\small Density profiles in terms of the weak potential $
  V_{\rm CC}=\sqrt{2}\,G_{\rm F}n_e$ at 1~ms and 1~s post bounce of
  the numerical SN models described in Ref.~\cite{Arcones:2006uq}
  (solid lines).  The dashed lines represent the simplified matter
  profile of Eq.~(\ref{eq:profile}) for $\lambda_0 = 4\times
  10^{6}$~km$^{-1}$ and $\lambda_0 = 5\times 10^{9}$~km$^{-1}$, used
  in our numerical calculations in Figure \ref{fig:evolution}. As
  horizontal bands we indicate the conditions $ V_{\mu\tau}=\Delta
  m^2_{\rm atm}/2E$, $ V_{\rm CC}=\Delta m^2_{\rm atm}/2E$, and $
  V_{\rm CC}=\Delta m^2_{\rm sol}/2E$ for a typical range of SN
  neutrino energies. The gray shaded range of radii corresponds to the
  region of collective neutrino transformations. Within the radius
  $r_{\rm syn}$ the collective oscillations are of the synchronized
  type~\cite{EstebanPretel:2007yq}.\label{fig:profiles_ch5}}
\end{center}
\end{figure}

Alternatively, the mu--tau matter effect can be important at the large
densities encountered by neutrinos streaming off a SN
core~\cite{Akhmedov:2002zj}. For $E=20~{\rm MeV}$ the condition $
V_{\mu\tau}=\Delta m^2_{\rm atm}/2E$ implies
$\rho\approx3\times10^7~{\rm g}~{\rm cm}^{-3}$. Numerical SN density
profiles~\cite{Arcones:2006uq} reveal that this occurs far beyond the
shock-wave radius during the accretion phase, but retracts close to
the neutrino sphere after the explosion has begun. To illustrate this
point we show in Fig.~\ref{fig:profiles_ch5} the same matter density
profiles as in Ref.~\cite{Arcones:2006uq} at 1~ms post bounce (red
line) and at 1~s post bounce (blue line). As a green horizontal band
we indicate the condition $ V_{\mu\tau}=\Delta m^2_{\rm atm}/2E$ for a
typical range of SN neutrino energies, whereas the yellow and
light-blue bands indicate the densities corresponding to the
$H$-resonance (driven by $\Delta m_{\rm atm}^2$) and the $L$-resonance
(driven by $\Delta m_{\rm sol}^2$). The $\nu_\mu$, $\nu_\tau$,
$\bar\nu_\mu$ and $\bar\nu_\tau$ fluxes from a SN are virtually
identical, leaving the $\mu\tau$-resonance moot, whereas the $H$- and
$L$-resonances cause well-understood consequences that are completely
described by the energy-dependent swapping probabilities for $\nu_e$
and $\bar\nu_e$ with some combination $\nu_x$ of the $\mu$ and $\tau$
flavor~\cite{Dighe:1999bi}, as discussed in
Chapter~\ref{chapter:oscillations}. Therefore, the traditional view
has been that genuine three-flavor effects play no role for SN
neutrino oscillations unless mu and tau neutrinos are produced with
different fluxes~\cite{Akhmedov:2002zj}.

It is now well known, however, that the traditional picture was not
complete. As discussed in Chapter~\ref{chapter:coll2flavors},
neutrino-neutrino interactions cause large collective flavor
transformations in the SN region out to a few 100~km (gray shaded
region in Fig.~\ref{fig:profiles_ch5}).

\section{Equations of motion}                         \label{sec:eoms}

Just as discussed in Chapter~\ref{chapter:coll2flavors}, mixed
neutrinos can be described by matrices of density $\rho_{\bf p}$ and
$\bar\rho_{\bf p}$ for each (anti)neutrino mode. The EOMs are the ones
given in Eqs.~(\ref{eq:gen_eoms})
\begin{equation}
\I\partial_t\varrho_{\bf p}=[{\sf H}_{\bf p},\varrho_{\bf p}]\,,
\label{eq:eoms}
\end{equation}
where the Hamiltonian is~\cite{Sigl:1992fn}
\begin{equation}
 {\sf H}_{\bf p}=\Omega_{\bf p}
 +{\sf V}+\sqrt{2}\,G_{\rm F}\!
 \int\!\frac{\D^3{\bf q}}{(2\pi)^3}
 \left(\varrho_{\bf q}-\bar\varrho_{\bf q}\right)
 (1-{\bf v}_{\bf q}\cdot{\bf v}_{\bf p})\,,
\label{eq:hamiltonian}
\end{equation}
and ${\bf v}_{\bf p}$ is the velocity, $\Omega_{\bf p}={\rm
  diag}(m_1^2,m_2^2,m_3^2)/2|{\bf p}|$ is the matrix of vacuum
oscillation frequencies and ${\sf V}=\sqrt{2}\,G_{\rm F}n_B\,{\rm
  diag}(Y_e,0,Y_\tau^{\rm eff})$ accounts for the matter effect.

In spherical symmetry the EOMs can be expressed as a closed set of
differential equations along the radial direction. We solve them
numerically as previously described, now using $3\times3$ matrices
instead of polarization vectors. The factor $(1-{\bf v}_{\bf
  q}\cdot{\bf v}_{\bf p})$ in the Hamiltonian implies multi-angle
effects for neutrinos moving on different
trajectories~\cite{Sawyer:2004ai, Sawyer:2005jk, Duan:2006an}.
However, for realistic SN conditions the modifications are small,
allowing for a single-angle approximation. We implement this
approximation by launching all neutrinos with $45^\circ$ relative to
the radial direction, see Chapter~\ref{chapter:coll2flavors}.

As a further simplification we use a monochromatic spectrum
($E=20~{\rm MeV}$), ignoring the spectral splits caused by collective
oscillation effects~\cite{Duan:2006an, Raffelt:2007cb, Raffelt:2007xt,
  Duan:2007fw, Fogli:2007bk}. Oscillation effects require
flavor-dependent flux differences. One expects $F^{R_\nu}_{\nu_e} >
F^{R_\nu}_{\bar\nu_e} > F^{R_\nu}_{\nu_\mu} = F^{R_\nu}_{\bar\nu_\mu}
= F^{R_\nu}_{\nu_\tau} = F^{R_\nu}_{\bar\nu_\tau}$.  The equal parts
of the fluxes drop out of the EOMs, so as initial condition we use
$F^{R_\nu}_{\nu_\mu,\bar\nu_\mu,\nu_\tau,\bar\nu_\tau}=0$ and
$F^{R_\nu}_{\nu_e} =(1+\epsilon) F^{R_\nu}_{\bar\nu_e}$ with
$\epsilon=0.25$.

For the neutrino parameters we use $\Delta m^2_{21}=\Delta m^2_{\rm
  sol}=7.6\times10^{-5}~{\rm eV}^2$, $\Delta m^2_{31}=\Delta m^2_{\rm
  atm}=2.4\times10^{-3}~{\rm eV}^2$, $\sin^2\theta_{12}=0.32$,
$\sin^2\theta_{13}=0.01$, and a vanishing Dirac phase
$\delta=0$, all consistent with
measurements~\cite{Maltoni:2004ei,Fogli:2005cq,GonzalezGarcia:2007ib}.
We consider the entire allowed range $0.35\leq\sin^2\theta_{23}\leq 0.65$
because our results depend sensitively on $\theta_{23}$.

We use a fixed matter profile of the form $\rho\propto r^{-3}$,
implying a radial variation of the weak potential~of
\begin{equation}
  V_{\rm CC}=Y_e\lambda_0\,\left(\frac{R_{\nu}}{r}\right)^3\,,
\label{eq:profile}
\end{equation}
where $R_{\nu}=10$~km is our nominal neutrino sphere radius and
$Y_e=0.5$. In Fig.~\ref{fig:profiles_ch5} we show this profile (dashed
lines) for two different values of $\lambda_0= 4\times
10^{6}$~km$^{-1}$ and $\lambda_0 = 5\times 10^{9}$~km$^{-1}$. For the
former case, the $H$-resonance is at $r^H_{\rm res}=1.9\times 10^3
$~km, the $L$-resonance at $r^L_{\rm res}=8.3\times 10^3$~km, and the
$\mu\tau$-resonance at $r^{\mu\tau}_{\rm res}=71$~km. For the latter
they are at $r^H_{\rm res}=2.0\times 10^4 $~km, $r^L_{\rm
  res}=9.0\times 10^4$~km, and $r^{\mu\tau}_{\rm
  res}=760$~km.\footnote{We loosely refer to the radius where $\Delta
  m_{\rm atm}^2/2E= V_{\mu\tau}$ as the $\mu\tau$-resonance, although
  this would be correct only for a small vacuum mixing angle in the
  23-subsystem.}

We remind the reader that the strength of the neutrino-neutrino
interaction can be parameterized by
\begin{equation}
\mu_0 = \sqrt{2}G_{\rm F}(F^{R_\nu}_{\bar\nu_e}-F^{R_\nu}_{\bar\nu_x})\,,
\end{equation}
where the fluxes are taken at the neutrino sphere radius~$R_\nu$. We
will again assume $\mu_0=7\times 10^5$~km$^{-1}$. In the single-angle
approximation where all neutrinos are launched with $45^\circ$
relative to the radial direction~\cite{EstebanPretel:2007ec}, the
radial dependence of the neutrino-neutrino interaction strength can be
explicitly written as
\begin{equation}
\mu(r) = \mu_0 \frac{R_\nu^4}{r^4}\frac{1}{2-R_\nu^2/r^2}\,.
\end{equation}
 While the $r^{-4}$ scaling of $\mu(r)$ for $r\gg R_\nu$ is
generic, the overall strength $\mu_0$ depends on the neutrino fluxes
and on their angular divergence, i.e., on the true radius of the
neutrino sphere. Our $R_\nu=10$~km is not meant to represent the physical
neutrino sphere, it is only a nominal radius where we fix the inner
boundary condition for our calculation.

The collective neutrino oscillations are of the synchronized type
within the synchronization radius. For our chosen $\mu_0$ and for the
assumed excess $\nu_e$ flux of 25\% we find $r_{\rm syn}\simeq
100$~km as indicated in Fig.~\ref{fig:profiles_ch5}. Collective flavor
transformations occur at $r > r_{\rm syn}$. Therefore, the $\mu\tau$
matter effect can be important only if it is sufficiently large for $r
> r_{\rm syn}$.

Figure~\ref{fig:profiles_ch5} illustrates that the region where the
$\mu\tau$-resonance takes place depends on the time after bounce. For
realistic values of the matter density profile and neutrino-neutrino
interaction, one expects $r^{\mu\tau}_{\rm res}$ to lie far beyond the
collective region at early times. This can be inferred from the
relative position of $r_{\rm syn}$ and the intersection of the 1~ms
profile and the green band. At later times though the proto neutron
star contracts and $r^{\mu\tau}_{\rm res}$ moves to smaller radii. Eventually
$r^{\mu\tau}_{\rm res}$ becomes smaller than $r_{\rm syn}$, at which point
$ V_{\mu\tau}$ becomes irrelevant.

In order to mimic these different situations we will use a simple
power-law matter profile of the form in Eq.~(\ref{eq:profile}). In
other words, we will use a mu-tau matter potential of the form
\begin{equation}
 V_{\mu\tau}=Y_{\tau}^{\rm eff}\lambda_0\,\left(\frac{R}{r}\right)^3\,,
\label{eq:mutaueff}
\end{equation}
with a fixed $Y_{\tau}^{\rm eff}$ given by Eq.~(\ref{eq:Ytau}) and a
variable coefficient $\lambda_0$. Therefore early and late times can
be reproduced by considering large and small values of $\lambda_0$,
respectively, as can be seen in Fig.~\ref{fig:profiles_ch5}. In other
words, we will always assume that the ordinary MSW resonances are far
outside of the collective neutrino region, whereas the $\mu\tau$
resonance can lie at smaller (vanishing $\mu\tau$ matter effect) or
larger (large $\mu\tau$ matter effect) radii than~$r_{\rm syn}$.

\section{Vanishing mu-tau matter effect}        \label{sec:smallmutau}

\begin{figure}[!ht]
\centering
\includegraphics[angle=0,width=0.98\textwidth]{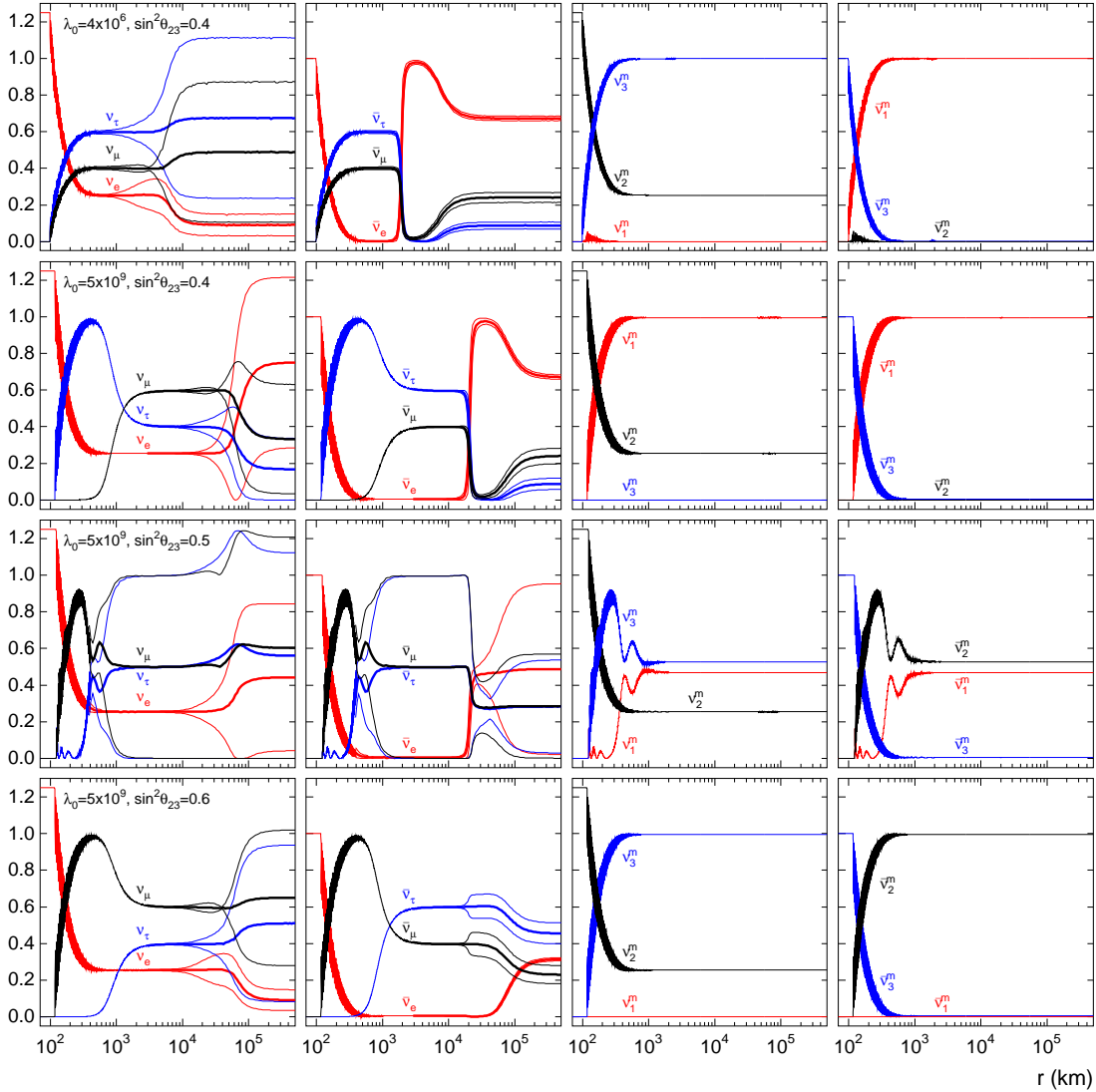}
\caption{\small Radial evolution of the neutrino fluxes, normalized to
  the initial $\bar{\nu}_e$ flux, for a fixed neutrino energy
  ($E_\nu=20$ MeV) and an inverted $\Delta m^2_{\rm atm}$. {}From left
  to right: neutrino weak eigenstates, antineutrino weak eigenstates,
  neutrino propagation eigenstates and antineutrino propagation
  eigenstates. In the first two columns, after bipolar conversions we
  show the average as thick lines and the envelopes of the
  fast-oscillating curves as thin lines. The top row shows the case of
  a vanishing $\mu\tau$ matter effect, while the three bottom rows use
  a large $\mu\tau$ effect with different values for the 23 mixing
  angle as
  indicated~\cite{EstebanPretel:2007yq}. \label{fig:evolution}}
\end{figure}

As a first case we consider the traditional assumption of a vanishing
$\mu\tau$ matter effect, which we account for using a value of
$\lambda_0=4\times 10^6$. We assume an inverted $\Delta m^2_{\rm atm}$
and use a non-maximal value $\sin^2\theta_{23}=0.4$.  Our numerical
calculations for this case are shown in the top row of
Fig.~\ref{fig:evolution}. The first two panels correspond to the
radial evolution of the fluxes of the weak interaction eigenstates of
neutrinos and antineutrinos, respectively, whereas in the last two
panels we show the evolution of the propagation eigenstates.  These
are the eigenstates of $\Omega_{\bf p} +{\sf V}$, i.e., of that part
of the Hamiltonian Eq.~(\ref{eq:hamiltonian}) that does not include
the neutrino-neutrino interactions. In the collective neutrino region,
we observe the usual pair conversion of the $\nu_e$ and $\bar\nu_e$
fluxes into the $\mu$ and $\tau$ flavors. Had we chosen a maximal $23$
mixing angle, the appearance curves for these flavors would be
identical.

For larger distances the evolution consists of ordinary MSW
transformations that are best pictured in the basis of instantaneous
propagation eigenstates in matter (last two
panels). Beyond the collective transformation region, all neutrinos
and antineutrinos stay fixed in their propagation eigenstates. In the
weak-interaction basis, on the other hand, this implies fast
oscillations because we have a fixed energy, preventing kinematical
decoherence between different energy modes. In the panels for neutrino
and antineutrino interaction states, for radii beyond the
dense-neutrino region we show as thick lines the average evolution as
well as the envelopes of the fast-oscillating flavor fluxes.

\begin{figure}
\begin{minipage}[b]{0.49\linewidth}
\centering
\includegraphics[angle=0,width=\textwidth]{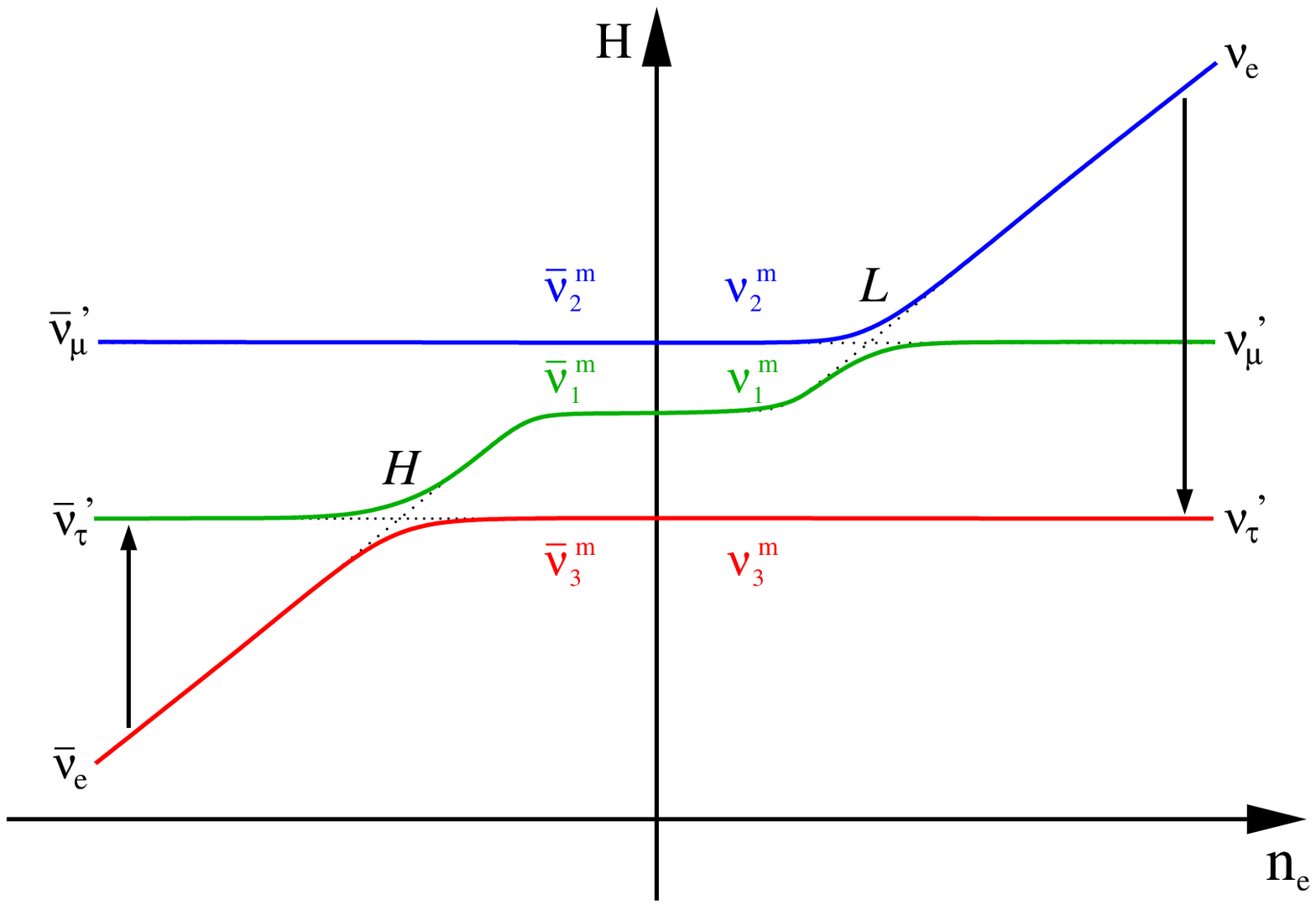}
\vspace{0.48cm}
\caption{\small Level crossing scheme of neutrino conversion for the
  inverted hierarchy in a medium with a vanishing $ V_{\mu\tau}$
  (upper-left panel) and a large $ V_{\mu\tau}$ with 23-mixing in the
  first octant (upper-right panel) or the second octant (bottom panel).
  The arrows indicate the transitions caused by collective flavor
  transformations~\cite{EstebanPretel:2007yq}. \label{fig:crossing}}
\end{minipage}
\begin{minipage}[b]{0.49\linewidth}
\centering
\includegraphics[angle=0,width=\textwidth]{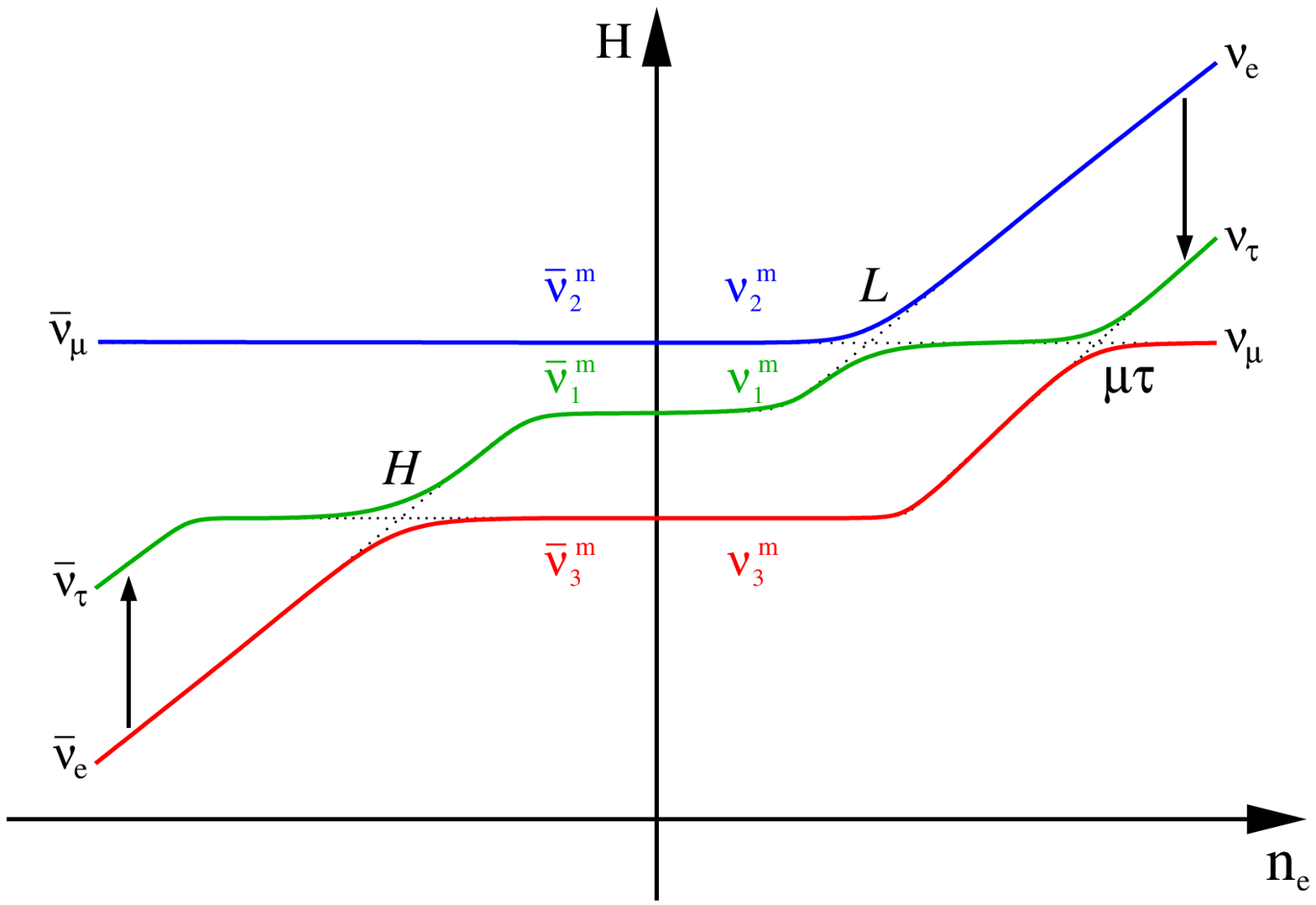}
\vskip12pt
\includegraphics[angle=0,width=\textwidth]{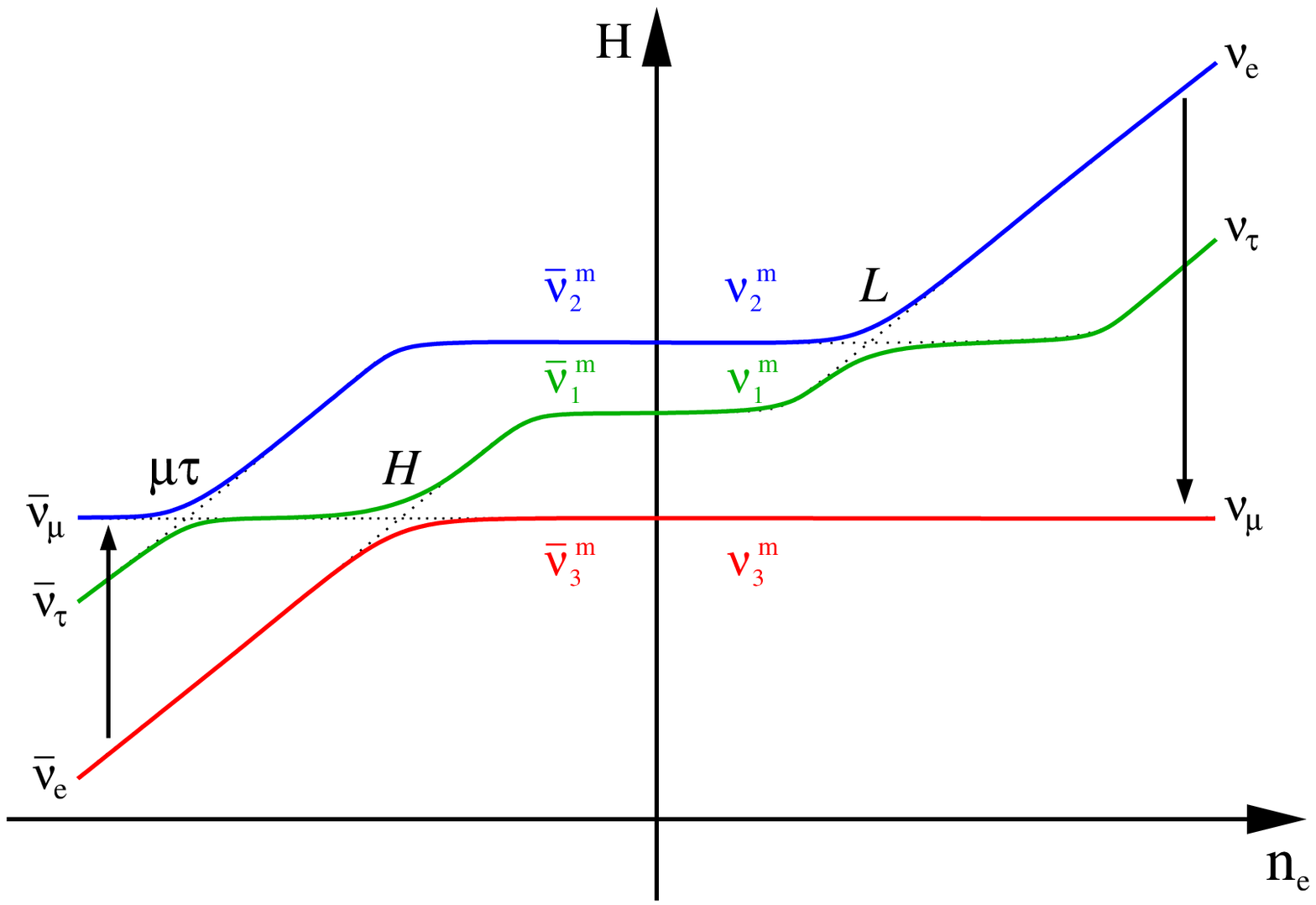}
\end{minipage}
\end{figure}

Another way of describing this evolution is by the level crossing
schemes of Fig.~\ref{fig:crossing}. The upper left panel represents
the case of vanishing $ V_{\mu\tau}$. The right panels represent the
case with large $ V_{\mu\tau}$ and a 23-mixing angle in the first
octant (upper right panel) or second octant (bottom panel). These
level crossing schemes are the same ones shown in
Fig.~\ref{fig:generic} of Chapter~\ref{chapter:oscillations}, but
adding the effect of collective transformations (vertical arrows). The
upper (blue) line corresponds to propagation eigenstate~2, the middle
(green) line to~1, and the bottom (red) line to~3, a scheme
representing the inverted hierarchy case.

While in vacuum the propagation eigenstates coincide with the mass
eigenstates, at large densities they correspond to weak interaction
eigenstates. For vanishing $ V_{\mu\tau}$ and at the low energies
relevant to our problem, the $\mu$ and $\tau$ flavor are not
distinguishable so that any convenient linear combination can be
chosen as interaction eigenstates. It is convenient to introduce the
states $\nu_\mu'$ and $\nu_\tau'$ that correspond to a vanishing
23-mixing angle, i.e., they diagonalize the 23-subsystem. If the small
13-mixing angle were to vanish, the 3-mass eigenstate would coincide
with $\nu_\tau'$. In the upper left panel of Fig.~\ref{fig:crossing}
and using the $(\nu_e,\nu_\mu',\nu_\tau')$ basis, the 2-state connects
adiabatically to $\nu_e$ and $\bar\nu_\mu'$, whereas the 3-state
connects adiabatically to $\bar\nu_e$ and $\nu_\tau'$.

At the neutrino sphere, the fluxes are prepared in $\nu_e$ and
$\bar\nu_e$ eigenstates, which in the case of inverted mass hierarchy
coincide with the propagation (or matter) eigenstates $\nu^m_2$
and $\bar\nu^m_3$, respectively.  In the absence of
neutrino-neutrino interactions, since the $L$-resonance is always
adiabatic, the $\nu_e$'s leave the star as $\nu_2$. In the case of
$\bar\nu_e$ the evolution depends on $\sin^2\theta_{13}$, see
Fig.~\ref{fig:adresHL}. For values larger than $10^{-3}$ they
propagate also adiabatically (MSW transformation) and escape as
$\bar\nu_3$, whereas for values smaller than $10^{-5}$ the transition
at the $H$-resonance is strongly non-adiabatic: there is a jump of
matter eigenstates from $\bar\nu^m_3$ to $\bar\nu^m_1$ and
the $\bar\nu_e$'s leave the star as $\bar\nu_1$. As a consequence, the
survival probability is $P(\nu_e\rightarrow \nu_e) \approx
\sin^2\theta_{12}$ and $P(\bar\nu_e\rightarrow \bar\nu_e) \approx
\sin^2\theta_{13}$ or $\cos^2\theta_{12}$ for large and small
$\theta_{13}$, respectively.

In the presence of neutrino-neutrino interactions, important
collective effects take place in the inner SN layers, where the
neutrino density is high.  We observe in the first two panels of
Fig.~\ref{fig:evolution} that collective pair transformations convert
the $\nu_e$ and $\bar\nu_e$ fluxes to $\nu_\tau'$ and $\bar\nu_\tau'$
as indicated by the arrows in the upper left panel of
Fig.~\ref{fig:crossing}.  The consequences for the subsequent
evolution are dramatic. In the case of $\nu_e$ a fraction equal to
$\epsilon F_{\bar\nu_e}$ stays in $\nu_2^m$ and evolves as in the
absence of neutrino-neutrino interactions, while the rest of $\nu_e$
are transformed to $\nu_3^m$. As a consequence, the final $\nu_e$
flux, normalized to the initial $\bar\nu_e$ one, is expected to be
approximately $\epsilon \sin^2\theta_{12}\simeq 0.08$, see thick line
in the upper left panel in Fig.~\ref{fig:evolution}.  In the case of
antineutrinos the effect of the collective pair conversion is to
interchange the eigenstates in which $\bar\nu_e$ and $\bar\nu_\tau'$
arrive at the $H$-resonance. Now $\bar\nu_e$ enters the resonance as
$\bar\nu_1^m$ instead of $\bar\nu_3^{\rm m}$. Therefore, for
$\sin^2\theta_{13}\gtrsim 10^{-3}$ the resonance is adiabatic and the
$\bar\nu_e$'s leave the star as $\bar\nu_1$, leading to a final
normalized flux of approximately $\cos^2\theta_{12}\simeq 0.68$, see
the thick line in the second panel in
Fig.~\ref{fig:evolution}. Instead, if $\sin^2\theta_{13}\lesssim
10^{-5}$ again there is a jump of matter eigenstates from
$\bar\nu^m_1$ to $\bar\nu^m_3$ at the $H$-resonance.  In this case
$\bar\nu_e$ leaves the star as $\bar\nu_3$, leading to a normalized
$\bar\nu_e$ flux equal to $\sin^2\theta_{13}$.

\begin{figure}
\begin{center}
\includegraphics[angle=0,width=0.4\textwidth]{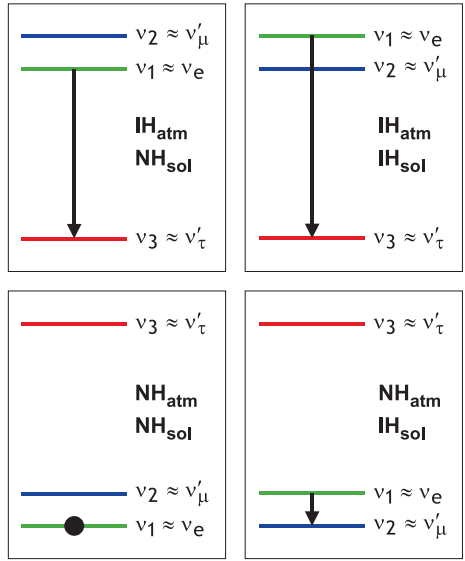}
\caption{\small Vacuum level diagram for all hypothetical combinations
  of atmospheric and solar mass hierarchies (normal or inverted). The
  12 and 13 mixing angles are assumed to be very small, mimicking the
  effect of ordinary matter. The effect of collective conversions is
  indicated by an arrow~\cite{EstebanPretel:2007yq}.\label{fig:levels}}
\end{center}
\end{figure}

The impact of collective effects is easier to understand if we do as
in the two-flavor system, where we have discussed that the impact of
ordinary matter can be transformed away by going into a rotating
reference frame for the polarization vectors. Collective conversions
proceed in the same way as they would in vacuum, except that the
effective mixing angle is reduced\footnote{For the sake of simplicity
  we will not take here into account the suppresion of collective
  neutrino transformations due to large matter densities, studied in
  Section~\ref{sec:densematter}.}. Therefore, assuming an inverted
hierarchy (IH) for the atmospheric mass splitting and a normal
hierarchy (NH) for the solar splitting, we should consider the level
scheme as in the upper left panel of Fig.~\ref{fig:levels}. The mass
eigenstates now approximately coincide with the interaction
eigenstates because the 23-mixing angle was removed by going to the
primed states, and the mixing angles involving $\nu_e$ are effectively
made small by the presence of matter. Of course, this level scheme
does not adiabatically connect to the true vacuum situation.

The initial state consists of $\nu_e$ and $\bar\nu_e$ and thus
essentially of $\nu_1$ and $\bar\nu_1$. Collective conversions
driven by $\Delta m^2_{\rm atm}$ then transform $\nu_1\bar\nu_1$
pairs to $\nu_3\bar\nu_3$ pairs in the familiar two-flavor way. If
both hierarchies are normal, we begin in the lowest-lying state and
nothing happens.
In the hypothetical case where both hierarchies are inverted (upper
right panel in Fig.~\ref{fig:levels}), we begin in the highest state
and $\Delta m^2_{\rm atm}$ drives us directly to the lowest state.
Finally, if the atmospheric hierarchy is normal and the solar one is
inverted (lower right panel in Fig.~\ref{fig:levels}), collective
transformations driven by $\Delta m^2_{\rm sol}$ take us to the
lowest state.

We have numerically solved the evolution of the three-flavor system
with a realistic SN matter profile and found that the results confirm
this simple picture. In a two-flavor treatment, the much smaller
$\Delta m^2_{\rm sol}$ leads to collective transformations at a much
larger radius than $\Delta m^2_{\rm atm}$. In a three-flavor
treatment, $\Delta m^2_{\rm atm}$ therefore acts first and takes us
directly to the lowest-lying state if the atmospheric hierarchy is
inverted. Otherwise only the hypothetical case of the lower-right
panel in Fig.~\ref{fig:levels} is an example where $\Delta m^2_{\rm
sol}$ plays any role. We have numerically verified that normal
$\Delta m_{\rm atm}^2$ combined with inverted $\Delta m_{\rm sol}^2$
is the only case where $\Delta m_{\rm sol}^2$ drives collective
transformations. Since $\Delta m_{\rm sol}^2$ is measured to be
normal, the previous two-flavor treatments based on $\Delta m_{\rm
atm}^2$ and $\theta_{13}$ fortuitously capture the full effect.

We conclude that in the limit of a vanishing $\mu\tau$ effect the
collective flavor transformations and the subsequent MSW evolution
factorize and that the collective effects are correctly treated in a
two-flavor picture. Of course, this situation may change if the matter
profile is so low that the ordinary MSW effects occur in the same
region as the collective phenomena~\cite{Duan:2007sh}.

\section{Large mu-tau matter effect}            \label{sec:largemutau}

\begin{figure}[t]
\centering
\includegraphics[angle=0,width=0.75\textwidth]{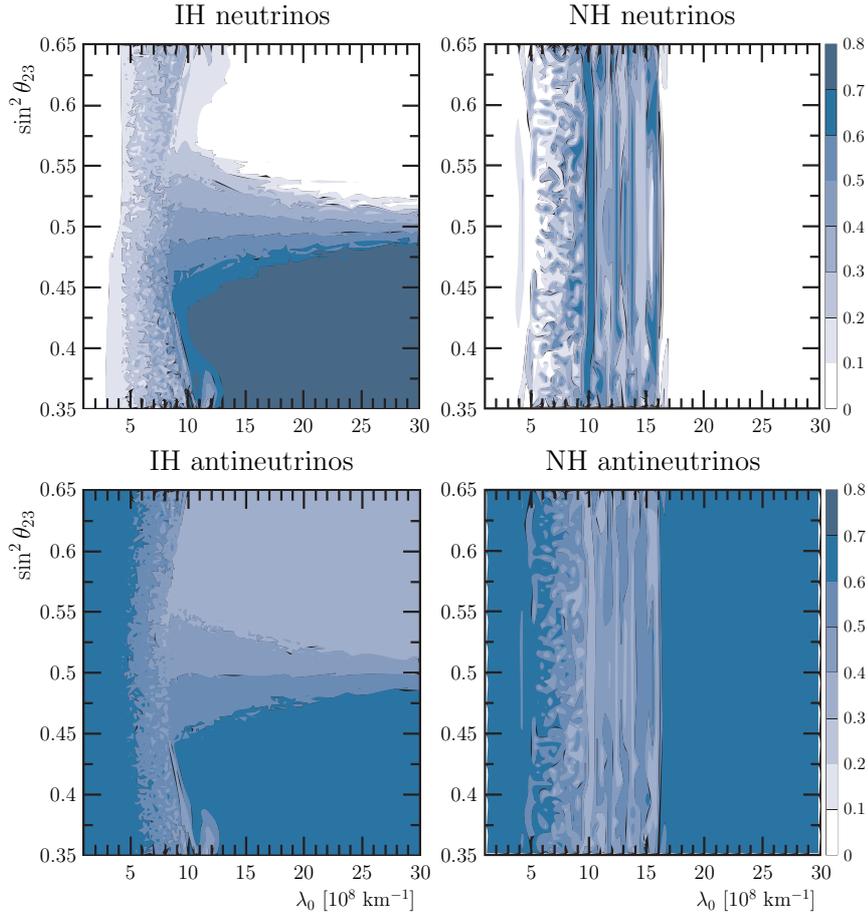}
\caption{\small Contours in the space of $\sin^2\theta_{23}$ and
  $\lambda_0$ for the $\nu_e$ (top) and $\bar\nu_e$ (bottom) fluxes
  emerging from the SN surface for both normal (right) and inverted
  (left) mass hierarchy. All fluxes are normalized to the initial
  $\bar{\nu}_e$ flux.  We show values averaged over fast vacuum
  oscillations~\cite{EstebanPretel:2007yq}. \label{fig:contours}}
\end{figure}

Next we calculate the flavor evolution for the same model, now
including a significant $ V_{\mu\tau}$, i.e. we assume a large
$\lambda_0$. In this case the flavor content of the neutrino and
antineutrino fluxes emerging from the SN surface depend on the
strength of $ V_{\mu\tau}$ as well as the choice of $\theta_{23}$, as
can be seen in the corresponding panels of
Fig.~\ref{fig:evolution}. In this figure, one can also notice that the
inclusion of $ V_{\mu\tau}$ delays the onset of the bipolar
conversions. We will discuss the latter effect in the next section,
while here we will concentrate on the former one. This dependence is
best illustrated with the help of the contour plot
Fig.~\ref{fig:contours} where we show the $\nu_e$ and $\bar\nu_e$
fluxes emerging from the SN, averaged over fast vacuum oscillations.

If $ V_{\mu\tau}$ is so large that the mu-tau effect is strong in the
region of collective neutrino oscillations, there are two stable
limiting cases, depending on the 23 mixing angle. If the mixing angle
is sufficiently non-maximal and in the first octant, the collective
oscillations transform the initially prepared $\nu_e$ and $\bar\nu_e$
fluxes to the propagation eigenstates as indicated by the arrows in
the upper right panel of Fig.~\ref{fig:crossing}, i.e., we observe
pair transformations to $\nu_\tau\bar\nu_\tau$.

This behavior is understood if we assume that in the $\mu\tau$ system
we can once more go to a rotating frame and now simply imagine that
the 23 mixing angle is effectively small by the impact of the
$\mu\tau$ matter effect. In this case $\nu_3\approx \nu_\tau$. Since
collective quasi-vacuum oscillations take us to the lowest-lying
state, the $\nu_3$ state in the inverted hierarchy, we are
effectively taken to $\nu_\tau\bar\nu_\tau$ pairs. Instead, if the 23
mixing angle is in the second octant, $\nu_\mu$ and $\nu_\tau$ switch
roles, explaining that now $\nu_3\approx\nu_\mu$ and
$\bar\nu_3\approx\bar\nu_\mu$.

\begin{figure}[t]
\begin{center}
\includegraphics[angle=0,width=0.49\textwidth]{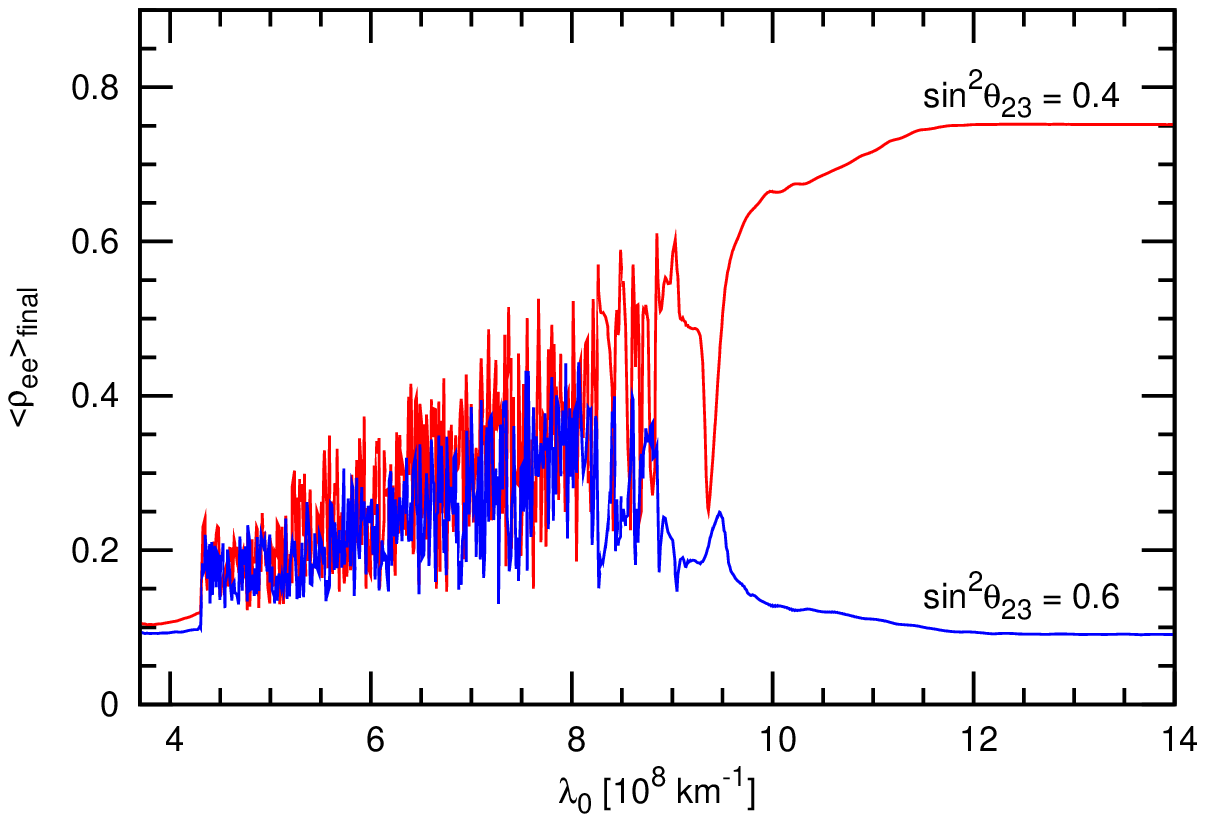}
\includegraphics[angle=0,width=0.49\textwidth]{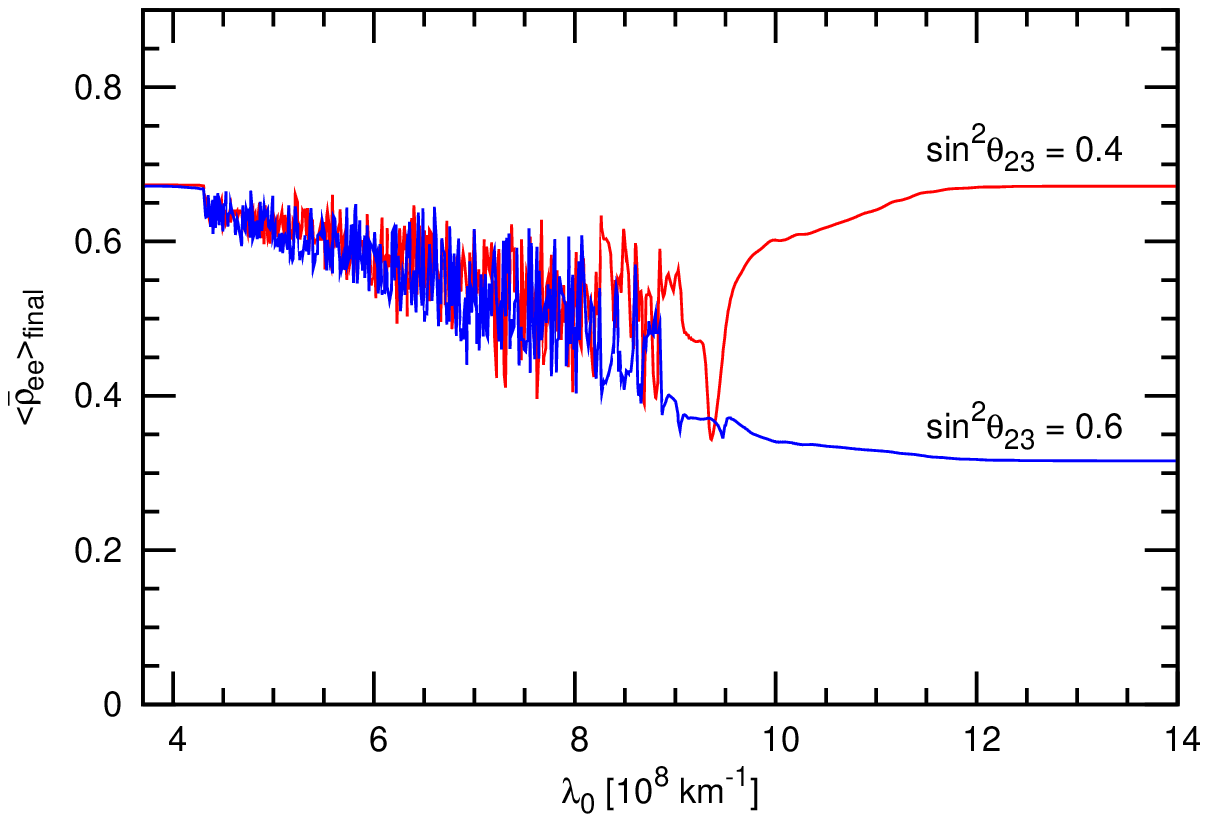}
\includegraphics[angle=0,width=0.49\textwidth]{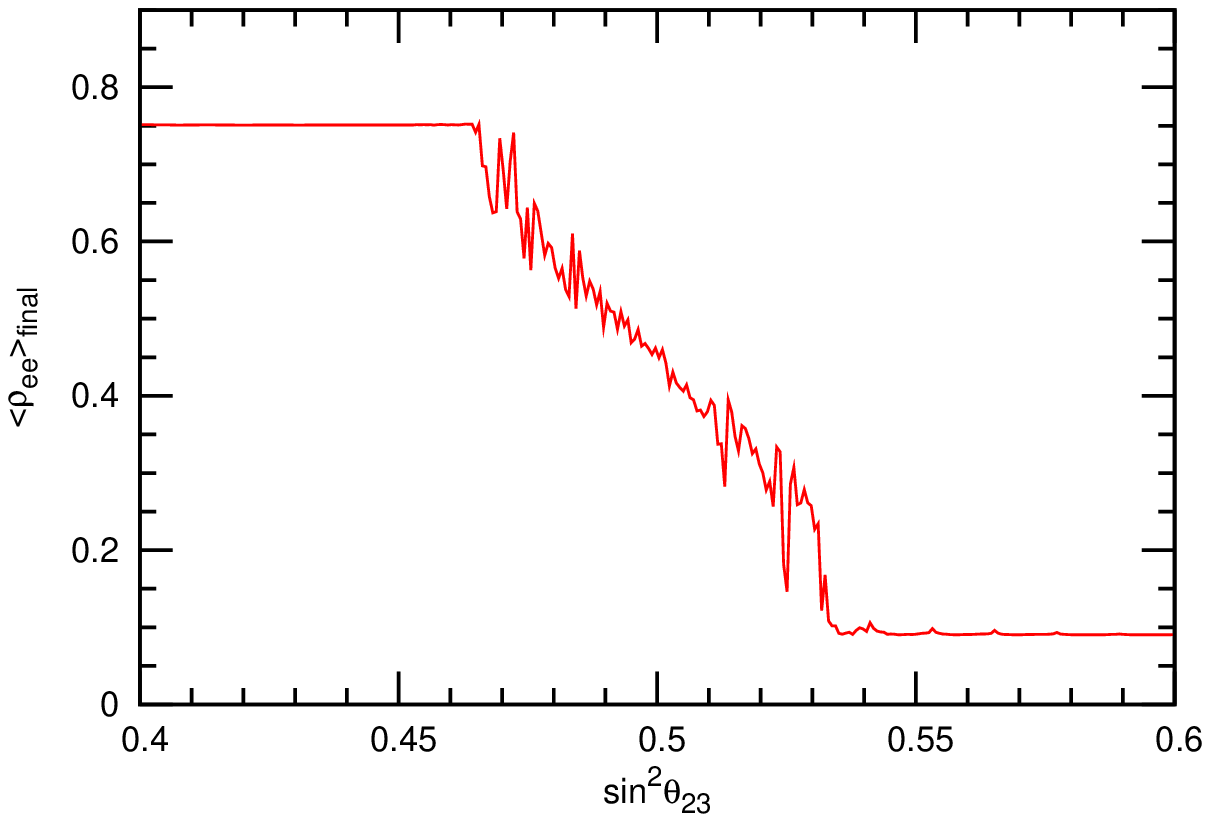}
\includegraphics[angle=0,width=0.49\textwidth]{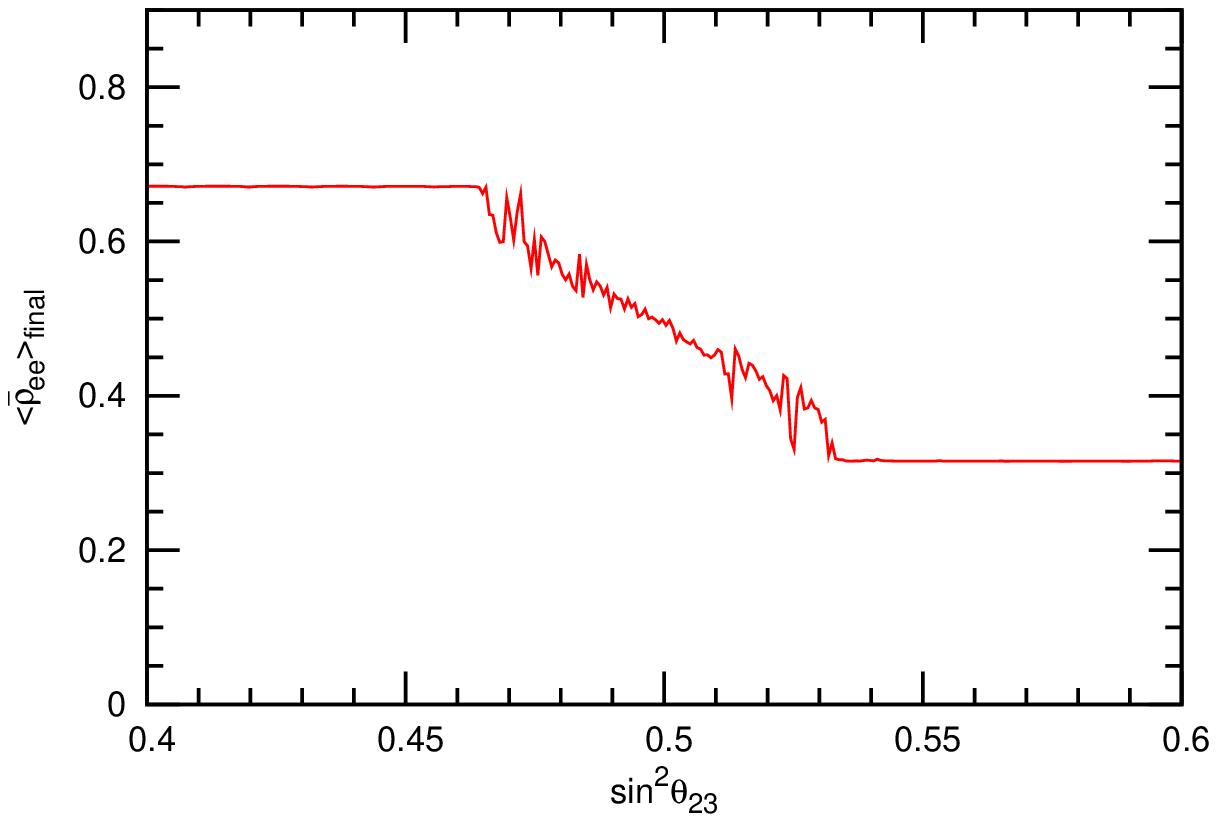}
\caption{\small Fluxes of $\nu_e$ (left column) and $\bar\nu_e$ (right
  column), normalized to the initial $\bar{\nu}_e$ flux, emerging from
  the SN. The first row is done as a function of $\lambda_0$ for a 23
  mixing angle in the first (red line) or second (blue line) octant,
  while the second row is represented as a function of
  $\sin^2\theta_{23}$ for $\lambda_0= 1.85\times
  10^9$~km$^{-1}$. These curves represent cuts through the inverted
  hierarchy contour plots of Fig.~\ref{fig:contours} at the indicated
  values of $\lambda_0$ and
  $\sin^2\theta_{23}$~\cite{EstebanPretel:2007yq}. \label{fig:cuts}}
\end{center}
\end{figure}

For intermediate values of $ V_{\mu\tau}$ and for 23 mixing angles
near maximal, the final fluxes depend sensitively on parameters. For
intermediate values of $ V_{\mu\tau}$, there are also nontrivial
effects for the normal hierarchy. The collective effects do not place
the ensemble into propagation eigenstates, preventing a simple
interpretation. The sensitive dependence for intermediate $
V_{\mu\tau}$ is also illustrated in the top panels of
Fig.~\ref{fig:cuts} where we show the emerging average $\nu_e$ and
$\bar\nu_e$ fluxes as functions of $\lambda_0$ for two values of
$\theta_{23}$, one in the first and the other in the second octant. In
the bottom panels of Fig.~\ref{fig:cuts} we show the same $\nu_e$ and
$\bar\nu_e$ fluxes as functions of $\sin^2\theta_{23}$ for $\lambda_0=
1.85\times 10^9$~km$^{-1}$. One can notice how the fall of
$\bar\rho_{ee}$ is not exactly centered at $\sin^2\theta_{23}=0.5$ but
slightly shifted to smaller values. This is due to second-order
corrections to the $\mu\tau$ resonance condition.

This dependence on the $\theta_{23}$ octant leads to a clear imprint
on the final survival probability.  Let us first consider the first
octant. In the case of $\nu_e$ a fraction equal to $\epsilon
F_{\bar\nu_e}$ stays in $\nu_2^m$.  However the presence of the
$\mu\tau$-resonance in the neutrino channel makes the rest of the
$\nu_e$ to be transformed to $\nu_1^m$. 
Their subsequent evolution would depend on the
adiabaticity of the
$\mu\tau$-resonance, but it has been shown to be always
adiabatic~\cite{Akhmedov:2002zj}. As a consequence, the 
final $\nu_e$ flux is expected to be approximately
$\cos^2\theta_{12}+\epsilon \sin^2\theta_{12} \simeq 0.76$, see thick
line in the left panel of the second row in
Fig.~\ref{fig:evolution}. In the case of antineutrinos the situation
is completely analogous to the case of vanishing $ V_{\mu\tau}$
so that $P(\bar\nu_e\rightarrow \bar\nu_e) \approx \cos^2\theta_{12}$
or $\sin^2\theta_{13}$, depending on the value of $\theta_{13}$.

If $\theta_{23}$ belongs to the second octant, then the
$\mu\tau$-resonance lies in the antineutrino channel. The crucial
point is that now all $\bar\nu_e$ are transformed to
$\bar\nu_\mu=\bar\nu_2^m$ before reaching the
$\mu\tau$-resonance, see the lower panel in Fig.~\ref{fig:crossing}.
Taking into account that $\bar\nu_2^m$ does not encounter the
$H$-resonance, the survival probability will be always
$P(\bar\nu_e\rightarrow \bar\nu_e) \approx \sin^2\theta_{12}$,
independently of the value of $\theta_{13}$. On the other hand
neutrinos do not feel the $\mu\tau$-resonance and therefore their
propagation is the same as in the vanishing $ V_{\mu\tau}$ case.

We present in Table~\ref{tab:barnueP} a summary of the cases discussed
so far. One can see the importance of the presence of collective
neutrino effects, as well as the dependence on the strength of the
mu-tau matter effect.
\begin{table}[t]
\begin{center}
\caption{\small Summary of the approximate values of the $\bar\nu_e$ survival
probability for an inverted hierarchy, including or not collective
effects. Here a small (large) mixing angle $\theta_{13}$
stands for $\sin^2\theta_{13}\lesssim 10^{-5}$
($\sin^2\theta_{13}\gtrsim 10^{-3}$), while a small (large) $
V_{\mu\tau}$ represents $r^{\mu\tau}_{\rm res}$ being smaller
(larger) than~$r_{\rm syn}$.
\label{tab:barnueP}}
\vskip10pt
\begin{tabular}{cccccc}
\hline
\hline
Collective & \multirow{2}{*}{$ V_{\mu\tau}$} &
\multirow{2}{*}{$\theta_{23}$} &
\multirow{2}{*}{$\theta_{13}$} &
$\bar\nu_e$ & \multirow{2}{*}{$P(\bar\nu_e\rightarrow \bar\nu_e)$}\\
effects & & & & leaves as & \\
\hline
no & any & any & small & $\bar\nu_1$ & $\cos^2\theta_{12}$\\
no & any & any & large & $\bar\nu_3$ & $\sin^2\theta_{13}$\\
yes & small & any & small & $\bar\nu_3$ & $\sin^2\theta_{13}$\\
yes & small  & any & large & $\bar\nu_1$ & $\cos^2\theta_{12}$\\
yes & large & $<\pi/4$ & small & $\bar\nu_3$ & $\sin^2\theta_{13}$\\
yes & large  & $<\pi/4$ & large & $\bar\nu_1$ & $\cos^2\theta_{12}$\\
yes & large  & $>\pi/4$ & any & $\bar\nu_2$ & $\sin^2\theta_{12}$\\
\hline
\hline
\end{tabular}
\end{center}
\end{table}

\section{Delay on the onset of bipolar conversions}\label{sec:onset}

Another interesting feature concerns the position of $r_{\rm syn}$
in the presence of a large $\mu\tau$ matter effect.  As can be seen
comparing the first two rows of Fig.~\ref{fig:evolution}, the radius
where collective neutrino transformations begin is slightly larger
($r_{\rm syn}\simeq 115$~km) when we include a significant $
V_{\mu\tau}$.

\subsection{Rotation-averaged mass-squared matrix}

As it was discussed in Sec.~\ref{sec:densematter}, when we include a
background medium the oscillation frequency of the system is
effectively modified, $\omega \to \omega \cos2\theta$. In the usual
two-flavor treatment of collective SN neutrino oscillations, driven by
the small mixing angle $\theta_{13}$, this has little practical
relevance because $\cos2\theta\approx1$ is a good approximation. In
the realistic situation of three flavors, however, two of the mixing
angles are large so that the projection effect of the squared masses
caused by a large matter effect can become non-negligible. One
particularly interesting case is the one we have been discussing in
this chapter, when matter is so dense that the second-order difference
between the $\nu_\mu$ and $\nu_\tau$ refractive effect is large. Here
the matter effect encoded in the matrix ${\sf V}$ of the Hamiltonian
Eq.~(\ref{eq:hamiltonian}) is large compared to the effect of the
vacuum oscillation matrix ${\sf\Omega}={\sf M}^2/2E$. Going to the
rotating three-flavor frame implies that all off-diagonal elements of
${\sf M}^2$ average to zero and we should think of all three
weak-interaction eigenstates $\nu_e$, $\nu_\mu$ and $\nu_\tau$ as
unmixed mass-eigenstate neutrinos with masses implied by the
rotation-averaged ${\sf M}^2$.

To determine these effective mass-squares of the weak-interaction states
we use the parametrization of the leptonic mixing matrix in terms of
the usual mixing angles $\theta_{12}$, $\theta_{13}$, $\theta_{23}$,
and the Dirac phase $\delta$. We here consider the inverted mass
hierarchy that is most relevant for our study. For the vacuum
mass-squares we use $m_3^2=0$, $m_1^2=\Delta m_{\rm atm}^2$ and
$m_2^2=(1+a)\,\Delta m_{\rm atm}^2$, where
\begin{equation}
a=\frac{\Delta m^2_\odot}{\Delta m_{\rm atm}^2}=\frac{1}{30}
\end{equation}
is the neutrino mass hierarchy parameter. We then find for the
diagonal elements of ${\sf M}^2$ in the weak-interaction basis
\begin{eqnarray}
  \frac{m^2_{\nu_e}}{\Delta m^2_{\rm atm}}&=&
  \left(1+a\,s_{12}^2\right)\,c_{13}^2\,,
  \nonumber\\*
  \frac{m^2_{\nu_\mu}}{\Delta m^2_{\rm atm}}&=&
  \left(1+a\,c_{12}^2\right)\,c_{23}^2-2a\,c_{12}\,c_{23}\,s_{12}\,s_{23}\,c_\delta\,s_{13}
  +\left(1+a\,s_{12}^2\right)\,s_{23}^2\,s_{13}^2\,,\\*
  \frac{m^2_{\nu_\tau}}{\Delta m^2_{\rm atm}}&=&
  \left(1+a\,c_{12}^2\right)\,s_{23}^2
  +2a\,c_{12}\,c_{23}\,s_{12}\,s_{23}\,c_\delta\,s_{13}
  +\left(1+a\,s_{12}^2\right)\,c_{23}^2\,s_{13}^2\,.
  \nonumber
\end{eqnarray}
In the absence of any
mixing, the r.h.s.\ would be 1, $1+a$, and 0 from top to bottom.

\begin{figure}[t]
\begin{center}
\includegraphics[angle=0,width=0.65\textwidth]{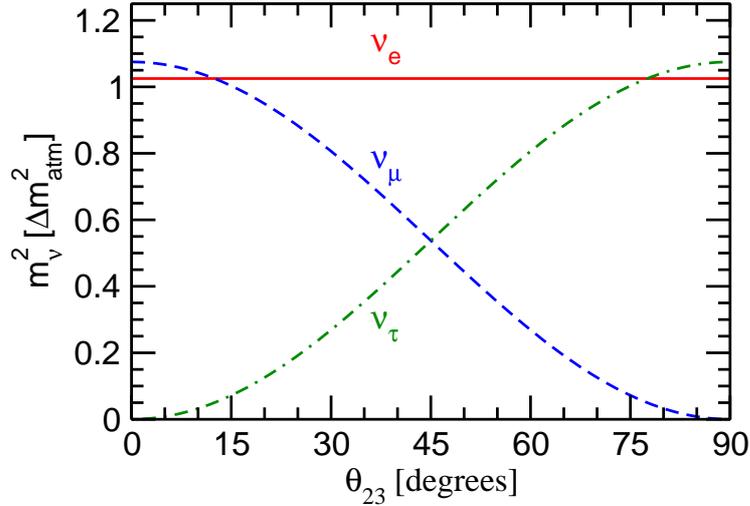}
\end{center}
\caption{\small The diagonal elements of ${\sf M}^2$ in the
  weak-interaction basis as a function of $\theta_{23}$, assuming
  $\theta_{13}=0$ and $\theta_{12}=\pi/6$. To construct a legible
  plot, the mass-hierarchy parameter was chosen as $a=1/10$, although
  in reality it is $a\approx1/30$.}
\label{fig:masses}
\end{figure}

The 13--mixing angle is known to be small so that we have two small
parameters, $s_{13}\ll1$ and $a\ll1$. In the second and third line
both the second terms are of second order in small quantities and
can be neglected. If in addition we use for the solar mixing angle
$\theta_{12}=\pi/6$, the mass-squares are
\begin{eqnarray}
 \frac{m^2_{\nu_e}}{\Delta m^2_{\rm atm}}&=&
 1+\frac{a}{4}\,,
 \nonumber\\*
 \frac{m^2_{\nu_\mu}}{\Delta m^2_{\rm atm}}&=&
 \left(1+\frac{3a}{4}\right)\,\cos^2\theta_{23}\,,\\*
 \frac{m^2_{\nu_\tau}}{\Delta m^2_{\rm atm}}&=&
 \left(1+\frac{3a}{4}\right)\,\sin^2\theta_{23}\,.\nonumber
\end{eqnarray}
The mass-squared spectrum, projected on the weak-interaction basis,
depends most crucially on the mixing angle $\theta_{23}$ that is
known to be nearly maximal. In principle, though, this mixing angle
can vary in the range $0\leq\theta_{23}\leq\pi/2$. We show the
projected mass spectrum as a function of $\theta_{23}$ in
Fig.~\ref{fig:masses}.

\subsection{Role of mu-tau matter effect}

\begin{figure}[t]
\begin{center}
\includegraphics[angle=0,width=0.65\textwidth]{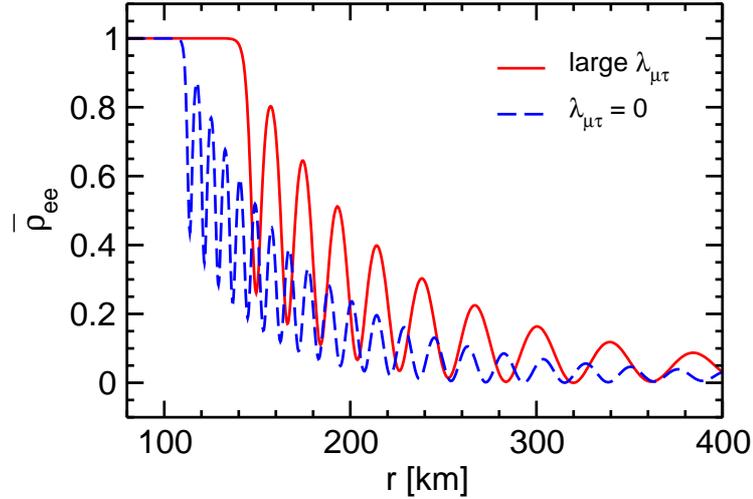}
\end{center}
\caption{\small Radial variation of the numerical $\bar\nu_e$ survival
  probability and the adiabatic limit. Dashed lines: No $\mu\tau$ matter
  effect. Solid lines: Large $\mu\tau$ matter effect.}
\label{fig:threeflavor}
\end{figure}

As a first case we return to the absence of a $\mu\tau$ matter effect.
In this case mu and tau neutrinos are not distinguishable in the SN
context. Therefore, we can re-define these flavors in the usual way as
$\nu_\mu'$ and $\nu_\tau'$. This amounts to using the mass basis in
the 23-system or rather, effectively to a situation of
$\theta_{23}=0$. Therefore, the relevant mass spectrum driving
collective oscillations is the one in Fig.~\ref{fig:masses} at
$\theta_{23}=0$ and corresponds to the usual inverted-hierarchy
spectrum, with the only modification that the splitting between the
upper two neutrinos is not given by $\Delta m_{\rm sol}^2$, but only
by half this amount because of the large 12--mixing angle. Corrections
of order $a$ aside, collective oscillations are driven by $\Delta
m^2_{\rm atm}$. This case corresponds essentially to the two-flavor
example studied in Chapter~\ref{chapter:coll2flavors}. In
Fig.~\ref{fig:threeflavor} we show an analogous case in terms of the
$\bar\nu_e$ survival probability, assuming $\omega=0.1~{\rm km}^{-1}$
and $\mu_0=10^5~{\rm km}^{-1}$, $\sin2\theta=0.01$, $\epsilon=0.4$ and
$\lambda=100~{\rm km}^{-1}$.

Next we consider the case of a large $\mu\tau$ matter effect where now
we need to use the physical weak-interaction basis and not the primed
states. In terms of Fig.~\ref{fig:masses} we now need to use
$\theta_{23}=\pi/4$ or a value nearby within experimental errors.
Therefore, we are now close to the cross-over between the $\nu_\mu$
and $\nu_\tau$ curves in Fig.~\ref{fig:masses}. Therefore, the
effective mass spectrum relevant for collective effects is now very
different from the vacuum mass spectrum. In the absence of the
$\mu\tau$ matter effect, the transitions were driven by
\begin{equation}
\omega=\frac{\Delta m^2_{\rm atm}}{2E}\,\left(1+\frac{a}{4}\right)\,.
\end{equation}
Now, using $\theta_{23}=\pi/4$, they should be driven by
\begin{equation}
\omega'=\frac{\Delta m^2_{\rm atm}}{2E}\,
\frac{1}{2}\left(1-\frac{a}{4}\right)\,.
\end{equation}
Therefore, the new oscillation frequency is predicted to be
\begin{equation}
\frac{\omega'}{\omega}=\frac{4-a}{8(1-a)}
=\frac{1}{2}+\frac{5}{8}\,a+{\cal O}(a^2)\,.
\end{equation}
According to Eq.~(\ref{eq:r_synch}) a change in the oscillation
frequency automatically means a change in $r_{\rm syn}$. This is
exactly what we obtain in Fig.~\ref{fig:threeflavor}, where we show
the numerical solution for the case with a strong $\mu\tau$ matter
effect together with the case of vanishing $\lambda_{\mu\tau}$.

Another test consists of returning to the case without a $\mu\tau$
matter effect and to modify $\omega\to\omega'$ by hand. We find
almost the same numerical curve as before, i.e., the impact of the
$\mu\tau$ matter effect indeed essentially amounts to a modified
mass-squared spectrum as predicted.

If we use $\theta_{23}<\pi/4$, the lowest effective mass eigenstate
corresponds to $\nu_\tau$, for $\theta_{23}>\pi/4$ it corresponds to
$\nu_\mu$. Collective oscillations always seem to go directly to the
lowest mass eigenstate. We can repeat the above exercise for
different values for $\theta_{23}$ and the corresponding $\omega'$
and find a behavior consistent with expectations.

\section{Competition between mu-tau and dense matter effects}\label{sec:competition}

We seem to have here a competition between two different effects, both
result of large matter densities. On the one hand, we have shown in
this chapter that the mu-tau matter effect could have important
consequences in neutrinos streaming off a SN core if $V_{\mu\tau}$ is
large in the collective region. On the other hand, in
Sec.~\ref{sec:densematter} we have studied the multi-angle matter
effects in the collective neutrino transformations, obtaining a
suppression of the bipolar conversions if the matter density is
sufficiently large. Therefore we want to know if these effects happen
at different density scales and will both be present, or on the
contrary the multi-angle effects will destroy the interesting mu-tau
matter effect.

In order to have an important mu-tau matter effect, we require
\begin{equation}
V_{\mu\tau}(r_{\rm bip}) \gtrsim \omega_{\rm H}
\end{equation}
where $r_{\rm bip}$ denotes the end of the bipolar conversion region,
and we recall that $\omega_{\rm H} = \Delta m^2_{\rm
  atm}/2E$. Applying here Eq.~(\ref{eq:mutaueff}) we obtain a lower
limit for $\lambda_0$ leading to mu-tau matter effect,
\begin{equation}
  \lambda_0 \gtrsim \frac{\omega_{\rm H}}{Y^{\rm eff}_\tau}\left(\frac{r_{\rm bip}}{R_{\nu}}\right)^3\,.\label{eq:lambdamtcond}
\end{equation}
For the typical values of the parameters we are assuming in our
simulations ($\omega_{\rm H} = 0.3$ km$^{-1}$, $Y^{\rm eff}_\tau =
2.7\times10^{-5}$ km$^{-1}$, $R_{\nu} = 10$ km, and $r_{\rm bip} =
330$ km) we obtain this limit to be $\lambda_0 \gtrsim 4 \times 10^8$
km$^{-1}$.

On the other hand, we have obtained in Eq.~(\ref{eq:potcondition}) that
the collective transformations will be suppressed until
\begin{equation}
  \lambda^*(r_{\rm syn})\lesssim \mu^*(r_{\rm syn})\,.\label{eq:potcondition2}
\end{equation}
Noting that $\lambda^* = (R_{\nu}^2/2r^2)\lambda$ (where $\lambda
\equiv V_{\rm CC}$) and $\mu^* \approx \mu_0(R_{\nu}^4/2r^4)$, we
obtain an upper limit on $\lambda_0$ in order for the bipolar
conversions to occur,
\begin{equation}
  \lambda_0 \lesssim \frac{\mu_0}{Y_e} \frac{r_{\rm syn}}{R_\nu}\,.\label{eq:lambdasuppcond}
\end{equation}
Applying our standard values ($\mu_0 = 7 \times 10^{-5}$ km$^{-1}$,
$Y_e = 0.5$, $R_{\nu} = 10$ km, and $r_{\rm syn} = 100$ km) we get
$\lambda_0 = 1.4 \times 10^7$ km$^{-1}$.

These simple arguments seem to show that both effects take place for
the same density scales, leaving no room for a coexistence between
them. The suppression of collective transformations due to multi-angle
effects in the dense matter region will destroy any possible effect
coming from the $\mu\tau$-resonance\footnote{In Chapter
  \ref{chapter:SN_NSI} we will show how this situation changes when
  including non-standard neutrino interactions.}.

\section{Conclusions}                           \label{sec:conclusion}

At the relatively low energies relevant for SN neutrinos, charged mu
and tau leptons cannot be produced so that mu- and tau-flavored
neutrinos are not distinguishable in the SN or in detectors. (In the
inner core of a SN the temperatures may be high enough to produce a
significant thermal muon density, but this would not affect the
emission from the neutrino sphere.) The impact of the small
second-order difference between the $\nu_\mu$ and $\nu_\tau$
refractive index does not produce observable effects as long as one
only considers the traditional MSW flavor
conversion~\cite{Akhmedov:2002zj}.

The picture changes if one includes the unavoidable effect of
collective neutrino transformations in the region above the neutrino
sphere. If the matter density is large enough that $
V_{\mu\tau}$ is comparable to or larger than $\Delta m_{\rm
atm}^2/2E$, the survival probability of $\nu_e$ and $\bar\nu_e$ can be
completely modified and depends sensitively on the mixing angle
$\theta_{23}$.

When it is important, the mu-tau matter effect adds one more layer of
complication to the already vexed problem of collective SN neutrino
oscillations. It was previously recognized that ``ordinary''
collective oscillations are almost completely insensitive to the
smallness of $\theta_{13}$ as long as it is not exactly zero. Here we
have found the opposite for the large mixing angle $\theta_{23}$ that
is often assumed to be maximal. Even small deviations from maximal
23-mixing can imprint themselves in the collective oscillation
effect. Both results are counter-intuitive and opposite to ordinary
flavor oscillations.

It seems though, that the density requirement for this $\mu\tau$ effect
to be important implies that the multi-angle matter effect can not be
avoided. In this sense, one complicated effect caused by a large
matter density annihilates another one.

\cleardoublepage

\pagestyle{normal}
\chapter{Non-Standard Neutrino Interactions in Supernova}\label{chapter:SN_NSI}

In this chapter we investigate the impact of non-standard neutrino
interactions on SN physics. We show how complementary
information on the NSI parameters could be inferred from the detection
of core-collapse SN neutrinos.

The motivation for the study is twofold. First, if a future SN
explosion takes place in our Galaxy the number of neutrino events
expected in the current or planned neutrino detectors would be
enormous, $\mathcal{O}(10^4$--$10^5)$~\cite{Scholberg:2007nu}.
Moreover, the extreme conditions under which neutrinos have to travel
since they are created in the SN core, in strongly deleptonized
regions at nuclear densities, until they reach the Earth, lead to
strong matter effects. In particular the effect of small values of the
NSI parameters can be dramatically enhanced, possibly leading to
observable consequences.

As we will later see, the inclusion of NSI to the SN scenario can lead
to strong effects in the same inner layers where collective
transformations take place. Therefore, in order to better understand
the physics of the problem, we will not consider the neutrino
background in the first part of this chapter, and focus on the NSI
side. In the second part, however, we will switch on the neutrino
self-interactions and study the interplay between these two effects.

\section{Previous literature}
\label{sec:nsi_sn}

According to the currently accepted SN paradigm, neutrinos are
expected to play a crucial role in SN dynamics. Moreover, many future
large neutrino detectors are currently being
discussed~\cite{Katsanevas:2006}.  The huge number of events,
$\mathcal{O}(10^4$--$10^5)$, that would be ``seen'' in these detectors
indicates that a future SN in our Galaxy would provide a very
sensitive probe of neutrino NSI effects. The presence of NSI can lead
to important consequences for SN neutrino physics both in the highly
dense core as well as in the envelope where neutrinos basically freely
stream.

The role of non-forward neutrino scattering processes on heavy nuclei
and free nucleons giving rise to flavor change within the SN core has
been recently analyzed in Ref.~\cite{Amanik:2004vm,Amanik:2006ad}. The
main effect found was a reduction in the core electron fraction $Y_e$
during core collapse. A lower $Y_e$ would lead to a lower homologous
core mass, a lower shock energy, and a greater nuclear
photon-disintegration burden for the shock wave. By allowing a maximum
$\Delta Y_e = -0.02$ it has been claimed that
$\varepsilon_{e\alpha}\lesssim 10^{-3}$, where
$\alpha=\mu,~\tau$~\cite{Amanik:2006ad}.

On the other hand it has been noted since long ago that the existence
of NSI plays an important role in the propagation of SN neutrinos
through the envelope leading to the possibility of a new resonant
conversion in the innermost layers. In contrast to the well known MSW
effect it would take place even for massless
neutrinos~\cite{Valle:1987gv}. Two basic ingredients are necessary:
universal and flavor changing NSI.  In the original scheme neutrinos
were mixed in the leptonic charged current and universality was
violated thanks to the effect of mixing with heavy gauge singlet
leptons~\cite{Schechter:1980gr,Mohapatra:1986bd}.
Such resonance would induce strong neutrino flavor conversion both for
neutrinos and antineutrinos simultaneously, possibly affecting the
neutrino signal of the SN1987A as well as the possibility of having
$r$-process nucleosynthesis.
This was first quantitatively considered within a two-flavor
$\nu_e$-$\nu_\tau$ scheme, and bounds on the relevant NSI parameters
were obtained using both arguments~\cite{Nunokawa:1996tg}.

One of the main features of such ``massless'' resonant conversion
mechanism is that it requires the violation of universality, its
position being determined only by the matter chemical composition,
namely the value of the electron fraction $Y_e$, and not by the
density.
In view of the experimental upper bounds on the NSI parameters such a
new resonance can only take place in the inner layers of the SN, near
the neutrino sphere, where $Y_e$ takes its minimum values (few per
cent), see Fig.~\ref{fig:snprofiles}.  In this region the values of
$Y_e$ are small enough to allow for resonance conversions to take
place in agreement with existing bounds on the strengths of
non-universal NSI parameters.

The SN physics implications of another type of NSI present in
supersymmetric R-parity violating models have also been studied in
Ref.~\cite{Nunokawa:1996ve}, again for a system of two neutrinos. For
definiteness NSI on $d$-quarks were considered, in two cases: (i)
massless neutrinos without mixing in the presence of flavor-changing
(FC) and non-universal (NU) NSIs, and (ii) neutrinos with eV masses
and FC NSI.
Different arguments have been used in order to constrain the parameters
describing the NSI, namely, the SN1987A signal, the possibility to get
successful $r$-process nucleosynthesis, and the possible enhancement
of the energy deposition behind the shock wave to reactivate it.

On the other hand, NSI could also affect the propagation of neutrinos
in the outer layers. This was considered in
Refs.~\cite{Mansour:1997fi,Bergmann:1998rg,Fogli:2002xj} in a
three-neutrino mixing scenario for the case $Y_e >0.4$, typical for
the outer SN envelope.  Together with the assumption that
$\varepsilon^{dV}_{\alpha\beta}\lesssim 10^{-2}$ this prevents the
appearance of internal resonances in contrast to previous references.
Motivated by supersymmetric theories without R-parity, in
Ref.~\cite{Mansour:1997fi} the authors considered the effects of
small-strength NSI with $d$-quarks.  Following the formalism developed
in Refs.~\cite{Kuo:1987qu,Bergmann:1997mr} they studied the
corrections that such NSI would have on the expressions for the
survival probabilities in the standard resonances MSW-H and MSW-L.
A similar analysis was performed in Ref.~\cite{Bergmann:1998rg}
assuming Z-induced NSI interactions originated by additional heavy
neutrinos.
A phenomenological generalization of these results was carried out in
Ref.~\cite{Fogli:2002xj}. The authors found an analytical compact
expression for the survival probabilities in which the main effects of
the NSI can be embedded through shifts of the mixing angles
$\theta_{12}$ and $\theta_{13}$. In contrast to similar expressions
found previously these directly apply to all mixing angles, and in the
case with Earth matter effects. The main phenomenological consequence
was the identification of a degeneracy between $\theta_{13}$ and
$\varepsilon_{e\alpha}$, similar to the analogous ``confusion''
between $\theta_{13}$ and the corresponding NSI parameter noted to
exist in the context of long-baseline neutrino
oscillations~\cite{Huber:2001de,Huber:2002bi}.


\section{Neglecting neutrino background}
\label{sec:no-background}


We here reconsidered the general three-neutrino mixing scenario with
NSI in the absence of a neutrino background, as it has been assumed in
all previous literature. This first approach will help us to better
understand the genuine effects of the NSI, and will be very useful for
the complete analysis when considering neutrino self-interactions. In
contrast to previous
work~\cite{Mansour:1997fi,Bergmann:1998rg,Fogli:2002xj}, we have not
restricted ourselves to large values of $Y_e$, discussing also small
values present in the inner layers.  This way our generalized
description includes both the possibility of neutrinos having the
``massless'' NSI-induced resonant conversions in the inner layers of
the SN envelope~\cite{Valle:1987gv,Nunokawa:1996tg,Nunokawa:1996ve},
as well as the ``outer'' oscillation-induced
conversions~\cite{Mansour:1997fi,Bergmann:1998rg,Fogli:2002xj}. However
we have confined ourselves to values of $\varepsilon_{e\alpha}$ small
enough not to lead to drastic consequences during the core collapse.


\subsection{Neutrino evolution}
\label{sec:neutrino-evolution}


In this section we describe the main ingredients of our analysis.  Our
emphasis will be on the use of astrophysically realistic SN density
and $Y_e$ profiles. Their details, in particular their time
dependence, are crucial in determining the way NSI affect the
propagation of neutrinos in the SN
medium.\vspace{0.5cm}\\
\textbf{A) Evolution equation}\vspace{0.2cm}\\
As discussed in Chapter~\ref{chapter:NSI}, in an unpolarized medium
the neutrino propagation in matter will be affected by the vector
coupling constant of the NSI\footnote{For the sake of simplicity we
  will omit the superindex $V$.}, $\varepsilon_{\alpha\beta}^{fV}=
\varepsilon_{\alpha\beta}^{fL} + \varepsilon_{\alpha\beta}^{fR}$.
The way the neutral current NSI modifies the neutrino evolution will
be parametrized phenomenologically through the effective low-energy
four-fermion operator described in Eq.~(\ref{eq:Lnsi}).
We also assume $\varepsilon^f_{\alpha\beta} \in\Re$, neglecting
possible $CP$ violation in the new interactions.

Under these assumptions the Hamiltonian describing the SN neutrino
evolution in the presence of NSI can be cast in the following
form
\begin{equation}
{\rm i}\frac{{\rm d}}{{\rm d}r}
\nu_\alpha
=
\left(
H_{\rm kin} + H_{\rm int} 
\right)_{\alpha\beta}
\nu_\beta
~,
\end{equation}
where we recall that $H_{\rm kin}$ stands for the kinetic term $H_{\rm
  kin} = U(M^2/2E)U^\dagger$, with $M^2={\rm
  diag}(m_1^2,m_2^2,m_3^2)$, and $U$ the three-neutrino lepton mixing
matrix~\cite{Schechter:1980gr} in the PDG
convention~\cite{Amsler:2008zzb} and with no $CP$ phases. The second
term of the Hamiltonian accounts for the interaction of neutrinos with
matter and can be split into two pieces,
\begin{equation}
H_{\rm int} = H^{\rm std}_{\rm int} + H^{\rm nsi}_{\rm int}~.
\end{equation}
Here, $ H^{\rm std}_{\rm int}$ describes the standard interaction with
matter and can be written as $H^{\rm std}_{\rm int}$ = diag $(V_{\rm
  CC},0,0)$ up to one loop corrections due to different masses of the
muon and tau leptons, unimportant for this analysis.
The standard matter potential for neutrinos is given by
Eq.~(\ref{eq:Vcc_Ye}):
\begin{eqnarray}\label{eq:Vcc_ch6}
  V_{\rm CC} & = & \sqrt{2}G_F N_e = V_0 \rho Y_e~,
\end{eqnarray}
where $V_0\approx 7.6\times 10^{-14}$~eV, the density is given in
${\rm g/cm}^3$, and $Y_e$ stands for the relative number of electrons
with respect to baryons.  For antineutrinos the potential is identical
but with the sign changed.

The term in the Hamiltonian describing the non-standard neutrino
interactions with a fermion $f$ can be expressed as,
\begin{equation}
(H_{\rm int}^{\rm nsi})_{\alpha\beta} = \sum_{f=e,u,d} (V_{\rm
    nsi}^{f})_{\alpha\beta}~,
\end{equation}
with $(V_{\rm nsi}^f)_{\alpha\beta} \equiv \sqrt{2}G_F
N_f\varepsilon^f_{\alpha\beta}$.
For definiteness and motivated by actual models, for example, those
with broken R-parity supersymmetry we take for $f$ the down-type
quark. However, an analogous treatment would apply to the case of NSI
on up-type quarks (see below). The existence of NSI with electrons
brings no drastic qualitative differences with respect to the pure
oscillation case.  Therefore the NSI potential can be expressed as
follows,
\begin{equation}
(V_{\rm nsi}^{d})_{\alpha\beta} 
=\varepsilon^{d}_{\alpha\beta}V_0\rho(2-Y_e)~. 
\end{equation}
From now on we will not explicitely write the superindex $d$.  In
order to further simplify the problem we will redefine the diagonal
NSI parameters so that $\varepsilon_{\mu\mu}=0$, as one can easily see
that subtracting a matrix proportional to the identity leaves the
physics involved in the neutrino oscillation unaffected.\vspace{0.5cm}\\
\textbf{B) Supernova matter profiles}\vspace{0.2cm}\\
Neutrino propagation depends on the SN matter and chemical profile
through the effective potential. As it was discussed in
Sec.~\ref{sec:SN_profiles}, this profile exhibits an important time
dependence during the explosion, see Fig.~\ref{fig:snprofiles}. We
will use the parameterization given there for the density $\rho(t,r)$
and the electron fraction $Y_e(t,r)$ profiles. Progenitor density
profiles can be roughly parameterized by the power-law function given
in Eq.~(\ref{eq:rho_prof}):
\begin{equation}\label{eq:rho_prof_ch6}
\rho(r) = \rho_0 \left(\frac{r_0}{r}\right)^n~,
\end{equation}
where $\rho_0 \sim 10^4$~g/cm$^3$, $r_0\sim 10^9$~cm, and $n\sim 3$.
While the electron fraction profile can be phenomenologically
approximated by Eq.~(\ref{eq:Ye}):
\begin{equation}\label{eq:Ye_ch6}
Y_e = a + b\arctan[(r-r_0)/r_s]~,
\end{equation}
where $a\approx 0.23$--$0.26$ and $b\approx 0.16$--$0.20$. We recall
that the parameters $r_0$ and $r_s$ describe where the rise takes
place and how steep it is, respectively.

\subsection{The two regimes}
\label{sec:two-regimes}


In order to study the neutrino propagation through the SN envelope we
will split the problem into two different regions: the inner envelope,
defined by the condition $V_{\rm CC}\gg \Delta m^2_{\rm atm}/(2E)$,
and the outer one, where $\Delta m^2_{\rm atm}/(2E) \gtrsim V_{\rm
  CC}$. From the upper panel of Fig.~\ref{fig:snprofiles} one can see
how the boundary roughly varies between $r\approx 10^8$~cm and
$10^9$~cm, depending on the time considered.
This way one can fully characterize all resonances that can take place
in the propagation of SN neutrinos, both the outer resonant
conversions related to neutrino masses and indicated as the upper
bands in Fig.~\ref{fig:snprofiles}, and the inner resonances that
follow from the presence of NSI.
Here we pay special attention to the use of realistic matter and
chemical SN profiles and three-neutrino flavors thus
generalising previous studies.\vspace{0.5cm}\\
\textbf{A) Neutrino evolution in the inner regions}\vspace{0.2cm}\\
Let us first write the Hamiltonian in the inner layers, where $H_{\rm
  int}\gg H_{\rm kin}$. In this case the Hamiltonian can be written as
\begin{equation}\label{eq:H_inner}
H\approx H_{\rm int} =
V_0\rho(2-Y_e)
\left(\begin{array}{ccc} \frac{Y_e}{2-Y_e}+\varepsilon_{ee} &
  \varepsilon_{e\mu} & \varepsilon_{e\tau} \\
\varepsilon_{e\mu} & 0 & \varepsilon_{\mu\tau} \\
\varepsilon_{e\tau} & \varepsilon_{\mu\tau} & \varepsilon_{\tau\tau}  
\end{array} \right)~.
\end{equation}

When the value of the $\varepsilon_{\alpha\beta}$ is of the same order
as the electron fraction $Y_e$ internal resonances can
arise~\cite{Valle:1987gv}.
Taking into account the current constraints on the $\varepsilon$'s
discussed in Chapter~\ref{chapter:NSI} one sees that small values
of $Y_e$ are required~\cite{Nunokawa:1996tg,Nunokawa:1996ve}.
As a result, these can only take place in the most deleptonised inner
layers, close to the neutrino sphere, where the kinetic terms of the
Hamiltonian are negligible.

Given the large number of free parameters $\varepsilon_{\alpha\beta}$
involved we consider one particular case where 
$|\varepsilon_{e\mu}|$ and $|\varepsilon_{e\tau}|$ are small enough to
neglect a possible initial mixing between $\nu_e$ and $\nu_\mu$ or
$\nu_\tau$. 
Barring fine tuning, this basically amounts to
$|\varepsilon_{e\mu}|,~|\varepsilon_{e\tau}|\ll 10^{-2}$. According to
the discussion of Chapter~\ref{chapter:NSI}, $\varepsilon_{e\mu}$
automatically satisfies the condition, whereas one expects that the
window $|\varepsilon_{e\tau}|\gtrsim 10^{-2}$ will eventually be
probed in future experiments.

Since the initial fluxes of $\nu_\mu$ and $\nu_\tau$ are expected to
be basically identical, it is convenient to redefine the weak basis by
performing a rotation in the $\mu$-$\tau$ sector:
\begin{equation}
\left(\begin{array}{c} \nu_e \\ \nu_\mu \\ \nu_\tau \end{array}\right)
= 
U(\theta_{23}') 
\left(\begin{array}{c} \nu_e \\ \nu_\mu' \\ \nu_\tau' \end{array}\right)
=
\left(\begin{array}{ccc} 1 & 0 & 0\\ 0 & c_{23'} &
  s_{23'} \\ 0 & -
  s_{23'} & c_{23'} 
\end{array}\right)
\left(\begin{array}{c} \nu_e \\ \nu_\mu' \\ \nu_\tau' \end{array}\right)~,
\end{equation}
where $c_{23'}$ and $s_{23'}$ stand for
$\cos(\theta_{23}')$ and
$\sin(\theta_{23}')$, respectively. 
The angle $\theta_{23}'$ can be written as
\begin{equation}
\tan(2\theta_{23}') \approx
\frac{2H_{23}}{H_{33}} =
  \frac{2\varepsilon_{\mu\tau}}{\varepsilon_{\tau\tau}}~.   
\end{equation}

The Hamiltonian becomes in the new basis
\begin{equation}
  H'_{\alpha\beta} = U^\dagger(\theta_{23}')H_{\alpha\beta}U(\theta_{23}') = 
  V_0\rho(2-Y_e)\left(\begin{array}{ccc}
      \frac{Y_e}{2-Y_e}+\varepsilon_{ee} & \varepsilon_{e\mu}' &
      \varepsilon_{e\tau}' \\
      \varepsilon_{e\mu}' & \varepsilon_{\mu\mu}' & 0 \\
      \varepsilon_{e\tau}' & 0 & \varepsilon_{\tau\tau}'
\end{array}\right)~,
\label{eq:HI1}
\end{equation}
where
\begin{eqnarray}
\varepsilon_{e\mu}' & = & \varepsilon_{e\mu}c_{23'}-
  \varepsilon_{e\tau}s_{23'}\\
\varepsilon_{e\tau}' & = & \varepsilon_{e\mu}s_{23'}+
  \varepsilon_{e\tau}c_{23'}\\
\varepsilon_{\mu\mu}' & = &
  (\varepsilon_{\tau\tau}-
  \sqrt{\varepsilon^2_{\tau\tau}+4\varepsilon^2_{\mu\tau}})/2~\\    
\varepsilon_{\tau\tau}' & = &
  (\varepsilon_{\tau\tau}+
\sqrt{\varepsilon^2_{\tau\tau}+4\varepsilon^2_{\mu\tau}})/2~.
\end{eqnarray}

With our initial assumptions on $\varepsilon_{e\alpha}$ one notices
that the new basis $\nu_\alpha'$ essentially diagonalizes the
Hamiltonian, and therefore coincides roughly with the matter
eigenstate basis.
A novel resonance can arise if the condition $H_{ee}'=H_{\tau\tau}'$
is satisfied, we call this $I$-resonance, $I$ standing for
``internal''. The alternative condition $H_{ee}' = H_{\mu\mu}'$ would
give rise to another internal resonance which can be studied using the
same method.  The corresponding resonance condition can be written as
\begin{equation}
\label{eq:Ye_resI}
Y_e^{I} = \frac{2\varepsilon^{I}}{1+\varepsilon^{I}}~,
\end{equation}
where
$\varepsilon^{I}$ is defined as
$\varepsilon_{\tau\tau}'- \varepsilon_{ee}$. 
\begin{figure}
  \begin{center}
    \includegraphics[width=0.45\textwidth]{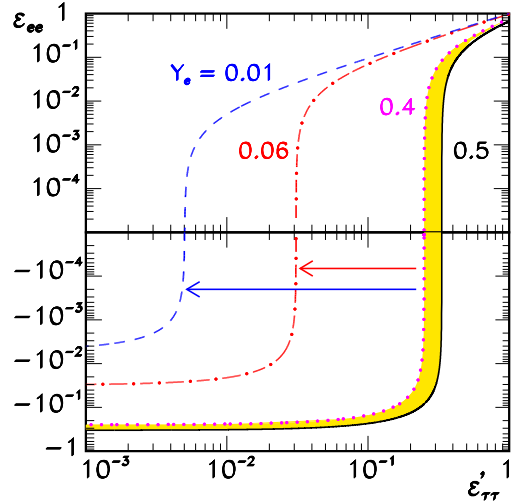}
  \end{center}
  \caption{\small Contours of $Y_e^{I}$ as function of
    $\varepsilon_{ee}$ and $\varepsilon_{\tau\tau}'$ according to
    Eq.~(\ref{eq:Ye_resI}) for different values of $Y_e$. The region
    in yellow represents the region of parameters that gives rise to
    $I$-resonance before the collapse. The arrows indicate how this
    region widens with time~\cite{EstebanPretel:2007yu}.}
  \label{fig:YeI-eI}
\end{figure}
In Fig.~\ref{fig:YeI-eI} we represent the range of $\varepsilon_{ee}$
and $\varepsilon_{\tau\tau}'$ leading to the $I$-resonance for an
electron fraction profile between different $Y_e^{\rm min}$'s and
$Y_e^{\rm max}=0.5$.  It is important to notice that the value of
$Y_e^{\rm min}$ depends on time, as discussed in
Sec.~\ref{sec:SN_profiles}. Right before the collapse the minimum
value of the electron fraction is around $0.4$. Hence the window of
NSI parameters that would lead to a resonance would be relatively
narrow, as indicated by the shaded (yellow) band in
Fig.~\ref{fig:YeI-eI}.
As time goes on $Y_e^{\rm min}$ decreases to values of the order of a
few \%, and as a result the region of parameters giving rise to the
$I$-resonance significantly widens.
For example, in the range $|\varepsilon_{ee}| \leq 10^{-3}$
possibly accessible to future experiments one sees that the
$I$-resonance can take place for values of $\varepsilon_{\tau\tau}'$
of the order of $\mathcal{O}(10^{-2})$. 
This indicates that the potential sensitivity on NSI parameters that
can be achieved in SN studies is better than that of the
current limits. 
One can see in Fig.~\ref{fig:snprofiles} in order to fulfill the
$I$-resonance condition for such small values of the NSI parameters
the values of $Y_e$ must indeed lie, as already stated, in the inner
layers.

\begin{figure}[th!]
  \begin{center}
  \includegraphics[width=0.45\textwidth]{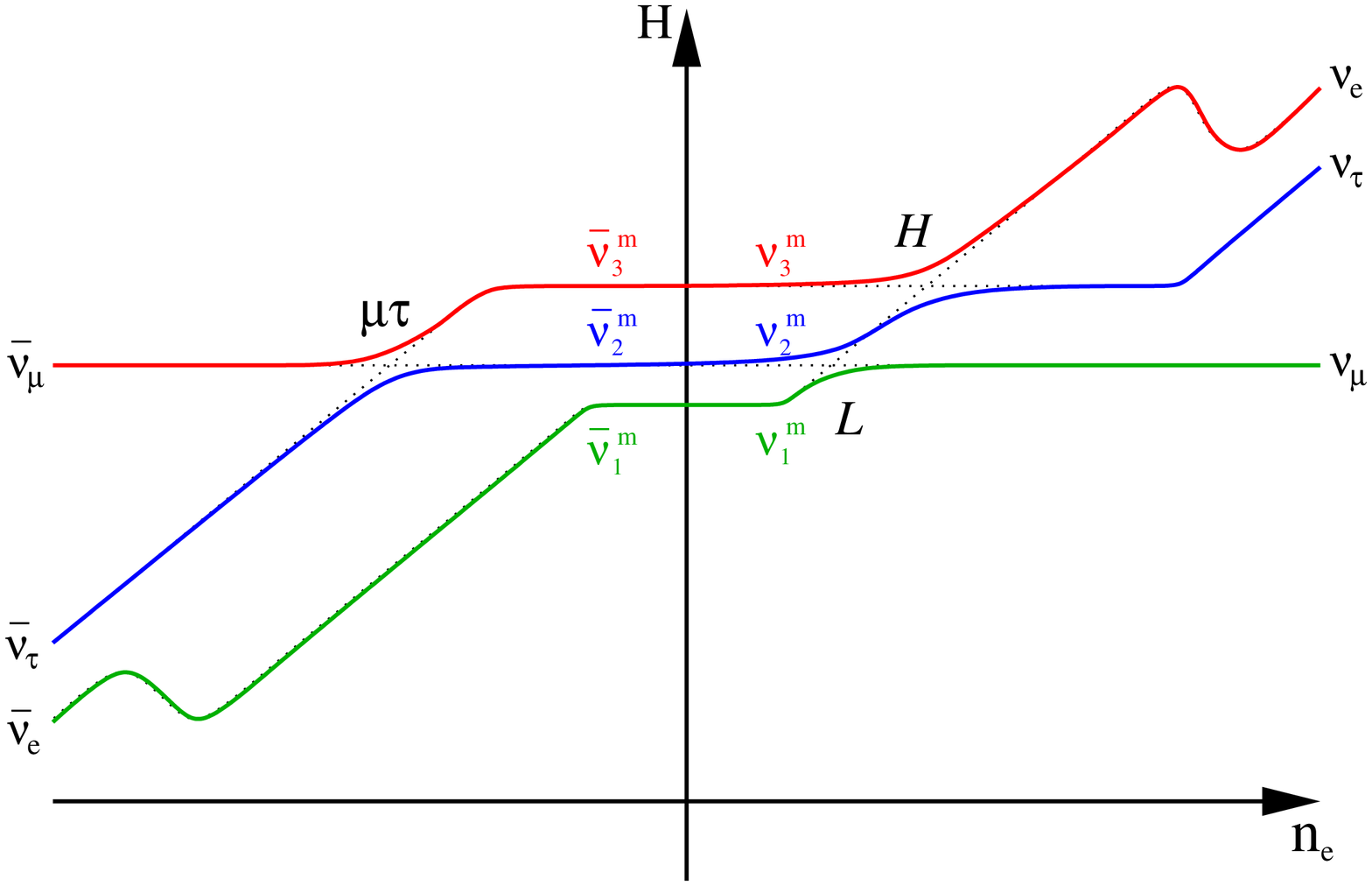}
\hskip6pt
  \includegraphics[width=0.45\textwidth]{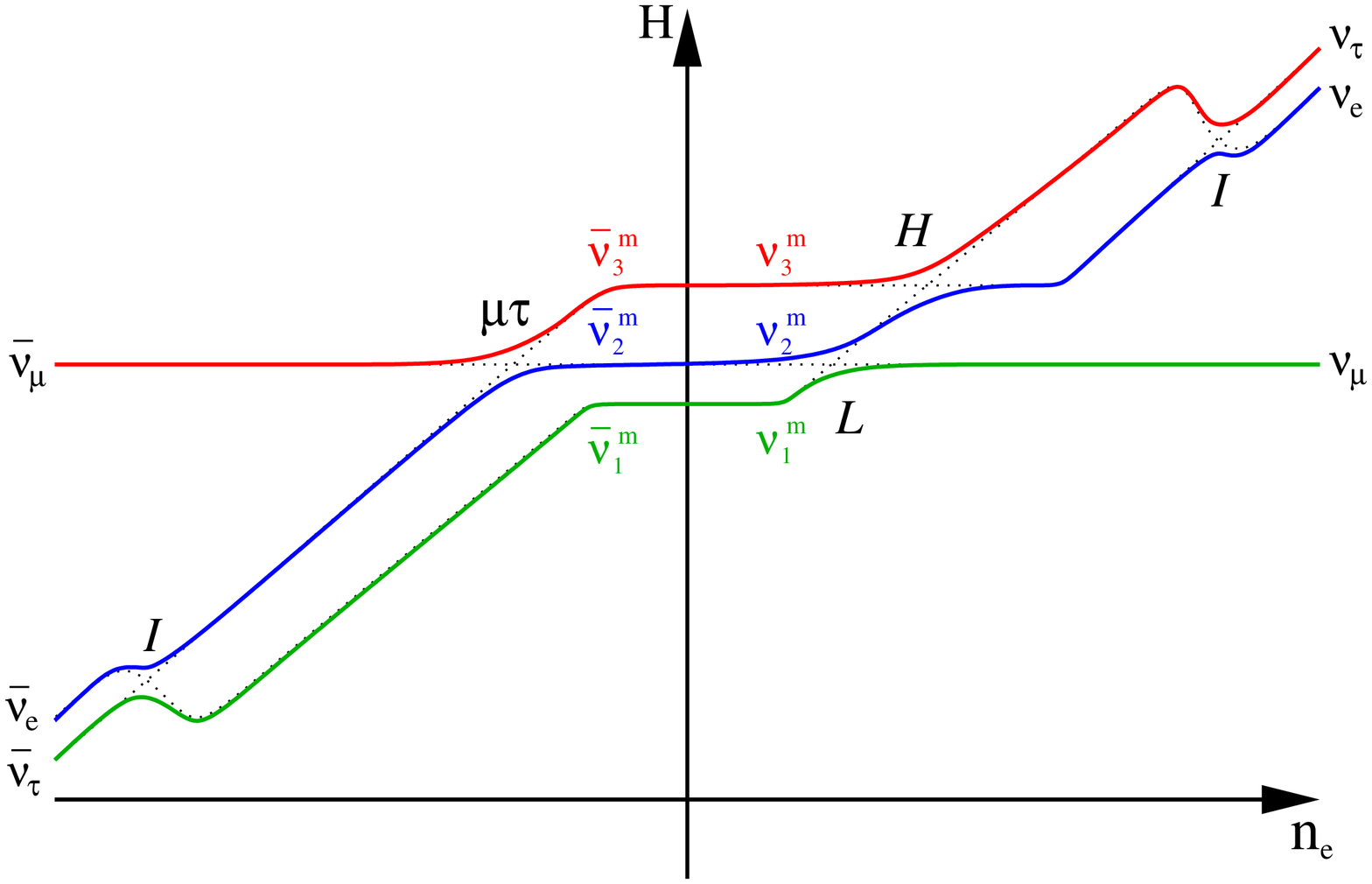}
\vskip6pt
  \includegraphics[width=0.45\textwidth]{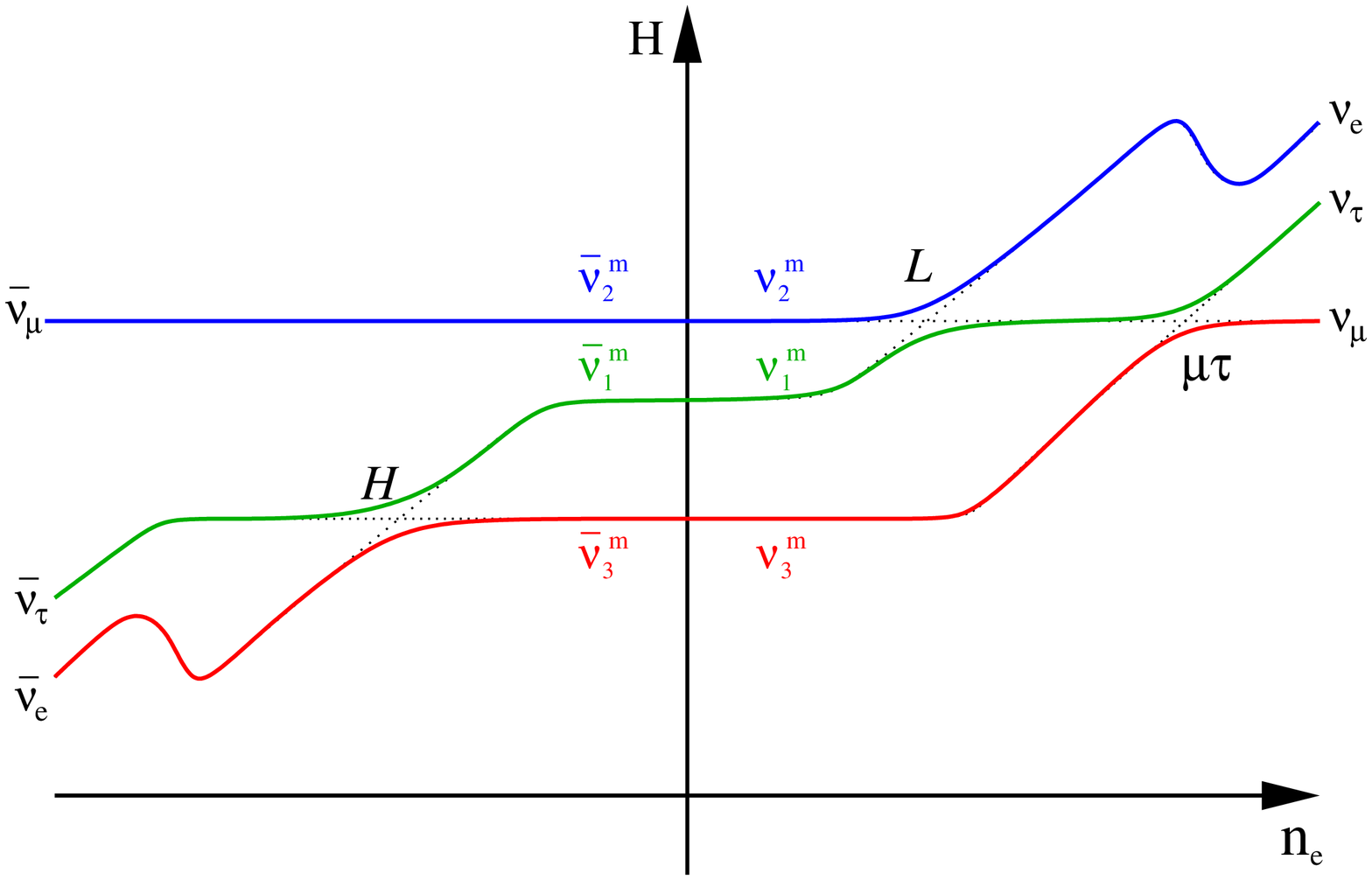}
\hskip6pt
  \includegraphics[width=0.45\textwidth]{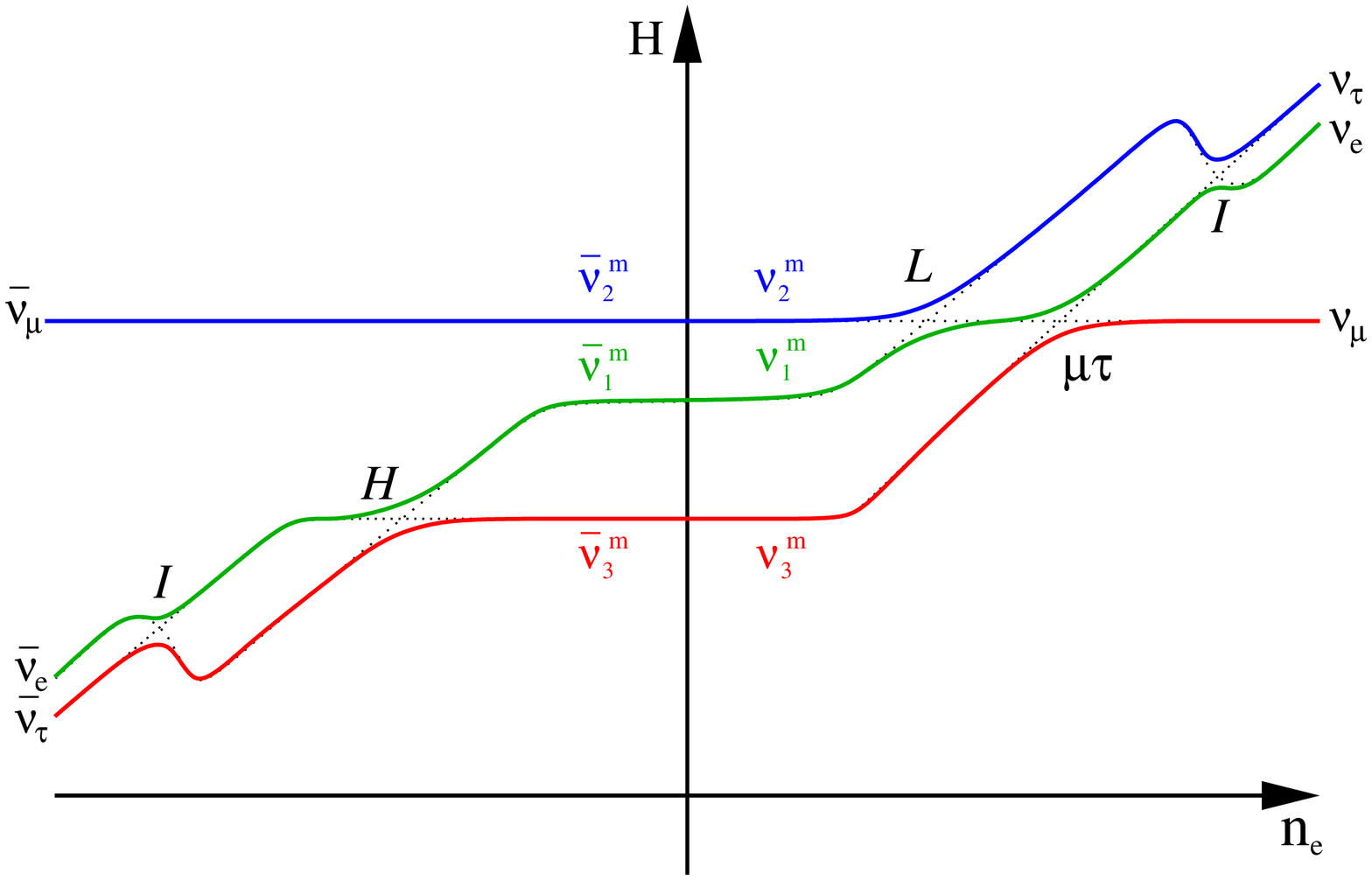}
  \end{center}
  \caption{\small Level-crossing schemes, first panel is for the case
    of normal hierarchy (oscillations only), the second includes the
    NSI effect.  The two lower panels correspond to the inverse
    hierarchy, oscillations only and oscillations + NSI,
    respectively.}
  \label{fig:scheme}
\end{figure}

Several comments are in order: First, in contrast to the standard $H$-
and $L$-resonances, related to the kinetic term, the density itself
does not explicitly enter into the resonance condition, provided that
the density is high enough to neglect the kinetic terms. Analogously
the energy plays no role in the resonance condition, which is
determined only by the electron fraction $Y_e$. Moreover, in contrast
to the standard resonances, the $I$-resonance occurs for both
neutrinos and antineutrinos simultaneously~\cite{Valle:1987gv}.
Finally, as indicated in Fig.~\ref{fig:scheme}
the $\nu_e$'s ($\bar\nu_e$) are not created as the heaviest (lightest)
state but as the intermediate state, therefore the flavor composition
of the neutrinos arriving at the $H$-resonance is exactly the opposite
of the case without NSI.

In order to calculate the hopping probability between matter
eigenstates at the $I$-resonance we use the Landau-Zener
approximation for two flavors, see Eq.~(\ref{eq:hop}),
\begin{equation}
P_{LZ}^{I} \approx e^{-\frac{\pi}{2}\gamma_{I}}~,
\label{eq:plz-i1}
\end{equation}
where $\gamma_{I}$ stands for the adiabaticity parameter, defined in
Eq.~(\ref{eq:adiab_param}), generally written as
\begin{equation}
\gamma_{I} = \left|\frac{E_2^{\rm m}-E_1^{\rm m}}{2\dot \theta^{\rm
    m}}\right|_{r_{I}}~,
\label{eq:gamma-i1-1}
\end{equation}
with $\dot \theta^{\rm m}\equiv {\rm d}\theta^{\rm m}/{\rm d}r$.  If
one applies this formula to the $e-\tau'$ box of Eq.~(\ref{eq:HI1})
assuming that $\tan 2\theta^{\rm m}_{I} =
2H_{e\tau}'/(H_{\tau\tau}'-H_{ee})$ and $E_2^{\rm m}-E_1^{\rm m} =
\left[ (H_{\tau\tau}'-H_{ee})^2 + 4H_{e\tau}' \right]^{1/2}$ one gets
\begin{equation}
\gamma_{I} = \left| \frac{4H_{e\tau}'^2}{(\dot H_{\tau\tau}'-\dot
    H_{ee})} \right|_{r_{I}} = \left| 
    \frac{16V_0\rho \varepsilon_{e\tau}'^2}{(1+\varepsilon^{I})^3
    \dot Y_e} \right|_{r_{I}} \approx
    4\times10^{9} r_{s,5}\rho_{11}
    \varepsilon_{e\tau}'^2 f(\varepsilon^{I})~,
\label{eq:gamma-i1-2}
\end{equation}
where the parametrization of the $Y_e$ profile has been defined as in
Eq.~(\ref{eq:Ye}) with $b=0.16$. The density $\rho_{11}$ represents the
density in units of $10^{11}$~g/cm$^3$, $r_{s,5}$ stands for $r_s$ in
units of $10^5$~cm, and $f(\varepsilon^{I})$ is a function whose value
is of the order $\mathcal{O}(1)$ in the range of parameters we are
interested in.
Taking all these factors into account it follows that the internal
resonance will be adiabatic, provided that
$\varepsilon_{e\tau}'\gtrsim 10^{-5}$. This value is well below the
current limits and in full numerical agreement with, e.g.,
Ref.~\cite{Nunokawa:1996ve}.

\begin{figure}
  \begin{center}
\includegraphics[width=0.33\textwidth,angle=-90]{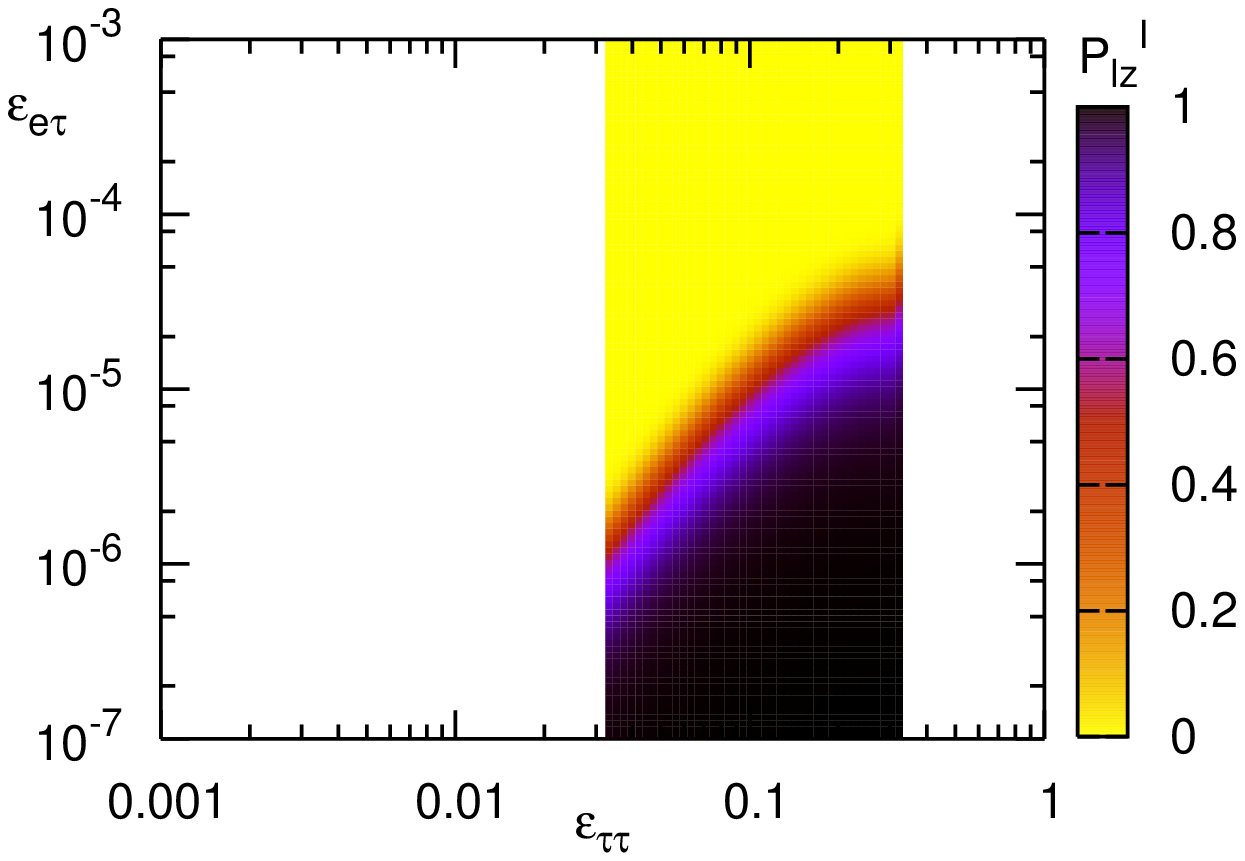}
\includegraphics[width=0.33\textwidth,angle=-90]{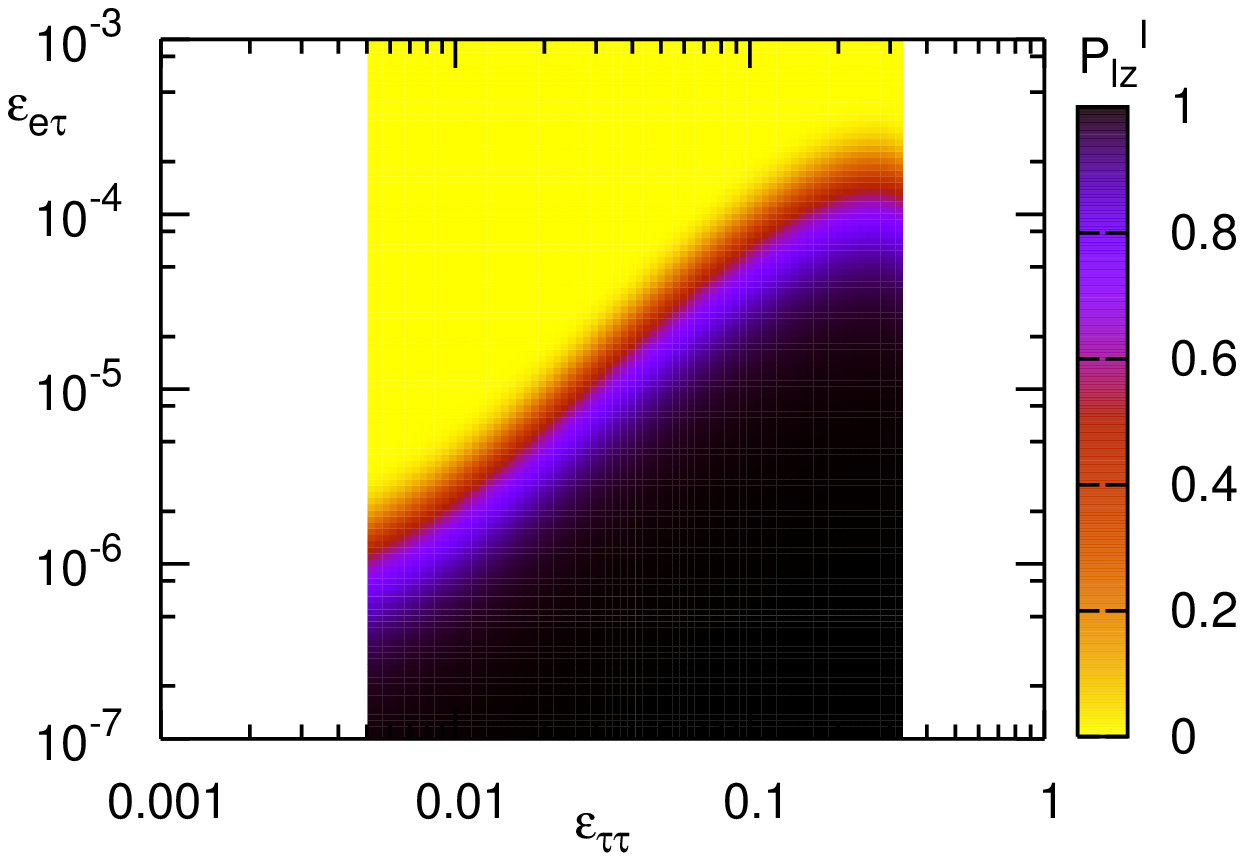}
  \end{center}
  \caption{\small Contours of constant jump probability at the
    $I$-resonance in terms of $\varepsilon_{\tau\tau}$ and
    $\varepsilon_{e\tau}$ for two profiles corresponding to
    Fig.~\ref{fig:snprofiles} at $2$~s with $a = 0.235$ and $b =
    0.175$ (left panel) and $15.7$~s with $a = 0.26$ and $b = 0.195$
    (right panel). For simplicity the other $\varepsilon$'s have been
    set to zero~\cite{EstebanPretel:2007yu}.}
  \label{fig:adiab_I}
\end{figure}

In Fig.~\ref{fig:adiab_I} we show the resonance condition as well as
the adiabaticity in terms of $\varepsilon_{\tau\tau}$ and
$\varepsilon_{e\tau}$ assuming the other
$\varepsilon_{\alpha\beta}=0$. In order to illustrate the dependence
on time we consider profiles inspired in the numerical profiles of
Fig.~\ref{fig:snprofiles} at $t=2$~s (left panel) and 15.7 s (right
panel). For definiteness we take $Y_e^{\rm min}$ as the electron
fraction at which the density has value of $5\times 10^{11}$g/cm$^3$.
For comparison with Fig.~\ref{fig:YeI-eI} we have assumed $Y_e^{\rm
  min}=10^{-2}$ in the case of 15.7~s. We observe how the border of
adiabaticity depends on $\varepsilon_{\tau\tau}$ through the value of
the density at $r_I$ which in turn depends on time.

Before moving to the discussion of the outer resonances a comment is
in order, namely, how does the formalism change for other non-standard
interaction models. First note that the whole treatment presented
above also applies to the case of NSI on up-type quarks, except that
the position of the internal resonance shifts with respect to the
down-quark case.  Indeed, in this case the NSI potential
\begin{equation}
(V_{\rm nsi}^{u})_{\alpha\beta} 
=\varepsilon^{u}_{\alpha\beta}V_0\rho(1+Y_e)~,
\end{equation}
would induce a similar internal resonance for the condition
$Y_e=\varepsilon^I/(1-\varepsilon^I)$. 

In contrast, for the case of NSI with electrons, the NSI potential is
proportional to the electron fraction, and therefore no internal
resonance would appear.\vspace{0.5cm}\\
\textbf{B) Neutrino evolution in the outer regions}\vspace{0.2cm}\\
As it has been extensively discussed, in the outer layers of the SN
envelope neutrinos can undergo important flavor transitions at those
points where the matter induced potential equals the kinetic terms.
In absence of NSI this condition can be expressed as $V_{\rm
  CC}\approx \Delta m^2/(2E)$.  Therefore, two different resonance
layers arise, the so-called $H$-resonance and the $L$-resonance,
corresponding to the atmospheric and solar squared mass diferences,
respectively.

The presence of NSI with values of $|\varepsilon_{\alpha\beta}|
\lesssim 10^{-2}$ modifies the properties of the $H$- and
$L$-transitions~\cite{Mansour:1997fi,Bergmann:1998rg,Fogli:2002xj}. In
particular one finds that the effects of the NSI can be described as
in the standard case by embedding the $\varepsilon$'s into effective
mixing angles~\cite{Fogli:2002xj}. An analogous ``confusion'' between
$\sin\theta_{13}$ and the corresponding NSI parameter
$\varepsilon_{e\tau}$ has been pointed out in the context of
long-baseline neutrino oscillations in
Refs.~\cite{Huber:2001de,Huber:2002bi}.

In this section we perform a more general and complementary study for
slightly higher values of the NSI parameters:
$|\varepsilon_{\alpha\beta}|\gtrsim {\rm few}\times 10^{-2}$, still
allowed by current limits, and for which the $I$-resonance could
occur.

The phenomenological assumption of a hierarchy in the squared mass
differences, $|\Delta m^2_{\rm atm}|\gg \Delta m^2_\odot$, allows, for
not too large $\varepsilon$'s, a factorization of the 3$\nu$ dynamics
into two 2$\nu$ subsystems roughly decoupled for the $H$- and
$L$-transitions~\cite{Kuo:1989qe}.
To isolate the dynamics of the $H$-transition, one usually rotates the 
neutrino flavor basis by $U^\dagger(\theta_{23})$, and extracts the
submatrix with indices (1,3)~\cite{Mansour:1997fi,Fogli:2002xj}.
Whereas this method works perfectly for small values of
$\varepsilon_{\alpha\beta}$ it can be dangerous for values above
$10^{-2}$. In order to analyze how much our case deviates from the
simplest approximation we have performed a rotation with the angle
$\theta_{23}''\equiv \theta_{23}-\alpha$ instead of just
$\theta_{23}$. By requiring that the new rotation diagonalizes the
submatrix (2,3) at the $H$-resonance layer one obtains the following
expression for the correction angle $\alpha$
\begin{eqnarray}
\label{eq:alfa}
\tan(2\alpha) & = & \left[\Delta_{\odot}s2_{12}s_{13} +
  V_{\tau\tau}^{nsi} s2_{23} - 2V_{\mu\tau}^{nsi}
  c2_{23}\right]/ \\ 
& & \left[(\Delta_{\rm atm}+ \frac{1}{2}\Delta_{\odot}) 
  c^2_{13}+\frac{1}{4}\Delta_{\odot} c2_{12}(-3+c2_{13})+ V_{\tau\tau}^{nsi} c2_{23} + 2 V_{\mu\tau}^{nsi}
  s2_{23}\right]~,\nonumber
\end{eqnarray}
where $\Delta_{\rm atm}\equiv \Delta m^2_{\rm atm}/(2E)$ and
$\Delta_{\odot}\equiv \Delta m^2_\odot/(2E)$. In our notation $s_{ij}$
and $s2_{ij}$ represent $\sin\theta_{ij}$ and $\sin(2\theta_{ij})$,
respectively. The parameters $c_{ij}$ and $c2_{ij}$ are
analogously defined. 
In the absence of NSI $\alpha$ is just a small
correction\footnote{Note that, in the limit of high densities one
  recovers the rotation angle obtained for the internal $I$-resonance
  $\theta_{23}'' \to \theta_{23}'$ after neglecting the kinetic terms.
} to $\theta_{23}$,
\begin{equation}
\tan(2\alpha)\approx
\Delta_{\odot}s2_{12}s_{13}/\Delta_{\rm atm}c^2_{13} \lesssim
\mathcal{O}(10^{-3})~. 
\end{equation}

In order to calculate $\alpha$ we need to know the $H$-resonance
point. To calculate it one can proceed as in the case without NSI,
namely, make the $\theta_{23}''$ rotation and analyze the submatrix
$(1,3)$. The new Hamiltonian $H_{\alpha\beta}''$ has now the form
\begin{eqnarray}
\label{eq:H'_h}
H_{ee}'' & = & V_0\rho [Y_e + \varepsilon_{ee}(2-Y_e)] + \Delta_{\rm
  atm}s^2_{13} + \Delta_{\odot}(c^2_{13}s^2_{12}+s^2_{13})~, \nonumber\\ 
H_{\tau\tau}'' & = & V_0\rho (2-Y_e)\varepsilon_{\tau\tau}'' + \Delta_{\rm
  atm}c^2_{13}c^2_{\alpha} + \Delta_{\odot}\left[c^2_{13}c^2_{\alpha} +
  (s_{\alpha}c_{12}+c_{\alpha}s_{12}s_{13})^2 \right]~,~~~~ \\ 
H_{e\tau}'' & = &  V_0\rho (2-Y_e)\varepsilon_{e\tau}'' +
\frac{1}{2}\Delta_{\rm   atm}s2_{13}c_{\alpha} + \frac{1}{2}\Delta_{\odot}(-c_{13}s_{\alpha}s2_{12}+c^2_{12}c_{\alpha}s2_{13})~.  \nonumber
\end{eqnarray} 
We have defined $\varepsilon_{\tau\tau}'' =
\varepsilon_{\tau\tau}c^2_{23-\alpha} +
\varepsilon_{\mu\tau}s2_{23-\alpha}$, and  $\varepsilon_{e\tau}'' =
\varepsilon_{e\tau}c_{23-\alpha} +
\varepsilon_{e\mu}s_{23-\alpha}$, where $s_{23-\alpha}\equiv
\sin(\theta_{23}-\alpha),~c_{23-\alpha}\equiv\cos(\theta_{23}-\alpha)$,
and
$s2_{23-\alpha}\equiv\sin(2\theta_{23}-2\alpha),
~c2_{23-\alpha}\equiv\cos(2\theta_{23}-2\alpha)$.  
The resonance condition for the $H$-transition,
$H_{ee}''=H_{\tau\tau}''$ can be then written as
\begin{eqnarray}
  & V_0 \rho^H [Y_e^H + 
  (\varepsilon_{ee}-\varepsilon_{\tau\tau}'')(2-Y_e^H)] =& \nonumber \\
  &  \Delta_{\rm atm}(c^2_{13} c^2_{\alpha} -s^2_{13})
  +\Delta_{\odot}[c^2_{12}(c^2_{13}-c^2_{\alpha}s^2_{13})- s^2_{\alpha}s^2_{12}+\frac{1}{2}s2_{\alpha}s2_{12} s_{13}]~. &
\label{eq:resH}
\end{eqnarray}
It can be easily checked how in the limit of
$\varepsilon_{\alpha\beta} \to 0$ one recovers the standard resonance
condition,
\begin{equation}
V_0 \rho^H Y_e^H\approx
\Delta_{\rm atm}c2_{13}~.
\end{equation}
In the region where the $H$-resonance occurs $Y_e^H\approx 0.5$. 

\begin{figure}[t]
  \begin{center}
    \includegraphics[angle=-0,width=0.45\textwidth]{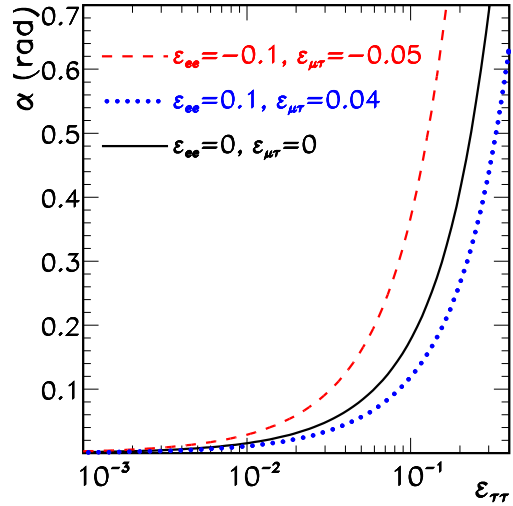}
  \end{center}
  \caption{\small Angle $\alpha$ as function of $\varepsilon_{\tau\tau}$ for
  different values of $\varepsilon_{ee}$ and $\varepsilon_{\mu\tau}$, 
  in the case of neutrinos of energy $10$~MeV, with normal mass hierarchy, and
  $s^2_{13}=10^{-5}$. The other NSI parameters take the following
  values: $\varepsilon_{e\mu}=0$ and  $\varepsilon_{e\tau}=10^{-3}$~\cite{EstebanPretel:2007yu}.
  \label{fig:alpha}}
\end{figure}

Taking into account Eqs.~(\ref{eq:alfa}) and~(\ref{eq:resH}) one can
already estimate how the value of $\alpha$ changes with the NSI
parameters. 
In Fig.~\ref{fig:alpha} we show the dependence of $\alpha$ on the
$\varepsilon_{\tau\tau}$ after fixing the value of the other NSI
parameters. One can see how for $\varepsilon_{\tau\tau}\gtrsim
10^{-2}$ the approximation of neglecting $\alpha$ significantly
worsens.  Assuming $\theta_{23}=\pi/4$ and a fixed value of
$\varepsilon_{\mu\tau}$ one can easily see that
$\varepsilon_{\tau\tau}$ basically affects the numerator in
Eq.~(\ref{eq:alfa}). Therefore one expects a rise of $\alpha$ as the
value of $\varepsilon_{\tau\tau}$ increases, as seen in
Fig.~\ref{fig:alpha}.  The dependence of $\alpha$ on
$\varepsilon_{\mu\tau}$ is correlated to the relative sign of the mass
hierarchy and $\varepsilon_{\mu\tau}$. For instance, for normal mass
hierarchy and positive values of $\varepsilon_{\mu\tau}$ the
dependence is inverse, namely, higher values of
$\varepsilon_{\mu\tau}$ lead to a suppression of $\alpha$.  Apart from
this general behavior, $\alpha$ also depends on the diagonal term
$\varepsilon_{ee}$ as seen in Fig.~\ref{fig:alpha}. This effect occurs
by shifting the resonance point through the resonance condition in
Eq.~(\ref{eq:resH}).

One can now calculate the jump probability between matter eigenstates
in analogy to the standard case by means of the Landau-Zener
approximation, see Eqs.~(\ref{eq:hop}) and (\ref{eq:gamma_H}),
\begin{equation}
P_{LZ}^{H} \approx e^{-\frac{\pi}{2}\gamma_{H}}~,
\label{eq:plz-h}
\end{equation}
where $\gamma_{H}$ represents the adiabaticity parameter at the
$H$-resonance, which can be written as
\begin{equation}
\gamma_{H} =  \left| \frac{4H_{e\tau}''^2}{(\dot H_{\tau\tau}''-\dot
    H_{ee}'')} \right|_{r_{H}}~,
\label{eq:gammaH}
\end{equation}
where the expressions for $H_{\alpha\beta}''$ are given in
Eqs~(\ref{eq:H'_h}).  

Let us first consider the case $|\varepsilon_{\alpha\beta}|\lesssim
10^{-2}$.  In this case $\alpha\approx 0$ and one can rewrite the
adiabaticity parameter as
\begin{equation}
\gamma_{H} \approx \frac{\Delta_{\rm
    atm}\sin^2(2\theta_{13}^{eff})}{\cos(2\theta_{13}^{eff}) |{\rm
    d}\ln V /{\rm d}r|_{r_H}}~, 
\label{eq:gammaH_lisi}
\end{equation}
where 
\begin{equation}
  \label{eq:deg}
  \theta_{13}^{eff} = \theta_{13} + \varepsilon_{e\tau}''
  (2-Y_e)/Y_e  
\end{equation}
in agreement with Ref.~\cite{Fogli:2002xj}.
\begin{figure}[t]
  \begin{center}
  \includegraphics[angle=-0,width=0.48\textwidth]{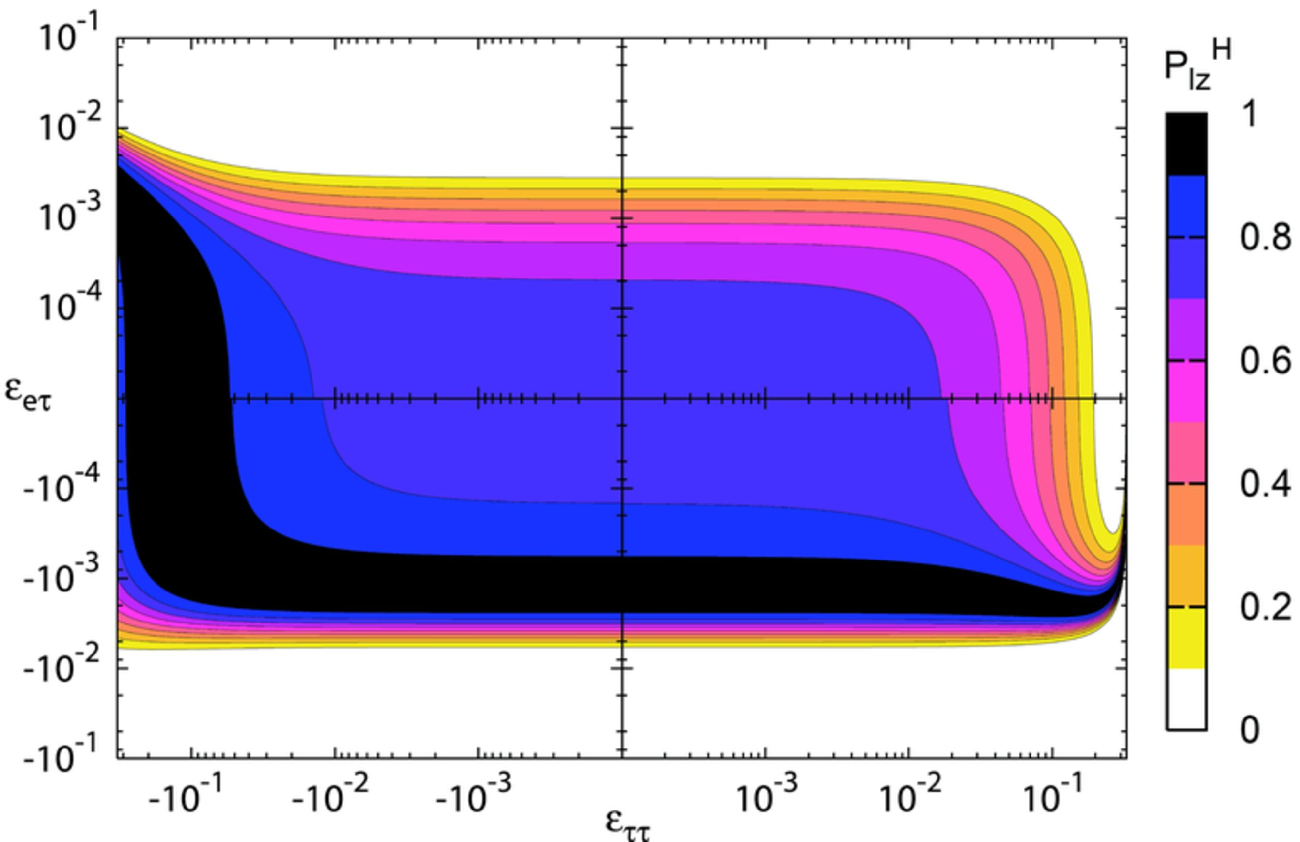} 
   \includegraphics[angle=-0,width=0.48\textwidth]{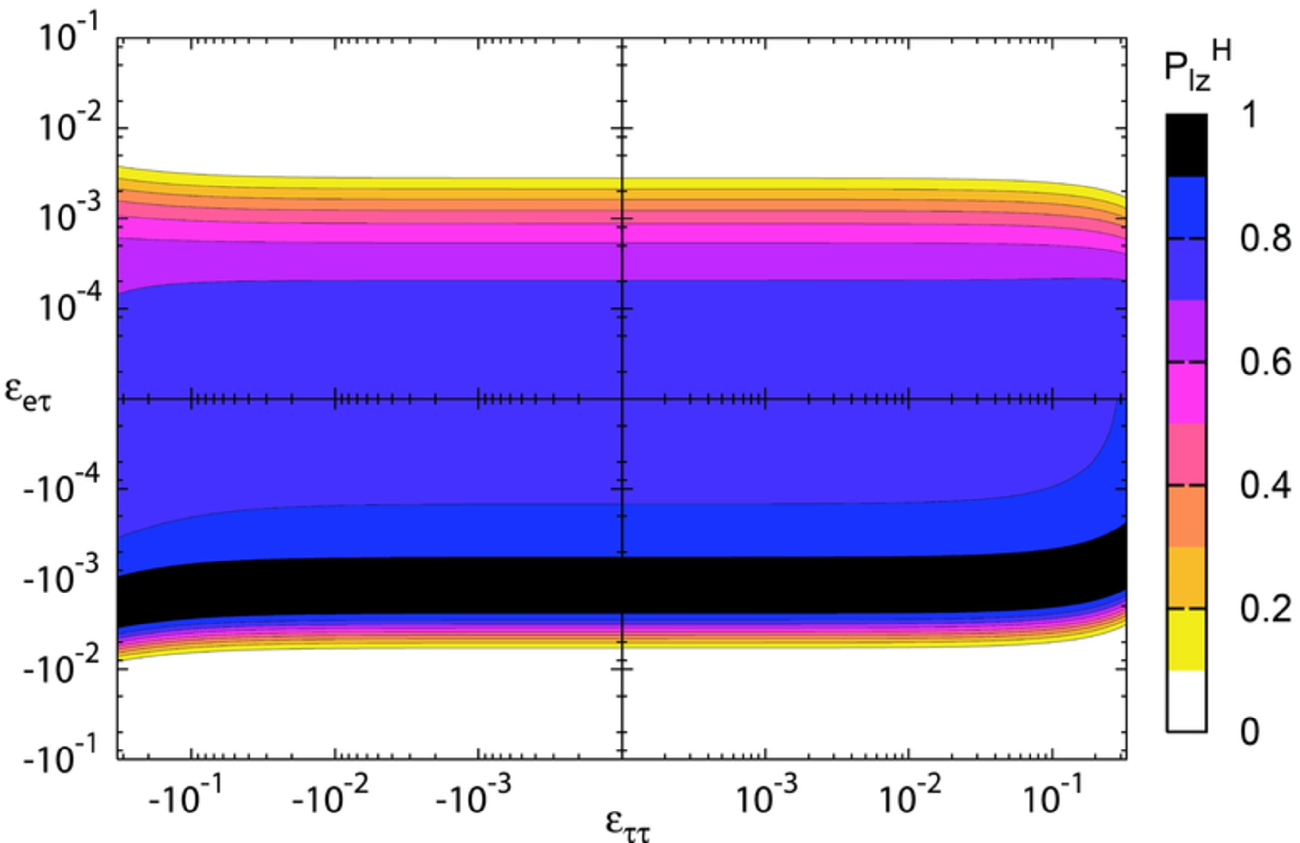} 
  \end{center}
  \caption{\small Landau-Zener jump probability isocontours at the
    $H$-resonance in terms of $\varepsilon_{e\tau}$ and
    $\varepsilon_{\tau\tau}$ for $10$~MeV antineutrinos in the case of
    inverted mass hierarchy. Left panel: $\alpha$ given by
    Eq.~(\ref{eq:alfa}). Right panel: $\alpha$ set to zero.  The
    remaining parameters take the following values:
    $\sin^2\theta_{13}=10^{-5},~\varepsilon_{ee}=\varepsilon_{e\mu}=\varepsilon_{\mu\tau}=0$. See
    text~\cite{EstebanPretel:2007yu}.  }
\label{fig:p-h}
\end{figure}
For slightly larger $\varepsilon$'s there can be significant
differences. In Fig.~\ref{fig:p-h} we show $P_{LZ}^{H}$ in the
$\varepsilon_{e\tau}$-$\varepsilon_{\tau\tau}$ plane for antineutrinos
with energy $10$~MeV in the case of inverse mass hierarchy, using
Eq.~(\ref{eq:plz-h}) with (left panel) and without (right panel) the
$\alpha$ correction.  The values of $\theta_{13}$ and
$\varepsilon_{e\tau}$ have been chosen so that the jump probability
lies in the transition regime between adiabatic and strongly non
adiabatic.
In the limit of small $\varepsilon_{\tau\tau}$, $\alpha$ becomes
negligible and therefore both results coincide. From
Eq.~(\ref{eq:gammaH_lisi}) one sees how as the value of
$\varepsilon_{e\tau}$ increases $\gamma_H$ gets larger and therefore
the transition becomes more and more adiabatic. For negative values of
$\varepsilon_{e\tau}$ there can be a cancellation between
$\varepsilon_{e\tau}$ and $\theta_{13}$, and as a result the
transition becomes non-adiabatic.

\begin{figure}[t]
  \begin{center}
\includegraphics[width=0.45\textwidth,height=7.cm,angle=0]{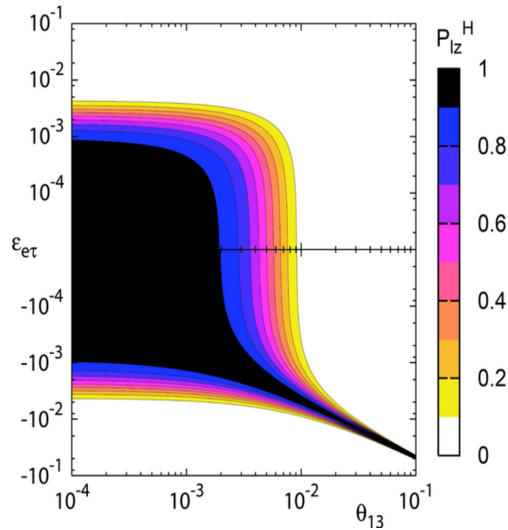}
  \end{center}
  \caption{\small Landau-Zener jump probability isocontours at the
    $H$-resonance in terms of $\varepsilon_{e\tau}$ and $\theta_{13}$
    for $\varepsilon_{\tau\tau}=10^{-4}$. Antineutrinos with energy
    $10$~MeV and inverted mass hierarchy has been
    assumed~\cite{EstebanPretel:2007yu}.}
  \label{fig:confusion}
\end{figure}

An additional consequence of Eq.~(\ref{eq:deg}) is that a degeneracy
between $\varepsilon_{e\tau}$ and $\theta_{13}$ arises. This is seen
in Fig.~\ref{fig:confusion}, which gives the contours of $P_{\rm
  LZ}^H$ in terms of $\varepsilon_{e\tau}$ and $\theta_{13}$ for
$\varepsilon_{\tau\tau}=10^{-4}$. One sees clearly that the same
Landau-Zener hopping probability is obtained for different
combinations of $\varepsilon_{e\tau}$ and $\theta_{13}$. This leads to
an intrinsic ``confusion'' between the mixing angle and the
corresponding NSI parameter, which can not be disentangled only in the
context of SN neutrinos, as noted in Ref.~\cite{Fogli:2002xj}.
 
We now turn to the case of $|\varepsilon_{\tau\tau}| \geq 10^{-2}$.
As $|\varepsilon_{\tau\tau}|$ increases the role of $\alpha$ becomes
relevant.  Whereas in the right panel of Fig.~\ref{fig:p-h}
$P_{LZ}^{H}$ remains basically independent of
$\varepsilon_{\tau\tau}$, one can see how in the left panel
$P_{LZ}^{H}$ becomes strongly sensitive to $\varepsilon_{\tau\tau}$
for $|\varepsilon_{\tau\tau}| \geq 10^{-2}$.

One sees that for positive values of $\varepsilon_{\tau\tau}$ it tends
to adiabaticity whereas for negative values to non-adiabaticity. This
follows from the dependence of $H_{e\tau}''$ on $\alpha$, essentially
through the term $-\Delta_\odot c_{13}s_\alpha s2_{12}$, see
Eq.~(\ref{eq:H'_h}).  For $|\varepsilon_{\tau\tau}| \geq 10^{-2}$ one
sees that $\sin\alpha$ starts being important, and as a result this
term eventually becomes of the same order as the others in
$H_{e\tau}''$.  At this point the sign of $\varepsilon_{\tau\tau}$,
and so the sign of $\sin\alpha$, is crucial since it may contribute to
the enhancement or reduction of $H_{e\tau}''$. This directly
translates into a trend towards adiabaticity or non-adiabaticity, seen
in Fig.~\ref{fig:p-h}.  Thus, for the range of
$\varepsilon_{\tau\tau}$ relevant for the NSI-induced internal
resonance the adiabaticity of the outer $H$-resonance can be affected
in a non-trivial way.

Turning to the case of the $L$-transition a similar expression can be
obtained by rotating the original Hamiltonian by
$U(\theta_{13})^\dagger U(\theta_{23})^\dagger$, as discussed in
Chapter~\ref{chapter:oscillations}. However, in contrast to the case
of the $H$-resonance, where the mixing angle $\theta_{13}$ is still
unknown, in the case of the $L$-transition the angle $\theta_{12}$ has
been shown by solar and reactor neutrino experiments to be
large~\cite{Maltoni:2004ei}. As a result, for the mass scale
$\Delta_\odot$ this transition will always be adiabatic irrespective
of the values of $\varepsilon_{\alpha\beta}$, and will affect only
neutrinos.

Summarizing, we have obtained that in the absence of collective flavor
transformations, the inclusion of NSI to the SN scenario mainly
affects the evolution of neutrinos in two ways. Firstly, for
sufficiently large diagonal NSI parameters, an internal resonance is
induced. Secondly, the usual MSW $H$- and $L$-resonances are modified,
essentially by a change in their position, but for some given values
of $\theta_{13}$ and the NSI parameters also affecting the
adiabaticity of the $H$-resonance.

\section{Including the neutrino background}
\label{sec:with-background}

After discussing the genuine effects that NSI could induce in the
evolution of SN neutrinos, let us now add to the picture the neutrino
self-interactions. The important point to have in mind is that both
ingredients may have drastic consequences in the same inner layers of
the SN. It is therefore crucial to analyze the interplay between these
two in principle coexisting effects.

\subsection{Equations of motion}                         
\label{sec:eoms}

As it has already been discussed, in the presence of neutrino
self-interactions it is convenient to make use of the density matrix
formalism. We therefore recover the EOMs for neutrinos given in
Eq.~(\ref{eq:gen_eoms}),
\begin{equation}
\I\partial_t\varrho_{\bf p}=[{\sf H}_{\bf p},\varrho_{\bf p}]\,,
\end{equation}
where $\varrho_{\bf p}$ and $\bar\varrho_{\bf p}$ represent the matrices of
density describing each (anti)neutrino mode and the Hamiltonian was
given in Eq.~(\ref{eq:hamiltonianrho}):
\begin{equation}
 {\sf H}_{\bf p}=\Omega_{\bf p}
 +{\sf V}+\sqrt{2}\,G_{\rm F}\!
 \int\!\frac{\D^3{\bf q}}{(2\pi)^3}
 \left(\varrho_{\bf q}-\bar\varrho_{\bf q}\right)
 (1-{\bf v}_{\bf q}\cdot{\bf v}_{\bf p})\,.
\label{eq:hamiltonian}
\end{equation}
For antineutrinos the only difference is $\Omega_{\bf
  p}\to-\Omega_{\bf p}$. 

\begin{figure}[t]
\begin{center}
\includegraphics[angle=0,width=0.55\textwidth]{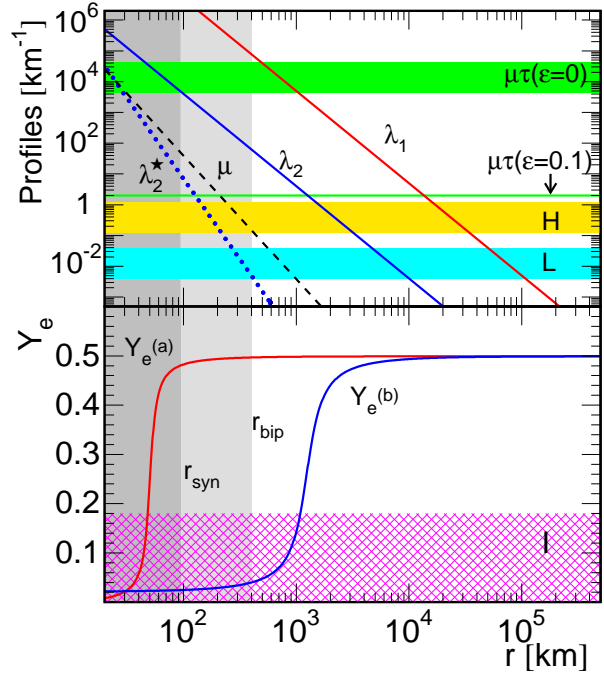}
\caption{\small Top panel: Density profiles, $\lambda (r)$, for
  $\lambda_0= 5\times 10^9$~km$^{-1}$ ($\lambda_1$) and $\lambda_0=
  4\times 10^6$~km$^{-1}$ ($\lambda_2$) in solid red and blue lines,
  respectively; $\lambda^\star(r)$ for $\lambda_2$ is shown in blue
  dotted lines; $\mu(r)$ for $\mu_0=7\times 10^5~$km$^{-1}$ in black
  dashed lines; the vacuum oscillation frequencies $\omega_{\rm H}$
  (yellow band), $\omega_{L}$ (cyan band), and $\omega_{\mu\tau}$ for
  $\varepsilon_{\tau\tau}=0$ (green band), for energies between 5 and
  50 MeV; $\omega^{\rm nsi}_{\mu\tau}$ for
  $\varepsilon_{\tau\tau}=0.1$ and $E=20$~MeV is also displayed (green
  solid line). The position of the synchronization and bipolar radii
  are also shown. Bottom: radial dependence of $Y_e$ for two set of
  parameters: $a=0.24,~b=0.165,~r_0=50~(1.2\times 10^3)$~km, and $r_s=
  5~(3\times 10^2)$~km, for $Y_e^{\rm a}~(Y_e^{\rm b})$. The
  horizontal magenta band represents the $Y_e^I$ leading to an
  internal $I$-resonance for $\varepsilon_{\tau\tau}\le
  0.1$~\cite{EstebanPretel:2009is}.  }
\label{fig:profiles_ch6}
\end{center}
\end{figure}

The kinetic term will be determined by the mass and mixing
parameters. We will use $\Delta m_{21}^2\equiv m_2^2-m_1^2 =
7.65\times 10^{-5}$~eV$^2$, $|\Delta m_{31}^2|\equiv |m_3^2-m_1^2| =
2.40\times 10^{-3}$~eV$^2$ and $\sin^2\theta_{12}=0.3$. We consider
also $\sin^2\theta_{13}=10^{-2}$ and three different values for
$\theta_{23}$ in the allowed range at 3$\sigma$, $\sin^2\theta_{23} =$
0.4, 0.5 and 0.6, because our results depend sensitively on
$\theta_{23}$. Given the values of $\Delta m^2$ we obtain the two
associated vacuum oscillation frequencies: $\omega_{\rm H}\equiv
\Delta m^2_{31}/2E$ and $\omega_{\rm L}\equiv \Delta m^2_{21}/2E$,
which in the case of neutrinos with $E=20$~MeV, lead to $\omega_{\rm
  H}=0.3$~km$^{-1}$ and $\omega_{\rm L}=0.01$~km$^{-1}$. In the top
panel of Fig.~\ref{fig:profiles_ch6} we represent $\omega_{\rm H}$ and
$\omega_{L}$ for energies typical in SNe, between 5 MeV and 50 MeV, as
yellow and light blue bands, respectively.

The only difference in Eq.~(\ref{eq:hamiltonian}) compared to the
previous chapters resides in the matter potential, where we have to
add the NSI contribution. Let us then concentrate the discussion in
this term. As we have seen, the interaction of neutrinos with matter
can be now split in two pieces:
\begin{equation}
{\sf V}= {\sf V}_{\rm std} + {\sf V}_{\rm nsi}~.
\end{equation}
The first term, ${\sf V}_{\rm std}$, describes the standard
interaction with matter and can be represented in the weak basis by
\begin{equation}
{\sf V}_{\rm std}=\lambda(r){\rm diag}(Y_e,0,Y_\tau^{\rm eff})\,, 
\label{eq:V}
\end{equation}
with 
\begin{equation}
\label{eq:lambda(r)}
  \lambda(r) =
    \lambda_0\,\left(\frac{R_\nu}{r}\right)^3 \,.
\end{equation} 
These expressions are equivalent to the ones in
Eqs.~(\ref{eq:Vcc_ch6}) and (\ref{eq:rho_prof_ch6}) in units of
km$^{-1}$. We have here left $Y_e$ outside the definition of $\lambda$
because of its special importance in the NSI effects. In the following
we assume $R_\nu= 10$~km.  In the top panel of
Fig.~\ref{fig:profiles_ch6} we show two $\lambda(r)$ profiles for
$\lambda_0 = 5\times 10^9$~km$^{-1}$ and $4\times 10^6$~km$^{-1}$
denoted by $\lambda_1$ and $\lambda_2$, corresponding to typical early
and late time profiles, respectively, as illustrated in
Fig.~\ref{fig:profiles_ch5}.

The first element in ${\sf V}_{\rm std}$ represents the charged
current potential and is proportional to the electron fraction,
$Y_e$. We parameterize $Y_e$ as in Eq.~(\ref{eq:Ye_ch6}) with $a=0.24$
and $b=0.165$. The parameters $r_0$ and $r_s$ describe where the rise
takes place and how steep it is, respectively.  In the bottom panel of
Fig.~\ref{fig:profiles_ch6} we show two $Y_e(r)$ profiles for two
different choices of these parameters.
The radius where $\lambda(r)Y_e(r)$ crosses the horizontal bands
$\omega_{\rm H}$ ($\omega_{\rm L}$) determines the well known $H$
($L$) MSW resonances. For the $\lambda$ and $Y_e$ profiles shown in
Fig.~\ref{fig:profiles_ch6} and energies typical in SNe the position
of both resonances $r^{\rm H}_{\rm res}$ and $r^{\rm L}_{\rm res}$ lie
above $10^3$~km.

The other non-zero element in ${\sf V}_{\rm std}$ arises from
radiative corrections to neutral-current $\nu_\mu$ and $\nu_\tau$
scattering, as discussed in
Chapter~\ref{chapter:oscillations}. Although there are no $\mu$ nor
$\tau$ leptons in normal matter, they appear as virtual states causing
a shift $\Delta V_{\mu\tau}= \sqrt{2}\,G_{\rm F}Y_\tau^{\rm eff}n_B$
between $\nu_\mu$ and~$\nu_\tau$ due to the difference in their
masses, with $Y_\tau^{\rm eff}$ given in Eq.~(\ref{eq:Ytau}). In the
top panel of Fig.~\ref{fig:profiles_ch6} we show
$\omega_{\mu\tau}\equiv \omega_{\rm H}/Y_\tau^{\rm eff}$ for energies
between 5 and 50 MeV, as a green band. Analogously to the $H$- and
$L$-resonances, the radius where $\lambda(r)\approx\omega_{\rm
  \mu\tau}$ defines the $\mu\tau$-resonance.

According to the description given in the previous section and in
Chapter~\ref{chapter:NSI}, the term in the Hamiltonian describing the
non-standard neutrino interactions with a fermion $f$ can be expressed
as,
\begin{equation}
({\sf V}_{\rm nsi})_{\alpha\beta} = \sum_{f=e,u,d} ({\sf V}_{\rm
    nsi}^f)_{\alpha\beta}=\sqrt{2}G_FN_f \varepsilon^f_{\alpha\beta} \,,
\end{equation}
where $N_f$ represents the fermion $f$ number density.
Again, we consider $\varepsilon^f_{\alpha\beta} \in\Re$, neglecting
possible $CP$ violation in the new interactions, and take for $f$ the
down-type quark.
Therefore the NSI potential can be expressed as follows,
\begin{equation}
\label{eq:Vnsi}
({\sf V}_{\rm nsi})_{\alpha\beta} = ({\sf V}_{\rm nsi})_{\alpha\beta} 
=\varepsilon_{\alpha\beta}\lambda(r)(2-Y_e)~.
\end{equation}
In principle at least five of the six independent
$\varepsilon_{\alpha\beta}$ parameters, after removing one of the
diagonals, should be taken into account. Nevertheless, the exhaustive
description developed in the previous section shows that all the
physics involved can be described in terms of
$\varepsilon'_{\tau\tau}$ and $\varepsilon'_{e\tau}$, which are just a
suitable combination of $\varepsilon$'s. This motivates us to
illustrate the interplay that could arise between collective effects
and NSI by only considering two non-zero NSI parameters:
$\varepsilon_{e\tau}$~and $\varepsilon_{\tau\tau}$, describing
flavor-changing (FC) processes and non-universality (NU),
respectively.
Therefore the term in the Hamiltonian responsible for the interactions
with matter can be written as
\begin{equation}
\label{eq:Vmatrix}
{\sf V}=
\lambda(r)(2-Y_e)
\left(\begin{array}{ccc} \frac{Y_e}{2-Y_e} &
  0 & \varepsilon_{e\tau} \\
0 & 0 & 0 \\
\varepsilon_{e\tau} & 0 & \varepsilon_{\tau\tau}+\frac{Y_\tau^{\rm eff}}{2-Y_e}  
\end{array} \right)~.
\end{equation}
This expresion is equivalent to Eq.~(\ref{eq:H_inner}), including the
$\mu\tau$-term to the potential and considering only the desired NSI
parameters. The range of values for the $\varepsilon$'s we consider is
for the off-diagonal term $10^{-5}\lesssim
|\varepsilon_{e\tau}|\lesssim {\rm few}\times 10^{-3}$. This prevents
any significant NSI-induced reduction of the electron fraction $Y_e$
during the core collapse.  For the diagonal term we assume
$|\varepsilon_{\tau\tau}|\lesssim 0.1$, allowed by the current
experimental constraints.

Finally, the third term in the Hamiltonian accounts for the collective
flavor transformations induced by neutrino-neutrino interaction, and
has been extensively described in Chapters~\ref{chapter:coll2flavors}
and \ref{chapter:coll3flavors}. We consider the single-angle
approximation by launching all neutrinos with $45^\circ$ relative to
the radial direction. This approximation has been shown to be valid
for realistic SN neutrino fluxes, provided that the neutrino density
exceeds the electron density. The radial dependence of the
neutrino-neutrino interaction strength can be
explicitly written in such a system as
\begin{equation}
\label{eq:mu(r)}
\mu(r) = \mu_0 \frac{R_\nu^4}{r^4}\frac{1}{2-R_\nu^2/r^2}\approx
\mu_0\frac{R_\nu^4}{2r^4}\,. 
\end{equation}
In the top panel of Fig.~\ref{fig:profiles_ch6} we show the typical
$\mu(r)$ profile we are using, with $\mu_0=7\times 10^5$~km$^{-1}$.
One final property of SN neutrinos with important consequences
  for our study is the hierarchy of fluxes obtained in SN models. 
The typical conditions of the proto-neutron star lead to the
following hierarchy of fluxes $F^{R_\nu}_{\nu_e} > F^{R_\nu}_{\bar\nu_e} >
F^{R_\nu}_{\nu_\mu} = F^{R_\nu}_{\bar\nu_\mu} = F^{R_\nu}_{\nu_\tau} =
F^{R_\nu}_{\bar\nu_\tau}$.  
As already discussed we express the lepton asymmetry with the
parameter
$\epsilon=(F^{R_\nu}_{\nu_e}-F^{R_\nu}_{\bar\nu_e})/(F^{R_\nu}_{\bar\nu_e}-F^{R_\nu}_{\bar\nu_x})$.
Throughout the analysis we shall consider $\epsilon=0.25$.
The equal parts of the fluxes drop out of the
EOMs, so as initial condition we use in the monoenergetic case
$F^R_{\nu_\mu,\bar\nu_\mu,\nu_\tau,\bar\nu_\tau}=0$ and $F^R_{\nu_e}
=(1+\epsilon) F^R_{\bar\nu_e}$.

\subsection{Competition between NSI and collective effects}
\label{sec:nsi-collective}

As it has already been discussed, in the absence of NSI and collective
effects the neutrino propagation through the SN envelope is basically
determined at the well-known MSW resonances.  The $L$-resonance occurs
always for neutrinos whereas the $H$-one takes place for
(anti)neutrinos for (inverted) normal mass hierarchy. For our matter
profiles and the values of $\theta_{12}$ and $\theta_{13}$ both
resonances are adiabatic, see Fig.~\ref{fig:adresHL}.  Moreover both
involve electron neutrino flavor and happen in the outer layers of the
SN envelope, see top panel of Fig.~\ref{fig:profiles_ch6}.

In addition, the $\mu\tau$-resonance is also adiabatic, but occurs
between the $\nu_\mu$ and $\nu_\tau$ or $\bar\nu_\mu$ and
$\bar\nu_\tau$ depending on the neutrino mass hierarchy and the
$\theta_{23}$ octant. However, when considering the neutrino
self-interaction this resonance can also cause significant
modifications of the overall $\nu_e$ and $\bar\nu_e$ survival
probabilities~\cite{EstebanPretel:2007yq}. According to the discussion
in Sec.~\ref{sec:eoms} the $\mu\tau$-resonance occurs at
\begin{equation}
\label{eq:rmutau}
r_{\mu\tau} \approx R_\nu~\left(\frac{\lambda_0 Y_\tau^{\rm eff}}{\omega_{\rm
    H}}\right)^{1/3}= R_\nu~\left(\frac{\lambda_0}{\omega_{\rm
    \mu\tau}}\right)^{1/3}\,.
\end{equation}
Due to the smallness of $ Y_\tau^{\rm eff}$ the $\mu\tau$-resonance
happens at deeper layers than the $H$- and $L$-resonances. In
particular, for $\omega_{\rm H}=0.3$~km$^{-1}$, $Y_e=0.5$, and $\lambda_0 =
5\times 10^9$~km$^{-1}$ ($4\times 10^6$~km$^{-1}$)
$r_{\mu\tau}=770$~km (71 km), see the intersection between the green
band and the profiles $\lambda_1(r)$ and $\lambda_2(r)$ in the top
panel of Fig.~\ref{fig:profiles_ch6}.
%


The consequence of the addition of an NSI term such as that of
Eq.~(\ref{eq:Vnsi}) is twofold. First, it will affect the MSW
resonances. For the values assumed here the main effect on the $H$-
and $L$-resonances will be just a slight shift in the resonance point,
discussed in Sec.~\ref{sec:two-regimes}. The consequences for the
$\mu\tau$-resonance can be more drastic. For sufficiently large values
of $|\varepsilon_{\tau\tau}|$ a negative sign can change the resonance
channel, from $\nu$ to $\bar\nu$ or viceversa, depending on the octant
of $\theta_{23}$. On the other hand, it can significantly modify the
position of the resonance. In the presence of NSI the
$\mu\tau$-resonance happens at
\begin{equation}\label{eq:rmutaunsi}
r_{\mu\tau} \approx R_\nu~\left(\frac{\lambda_0 Y_{\tau,{\rm nsi}}^{\rm
    eff}}{\omega_{\rm H}}\right)^{1/3}=
R_\nu~\left(\frac{\lambda_0}{\omega_{\rm \mu\tau}^{\rm nsi}}\right)^{1/3}\,,
\end{equation}
where we have defined
\begin{eqnarray}
Y_{\tau,{\rm nsi}}^{\rm eff} & \equiv & Y_\tau^{\rm
  eff}+(2-Y_e)\varepsilon_{\tau\tau} \,,\\   
\omega^{\rm  nsi}_{\mu\tau} & \equiv &\omega_{\rm H}/Y_{\tau,{\rm nsi}}^{\rm
  eff} \,.
\end{eqnarray}
In particular, for $|\varepsilon_{\tau\tau}|>Y_\tau^{\rm eff}/(2-Y_e)$
the value of $\omega_{\mu\tau}^{\rm nsi}$ decreases, and therefore
$r_{\mu\tau}$ is pushed outwards with respect to the standard case.
In the top panel of Fig.~\ref{fig:profiles_ch6} we show the value of
$\omega^{\rm nsi}_{\mu\tau}$ in the presence of
$\varepsilon_{\tau\tau}=0.1$ and $E=20$~MeV. For such a choice of
parameters and the matter profile corresponding to $\lambda_0=4\times
10^6$~km$^{-1}$, we can see how the position of the
$\mu\tau$-resonance moves out to a radius of $r_{\mu\tau}\approx
1.3\times 10^3$~km. 

The second important consequence is that the new NSI terms can induce
additional resonances in the inner layers, as described in
Sec.~\ref{sec:no-background}. The condition required for this
$I$-resonance to take place was given in Eq.~(\ref{eq:Ye_resI}) and
can be written, for our simplified system, as
\begin{equation}
\label{eq:Ye_resI}
Y_e^{I} = \frac{2\varepsilon_{\tau\tau}}{1+\varepsilon_{\tau\tau}}~.
\end{equation}
In the bottom panel of Fig.~\ref{fig:profiles_ch6} we show as a
horizontal band the range of $Y_e^{I}$ required for the $I$-resonance
to take place for $\varepsilon_{\tau\tau}\le 0.1$. For typical values
of $Y_e$ one expects to have the $I$-resonance for
$\varepsilon_{\tau\tau}\gtrsim 10^{-2}$. Moreover the $Y_e(r)$ and
$|\varepsilon _{e\tau}|$ considered guarantee the adiabaticity.


At the same time, also in the internal region, the neutrino flux
emerging from the SN core is so dense that, neutrino-neutrino
refraction can cause nonlinear flavor oscillation phenomena. For the
hierarchy of neutrino fluxes assumed the induced pair-wise flavor
transformation occurs only in the case that the neutrino mass
hierarchy is inverted.
Collective flavor transformations start after the synchronization
phase, where $\mu(r_{\rm syn}) \approx 2\omega_{\rm
  H}/(1-\sqrt{1+\varepsilon})^2$, and extends a few hundred km in the
so-called bipolar regime until $\mu(r_{\rm bip})\approx \omega_{\rm
  H}$. At larger radii $\mu(r)<\omega_{\rm H}$ and the neutrino
self-interaction becomes negligible. For our chosen $\mu_0$, an excess
$\nu_e$ flux of 25\%, and $\omega_{\rm H}=0.3$~km$^{-1}$, we find a
synchronization and bipolar radius of $r_{\rm syn}\simeq 100$~km and
$r_{\rm bip}\simeq 330$~km, as indicated in
Fig.~\ref{fig:profiles_ch6} by dark and light vertical gray bands,
respectively.
One important consequence of this flavor transformation in the context
of three neutrino flavors is its potential sensitivity to deviations
of $\theta_{23}$ from maximal mixing. As discussed in
Chapter~\ref{chapter:coll3flavors}, in the particular case that the
$\mu\tau$-resonance takes place outside the synchronization radius the
final $\nu_e$ and $\bar\nu_e$ survival probability depend crucially on
the octant of $\theta_{23}$. In Sec.~\ref{sec:competition}, we argued
that this potential effect would most likely be suppressed by
multi-angle decoherence induced by dense matter. Let us now analyze
how the picture changes when considering NSI. According to
Eqs.~(\ref{eq:rmutaunsi}) and~(\ref{eq:r_synch}) the condition for this
{\em $\mu\tau$ effect} to happen is given by
%
\begin{equation}
\label{eq:rmutausyn}
\lambda_0 Y_{\tau,{\rm nsi}}^{\rm eff} \gtrsim \left(\frac{\sqrt{1+\epsilon}-1}{2}\right)^{3/2}\mu_0^{3/4}\omega_{\rm H}^{1/4}\sim 3\times 10^2~{\rm 
  km}^{-1}\,. 
\end{equation}
Although this is a minimum requirement, the possibility to discern the
$\theta_{23}$ octant becomes cleaner when the $\mu\tau$-resonance
happens outside the bipolar radius, $r_{\mu\tau}>r_{\rm bip}$. The
condition we obtain according to Eqs.~(\ref{eq:rmutaunsi})
and~(\ref{eq:r_bip}) is, therefore,
\begin{equation}
\label{eq:rmutaubip}
\lambda_0 Y_{\tau,{\rm nsi}}^{\rm eff} \gtrsim
\left(\frac{\mu_0}{2}\right)^{3/4}\omega_{\rm H}^{1/4}\sim 10^4~{\rm km}^{-1}\,.
\end{equation}
In the standard case this possibility only occurs for
large density profiles, $\lambda_0\gtrsim 4\times 10^8$~km$^{-1}$,
i.e.~at early times. This situation would correspond to the
$\lambda_1(r)$ profile in top panel of Fig.~\ref{fig:profiles_ch6}, but
not to $\lambda_2(r)$. However the presence of NSI terms in the
Hamiltonian may shift the $\mu\tau$-resonance to outer layers, making
this condition more flexible. For instance, for
$\varepsilon_{\tau\tau}=0.1$ the previous condition requires only
$\lambda_0\gtrsim 1.3\times 10^5$~km$^{-1}$, see green solid line in the
top panel of Fig.~\ref{fig:profiles_ch6}. Therefore the presence of NSI
could keep the possibility to distinguish between the two
$\theta_{23}$ octants for several seconds.


The self-induced flavor transformations however do not occur for
arbitrarily large density profiles. If the electron density $n_e$
significantly exceeds the neutrino density $n_\nu$ in the conversion
region they can be suppressed by matter.
This is a consequence of neutrinos traveling on different trajectories
when streaming from a source that is not point-like. This multi-angle
matter effect can be neglected if in the collective region, prior
  to  the synchronization radius, we have
\begin{equation}
\label{eq:lambda*}
\lambda^{\star}(r)\equiv Y_e(r)\lambda(r)\frac{R_\nu^2}{2r^2} \ll \mu(r)\,.
\end{equation}
The limiting condition can be determined by imposing
  Eq.~(\ref{eq:lambda*}) at $r_{\rm syn}$.  Taking into
account Eqs. (\ref{eq:lambda(r)}) and (\ref{eq:mu(r)}) we obtain
\begin{equation}
\label{eq:lambda*2}
Y_e(r_{\rm syn})\lambda_0\frac{R_\nu}{r_{\rm syn}} \ll \mu_0\,.
\end{equation}
Assuming $Y_e=0.5$ and $r_{\rm syn}=100$~km, this condition amounts to
$\lambda_0\ll 1.4\times 10^7$~km$^{-1}$.
In the top panel of Fig.~\ref{fig:profiles_ch6} we show
$\lambda_2^\star(r)$, corresponding to a $\lambda_0$ smaller than
$\lambda_0\ll 1.4\times 10^7$~km$^{-1}$. The condition
$\lambda_2^\star(r)\ll \mu(r)$ is then satisfied in the bipolar
region, and collective effects are not matter suppressed. This is not
the case of $\lambda_1(r)$.
In the standard case the limiting $\lambda_0=1.4\times
10^7$~km$^{-1}$, above which multiangle matter effects suppress the
collective effects, is though smaller than the minimum
$\lambda_0=4\times 10^8$~km$^{-1}$ required for the $\mu\tau$ effect
to be important.  The situation could drastically change in the
presence of NSI. Non-zero NSI diagonal parameters could help moving
the $\mu\tau$ resonance out of the $r_{\rm syn}$ even for $\lambda_0$
smaller than $1.4\times 10^7$~km$^{-1}$. The consequence is that large
enough $|\varepsilon_{\tau\tau}|$ would make the neutrino propagation
through the SN envelope highly sensitive to the $\theta_{23}$ octant.

\subsection{Classification of regimes}
\label{sec:classification}

\begin{figure}[t]
\begin{center}
\includegraphics[angle=0,width=0.55\textwidth]{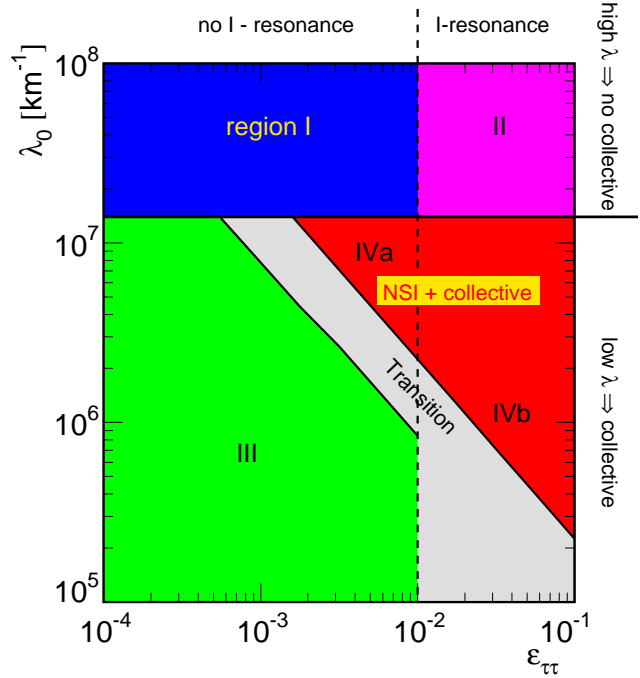}
\caption{\small Different regimes of the neutrino propagation
  depending on the value of $\lambda_0$ and $\varepsilon_{\tau\tau}$,
  as described in the text~\cite{EstebanPretel:2009is}.  }
\label{fig:regions}
\end{center}
\end{figure}

In this section we summarize all the information formerly
introduced. Taking into account the conditions given in
Eqs.~(\ref{eq:Ye_resI}), (\ref{eq:rmutausyn}),
(\ref{eq:rmutaubip}), (\ref{eq:lambda*}), and (\ref{eq:lambda*2}) we
can roughly identify four different regimes of the neutrino
propagation in terms of $\lambda_0$ and $\varepsilon_{\tau\tau}$. This
scheme is displayed in Fig.~\ref{fig:regions}.

To first approximation the four regions can be defined in terms of
matter suppression (or not) of collective effects and presence (or
not) of the internal $I$-resonance.  Equation~(\ref{eq:lambda*2}) is depicted as a horizontal solid line at
$\lambda_0=1.4\times 10^7$~km$^{-1}$. For higher $\lambda_0$ matter
suppresses collective effects whereas for smaller densities collective
effects are present. For intermediate values,
$\lambda^\star(r)\sim\mu(r)$, there would be a matter induced
decoherence~\cite{EstebanPretel:2008ni}. To make the discussion as
simple as possible we will only consider the extreme cases.
On the other hand the vertical dashed line at
$\varepsilon_{\tau\tau}=10^{-2}$ indicates the presence (right) or
absence (left) of the NSI-induced $I$-resonance.\vspace{0.5cm}\\
This simple scheme becomes further complicated if one adds the
possibility that the $\mu\tau$-resonance lies outside the bipolar
region. In the next subsections we analyze in detail the different
possibilities.

\textbf{A) Region I}\vspace{0.2cm}\\
On the upper left corner we have the region I, defined by $\lambda_0
\gtrsim 1.4\times 10^7$~km$^{-1}$ and $\varepsilon_{\tau\tau}\lesssim
10^{-2}$. According to the previous discussion, this range of
parameters leads to no collective effects, since they are suppressed
by matter, and no $I$-resonance. Assuming that the $L$- and
$H$-resonances are adiabatic the $\nu_e$ and $\bar\nu_e$ survival
probability is then only fixed by the mass hierarchy.  The NSI terms
will lead at most to a small shift in its position.

\begin{figure}[th!]
\begin{center}
\includegraphics[angle=0,width=0.45\textwidth]{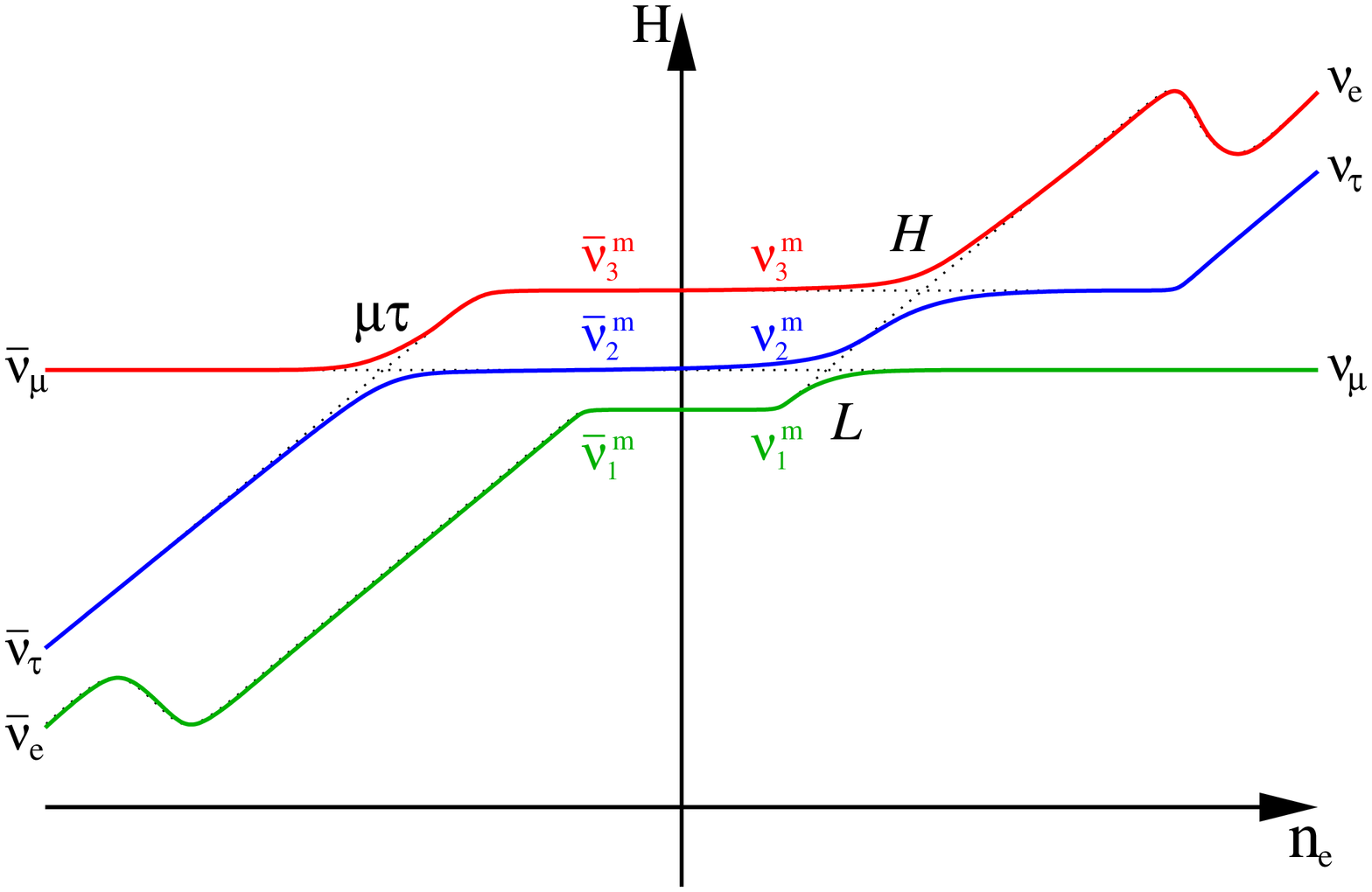}
\hskip6pt
\includegraphics[angle=0,width=0.45\textwidth]{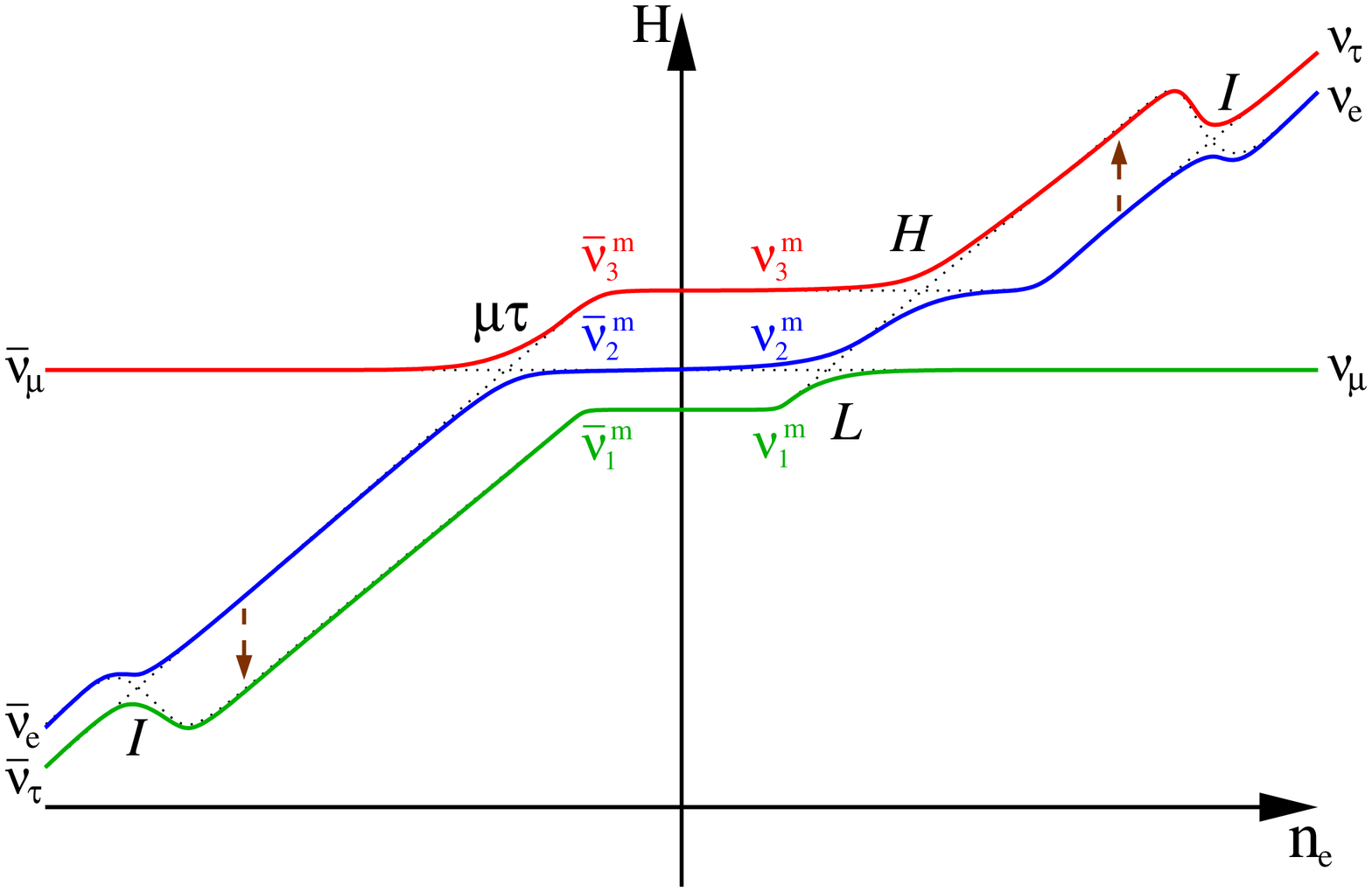}
\vskip6pt
\includegraphics[angle=0,width=0.45\textwidth]{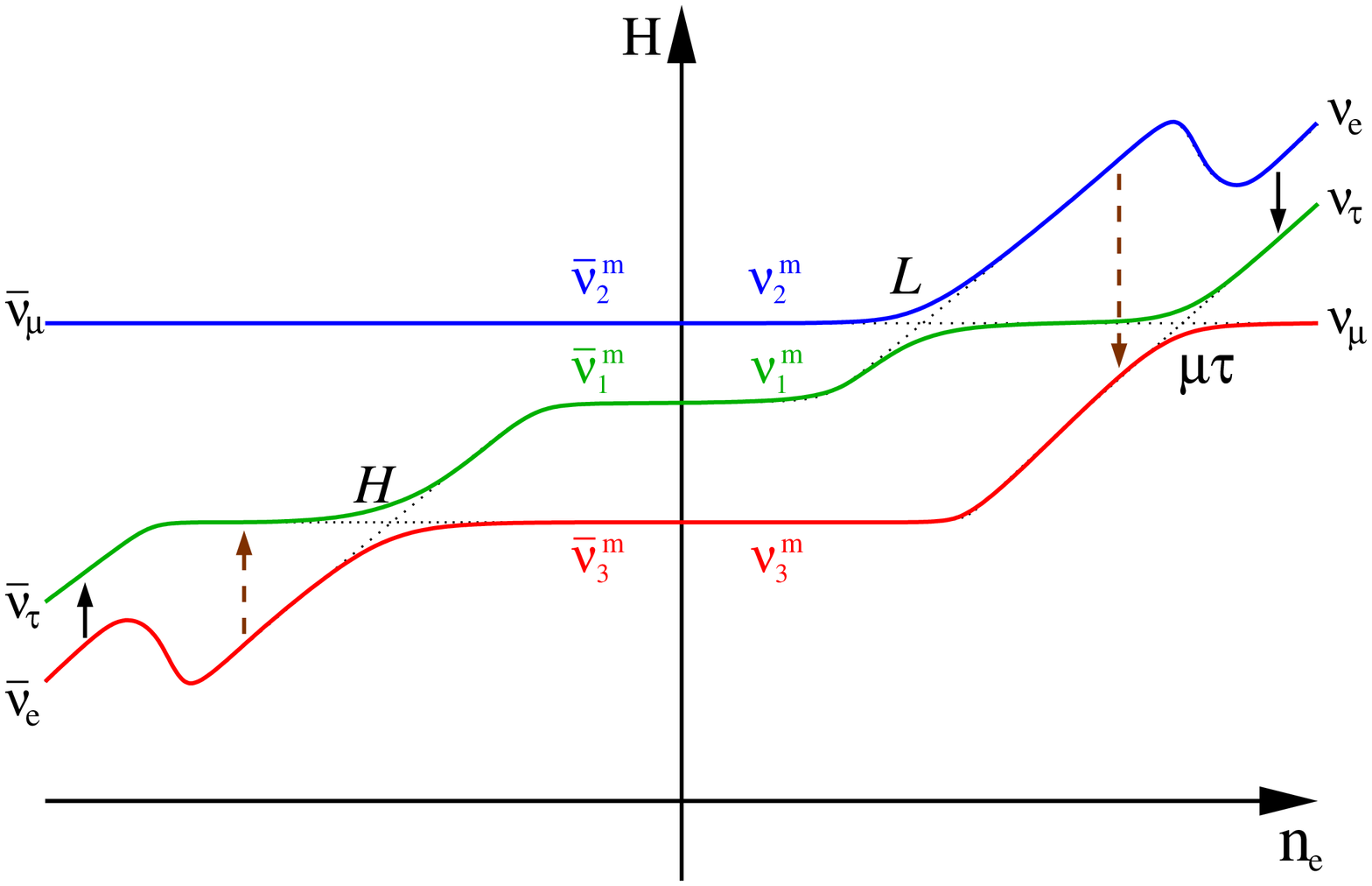}
\hskip6pt
\includegraphics[angle=0,width=0.45\textwidth]{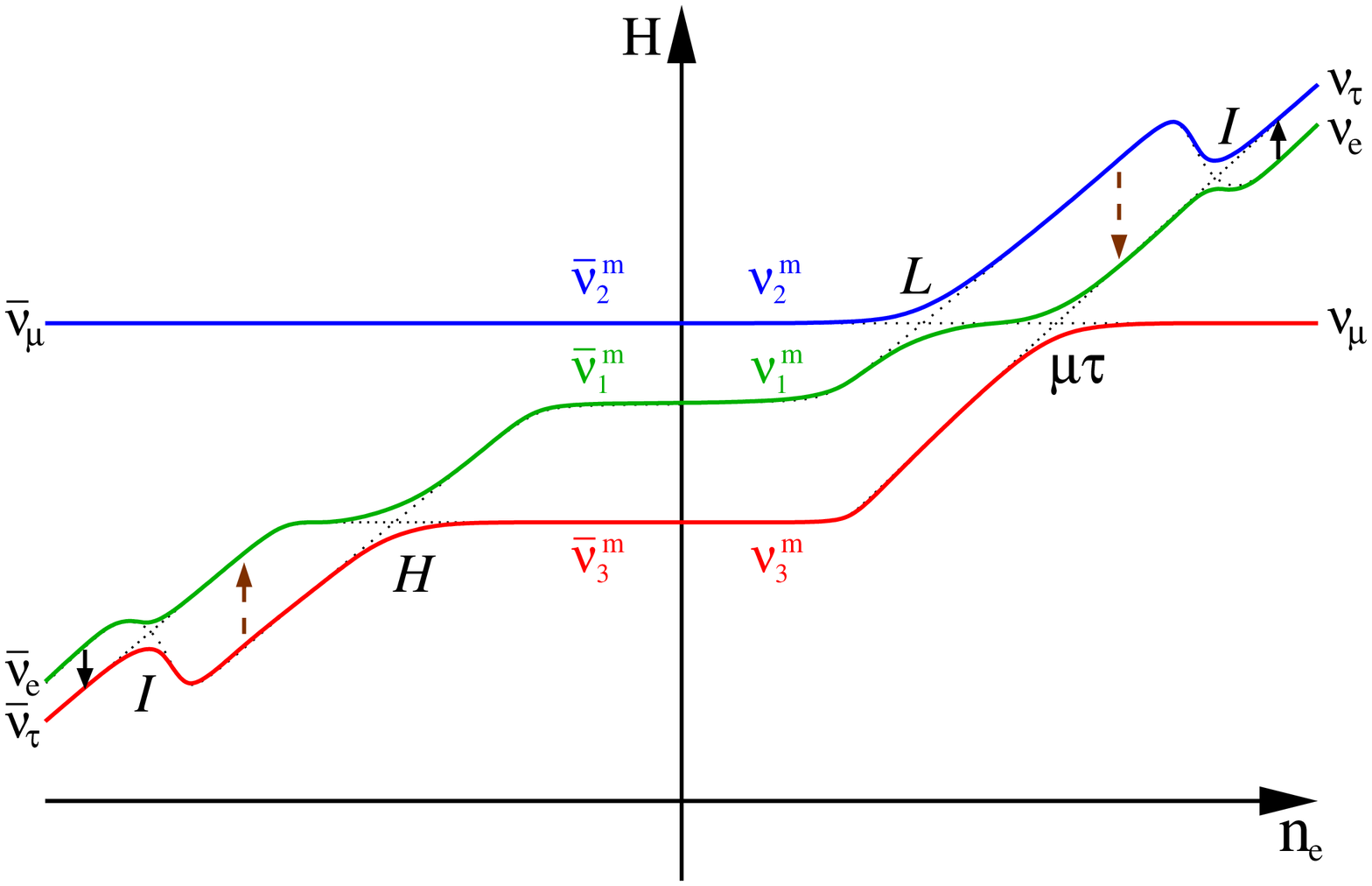}
\vskip6pt
\includegraphics[angle=0,width=0.45\textwidth]{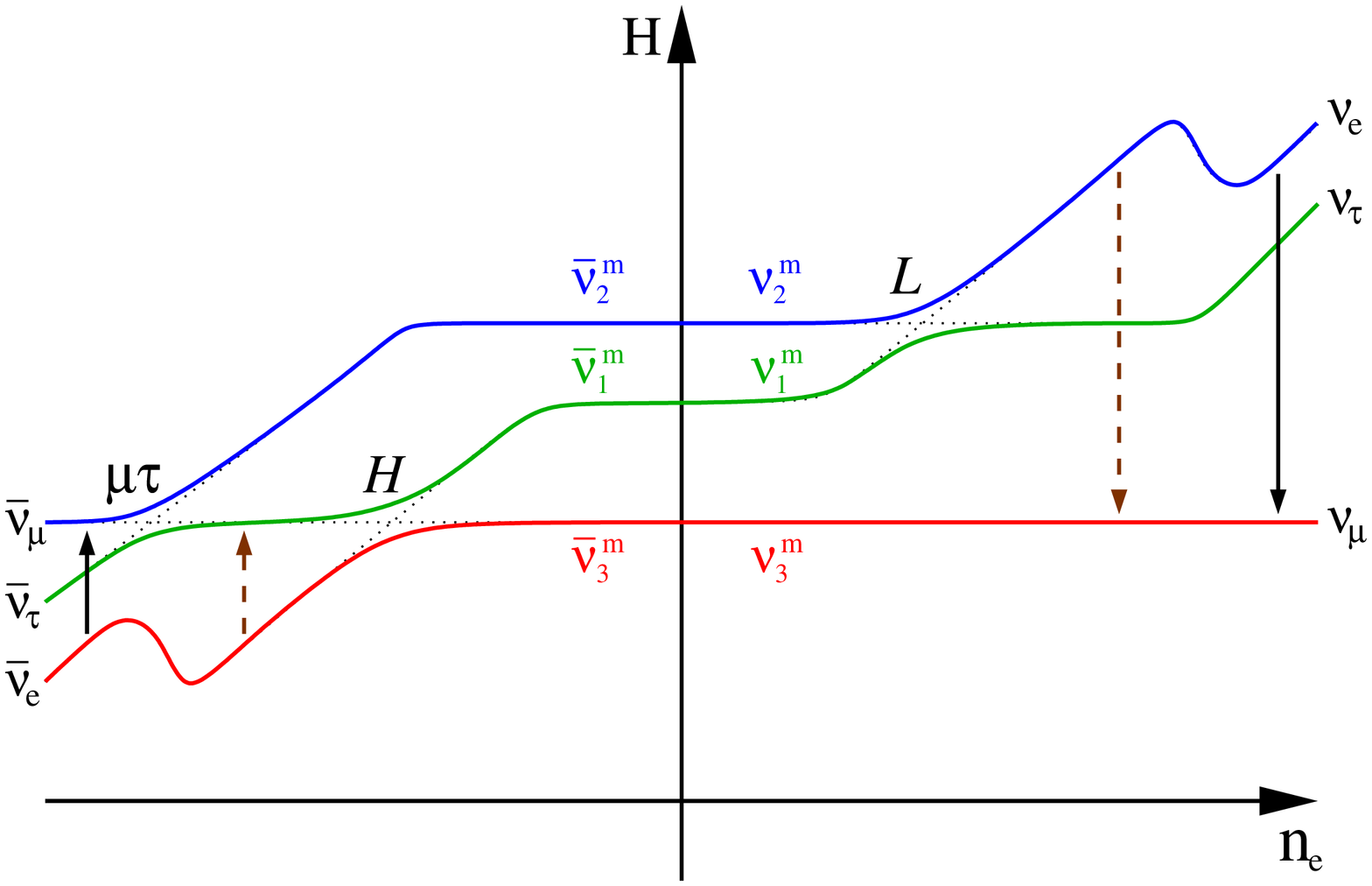}
\hskip6pt
\includegraphics[angle=0,width=0.45\textwidth]{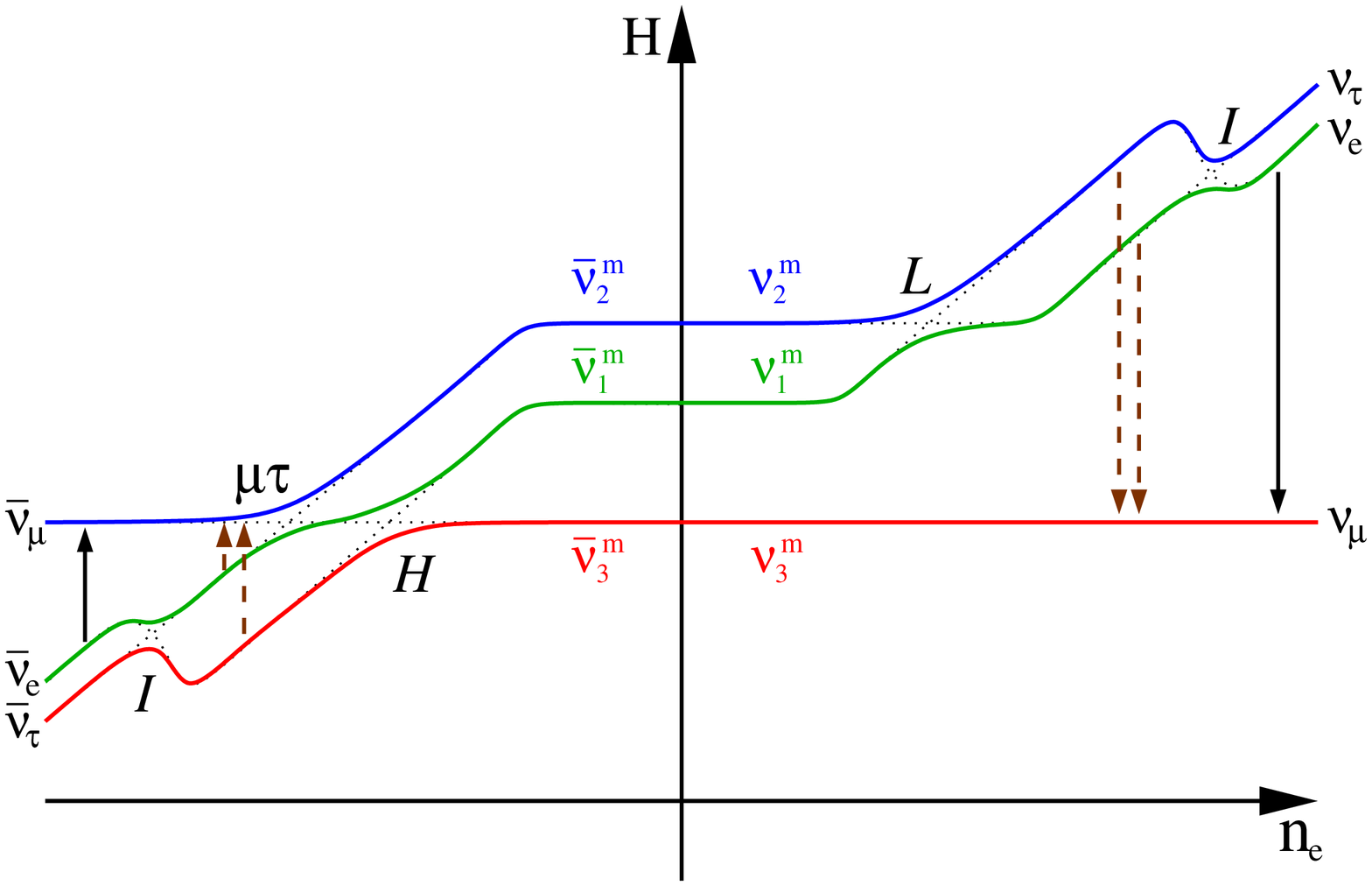}
\caption{\small Level crossings in the absence (left) and presence
  (right) of $I$-resonance for normal (top), and inverted mass
  hierarchy for $\sin^2\theta_{23} = 0.4$ (middle), and
  $\sin^2\theta_{23} = 0.6$ (bottom). The dashed and solid arrows in
  the middle and bottom indicate the pair transformations due to
  collective effects happening after (dashed) or before (solid) the
  $\mu\tau$-resonance~\cite{EstebanPretel:2009is}.  }
\label{fig:levcros_I-noI}
\end{center}
\end{figure}

\begin{figure}[t]
\begin{center}
\includegraphics[angle=0,width=0.45\textwidth]{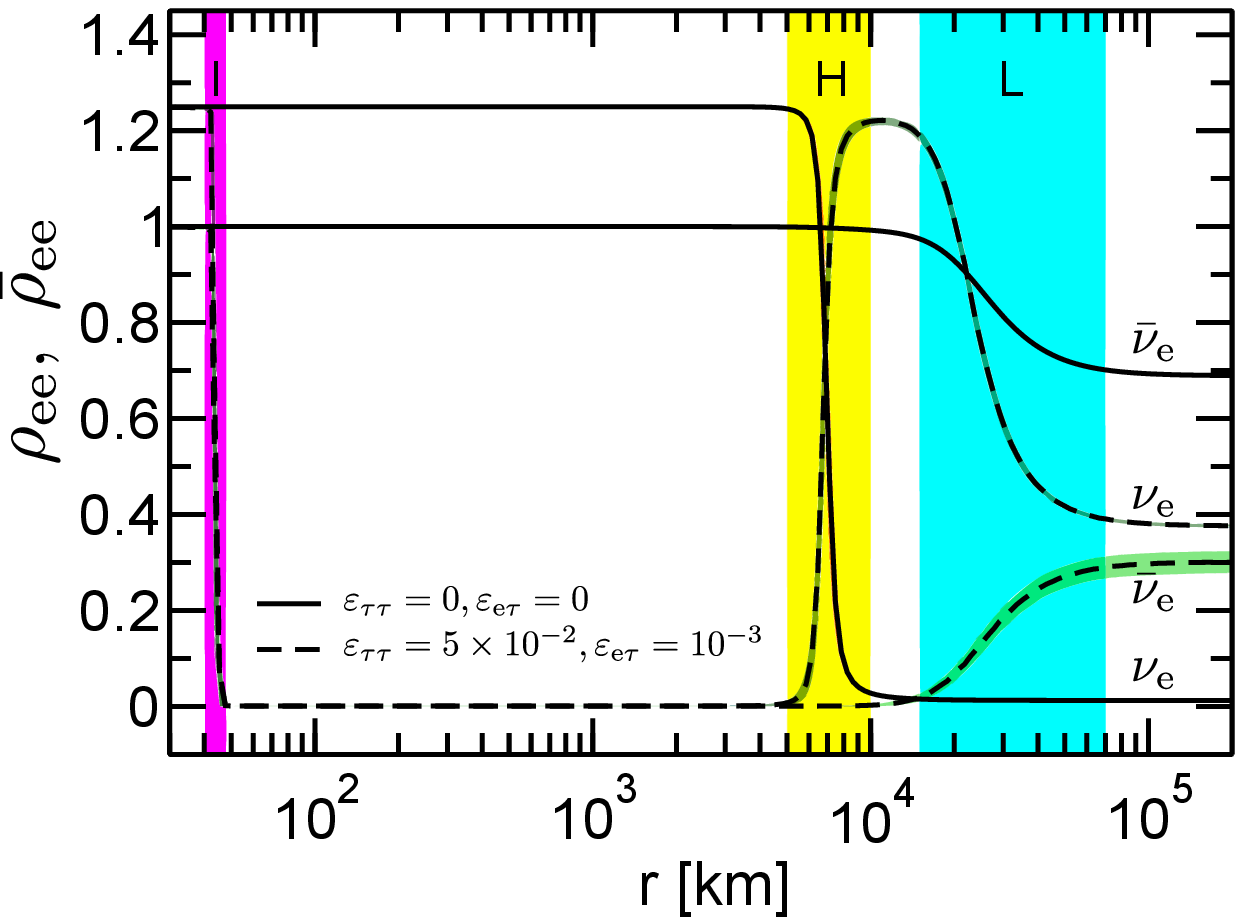}
\includegraphics[angle=0,width=0.45\textwidth]{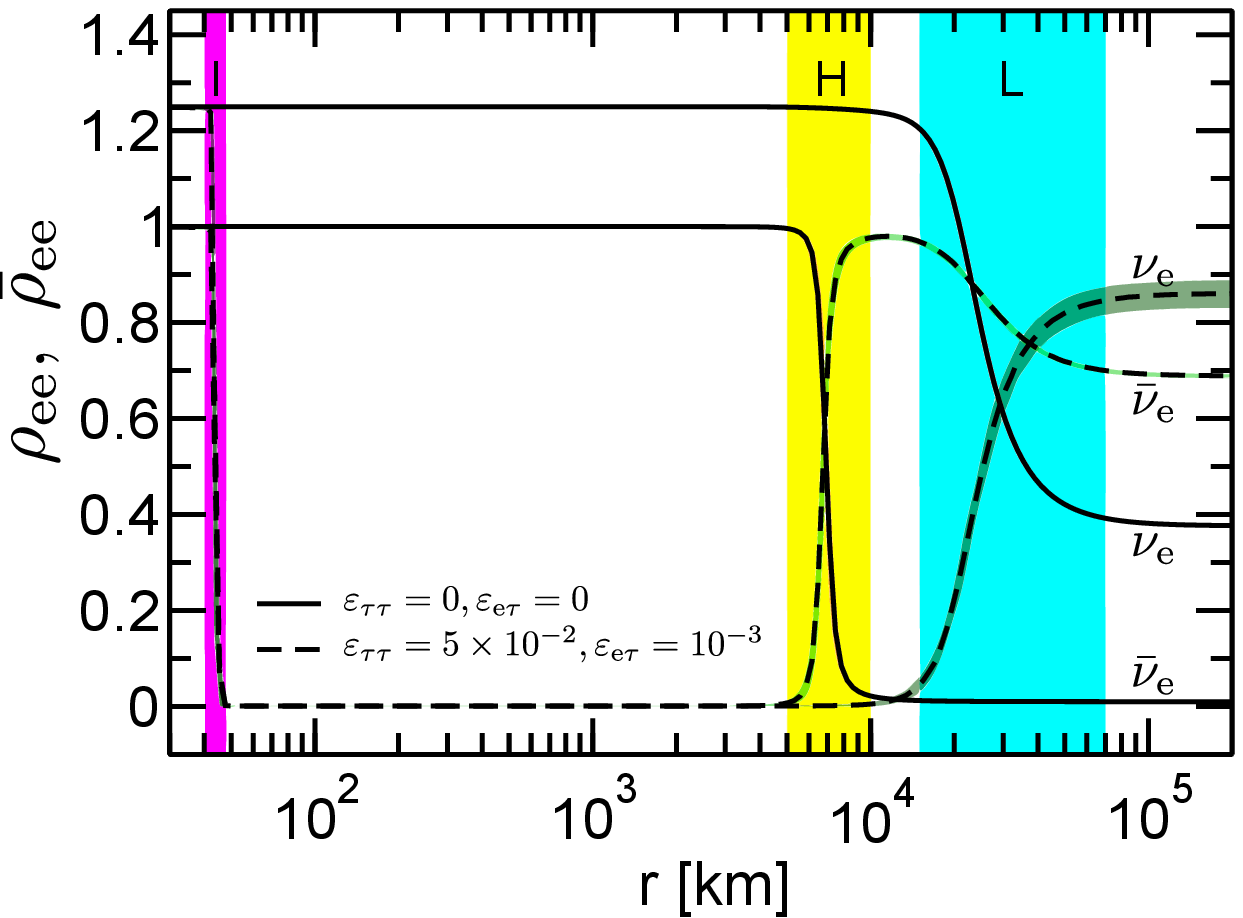}
\caption{\small Radial dependence of $\rho_{ee}$ and $\bar\rho_{ee}$
  corresponding to regions I (solid) and II (dashed) in
  Fig.~\ref{fig:regions}.  Left panel represents normal mass hierarchy
  and right panel inverted mass hierarchy. We assume
  $\lambda_0=10^8$~km$^{-1}$, $\omega_{\rm H}=0.3$~km$^{-1}$, and
  $\sin^2\theta_{23}=0.5$. Vertical bands indicate regions where
  resonances take place~\cite{EstebanPretel:2009is}. }
\label{fig:rho_regionI-II}
\end{center}
\end{figure}

In Fig.~\ref{fig:levcros_I-noI} we show the well known level crossing
schemes for normal (top), and inverted mass hierarchy for
$\sin^2\theta_{23} < \pi/4 $ (middle), and $\sin^2\theta_{23} > \pi/4$
(bottom), with (left column) and without (right column) $I$-resonance,
where we have now added arrows representing the transitions caused by
the collective effects. The arrows must therefore be ignored when
these are not present. In the normal hierarchy case $\nu_e$ and
$\bar\nu_e$ leave the SN as $\nu_3$ and $\bar\nu_1$, whereas for
inverted mass hierarchy they escape as $\nu_2$ and $\bar\nu_3$ for any
octant. The survival probabilities can then be written as
$P(\nu_e\rightarrow\nu_e)\approx
\sin^2\theta_{13}~(\sin^2\theta_{12})$ and
$P(\bar\nu_e\rightarrow\bar\nu_e)=
\cos^2\theta_{12}~(\sin^2\theta_{13})$ for normal (inverted) mass
hierarchy.
Figure~\ref{fig:rho_regionI-II} represents in solid lines the radial
evolution of $\rho_{ee}$ and $\bar\rho_{ee}$ assuming
$\lambda_0=10^8$~km$^{-1}$, $\omega_{\rm H}=0.3$~km$^{-1}$,
$\sin^2\theta_{23}=0.5$, and $\epsilon=0.25$. The vertical bands
indicate where the resonance conversions take place. In order to
perform the plot we have artificially set $\mu_0=0$.  We want to
recall here that both $\rho_{ee}$ and $\bar\rho_{ee}$ are normalized
to the $\bar\nu_e$ flux, and therefore, while $\bar\rho_{ee}$
corresponds directly to $\bar\nu_e$ survival probability, $\rho_{ee}$
must be corrected by a factor $(1+\epsilon)$ in order to obtain the
corresponding survival probability,
$\rho_{ee}=P(\nu_e\rightarrow\nu_e)(1+\epsilon)$. This region,
therefore, corresponds to the physics of oscillation discussed in
Chapter~\ref{chapter:oscillations}.\vspace{0.5cm}\\
\textbf{B) Region II}\vspace{0.2cm}\\
The region II, on the upper right corner, is defined by $\lambda_0
\gtrsim 1.4\times 10^7$~km$^{-1}$ and $\varepsilon_{\tau\tau}\gtrsim
10^{-2}$. As in I the matter density is so high that prevents
neutrinos from undergoing collective effects. However, the values of
the diagonal NSI terms in this region are large enough to fulfill
Eq.~(\ref{eq:Ye_resI}), causing the $I$-resonance to appear.
In contrast to the previous case $\nu_e$ and $\bar\nu_e$ are now
created as $\nu_2^{\rm m}$ ($\nu_1^{\rm m}$) and $\bar\nu_2^{\rm m}$
($\bar\nu_1^{\rm m}$) for normal (inverted) mass hierarchy, cross
adiabatically all resonances and leave the SN as $\nu_2$ ($\nu_1$) and
$\bar\nu_2$ ($\bar\nu_1$) for normal (inverted) mass
hierarchy~\cite{EstebanPretel:2007yu}. The survival probabilities are
now $P(\nu_e\rightarrow\nu_e)=P(\bar\nu_e\rightarrow\bar\nu_e)\approx
\sin^2\theta_{12}~(\cos^2\theta_{12})$ for normal (inverted) mass
hierarchy.
The black dashed lines in Fig.~\ref{fig:rho_regionI-II} show
the expected radial evolution of $\rho_{ee}$ and $\bar\rho_{ee}$,
respectively, when neutrinos and antineutrinos undergo an adiabatic
$I$-resonance. The green band represents the presence of
phases.
As for region I, we have made the calculation assuming
$\mu_0=0$. However, we have analyzed the single energy and multiangle
case within two-flavor framework for the range of parameters here
discussed, and have verified that collective effects are indeed
suppressed and the $I$-resonance is present for both normal and
inverted hierarchies. That means that the behavior in region II
corresponds indeed to the case discussed in Sec.~\ref{sec:no-background}. \vspace{0.5cm}\\
\textbf{C) Region III}\vspace{0.2cm}\\
Let us now consider the lower part of Fig.~\ref{fig:regions},
i.e.~when $\lambda_0\lesssim 1.4\times 10^7$~km$^{-1}$. The main
characteristic of this scenario is the presence of collective
effects. As it was discussed in Ref.~\cite{EstebanPretel:2007yq},
and here reviewed, these in turn depend on the relative position
of the $\mu\tau$-resonance with respect to the synchronization and
bipolar radius. We can then distinguish two different regimes:
On the bottom left corner we define the region III by 
the condition $r_{\mu\tau}\lesssim r_{\rm syn}$, and on the bottom
right corner we have region IV defined by $r_{\mu\tau}\gtrsim r_{\rm
  bip}$. In the middle of both there is a transition region where
$r_{\rm bip}\gtrsim r_{\mu\tau}\gtrsim r_{\rm syn}$, which we will not
consider here.

\begin{figure}[t]
\begin{center}
\includegraphics[angle=0,width=0.45\textwidth]{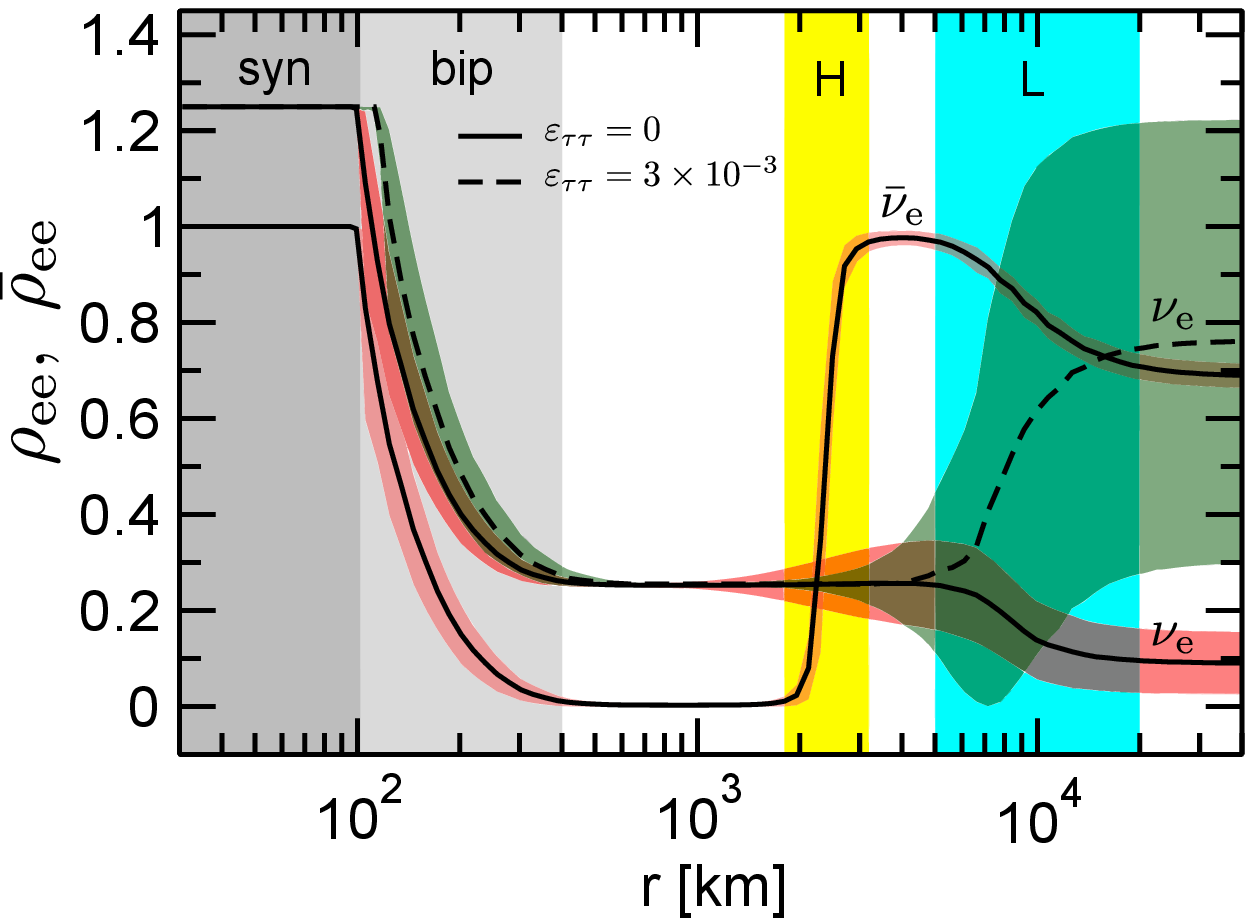}
\includegraphics[angle=0,width=0.45\textwidth]{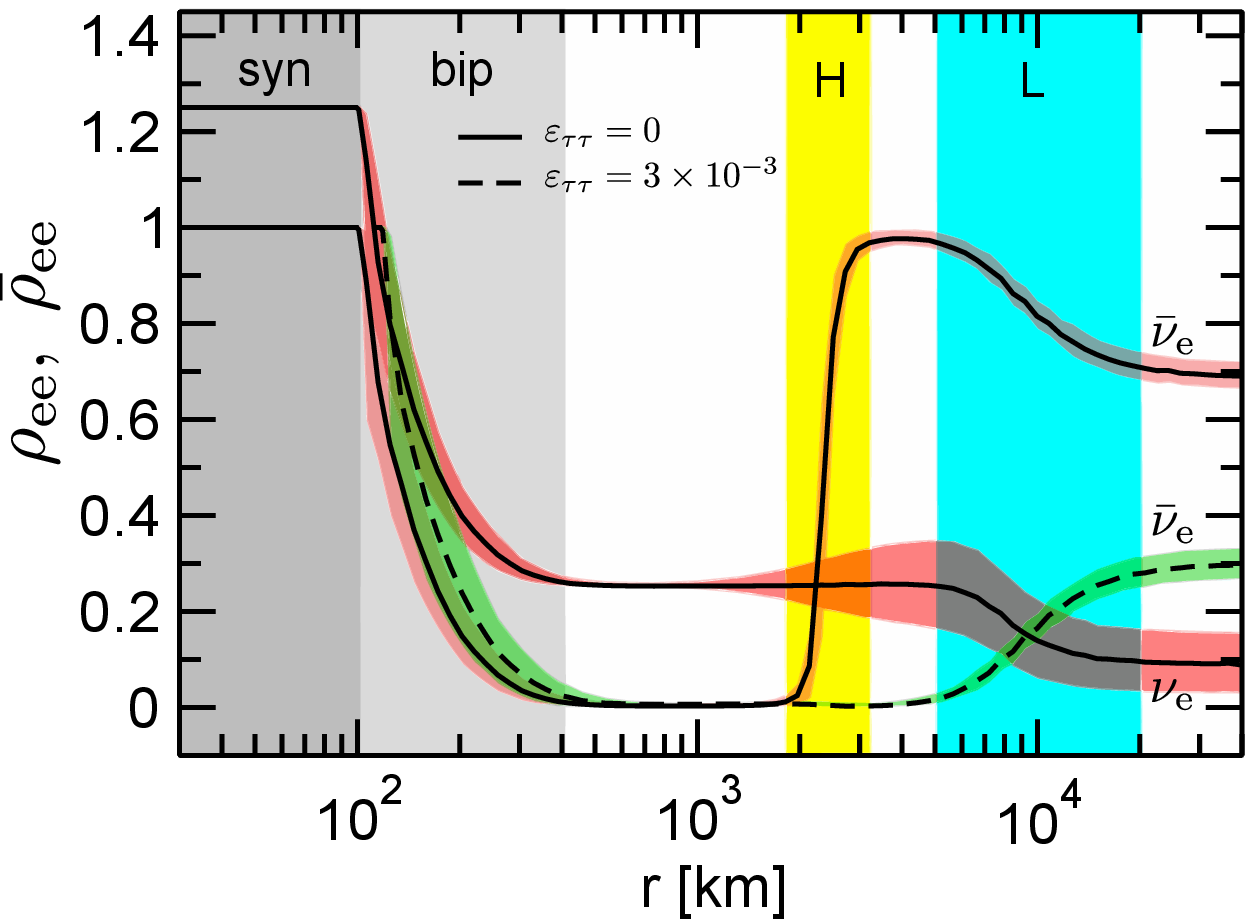}
\caption{\small Radial dependence of $\rho_{ee}$ and $\bar\rho_{ee}$
  for region III with $\varepsilon_{\tau\tau}=0$ (solid) and IVa
  (dashed) with $\varepsilon_{\tau\tau}= 3\times 10^{-3}$ for inverted
  mass hierarchy, and $\sin^2\theta_{23}=0.4~(0.6)$ in the left (right
  panel). In both cases $\lambda_0=4\times 10^6$~km$^{-1}$ and
  $\varepsilon_{e\tau}=0$. The bands around the lines represent
  modulations. Vertical gray bands stand for synchronized (dark) and
  bipolar (light) regime. Resonance regions are also
  displayed~\cite{EstebanPretel:2009is}.}
\label{fig:rho_regionIII-IVa}
\end{center}
\end{figure}

Let us now discuss region III. According to Eq.~(\ref{eq:rmutaubip})
this range of parameters satisfies the condition
\begin{equation}
\lambda_0[Y_\tau^{\rm eff} + (2-Y_e)\varepsilon_{\tau\tau}]\lesssim
10^4~{\rm km}^{-1}\,,
\end{equation}
which, for $Y_e=0.5$, roughly amounts to
$\lambda_0\varepsilon_{\tau\tau}\lesssim 7.3\times 10^3$~km$^{-1}$,
see Fig.~\ref{fig:regions}.  This situation can be reduced to the
standard two-flavor scenario previously analyzed in this thesis. In
order to better understand the consequences of collective effects it
is convenient to recall that the impact of ordinary matter can be
transformed away by going into a rotating reference frame for the
polarization vectors. Collective conversions proceed in the same way
as they would in vacuum, except that the effective mixing angle is
reduced.  The connection between flavor $\nu_\alpha$ and vacuum
$\nu_i$ eigenstates can be done in the level crossing schemes by
propagating the former ones from regions at high density to vacuum
crossing all resonances non-adiabatically.  The initial states $\nu_e$
and $\bar\nu_e$ can therefore be identified with $\nu_1$ and
$\bar\nu_1$, respectively.  If the neutrino mass hierarchy is normal,
we begin in the lowest-lying state and nothing happens.  The situation
is then similar to that in region I, i.e.~without collective effects,
see left panel of Fig.~\ref{fig:rho_regionI-II}.
However in the case of inverted mass hierarchy both $\nu_1$ and
$\bar\nu_1$ correspond to the intermediate state. The effect of
the self-interaction is to drive them to the lowest-lying states, which
in this case are $\nu_3$ and $\bar\nu_3$. This is shown by dashed
arrows in the middle and bottom left panels of
Fig.~\ref{fig:levcros_I-noI}. 
In the case of $\nu_e$ a fraction equal to $\epsilon F_{\bar\nu_e}$ is
not transformed and stays in $\nu_2^{\rm m}$ and evolves as in the
absence of neutrino-neutrino interactions, i.e.~adiabatically through
the $L$-resonance. The rest of $\nu_e$ are transformed to $\nu_3^{\rm
  m}$. As a consequence, the final $\nu_e$ flux, normalized to the
initial $\bar\nu_e$ one, is expected to be approximately $\rho^{\rm
  final}_{ee}=\epsilon \sin^2\theta_{12}+\sin^2\theta_{13}\simeq
0.08$. On the other hand, after the pair transformation $\bar\nu_e$
cross the $H$-resonance adiabatically and leave the star as $\nu_1$,
leading to a final normalized flux of approximately $\bar\rho^{\rm
  final}_{ee}=\cos^2\theta_{12}\simeq 0.7$.
This can be seen in Fig.~\ref{fig:rho_regionIII-IVa}, where we show in
solid lines the radial evolution of $\nu_e$ and $\bar\nu_e$ for
inverted mass hierarchy assuming
$\lambda_0=4\times 10^6$~km$^{-1}$, $\omega_{\rm H}=0.3$~km$^{-1}$,
and $\sin^2\theta_{23}=0.4$ (left) and 0.6 (right).
As can be seen in the figure, the result is independent of the
$\theta_{23}$ octant. The evolution of neutrinos in this region of
parameters therefore corresponds to the description given in
Chapters~\ref{chapter:coll2flavors} and \ref{chapter:coll3flavors}.\vspace{0.5cm}\\
\textbf{D) Region IV}\vspace{0.2cm}\\
Finally, neutrinos with parameters in the right bottom corner (region
IV) will feel both collective and NSI effects. This region of
parameters is defined by the condition that the $\mu\tau$-resonance
lies outside the bipolar region. According to Eq.~(\ref{eq:rmutaubip})
this amounts to 
\begin{equation}
\lambda_0[Y_\tau^{\rm eff} + (2-Y_e)\varepsilon_{\tau\tau}]\gtrsim
2\times 10^4~{\rm km}^{-1}\,.
\end{equation}
As discussed above, for the standard case and $Y_e=0.5$ this is
satisfied for $\lambda_0\gtrsim 7\times 10^8$~km$^{-1}$, which implies
a strong matter suppression of the collective effects, see
Fig.~\ref{fig:regions}. However, if NSI diagonal parameters are of the
order of $|\varepsilon_{\tau\tau}|\gtrsim 5.3\times 10^3/\lambda_0~({\rm
  km}^{-1})$ then one can avoid the matter suppression condition.
Therefore the first NSI effect is to increase the value of
$Y_{\tau,{\rm nsi}}^{\rm eff}$ so that the $\lambda_0$ required to
have the $\mu\tau$-resonance outside $r_{\rm bip}$ is still consistent
with the presence of collective effects.
On top of that, if $\varepsilon_{\tau\tau}$ is of the order of a few
\% the condition given in Eq.~(\ref{eq:Ye_resI}) is fulfilled for the
typical values of $Y_e$ found in SNe. Thus, in the region IV we can
distinguish two subsets of parameters denoted by IVa and IVb defined
by the absence or presence of the $I$-resonance, respectively. For
defineteness we set the boundary at $\varepsilon_{\tau\tau}=10^{-2}$.

Let us first consider the IVa region. Depending on $\lambda_0$,
i.e.~on the instant considered, this range of parameters implies
values of $|\varepsilon_{\tau\tau}|$ from $10^{-3}$ to
$\sim 10^{-2}$.  Although these values are not high enough to induce
the $I$-resonance they are sufficiently large to push the
$\mu\tau$-resonance outside the bipolar region.  The situation is
therefore analogous to the one described in
Chapter~\ref{chapter:coll3flavors}. That means a flavor pair
transformation $\nu_e\bar\nu_e\rightarrow \nu_x\bar\nu_x$ due to
collective effects only for inverted neutrino mass hierarchy, like in
region III.  However, the final matter eigenstates depend on the
$\theta_{23}$ octant. In the middle left panel of
Fig.~\ref{fig:levcros_I-noI} we show with solid lines the pair
conversion for $\theta_{23}$ in the first octant. In terms of matter
eigenstates, $\nu_e$ and $\bar\nu_e$ are transformed into $\nu_1^{\rm
  m}$ and $\bar\nu_1^{\rm m}$, respectively. The presence of the
$\mu\tau$-resonance in the neutrino channel leads to a difference of
$\nu_e$ with respect to region III. In the left panel of
Fig.~\ref{fig:rho_regionIII-IVa} we show with dashed lines the
evolution of $\nu_e$ as function of the distance, for
$\varepsilon_{\tau\tau}=3\times 10^{-3}$ and $\sin^2\theta_{23}=0.4$.
In the collective bipolar conversions, the excess $\epsilon$ of
$\nu_e$ over $\bar\nu_e$ remains as $\nu_2^{\rm m}$ whereas the rest
will be transformed to $\nu_1^{\rm m}$. As a consequence, the original
$\nu_e$ flux leaving the star can be written as $\rho^{\rm
  final}_{ee}=\varepsilon\sin^2\theta_{12}+\cos^2\theta_{12}$, which
in our particular case amounts to roughly 0.75.
If $\theta_{23}$ belongs to the second octant the $\mu\tau$-resonance
takes place in the antineutrino channel, see bottom left panel of
Fig.~\ref{fig:levcros_I-noI}. The pair $\nu_e$ and $\bar\nu_e$ is driven
to the lowest-lying states, which in this case are $\nu_3^{\rm m}$ and
$\bar\nu_2^{\rm m}$, for neutrinos and antineutrinos,
respectively. Therefore, for $\nu_e$ the situation is completely
analogous to that in region III, whereas $\bar\nu_e$ leave the star as
$\bar\nu_2$. In the right panel of Fig.~\ref{fig:rho_regionIII-IVa}
it is displayed with dashed lines the radial evolution of $\bar\nu_e$
for $\varepsilon_{\tau\tau}=3\times 10^{-3}$ and
$\sin^2\theta_{23}=0.6$.  It is remarkable that neither $\nu_e$ nor
$\bar\nu_e$ undergo the H resonance, and therefore are blind to the
possible effect of the outwards propagating shock
wave~\cite{Schirato:2002tg,Fogli:2003dw,Tomas:2004gr}.

It is important to notice that the same effect observed for the
different octants of $\theta_{23}$ can be obtained by fixing the
octant and changing the sign of $\varepsilon_{\tau\tau}$. This
  can be easily understood if we study the $\mu\tau$-resonance
  condition,
\begin{equation}
\label{eq:mutau_rescond}
\lambda(r)[Y^{\rm
    eff}_{\tau}+(2-Y_e)\varepsilon_{\tau\tau}] \simeq -\omega_{\rm H}
  \cos^2{\theta_{13}}\cos{2\theta_{23}}\,, 
\end{equation}
where we have neglected subleading solar terms. This condition
dictates the channel where the resonance takes place. In the
standard case this only determined by the hierarchy of neutrino masses
and the octant of $\theta_{23}$. In the presence of NSI, though, the
sign of left-hand side of the equation depends on that of
$\varepsilon_{\tau\tau}$, what therefore affects directly to the
resonance condition. As a consequence, the same result of
Fig.~\ref{fig:rho_regionIII-IVa} is obtained by changing the sign of
$\varepsilon_{\tau\tau}$ and the octant of $\theta_{23}$,
i.e.~$\varepsilon_{\tau\tau}=-3\times 10^{-3}$ and second octant for
the panel on the left and first octant in the right panel.
The bottom line is that, in the presence of NSI parameters such that
$|\varepsilon_{\tau\tau}|\gtrsim$ a few$\times 10^{-4}$, SN neutrinos
are sensitive to the octant of $\theta_{23}$, the absolute value of
the NSI diagonal parameter as well as its sign. In this region, the
propagation of neutrinos is analogous to the one described in
Chapter~\ref{chapter:coll3flavors}.

Finally, for higher values of the NSI diagonal parameters,
$\varepsilon_{\tau\tau}\gtrsim 10^{-2}$ (region IVb), the internal
$I$-resonance will arise. In this case one has to analyze the
interplay between collective effects and the $I$-resonance. This is
discussed in the next section.

\subsection{Collective effects and NSI-induced $I$-resonance}
\label{sec:nsi-col}

In this section we analyze region IVb, defined by $\lambda_0\lesssim
1.4\times 10^7$~km$^{-1}$ and $\varepsilon_{\tau\tau}\gtrsim 10^{-2}$,
where both an adiabatic $I$-resonance and collective effects are
present. If the $I$-resonance is not adiabatic then neutrinos within
region IVb evolve exactly as in IVa.

As can be inferred from Fig.~\ref{fig:profiles_ch6}, one of the main
features of this scenario is that both effects happen nearly in the
same region, namely the deepest layers right above the neutrino sphere.
That means that the final result will also depend on the relative
position between the bipolar region and the location of the
$I$-resonance. Schematically two extreme scenarios can be
identified. In one case the rise  in the $Y_e$, and consequently
the $I$-resonance, takes place before the bipolar conversion region,
see $Y_e^{\rm a}$ in the bottom panel of Fig.~\ref{fig:profiles_ch6}.
In the second scenario one has first the bipolar conversion and then
neutrinos traverse the $I$-resonance, see $Y_e^{\rm b}$ in the bottom
panel of Fig.~\ref{fig:profiles_ch6}.\vspace{1.0cm}\\
\textbf{A) First NSI $I$-resonance}\vspace{0.2cm}\\
Let us analyze here the case where the $I$-resonance happens in deeper
layers than collective effects. This situation corresponds to the
$Y_e^{\rm a}$ in the bottom panel of Fig.~\ref{fig:profiles_ch6} and,
according to SN numerical simulations it is the most likely situation.

The main consequence of an $I$-resonance in the most inner layers,
right after the neutrino sphere, is an inversion of the neutrino
fluxes entering the bipolar region. For the initial flux pattern
assumed this implies the following new pattern after the
$I$-resonance: $F_{\nu_e} = F_{\bar\nu_e} = 0$,
$F_{\nu_\tau}=1+\epsilon$, and $F_{\bar\nu_\tau}=1$ normalized to
$F_{\bar\nu_e}$. Contrarily to the standard case, under this condition
collective effects arise in the case of normal mass hierarchy.

\begin{figure}[t]
\begin{center}
\includegraphics[angle=0,width=0.45\textwidth]{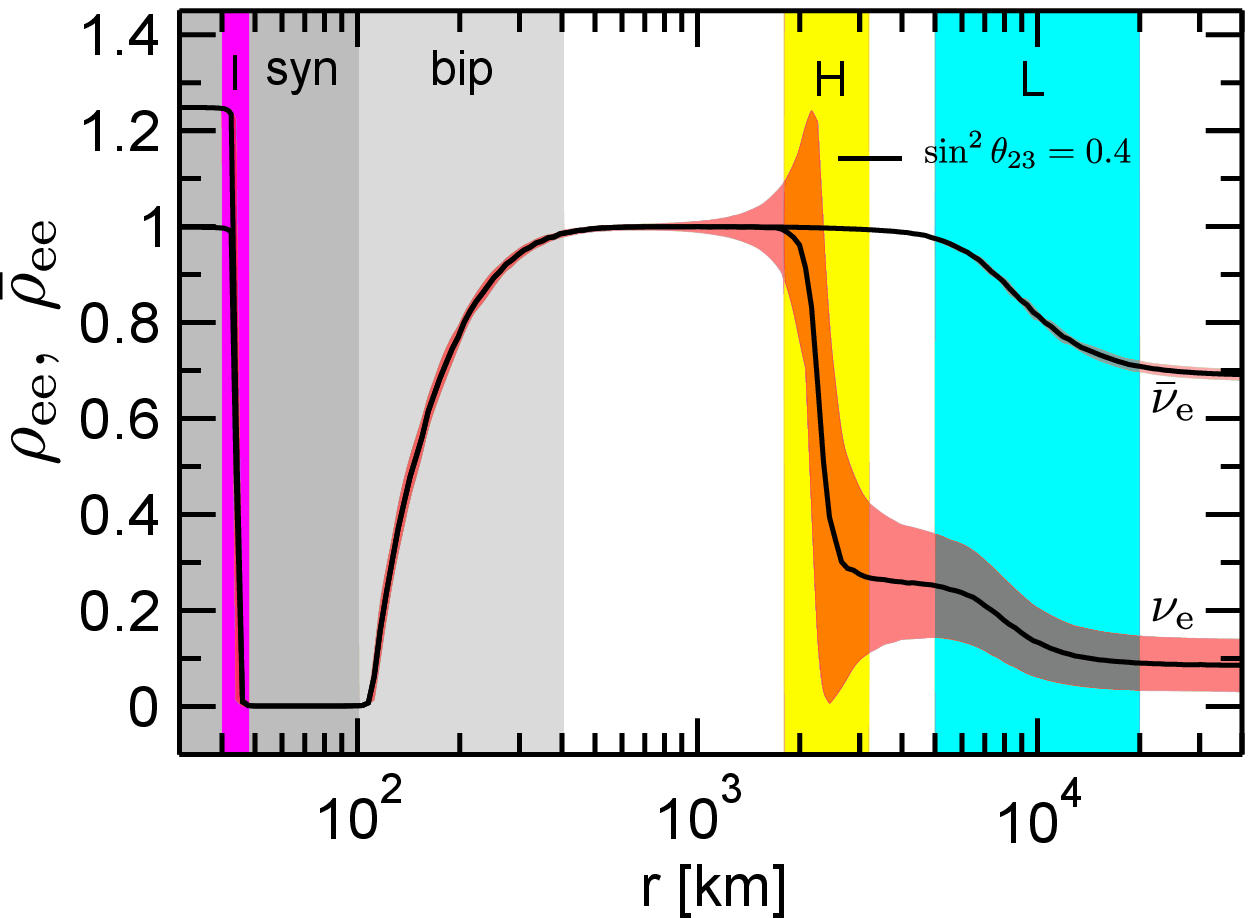}
\includegraphics[angle=0,width=0.45\textwidth]{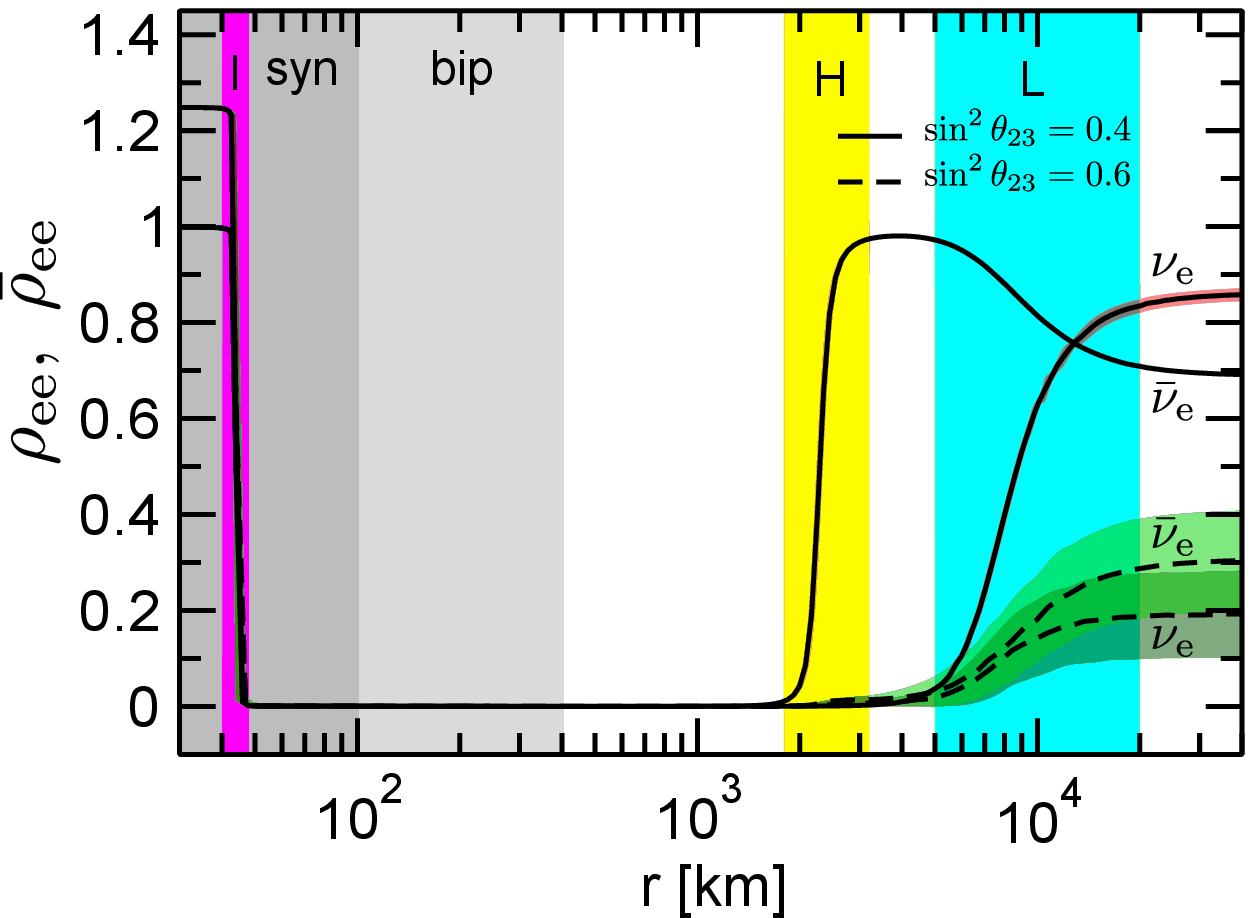}
\caption{\small Same as Fig.~\ref{fig:rho_regionI-II} for region IVb,
  for normal (left panel) and inverted (right panel) mass
  hierarchy. In the left panel it is shown the case
  $\sin^2\theta_{23}=0.4$, while in the right panel the case
  $\sin^2\theta_{23}=0.6$ is also displayed.  In both cases
  $\lambda_0=4\times 10^6$~km$^{-1}$ and
  $\varepsilon_{\tau\tau}=5\times 10^{-2}$ and
  $\varepsilon_{e\tau}=10^{-3}$. The $I$-resonance is assumed to occur
  before the collective effects~\cite{EstebanPretel:2009is}.}
\label{fig:rho_regionIVb}
\end{center}
\end{figure}

This can be understood using the pendulum analogy in the corresponding
reduced two flavor scenario, and keeping in mind that the bipolar
conversion drives the neutrinos to the lowest-lying states.
In the normal hierarchy the system is already created near the minimum
of the potential. Thus, in absence of the $I$-resonance collective
effects are not present. However once an adiabatic $I$-resonance is
switched on the neutrino flavor is swapped and the system is driven to
the maximum of the potential. In this situation bipolar effects act
leading $\nu_e$ and $\bar\nu_e$ back to the lowest-lying states,
i.e.~to $\nu_3^{\rm m}$ and $\bar\nu_1^{\rm m}$, respectively. See
dashed arrows in the top right panel of Fig.~\ref{fig:levcros_I-noI}.
As a consequence both the $I$-resonance and the induced collective
effects basically cancel each other. In the left panel of
Fig.~\ref{fig:rho_regionIVb} we show the radial evolution of $\nu_e$
and $\bar\nu_e$. This cancellation between the $I$-resonance and
collective effects is complete for $\bar\nu_e$, which leave as
$\bar\nu_1^{\rm m}$, but not for $\nu_e$: its excess $\epsilon$ over
$\bar\nu_e$ is not transformed back to $\nu_3^{\rm m}$ but remains as
$\nu_2^{\rm m}$. Therefore in the case of a monochromatic neutrino
flux we obtain $\rho^{\rm
  final}_{ee}=\epsilon\sin^2\theta_{12}+\sin^2\theta_{13}\approx
0.08$, see left panel of Fig.~\ref{fig:rho_regionIVb}, instead of
simply $\sin^2\theta_{13}=10^{-2}$ as in left panel of
Fig.~\ref{fig:rho_regionI-II}.  This result does not depend on the
$\theta_{23}$ octant since the collective effects bring the
$\nu_e\bar\nu_e$ pair to $\nu_3^{\rm m}\bar\nu_1^{\rm m}$ in both
cases. Hence, by comparing the left panels of
Figs.~\ref{fig:rho_regionI-II} and~\ref{fig:rho_regionIVb} one
realizes that, except for the excess $\epsilon$, the situation for
normal mass hierarchy is basically the same as in regions I, III, and
IVa.

\begin{figure}
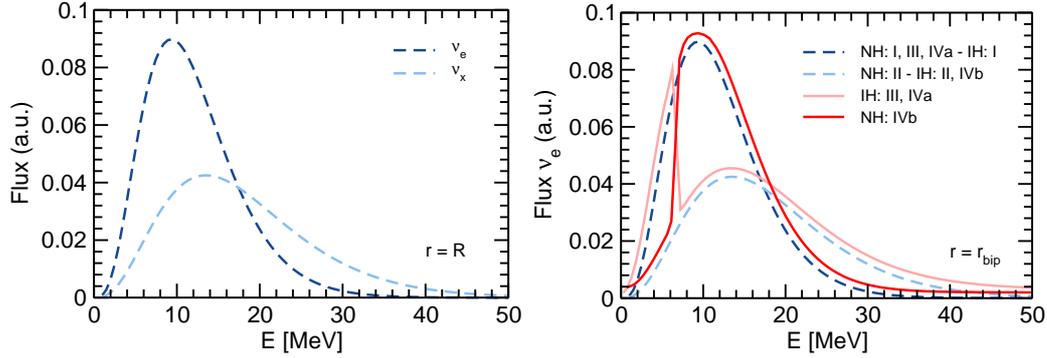

\begin{center}
\includegraphics[angle=0,width=0.45\textwidth]{./cap6/figures/sn_nsi_col/KE_inicial.eps}
\includegraphics[angle=0,width=0.45\textwidth]{./cap6/figures/sn_nsi_col/KE_final.eps}
\caption{\small Left: $\nu_e$ and $\nu_x$ fluxes as emitted at the
  neutrino sphere. Right: $\nu_e$ flux after the bipolar region for
  different cases. In the region IVb it is assumed that the
  $I$-resonance occurs before than the bipolar
  conversion~\cite{EstebanPretel:2009is}.  }
\label{fig:spectralsplit}
\end{center}
\end{figure}

While this is true in the monoenergetic case
a specific signature can be observed if we do not restrict
ourselves to that case but consider the whole energy
spectrum. The left panel of Fig.~\ref{fig:spectralsplit} displays the 
$\nu_e$ and $\nu_x$ fluxes at the neutrino sphere, $f^{\rm R}_{\nu_e}$
and $f^{\rm R}_{\nu_x}$. We have assumed the parameterization given in
Ref.~\cite{Keil:2002in}, 
\begin{eqnarray}
  f^{R_\nu}_{\nu_\alpha}(E) = C_{\nu_\alpha}
  \left(\frac{E}{\langle{E_{\nu_\alpha}}\rangle}\right)^{\beta_{\nu_\alpha}-1} 
  \exp\left(-\beta_{\nu_\alpha}\frac{E}{\langle{E_{\nu_\alpha}}\rangle}\right) \,,
\label{eq:flux-Gal}
\end{eqnarray}
with $\langle E_{\nu_e}\rangle = 12$~MeV, $\langle
E_{\nu_e}\rangle=15$~MeV, $\langle E_{\nu_x}\rangle=18$~MeV,
$\beta_{\nu_e}=5,~\beta_{\bar\nu_e}=4.5$ and $\beta_{\nu_x}=4$. The
normalization $C_{\nu_\alpha}$ has been chosen such that $F^{R_\nu}_{\bar\nu_e}\equiv \int f^{R_\nu}_{\bar\nu_e}(E){\rm d}E=1$,  $F^{R_\nu}_{\nu_e} = 1+\kappa\epsilon$ and $F^{R_\nu}_{\nu_x} = 1-\kappa$,
with $\kappa=0.15$.

As in standard case for inverted mass hierarchy, the excess $\epsilon$
of $\nu_e$ translates into a spectral split. However, in contrast to
the standard case this excess concentrates at high energies. In the
right panel of Fig.~\ref{fig:spectralsplit} we show the $\nu_e$ fluxes
after the bipolar region. In solid dark red lines is represented the
case under discussion: normal mass hierarchy in region IVb. By
comparing the two panels one sees how the conversion
$\nu_e\rightarrow\nu_x$ takes place only at low energies. This is
exactly the contrary to what occurs for the standard case (inverted
mass hierarchy in region III), shown as solid light red lines, where
the untransformed flux concentrates at low energies.
For completeness we show also the other cases. In the region of
parameters I, III and IVa there is neither collective effects nor
$I$-resonance for normal hierarchy, then the fluxes after the
bipolar region coincide with the initial ones, $f_{\nu_e}=f_{\nu_e}^{R_\nu}$
and $f_{\nu_x}=f_{\nu_x}^{R_\nu}$. In region II, the $I$-resonance implies a
complete conversion $\nu_e\rightarrow \nu_x$, what leads to a spectral
swap, $f_{\nu_e}=f_{\nu_x}^{R_\nu}$ and $f_{\nu_x}=f_{\nu_e}^{R_\nu}$. 

The case of inverted mass hierarchy is more subtle. According to the
previous discussion one would expect no collective effects after
neutrinos traverse the $I$-resonance.  The system starts its evolution
near the maximum of the potential and, in the absence of NSI, the bipolar
conversions would take it to the minimum. What the $I$-resonance is
doing in this language by swapping the flavor eigenstates is to take
the system to the minimum of the potential before any collective
effects can arise. The new stable situation prevents bipolar
conversions, leaving the system unchanged until the outer resonances
are reached. 
And this is indeed what happens if $\theta_{23}$ lies in the first
octant, see middle right panel of Fig.~\ref{fig:levcros_I-noI}. After
the $I$-resonance the original $\nu_e$ and $\bar\nu_e$ are already in
the state of ``minimum energy'', which in this case corresponds to
$\nu_1^{\rm m}$ and $\bar\nu_1^{\rm m}$ ($\nu_3$ and $\bar\nu_3$ after
rotating the matter term away).  Hence no collective effects take
place, and $\nu_e$ and $\bar\nu_e$ leave the star as $\nu_1$ and
$\bar\nu_1$, respectively, see solid lines in the right panel of
Fig.~\ref{fig:rho_regionIVb}.

However if $\theta_{23}$ belongs to the second octant things are
different. The $I$-resonance drives now $\nu_e$ and $\bar\nu_e$ to the
states of ``maximum energy'', $\nu_2$ and $\bar\nu_2$ after rotating
the matter term away, see bottom right panel of
Fig.~\ref{fig:levcros_I-noI}. That means, that in contrast to the
first-octant case, when the neutrinos traverse the bipolar regime they
will be driven to the lowest-lying states,
i.e.~$\nu_2\bar\nu_2\rightarrow\nu_3\bar\nu_3$, see dashed arrows in
the figure. In terms of matter eigenstates that means a pair
conversion from $\nu_1^{\rm m}\bar\nu_1^{\rm m}$ into $\nu_3^{\rm
  m}\bar\nu_2^{\rm m}$.  In the right panel of
Fig.~\ref{fig:rho_regionIVb} we show in dashed lines the evolution of
$\nu_e$ and $\bar\nu_e$ as function of distance for
$\sin^2\theta_{23}=0.6$. The bipolar conversion can not be seen as it
occurs between $\nu_1^{\rm m}\bar\nu_1^{\rm m}$ and $\nu_3^{\rm
  m}\bar\nu_2^{\rm m}$, while $\nu_e$ and $\bar\nu_e$ coincide in that
region with $\nu_3^{\rm m}$ and $\bar\nu_2^{\rm m}$, respectively. At
the end, $\bar\nu_e$ leave the SN as $\bar\nu_2$, see right panel of
Fig.~\ref{fig:rho_regionIVb}. In the case of $\nu_e$ the excess
$\epsilon$ over $\bar\nu_e$ remains as $\nu_1^{\rm m}$ whereas the
rest is transformed to $\nu_3^{\rm m}$. Therefore for a monoenergetic
flux we find $\rho^{\rm final}_{ee} = \epsilon
\cos^2\theta_{12}+\sin^2\theta_{13}\approx 0.2$. By comparing the
right panel of Figs.~\ref{fig:rho_regionIII-IVa}
and~\ref{fig:rho_regionIVb} one realizes that this case is analogous
to IVa.
Since the collective effects do not affect $\nu_e$ and $\bar\nu_e$
directly, considering neutrinos with an energy spectrum, one
expects simply a complete swap of spectra, $f_{\nu_e}=f_{\nu_x}^R$ and
$f_{\nu_x}=f_{\nu_e}^R$, like in scenario II, see right panel of
Fig.~\ref{fig:spectralsplit}.

The final conclusion is that the propagation of neutrinos with
parameters in region IV is practically independent of the presence
(IVb) or not (IVa) of the $I$-resonance. The main consequence of the
$I$-resonance is to remove the spectral split expected in the inverted
mass hierarchy, and create an inverted spectral split for $\nu_e$ for
normal mass hierarchy. \vspace{0.5cm}\\
\textbf{B) First Collective}\vspace{0.2cm}\\
For completeness we have also considered the possibility that the
bipolar conversion takes place before neutrinos traverse the
$I$-resonance.  This situation corresponds schematically to the
$Y_e^{\rm b}$ profile in the bottom panel of Fig.~\ref{fig:profiles_ch6}.

\begin{figure}
\begin{center}
\includegraphics[angle=0,width=0.45\textwidth]{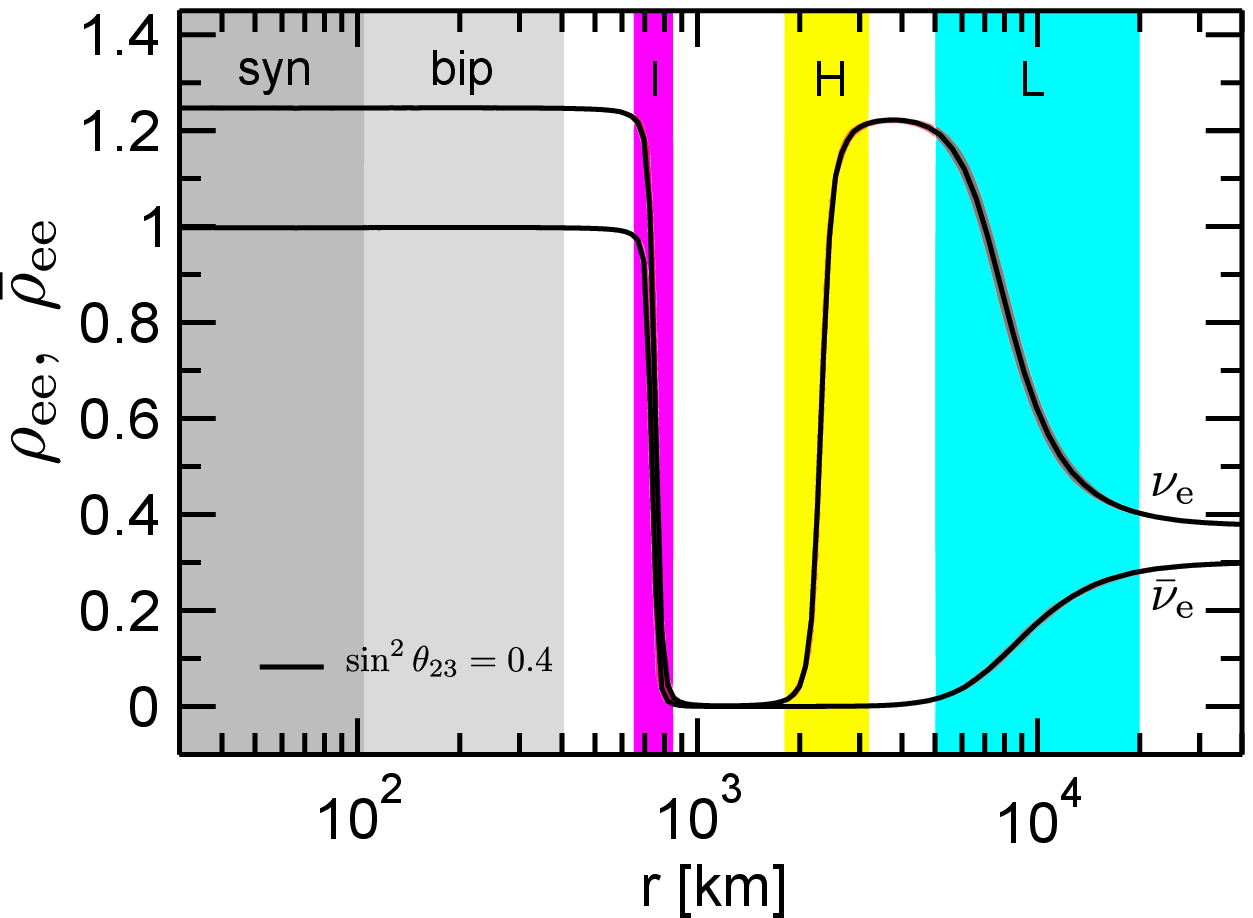}
\includegraphics[angle=0,width=0.45\textwidth]{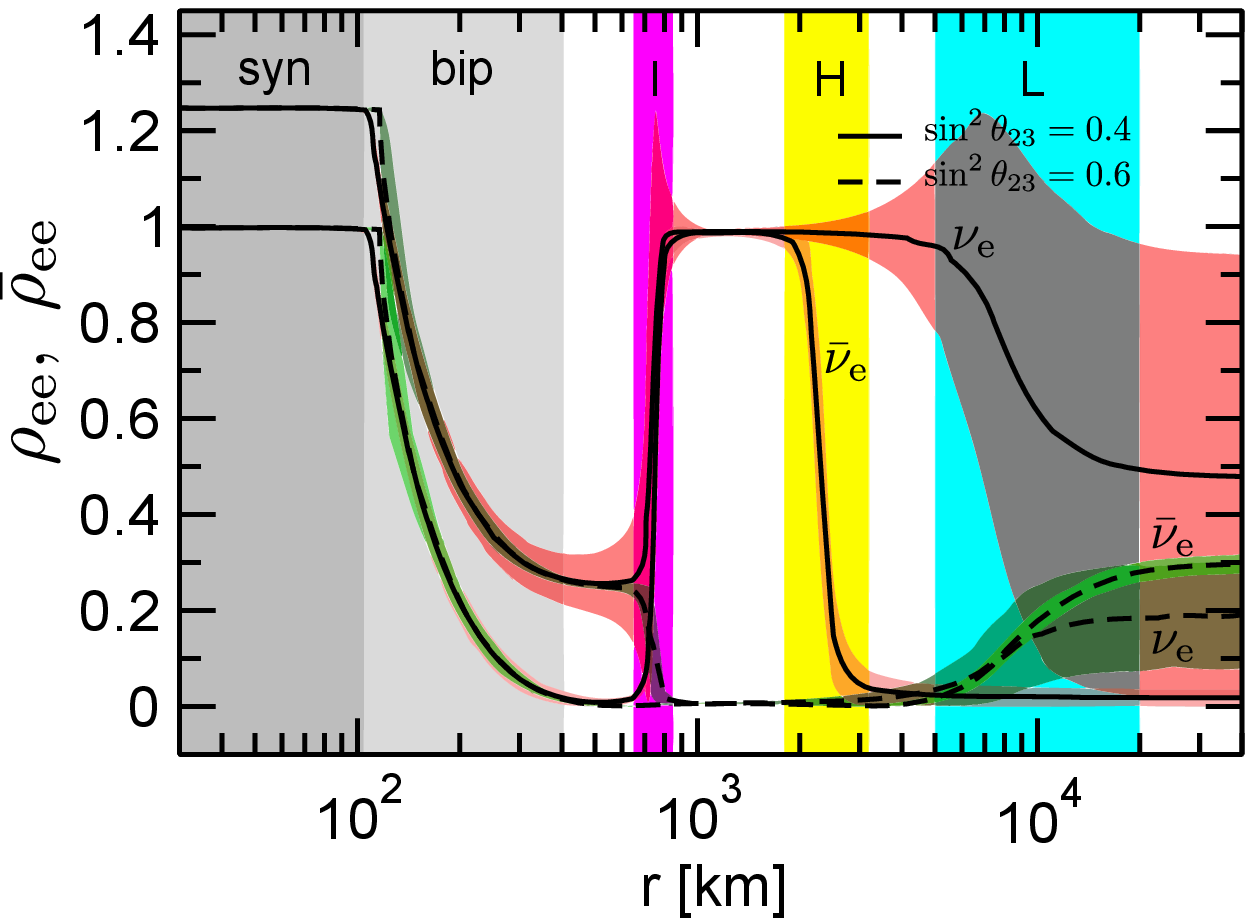}
\caption{\small Same as Fig.~\ref{fig:rho_regionIVb} but assuming that
  the collective effects take place before the
  $I$-resonance~\cite{EstebanPretel:2009is}.}
\label{fig:rho_regionIVbbis}
\end{center}
\end{figure}

The case of normal mass hierarchy is completely analogous to the one
of region II, that is, absence of collective effects and
$I$-resonance. The $\nu_e$ and $\bar\nu_e$ are created as $\nu_2^{\rm
  m}$ and $\bar\nu_2^{\rm m}$, respectively. Therefore, if all
resonances involved, $I,~\mu\tau,~H,$ and $L$, are adiabatic then they
leave the SN as $\nu_2$ and $\bar\nu_2$, respectively, see left panel
of Fig.~\ref{fig:rho_regionIVbbis}. The result is identical to the one
shown with dashed lines in the left panel of
Fig.~\ref{fig:rho_regionI-II}.

The situation with inverted mass hierarchy depends significantly on
the $\theta_{23}$ octant. Rotating the matter term away $\nu_e$ and
$\bar\nu_e$ are created as the intermediate states $\nu_1$ and
$\bar\nu_1$. Collective effects drive them to the lowest-lying states
$\nu_3$ and $\bar\nu_3$. However the corresponding matter eigenstates
are different depending on whether $\theta_{23}$ belongs to the first
or to the second octant. In the first case, most of $\nu_e$ and
$\bar\nu_e$ end up as $\nu_2^{\rm m}$ and $\bar\nu_3^{\rm m}$ before
crossing the $I$-resonance, see solid arrows in the middle right panel
of Fig.~\ref{fig:levcros_I-noI}. The excess $\epsilon$ of $\nu_e$
stays as $\nu_1^{\rm m}$. As a consequence, the final $\nu_e$ and
$\bar\nu_e$ fluxes, normalized to the initial $\bar\nu_e$ one, are
$\rho^{\rm final}_{ee}=\epsilon\cos^2\theta_{12}+\sin^2\theta_{12}$
and $\bar\rho^{\rm final}_{ee}=\sin^2\theta_{13}$, respectively. See
solid lines in the right panel of
Fig.~\ref{fig:rho_regionIVbbis}. Except for the excess $\epsilon$ in
$\nu_e$ the net result is a cancellation of the collective effects and
the $I$-resonance, leading to a similar result as in region I (solid
lines in right panel of Fig.~\ref{fig:rho_regionI-II}).
Qualitatively, the main difference shows up in the $\nu_e$
spectrum. The initial collective effects induces a ``standard''
spectral split, i.e.~spectral swap only at high
energies. Nevertheless, as neutrinos cross the $I$-resonance this
split turns into an inverse one, with a swap at low energies. The
final result right after the $I$-resonance is analogous to the case of
normal mass hierarchy and the $I$-resonance happening first, displayed
with dark red solid lines in the right panel of
Fig.~\ref{fig:spectralsplit}.

If $\theta_{23}$ lies in the second octant then most of $\nu_e$ and
$\bar\nu_e$ end up as $\nu_3^{\rm m}$ and $\bar\nu_2^{\rm m}$. As can
be seen in the solid lines in the bottom right panel of
Fig.~\ref{fig:levcros_I-noI} these neutrinos will not traverse the
$I$-resonance, except the excess $\epsilon$ of $\nu_e$, which stays as
$\nu_1^{\rm m}$. These neutrinos will be basically blind to the $I$-,
$H$-, and $L$-resonances. The final fluxes will be therefore
$\rho^{\rm final}_{ee}=\sin^2\theta_{13}+\epsilon\cos^2\theta_{12}$
and $\bar\rho^{\rm final}_{ee}=\sin^2\theta_{12}$.  This case is
represented with dashed lines in the bottom right panel of
Fig.~\ref{fig:rho_regionIVbbis}.
In the end the final evolution turns out to be similar to that in
region IVb.

\subsection{Discussion}
\label{sec:discussion}

In the previous sections we have studied the consequences of NSI on
the neutrino propagation through the SN envelope 
taking into account the presence of a neutrino background.
We have analyzed the different situations in terms of the non-universal
NSI parameter $\varepsilon_{\tau\tau}$ and the density at the
neutrino sphere $\lambda_0$. Depending on their values we were able to
identify four extreme regions of the parameters where the evolution of
the neutrinos have a specific pattern. 

\begin{figure}
\begin{center}
\includegraphics[angle=0,width=0.98\textwidth]{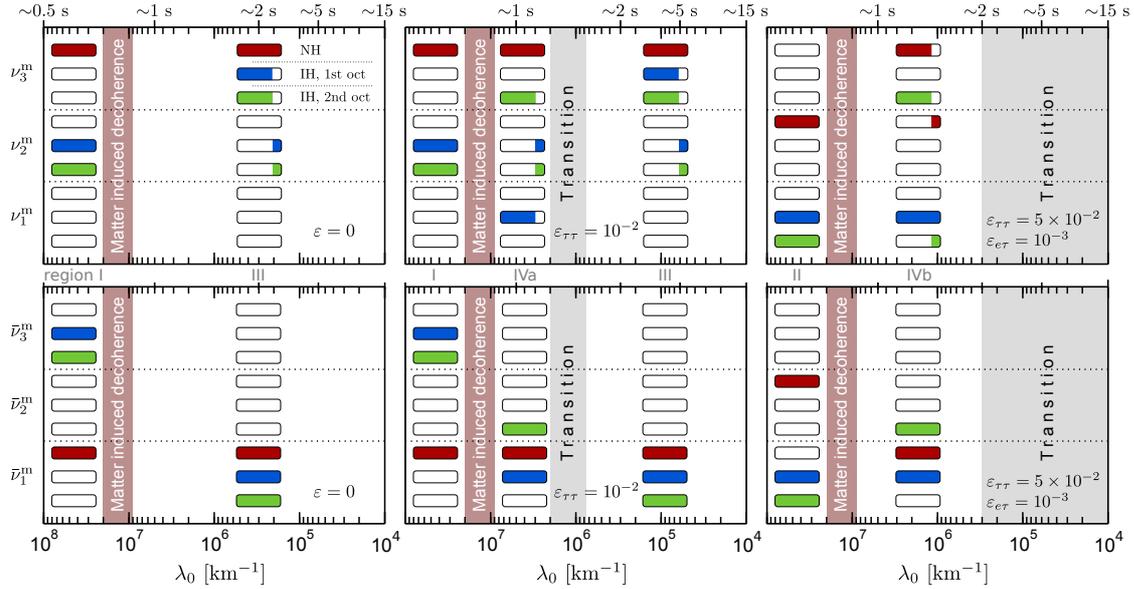}
\caption{\small Relationship between $\nu_e$ (top panels) and
  $\bar\nu_e$ (bottom panels) and the matter eigenstates at the SN
  surface as function of time for different neutrino mass and mixing
  schemes and for given values of $\varepsilon_{\tau\tau}=0$ (left),
  $3\times 10^{-2}$ (middle), and $5\times 10^{-2}$ (right).  }
\label{fig:finalscheme}
\end{center}
\end{figure}

In a realistic situation, though, we expect to find a combination
of these situations depending on the instant considered. As 
mentioned in Sec.~\ref{sec:eoms} one expects the value of $\lambda_0$
to decrease with time as the explosion goes on.
Therefore it is important to look for a time dependence in the
neutrino propagation for given values of the NSI parameters.  In
Fig.~\ref{fig:finalscheme} we show the relationship between $\nu_e$
(top panels) and $\bar\nu_e$ (bottom panels) and the matter
eigenstates at the SN surface as function of time for different
neutrino mass and mixing schemes and for a given value of
$\varepsilon_{\tau\tau}$.  The evolution in time shown in each panel
is equivalent to consider Fig.~\ref{fig:regions}, fix a value in the x
axis corresponding to some $\varepsilon_{\tau\tau}$ and following
vertically towards lower values of $\lambda_0$.  Depending on
$\varepsilon_{\tau\tau}$ and the instant considered one can
distinguish different regions separated by vertical bands denoting
transition phases. The position and the size of these transition bands
are not constant but depend on time, as $\mu_0$ and $\epsilon$
do. Nevertheless, unless multiangle decoherence is triggered (by
e.g. a strong reduction of $\epsilon$~\cite{EstebanPretel:2007ec}),
the sequence of different regimes undergone by the neutrinos is not
expected to change drastically.

Let us first discuss the antineutrino case, because they would give
rise to most of the signal.  The bottom left panel corresponds to the
standard case, i.e.~$\varepsilon_{\tau\tau}=0$. On the left we have
early times (or large $\lambda_0$), which corresponds to the region I in
Fig.~\ref{fig:regions}. From the previous discussion, we know that for
such a case and normal mass hierarchy (red box) $\bar\nu_e$ leaves as
$\bar\nu_1$ whereas in inverted mass hierarchy they escape as
$\bar\nu_3$ due to the adiabatic $H$-resonance, see
Figs.~\ref{fig:levcros_I-noI} and~\ref{fig:rho_regionI-II}.
At later times, $\lambda_0$ becomes smaller and matter can not
suppress collective effects any longer, i.e.~neutrinos enter region
III. These affect only in the inverted mass hierarchy case
``canceling'' the $H$-resonance conversion and making $\bar\nu_e$ to
escape as $\bar\nu_1$. There is then a time dependence in the survival
$\bar\nu_e$ probability for inverted mass hierarchy but not for the
normal one. As can be seen in the panel this behavior does not depend
on the $\theta_{23}$ octant, see Figs.~\ref{fig:levcros_I-noI}
and~\ref{fig:rho_regionIII-IVa}. In terms of $\bar\nu_e$ survival
probabilities there is then a transition from
$\sin^2\theta_{13}\approx 0$ at early times to
$\cos^2\theta_{12}\approx 0.7$ at later times, the details depending
on the specific time evolution of $\lambda(r)$.

Let us now take the bottom middle panel, with
$\varepsilon_{\tau\tau}=3\times 10^{-3}$. The situation at early times
is the same as in the previous panel, described by the region
I. However at intermediate times the situation changes in the case of
inverted mass hierarchy. Now the NSI parameters make the evolution go
through region IVa before entering eventually region III. The
$\mu\tau$-resonance is pushed outside the bipolar region and then the
degeneracy between the two $\theta_{23}$ octants is broken: for
$\theta_{23}$ in the first octant (blue box) $\bar\nu_e$ leaves as
$\bar\nu_1$ whereas for the second octant (green box) they escape as
$\bar\nu_2$, see Figs.~\ref{fig:levcros_I-noI}
and~\ref{fig:rho_regionIII-IVa}. At later times $\lambda_0$ further
decreases and the $\mu\tau$-resonance contracts to deeper layers
within $r_{\rm syn}$. That means neutrinos cross to region III and the
$\theta_{23}$ octant degeneracy is restored.  Concerning the
$\bar\nu_e$ survival probability, as before there is a transition from
$\sin^2\theta_{13}\approx 0$ directly to $\cos^2\theta_{12}\approx
0.7$ for $\theta_{23}$ in the first octant, and from
$\sin^2\theta_{13}\approx 0$ through $\sin^2\theta_{12}\approx 0.3$
until $\cos^2\theta_{12}\approx 0.7$ if $\theta_{23}$ lies in the
second octant. As it was discussed an analogous effect would arise for
a fixed $\theta_{23}$ and different signs of $\varepsilon_{\tau\tau}$.

Finally, we consider the case where the NSI parameters are large
enough, $\varepsilon_{\tau\tau}=5\times 10^{-2}$, to induce the
$I$-resonance. Now, at early times neutrino propagation follows the
prescription given in region II. For normal mass hierarchy
$\bar\nu_e$ leave as $\bar\nu_2$, whereas for inverted they escape the
star as $\bar\nu_1$ for both octants, see Figs.~\ref{fig:levcros_I-noI}
and~\ref{fig:rho_regionIVb}. After this phase neutrinos enter the
region IVb. That means that collective effects arise and, as before,
they break the degeneracy of the $\theta_{23}$ octant for inverted
mass hierarchy.

  The bottom line is that if $|\varepsilon_{\tau\tau}|\gtrsim$ a
  $10^{-3}$ neutrinos cross the region IV during some seconds, and
  this could help disentangle the $\theta_{23}$ octant. If the octant
  were known one could obtain information about the sign of the
  non-universal NSI parameters.

In the upper panels we show the same kind of plots but for
neutrinos. The main difference with respect to antineutrinos is that
in the presence of collective effects $\nu_e$ are not fully converted
like $\bar\nu_e$. Some fraction of them, corresponding to the excess
over $\bar\nu_e$, remains unaffected. This excess is represented in
Fig.~\ref{fig:finalscheme} with the small colored portion at the right
hand side of the corresponding boxes.  As it has been discussed, this
excess of $\nu_e$ is translated into a spectral split. That means that
the flavor spectral swap happens only for some energies. Whether these
correspond to the low-energy tail or high-energy tail of the initial
spectrum depends on the neutrino properties, see right panel of
Fig.~\ref{fig:spectralsplit}.  
Therefore one could hope to use this additional information to break possible
degeneracies between different mass and mixing schemes and different
values of the NSI parameters.

Throughout the analysis we have limited ourselves to the case of NSI
with $d$ quarks. Nevertheless most of the results here presented can
be generalized to the case of $u$ quarks and $e$. In the first case
the only effect is to shift the position of the $I$-resonance, since
the resonance condition is modified to
$Y_e^I=\varepsilon_{\tau\tau}/(1-\varepsilon_{\tau\tau})$
~\cite{EstebanPretel:2007yu}. In the case of $e$ the $I$-resonance is
absent. But nevertheless its contribution to increase the value of
$Y^{\rm eff}_{\rm \tau,nsi}$ would also make the neutrino propagation
highly sensitive to the $\theta_{23}$ octant and to its own sign,
exactly as in the case of $d$ quark.

Last but not least we briefly comment on the possibility to observe
the different regimes analyzed. This possibility will be hampered by
several uncertainties inherent in SN neutrinos. One is the lack of
knowledge on the exact matter profile traversed by the outgoing
neutrinos. In our study we assumed a simple power law given in
Eq.~(\ref{eq:lambda(r)}). This density profile will be significantly
distorted by the passage of the shock wave responsible for ejecting
the whole SN
envelope~\cite{Schirato:2002tg,Fogli:2006xy,Friedland:2006ta}. One of
the main effect will be 
to destroy the adiabaticity of the $H$-resonance, which was assumed in
the study. This effect, though, is not always present but depends on
the neutrino properties.
Therefore, far from being a problem, the time and energy dependence
modulation introduced in the spectra could further help disentangle
between the different scenarios here
considered~\cite{Fogli:2003dw,Tomas:2004gr,EstebanPretel:2007yu}. 

Another important source of uncertainties is our ignorance of the
exact initial fluxes $f^{\rm R}_{\nu_\alpha}(E)$. Although the initial
fluxes during the first stage of the explosion, the neutronization
burst, are rather model independent~\cite{Kachelriess:2004ds}, the
expected number of events is very low. Most of the signal is generated
later, during the accretion and cooling phases. The spectral features
observed in the numerical simulations depends strongly on the
properties of the SN.
It is therefore necessary to set up strategies combining different
observables to be able to pin down the underlying neutrino properties
independently of the initial fluxes. These include among others to
analyze the spectral modulations expected if neutrinos cross the Earth
before being
detected~\cite{Lunardini:2001pb,Dighe:2003be,Dighe:2003jg,Dighe:2003vm},
or benefit from the time dependence of the matter profiles
$\lambda(r)$ and $Y_e(r)$ themselves~\cite{EstebanPretel:2007yu}.

\section{Summary}
\label{sec:summary}

We have reexamined the effect of NSI on the neutrino propagation
through the SN envelope within a three-neutrino framework, first in the
absence and later in the presence of a neutrino background.
We have found that the small values of the electron fraction, typical
of the more deleptonized inner layers, allow for internal NSI-induced
resonant conversions, in addition to the standard MSW-H and MSW-L
resonances of the outer envelope. These new flavor conversions take
place for a relatively large range of NSI parameters, namely
$|\varepsilon_{\alpha\alpha}|$ between $10^{-2}-10^{-1}$, and
$|\varepsilon_{e\tau}|\gtrsim {\rm few}\times 10^{-5}$, currently
allowed by experiment.  For this range of strengths, in particular
$\varepsilon_{\tau\tau}$, NSI can significantly affect the
adiabaticity of the $H$-resonance. We have also obtained that when
including neutrino self-interaction to the system and considering
$|\varepsilon_{\tau\tau}|\gtrsim$ few$\times 10^{-4}$ the neutrino
propagation becomes for some time sensitive to the $\theta_{23}$
octant and the sign of $\varepsilon_{\tau\tau}$.
Furthermore, the coexistence of collective effects and the
$I$-resonance may lead to an exchange of the neutrino fluxes entering
the bipolar regime. The main consequences being a bipolar conversion
happening for normal mass hierarchy and an inverted spectral split.

\cleardoublepage

\pagestyle{normal}
\chapter{Conclusions}\label{chapter:conclusions}

Neutrino physics has reached a point where one can start talking about
precision physics. In the last ten years the situation has changed
from having no experimental evidence of massive neutrinos to measuring
the oscillation parameters within errors of 5\% to 10\%. The
conditions are, therefore, ideal for the study of neutrino
non-standard properties while we improve the precision in the
measurement of their oscillation parameters. These two research lines,
together with the determination of the Dirac or Majorana nature of
neutrinos, are nowadays centering the effort of the neutrino
physicists community. The thesis here presented has followed this
philosophy. We have analyzed various aspects of neutrino phenomenology
in two different scenarios: accelerator and reactor terrestrial
experiments, and neutrinos in a supernova (SN) environment.

In this way, we have started with a series of introductory chapters,
where we have reviewed, in a general form, the physics of SN
explosions, neutrino oscillations and the effect of non-standard
interaction (NSI). With these important concepts already clarified, we
have discussed whether OPERA could help in constraining neutrino NSI,
improving the results obtained from MINOS and Double Chooz. The
motivation for this study is twofold. Firstly, OPERA will measure for
the first time the oscillation $\nu_\mu \to \nu_\tau$, detecting
directly the $\nu_\tau$; secondly, the distance-energy ($L/E$)
relation is very different for MINOS and OPERA. Both conditions could
help in distinguishing NSI from standard oscillations. The inclusion
of Double Chooz into the analysis limits $\theta_{13}$ in an
independent way, since it is not sensitive to NSI because of the short
distance involved and the low energy of its neutrinos. In our study we
have obtained that the main improvement coming from OPERA is due to
the different $L/E$ relation with respect to MINOS. However, the
limits we got from their combination do not increase the precision
obtained with atmospheric neutrino experiments. Furthermore, the
expected $\nu_\tau$ signal in OPERA is too low to be of statistical
significance, helping only for large values of $\theta_{13}$.

Regarding the main neutrino scenario discussed in this thesis,
i.e. core collapse SN, we have essentially focused on two effects. On
the one hand, we have analyzed neutrino self-interaction in the SN
envelope, recently confirmed to be of special importance in their
evolution. This piece of the system had been underestimated for a long
time, and only very recently its relevance has been recognized,
drastically changing SN neutrino's evolution paradigm. Before this
discovery, their evolution was basically determined by the
adiabaticities of the MSW $H$ and $L$-resonances. In contrast, it has
been shown that neutrino collective conversion effects can occur in
the most inner layers of the star, given the appropriate astrophysical
(matter and neutrino densities, hierarchy of fluxes$\dots$) and
neutrino (inverted mass hierarchy) conditions. The nature of these
transformations is very different from that of the standard
resonances.

In this thesis we have reviewed the collective transformation
phenomenon. We have payed special attention to the most relevant
effects, mainly the inversion of the neutrino fluxes entering the MSW
resonances, and the spectral split. Keeping this in mind, we have
studied, in a two flavor scenario, the consequences that could result
from the multi-angular nature of the system, being neutrinos emitted
from a spherical surface, the neutrino sphere. This condition can
induce kinematical decoherence, since different angular modes feel
different refraction index in the medium. We have identified two
possible sources of decoherence, one related to the self-interaction
term, and another to the interaction with matter, which will have
different magnitude depending on the emission angle. From our study we
conclude that the first one does not involve any risk, given the
characteristics obtained in the simulations of this kind of
explosions. The second source, in contrast, translates into a time
dependence of the expected signal. At the initial moments of the
explosion, when the matter density near the neutrino sphere is much
larger than the neutrino one, a suppression of the collective
phenomenon is obtained. As time goes on, the matter density is reduced
becoming comparable to the neutrino one. We then obtain a transition
region marked by the decoherence. Nevertheless, the matter density
will quickly become lower than that of neutrinos, recovering the
collective effects. After having clarified the decoherence topic, we
have studied the existence of possible characteristic three-flavor
effects. In summary, we have obtained that in the absence of second
order corrections to the potentials affecting $\nu_\mu$ and
$\nu_\tau$, the system can be always reduced to the two flavor
case. However, if we take into account the radiative corrections, the
$\nu_e$ and $\bar\nu_e$ survival probabilities can be modified, and
depend on the $\theta_{23}$ octant. This effect however, requires a
large matter density, which in turn would suppress the collective
effect.

The second important point we have discussed in the SN scenario is the
effect that the inclusion of the possible NSI would induce in the
neutrino propagation. This effect occur in the same region where
collective neutrino transformation phenomena take place. Therefore, we
have first clarified the genuine NSI effects by initially neglecting
any self-interaction effect. The most important consequence we have
obtained in our study is the appearance of a new resonance ($I$) in
the internal regions of the SN. It is related to the low values of the
electron fraction in the most deleptonized layers of the star, near
the neutrino sphere. Furthermore, we obtain a modification in the $H$
and $L$-resonance conditions, basically affecting their positions. The
interplay of NSI and neutrino self-interactions has very interesting
consequences for the propagation of SN neutrinos. First, the time
dependence of the collective effects is modified in the presence of
NSI. For some time, the evolution of neutrinos becomes sensitive to
the octant of $\theta_{23}$ and the sign of $\varepsilon_{\tau\tau}$,
provided that $|\varepsilon_{\tau\tau}|\gtrsim 10^{-4}$. This is an
analogous effect to that obtained as a consequence of the radiative
corrections to the potentials. The difference here is that the
presence of NSI lowers the density required to induce the effect, and
the tension with the suppression of the collective effects is
therefore relaxed. On the other hand, if
$|\varepsilon_{\tau\tau}|\gtrsim 10^{-2}$ the $I$-resonance comes into
play, which in the case of being adiabatic will transform the neutrino
fluxes entering the bipolar region. As a consequence, we will obtain
bipolar conversion for normal mass hierarchy case and an inverted
spectral split.

\cleardoublepage

\chapter*{Conclusions}

La f\'{i}sica de neutrins ha arribat a un punt en qu\`e pot comen\c
car a parlar-se de f\'{i}sica de precisi\'o. En els darrers deu anys
s'ha passat de no tindre cap evid\`encia experimental de la massa dels
neutrins a la determinaci\'o dels par\`ametres d'oscil$\cdot$laci\'o
amb errors entre el 5\% i el 10\%. Es donen, doncs, les condicions per
a buscar possibles propietats no est\`andard dels neutrins alhora que
s'incrementa la precisi\'o en la mesura dels seus
par\`ametres. Aquestes dues l\'{i}nies d'investigaci\'o, junt a la
determinaci\'o de la naturalesa de Dirac o Majorana dels neutrins,
centren actualment els esfor\c cos de la comunitat de f\'{i}sics de
neutrins. La tesi ac\'{i} presentada ha seguit aquesta filosofia. Hem
fet una an\`alisi de diversos aspectes de fenomenologia de neutrins en
dos escenaris diferents: per un costat experiments terrestres
d'accelerador i reactor i per un altre neutrins emesos per una
supernova (SN).

Aix\'{i} doncs, hem comen\c cat amb una s\`erie de cap\'itols
introductoris on hem revisat de forma general la f\'isica de les
explosions de SN, d'oscil$\cdot$laci\'o de neutrins i les seues
possibles interaccions no est\`andard (NSI). Amb aquests conceptes
clarificats, hem discutit la possibilitat de millorar els l\'{i}mits
sobre les NSI obtinguts als experiments MINOS i Double Chooz amb la
inclusi\'o d'OPERA.
La motivaci\'o d'aquesta an\`alisi \'es doble. En primer lloc, OPERA
\'es el primer experiment capa\c c de mesurar l'oscil$\cdot$laci\'o
$\nu_\mu \to \nu_\tau$ detectant directament els $\nu_\tau$; en segon
lloc, la relaci\'o dist\`ancia-energia ($L/E$) \'es molt diferent
entre MINOS i OPERA. Totes dues condicions podrien ajudar a distingir
entre oscil$\cdot$lacions est\`andard i NSI. La inclusi\'o de Double
Chooz en l'an\`alisi permet limitar de forma independent
$\theta_{13}$, ja que no \'es sensible a les NSI degut a la curta
dist\`ancia involucrada i la baixa energia dels neutrins. Al nostre
estudi hem trobat que la principal millora que OPERA aporta ve de la
seua difer\`encia en $L/E$ respecte a MINOS. Tot i aix\`o, els
l\'{i}mits que s'obtenen no augmenten la precisi\'o aconseguida en
experiments amb neutrins atmosf\`erics. D'altra banda, el senyal de
$\nu_\tau$ esperat en OPERA \'es massa baix per a ser
estad\'{i}sticament significatiu, ajudant nom\'es per a valors grans
de $\theta_{13}$.

Respecte al principal escenari de neutrins discutit en aquesta tesi,
les explosions de SN per col$\cdot$lapse gravitatori del nucli, ens
hem centrat essencialment en dos efectes. D'un costat, hem analitzat
l'autointeracci\'o dels propis neutrins en la seua propagaci\'o a
trav\'es de la SN, recentment constatat com d'especial import\`ancia
en la seua evoluci\'o. Aquesta pe\c ca del sistema havia sigut
menyspreada durant molt de temps, i nom\'es als darrers anys s'ha
reconegut la seua rellev\`ancia, canviant dr\`asticament el paradigma
d'evoluci\'o de neutrins de SN. Abans d'aquest descobriment, la seua
evoluci\'o venia determinada b\`asicament per l'adiabaticitat de les
resson\`ancies MSW: $H$ i $L$. Per contra, s'ha vist que en la regi\'o
interna de l'estel, donades les condicions apropiades tant
astrof\'{i}siques (densitat de mat\`eria i de neutrins, jerarquia de
fluxos per als diferents sabors de neutrins$\dots$) com dels propis
neutrins (jerarquia inversa de massa), poden oc\'orrer efectes
col$\cdot$lectius de conversi\'o de sabor en neutrins i antineutrins,
de naturalesa molt diferent a la de les resson\`ancies est\`andard.

En aquesta tesi hem fet, doncs, una revisi\'o del fenomen parant
especial atenci\'o als efectes m\'es destacats, que principalment
s\'on la inversi\'o del flux de neutrins que entren en les
resson\`ancies MSW i el ``trencament de l'espectre'' (spectral
split). Partint d'aquesta base, hem estudiat, en un escenari de dos
sabors, les conseq\"u\`encies que la naturalesa multiangular del
sistema pot tindre en la seua evoluci\'o, per ser els neutrins emesos
des d'una superf\'{i}cie esf\`erica, la neutrinosfera. Aquesta
condici\'o pot donar lloc a una decoher\`encia cinem\`atica en el
sistema, ja que els diferents modes angulars senten diferent \'{i}ndex
de refracci\'o amb el medi. Hem identificat dues possibles fonts de
decoher\`encia, una deguda al terme d'autointeracci\'o dels neutrins,
i altra a la interacci\'o amb la mat\`eria, que tindr\`a diferent
intensitat segons l'angle d'emissi\'o. Del nostre estudi concloem que
la primera no suposa un perill per a l'estudi de neutrins de SN,
donades les caracter\'{i}stiques que t\'{i}picament s'obtenen en les
simulacions d'aquest tipus d'explosions. La segona font, per contra,
comporta una depend\`encia temporal del senyal esperat. En els
instants inicials de l'explosi\'o, quan la densitat de mat\`eria prop
de la neutrinosfera \'es molt gran comparada amb la de neutrins,
s'obt\'e una supressi\'o del fenomen col$\cdot$lectiu. Conforme passa
el temps la densitat de mat\`eria disminueix fent-se comparable a la
de neutrins. S'obt\'e, aleshores, una regi\'o de transici\'o marcada
per la decoher\`encia i per tant p\`erdua de la informaci\'o de
l'espectre incial dels neutrins. No obstant, la densitat de mat\`eria
r\`apidament es far\`a inferior a la de neutrins, recuperant l'efecte
col$\cdot$lectiu. Una vegada aclarit el tema de la decoh\`erencia, hem
estudiat l'exist\`encia de possibles efectes caracter\'{i}stics de
tres sabors. En resum, hem obtingut que en abs\`encia de correccions
de segon ordre als potencials que afecten els $\nu_\mu$ i $\nu_\tau$,
el sistema pot ser redu\"{i}t sempre al cas de dos sabors. No obstant,
si tenim en compte les correccions radiatives als potencials,
les probabilitats de superviv\`encia dels $\nu_e$ i $\bar\nu_e$ es
poden veure dr\`asticament modificades, depenent sensiblement de
l'octant de $\theta_{23}$. Aquest efecte per\`o, precisa d'una alta
densitat de mat\`eria, i possiblement estiga amagat sota la
supressi\'o del fenomen col$\cdot$lectiu que aix\`o comporta.


El segon punt important que hem tractat dins d'un escenari de SN \'es
l'efecte que la inclusi\'o de possibles NSI tindria en la propagaci\'o
dels neutrins. Aquestes afecten en la mateixa regi\'o interna de la SN
on tenen lloc els fen\`omens de transformaci\'o col$\cdot$lectiva de
neutrins. Per tant, hem volgut clarificar primer els efectes
genu\"{i}ns de les NSI, analitzant inicialment el sistema en
abs\`encia d'autointeracci\'o dels neutrins. La conseq\"u\`encia m\'es
important d'aquest estudi \'es l'aparici\'o d'una nova resson\`ancia
($I$) en les regions internes de la SN, relacionada amb valors
xicotets de la fracci\'o electr\`onica en la zona m\'es deleptonizada
de l'estel, prop de la neutrinosfera. A m\'es a m\'es, s'obt\'e una
modificaci\'o en les condicions de les resson\`ancies $H$ i $L$,
b\`asicament afectant a la seua posici\'o. L'estudi conjunt de les NSI
i l'autointeracci\'o dels neutrins t\'e conseq\"u\`encies molt
interessants per a la propagaci\'o dels neutrins de SN. En primer
lloc, la depend\`encia temporal dels fen\`omens col$\cdot$lectius es
veu modificada per la pres\`encia de les NSI. Durant un cert temps la
propagaci\'o dels neutrins es torna sensible a l'octant de
$\theta_{23}$ i al signe de $\varepsilon_{\tau\tau}$, sempre i quan
$|\varepsilon_{\tau\tau}|\gtrsim 10^{-4}$. Aquest efecte \'es an\`aleg
al que obten\'{i}em degut a les correccions radiatives, amb la
difer\`encia que la pres\`encia de NSI disminueix la densitat
requerida per a induir l'efecte. Es relaxa aix\'{i} la tensi\'o amb la
supressi\'o del fenomen col$\cdot$lectiu deguda a l'alta densitat de
mat\`eria. D'un altre costat, si $|\varepsilon_{\tau\tau}|\gtrsim
10^{-2}$ la resson\`ancia $I$ entrar\`a en joc, la qual, en cas de ser
adiab\`atica, transformar\`a els fluxos de neutrins que entren a la
regi\'o bipolar. Com a conseq\"u\`encia obtindrem conversi\'o bipolar
per a jerarquia normal de masses i un trencament de l'espectre invers.


\cleardoublepage

\pagestyle{appendixa}
\appendix
\chapter*{Appendix A \newline \newline Equations of motion}
\addcontentsline{toc}{chapter}{Appendix A: Equations of motion}
\newcounter{alpha}
\renewcommand{\thesection}{\Alph{alpha}}
\renewcommand{\theequation}{\Alph{alpha}.\arabic{equation}}
\renewcommand{\thetable}{\Alph{alpha}}
\setcounter{alpha}{1}
\setcounter{equation}{0}
\setcounter{table}{0}

\subsection{Temporal evolution}

A homogeneous ensemble of unmixed neutrinos is represented by the
occupation numbers $f_{\bf p}=\langle a^\dagger_{\bf p} a_{\bf
p}\rangle$ for each momentum mode ${\bf p}$, where $a^\dagger_{\bf
p}$ and $a_{\bf p}$ are the relevant creation and annihilation
operators and $\langle\ldots\rangle$ is the expectation value. A
corresponding expression can be defined for the antineutrinos,
${\bar f}_{\bf p}=\langle {\bar a}^\dagger_{\bf p} {\bar a}_{\bf
p}\rangle$, where overbarred quantities always refer to
antiparticles. In a multiflavor system of mixed neutrinos, the
occupation numbers are generalised to density matrices in flavor
space~\cite{Dolgov:1980cq,Sigl:1992fn,McKellar:1992ja}
\begin{equation}\label{eq:densitymatrixdefinition}
 (\varrho_{\bf p})_{ij}=
 \langle a^\dagger_{i} a_{j}\rangle_{\bf p}
 \hbox{\quad and\quad}
 (\bar \varrho_{\bf p})_{ij}=
 \langle \bar a^\dagger_{j}\,\bar a_{i}\rangle_{\bf p}\,.
\end{equation}
The reversed order of the flavor indices $i$ and $j$ in the
right-hand side for antineutrinos assures that $\varrho_{\bf p}$ and
$\bar\varrho_{\bf p}$ transform identically under a flavor
transformation.

Flavor oscillations of an ensemble of neutrinos and antineutrinos
are described by~\cite{Dolgov:1980cq, Sigl:1992fn, McKellar:1992ja}
\begin{equation}\label{eq:eom1}
 \I\partial_t\varrho_{\bf p}=[{\sf H}_{\bf p},\varrho_{\bf p}]
 \hbox{\quad and\quad}
 \I\partial_t\bar\varrho_{\bf p}=
 [\bar {\sf H}_{\bf p},\bar\varrho_{\bf p}]\,,
\end{equation}
where $[{\cdot},{\cdot}]$ is a commutator. The ``Hamiltonian'' for
each mode is
\begin{equation}\label{eq:ham1}
 {\sf H}_{\bf p}=\Omega_{\bf p}
 +\lambda {\sf L}+\sqrt{2}\,G_{\rm F}
 \int\!\frac{\D^3{\bf q}}{(2\pi)^3}
 \left(\varrho_{\bf q}-\bar\varrho_{\bf q}\right)
 (1-{\bf v}_{\bf q}\cdot{\bf v}_{\bf p})\,,
\end{equation}
where $G_{\rm F}$ is the Fermi constant. The matrix of vacuum
oscillation frequencies for relativistic neutrinos is in the mass
basis $\Omega_{\bf p}={\rm diag}(m_1^2,m_2^2,m_3^2)/2p$ with
$p=|{\bf p}|$. The matter effect is represented by
$\lambda=\sqrt{2}\,G_{\rm F}(n_{e^-}-n_{e^+})$ and ${\sf L}={\rm
diag}(1,0,0)$, given here in the weak interaction basis. We ignore
the possible presence of other charged-lepton flavors. The
Hamiltonian for antineutrinos $\bar {\sf H}_{\bf p}$ is the same
with $\Omega_{\bf p}\to-\Omega_{\bf p}$, i.e., in vacuum
antineutrinos oscillate ``the other way round.''

The factor $(1-{\bf v}_{\bf q}\cdot{\bf v}_{\bf
p})=(1-\cos\theta_{\bf pq})$ represents the current-current nature
of the weak interaction where ${\bf v}_{\bf p}={\bf p}/p$ is the
velocity. The angular term averages to zero if the gas is isotropic.
We ignore a possible net flux of charged leptons lest the ordinary
matter effect also involves an angular factor.

If the system is axially symmetric relative to some direction, the
angular factor simplifies after an azimuthal integration
to~\cite{Duan:2006an, Raffelt:2007yz}
\begin{equation}\label{eq:axial}
 (1-{\bf v}_{\bf q}\cdot{\bf v}_{\bf p})
 \to(1-{v}_{\bf q}{v}_{\bf p})\,,
\end{equation}
where the velocities are along the symmetry axis.

\subsection{Spatial evolution in spherical symmetry}

Instead of a homogeneous system that evolves in time we consider a
stationary system that evolves in space. The occupation numbers
become Wigner functions, which depend both on spatial coordinates
and on momenta, but there is no conceptual problem as long as we
consider spatial variations that are slow on the scale of the
inverse neutrino momenta.

Since multi-angle effects are at the focus of our problem, we cannot
reduce the equations to plane waves moving in the same direction.
Motivated by the SN application, however, we can take advantage of
global spherical symmetry, implying that the ensemble is represented
by matrices that depend on a radial coordinate $r$, the zenith angle
relative to the radial direction, and the energy $E$ which in the
relativistic limit is identical with $p=|{\bf p}|$.

We ignore gravitational deflection near the SN core and assume that
neutrinos move on straight lines after being launched at a radius
$R$ that we call the neutrino sphere. Consider a neutrino that was
launched at an angle $\vartheta_R$ relative to the radial direction.
Its radial velocity is
\begin{equation}
v_R=\cos\vartheta_R\,.
\end{equation}
At $r>R$ the trajectory's angle relative to the radial
direction is implied by simple geometry to be~\cite{Duan:2006an} (see
e.g. their Fig. 1)
\begin{equation}
R\sin\vartheta_R=r\sin\vartheta_r\,.
\end{equation}
Therefore, the radial velocity at $r$ is
\begin{equation}\label{eq:costheta}
v_{u,r}=\cos\vartheta_r=\sqrt{1-\frac{R^2}{r^2}\,u}
\end{equation}
where we have introduced
\begin{equation}\label{eq:udef}
u=1-v^2_R=\sin^2\vartheta_R\,.
\end{equation}
It is convenient to label the angular modes with $u$. The physical
zenith angles change with distance so that the equations would be
more complicated.

The density matrices $\varrho_{p,u,r}$ are not especially useful to
describe a spherically symmetric system because they vary with $r$
even in the absence of oscillations. (Note that we often write the
dependence of a quantity on a variable as an subscript.) A quantity
that is conserved in the absence of oscillations is the total flux
matrix
\begin{equation}
  {\sf J}_r=\frac{r^2}{R^2}
  \int\frac{\D^3{\bf p}}{(2\pi)^3}\,
  \varrho_{{\bf p},r}\,v_{{\bf p},r}\,.
\end{equation}
To express the integral in co-moving variables we observe that
$\D^3{\bf p}$ in spherical coordinates is $p^2\D
p\,\D\varphi\,\D\cos\vartheta_r$ and that Eq.~(\ref{eq:costheta})
implies
\begin{equation}\label{eq:transformation}
 \left|\frac{\D\cos\vartheta_r}{\D u}\right|=
 \frac{1}{2v_{u,r}}\,\frac{R^2}{r^2}\,.
\end{equation}
Therefore, we finally define the differential flux matrices
\begin{equation}\label{eq:Judef}
 {\sf J}_{p,u,r}=
 \frac{p^2\varrho_{p,u,r}}{2\,(2\pi)^2}\,,
\end{equation}
where we have used $\int\D\varphi=2\pi$ for axial symmetry. The
normalization is
\begin{equation}
{\sf J}_r=\int_0^1\D u\int_0^\infty\D p\; {\sf J}_{p,u,r}\,.
\end{equation}
In the absence of oscillations the total and differential fluxes are
conserved, $\partial_r {\sf J}_{r}=0$ and $\partial_r {\sf
J}_{p,u,r}=0$.

To include oscillations, we note that the radial velocity along a
neutrino trajectory is $v_{u,r}=\D r_{u}/\D t=\cos\vartheta_{u,r}$.
Therefore, if we wish to express the temporal evolution of the
neutrino density matrix along its trajectory in terms of an
evolution expressed in terms of the radial coordinate~$r$, we
substitute $\partial_t\to v_{u,r}\partial_r$ in Eq.~(\ref{eq:eom1})
so that
\begin{equation}
 {\rm i}\partial_r{\sf J}_{p,u,r}=
 \frac{[{\sf H}_{p,u,r},{\sf J}_{p,u,r}]}{v_{u,r}}
 \,,
\end{equation}
and analogous for antineutrinos. In other words, we project the
evolution along a given trajectory to an evolution along the radial
direction. For vacuum oscillations this has the effect of
``compressing'' the oscillation pattern for non-radial modes, i.e.,
even for monochromatic neutrinos, the effective vacuum oscillation
frequency depends on both $r$ and $u$.

The vacuum-oscillation and ordinary-matter contributions to ${\sf
H}_{p,u,r}$ were given in Eq.~(\ref{eq:ham1}), whereas the self-term
must be made explicit. To this end we introduce the matrix of number
densities
\begin{equation}
{\sf N}_{p,u,r}=v_{u,r}^{-1}\,{\sf J}_{p,u,r}
\end{equation}
and its integral as
\begin{equation}
 {\sf N}_{r}=\int_0^1 \D u\int_0^\infty\D p\;{\sf N}_{p,u,r}
 =\int_0^\infty\D u\int_0^\infty\D p\;
 \frac{{\sf J}_{p,u,r}}{v_{u,r}}\,.
\end{equation}
Collecting all terms and taking advantage of Eq.~(\ref{eq:axial})
for axial symmetry, we find
\begin{eqnarray}\label{eq:eom0}
 \I\partial_r{\sf J}_{p,u,r}&=&+\bigl[\Omega_p,{\sf N}_{p,u,r}\bigr]
 +\lambda_r\bigl[{\sf L},{\sf N}_{p,u,r}\bigr]\nonumber\\*
 &&+\sqrt{2}G_{\rm F}\,\frac{R^2}{r^2}\,
 \Bigl(\bigl[{\sf N}_r-\bar{\sf N}_r,{\sf N}_{p,u,r}\bigr]
 -\bigl[{\sf J}_r-\bar{\sf J}_r,{\sf J}_{p,u,r}\bigr]\Bigr)\,,
 \nonumber\\*
 \I\partial_r\bar{\sf J}_{p,u,r}&=&
 -\bigl[\Omega_p,\bar{\sf N}_{p,u,r}\bigr]
 +\lambda_r\bigl[{\sf L},\bar{\sf N}_{p,u,r}\bigr]\nonumber\\*
 &&
 +\sqrt{2}G_{\rm F}\,\frac{R^2}{r^2}\,
 \Bigl(\bigl[{\sf N}_r-\bar{\sf N}_r,\bar{\sf N}_{p,u,r}\bigr]
 -\bigl[{\sf J}_r-\bar{\sf J}_r,\bar{\sf J}_{p,u,r}\bigr]\Bigr)\,,
\end{eqnarray}
where the electron density's radial variation is included
in~$\lambda_r$.

\subsection{Angular emission characteristics}

In a numerical simulation we need to specify the fluxes at the
neutrino sphere $r=R$. For our usual multi-angle simulations we
assume that the neutrino radiation field is ``half isotropic''
directly above the neutrino sphere, i.e., that all outward-moving
angular modes are equally occupied as behooves a thermal radiation
field. Therefore, the occupation numbers are distributed as $\D
n/\D\cos\vartheta_R={\rm const.}$, implying that the radial fluxes
are distributed as $\D j/\D\cos\vartheta_R=v_R\D
n/\D\cos\vartheta_R\propto\cos\vartheta_R$ because
$v_R=\cos\vartheta_R$. Expressed in the angular variable $u$ this
implies $\D j/\D u={\rm const.}$ because of
Eq.~(\ref{eq:transformation}). In other words, a blackbody radiation
field at the neutrino sphere implies that
\begin{equation}
{\sf J}_u={\rm const.}
\end{equation}
in the interval $0\leq u\leq1$.

To avoid multi-angle effects one may sometimes wish to use a single
angular bin. To represent a uniform ${\sf J}_u$ distribution, the
natural choice is $u=1/2$, corresponding to a launch angle
$\vartheta_R=45^\circ$. Our numerical single-angle examples always
correspond to this choice in an otherwise unchanged numerical code.

In this case the radial velocity of all neutrinos as a function of
radius is
\begin{equation}
v_r=\sqrt{1-\frac{R^2}{2r^2}}\,.
\end{equation}
For a monochromatic spectrum, the remaining flavor matrices are
simply the total ${\sf J}_r$ (corresponding to the single $u=1/2$)
and ${\sf N}_r={\sf J}_r/v_r$. Ignoring the trivial ordinary matter
term, the equations of motion are
\begin{equation}
 \I\partial_r{\sf J}_{r}=
 \frac{\bigl[\Omega,{\sf J}_{r}\bigr]}{v_r}
 +\sqrt{2}G_{\rm F}\,\frac{R^2}{r^2}\,
 \left(\frac{1}{v_r^2}-1\right)
 \bigl[{\sf J}_r-\bar{\sf J}_r,{\sf J}_{r}\bigr]
\end{equation}
and analogous for the antineutrinos. The coefficient of the
neutrino-neutrino term is explicitly
\begin{equation}\label{eq:muvariation}
 \sqrt{2}G_{\rm F}\,\frac{R^4}{r^4}\,\frac{1}{2-R^2/r^2}\,.
\end{equation}
At the neutrino sphere this expression becomes equal to
$\sqrt{2}G_{\rm F}$, whereas at large distances it is $(\sqrt{2}G_{\rm
  F}/2)\,R^4/r^4$. As observed in the previous literature, the
neutrino-neutrino term dies out at large distances as $r^{-4}$.

One can define a ``single-angle case'' somewhat differently.
Assuming all angular modes evolve coherently, we can integrate the
equations of motion over $\int_0^1 \D u$ and study the evolution of
the quantities $J_{p,r}=\int_0^1\D u J_{p,u,r}$. To write the
equations in a compact form we introduce the notation
\begin{equation}
 \frac{1}{v_r^*}\equiv\frac{1}{{\sf J}_r}
 \int_0^\infty\!\D p\int_0^1\D u\;\frac{{\sf J}_{p,u,r}}{v_{u,r}}\,.
\end{equation}
The full equation of motion Eq.~(\ref{eq:eom0}) for neutrinos
becomes
\begin{eqnarray}\label{eq:eom0a}
 \I\partial_r{\sf J}_{p,r}&=&
 \frac{\bigl[\Omega_p,{\sf J}_{p,r}\bigr]}{v_r^*}
 +\lambda_r\,\frac{\bigl[{\sf L},{\sf J}_{p,r}\bigr]}{v_r^*}
 \nonumber\\
 &&{}+\sqrt{2}G_{\rm F}\,\frac{R^2}{r^2}\,
 \left(\frac{1}{(v_r^*)^2}-1\right)
 \bigl[{\sf J}_r-\bar{\sf J}_r,{\sf J}_{p,r}\bigr]
\end{eqnarray}
and analogous for antineutrinos with $\Omega_p\to-\Omega_p$.

At large distances we have $1/v_r^*=1+\frac{1}{2}(R/r)^2\langle
u\rangle$ where $\langle u\rangle$ is the average of $u$ at
emission. For the vacuum and matter terms, we only need the leading
terms so that we recover the familiar plane-wave form of the
equations of motion. The coefficient of the neutrino-neutrino term,
on the other hand, becomes
\begin{equation}
\sqrt{2}G_{\rm F}\,\frac{R^4}{r^4}\,\langle u\rangle\,.
\end{equation}
Both for half-isotropic emission and for our single-angle case we
have $\langle u\rangle=\frac{1}{2}$, in agreement with
Eq.~(\ref{eq:muvariation}).

\cleardoublepage

\pagestyle{bibliografia}

\pagestyle{agraiments}
\chapter*{Agra\"{i}ments/Acknowledgements}

\addcontentsline{toc}{chapter}{Agra\"{i}ments/Acknowledgements}

En primer lloc vull agrair als meus directors de tesi, Jose W. Furtado
Valle i Ricard Tom\`as Bayo, perqu\`e ells han fet possible la
realitzaci\'o d'aquesta tesi doctoral, cadascun amb la seua
responsabilitat.

I would also like to thank everyone whom I have had the opportunity to
work with: Alessandro Mirizzi, Georg G.~Raffelt, G\"unter Sigl,
Pasquale D.~Serpico, Patrick Huber and Sergio Pastor. It has been a
pleasure to be your coauthor. I have learnt a lot from all of you.

Dejando de lado la f\'{i}sica, pero sin salir del IFIC, quiero
agradecer a toda la gente que ha hecho que estos cuatro a\~nos y pico
de doctorado me hayan dejado tan buen recuerdo. Muchas gracias a toda
la gente que est\'a, o ha pasado por el grupo AHEP: Diego
Aristiz\'abal, Tegua, Albert, Mariam, Urbano, Avelino, Martin, Diego
Restrepo, Massimiliano, Timur, Federica, Satoru, Antonio, Stefano,
Werner, Celio, Omar, Juan, Thomas, Sofiane, Kiko, Laslo$\ldots$ Muchas
gracias tambi\'en a todos los amigotes del IFIC: Mart\'{i}n, Diego
Milan\'es, Paola, Alberto, Fabio, Paula, Natxo, la gente del
futbito$\ldots$ y a los de la universidad: Iv\'an, Enrique, Jacobo,
Ana, V\'{i}ctor$\ldots$

Doble agra\"{i}ment mereix Ricard, perqu\`e no nom\'es m'has guiat
durant tot el meu doctorat sin\'o que a m\'es has fet que el m\'on de
la f\'{i}sica prengu\'es un altre sentit, sempre amb eixa visi\'o tan
particular$\ldots$ tu m'entens.

Por supuesto quiero agradecer a mi familia, que siempre est\'a ah\'{i}
y me ha sufrido desde el principio de los d\'ias, porque aunque hace
tiempo que se fue ``Sor Oxidaci\'on de los Clavos de la Cruz'', tengo
mis momentos. Much\'{i}simas gracias a mis padres, Juli\'an y
Mar\'{i}a, a mis hermanos Julen y Guille, a mis cu\~nadas Jean y Ana,
a mi futuro sobrinito Pau, a Cecilia y por supuesto a Julia (la chica
guapa del IFIC), donde empieza mi nueva familia.

Por \'ultimo, quiero agradecer a mis amigos de la ``Pandilla led
Infiern''. Aunque nunca os hay\'ais cre\'{i}do el rollo de los
neutrinos y mucho menos que sirvan para nada, os merec\'eis un hueco
especial en esta tesis. Nos vemos en el TeOdio: Aida y Pepetoni,
Ferran y Alba, Carlos, Germ\'an, Goyo y Mar\'{i}a, Feli y Alex, Ariana
e Iv\'an.

\cleardoublepage

\end{document}